\documentclass[a4paper,amsmath,amssymb,nofootinbib,superscriptaddress,rmp,twocolumn,aps]{revtex4-1}

\usepackage[pdftex]{graphicx}
\usepackage{color}
\usepackage{hyperref}
 
\newcommand{\av}[1]{\langle {#1} \rangle}

\newcommand{\fluc}[1]{\langle {#1}^2\rangle}

\newcommand{\ki}{k^\textrm{in}}
\newcommand{\ko}{k^\textrm{out}}

\newcommand{\be}{\begin{equation}}
\newcommand{\ee}{\end{equation}}
\newcommand{\avk}{\langle k \rangle}
\newcommand{\AV}{\av{k^2}/\av{k}}

\begin{document}

\title{Epidemic processes in complex networks}
\author{Romualdo Pastor-Satorras}

\affiliation{\mbox{Departament de F\'{\i}sica i Enginyeria Nuclear,
    Universitat Polit\`ecnica de Catalunya, Campus Nord B4, 08034
    Barcelona, Spain}}

\author{Claudio Castellano}

\affiliation{\mbox{Istituto dei Sistemi Complessi (ISC-CNR), via dei
    Taurini 19, I-00185 Roma, Italy}}

\affiliation{\mbox{Dipartimento di Fisica, ``Sapienza'' Universit\`a
    di Roma, P.le A. Moro 2, I-00185 Roma, Italy}}

\author{Piet Van Mieghem}

\affiliation{\mbox{Delft University of Technology, Mekelweg 4, 2628 CD
    Delft, The Netherlands}}

\author{Alessandro Vespignani}

\affiliation{\mbox{Laboratory for the Modeling of Biological and
    Socio-technical Systems, Northeastern University, Boston MA 02115
    USA}}

\affiliation{\mbox{Institute for Scientific Interchange Foundation,
    Turin 10133, Italy}}

\date{\today}

\begin{abstract} 
In recent years the research community has accumulated overwhelming
evidence for the emergence of complex and heterogeneous connectivity
patterns in a wide range of biological and sociotechnical systems. The
complex properties of real-world networks have a profound impact on
the behavior of equilibrium and nonequilibrium phenomena occurring in
various systems, and the study of epidemic spreading is central to our
understanding of the unfolding of dynamical processes in complex
networks. The theoretical analysis of epidemic spreading in
heterogeneous networks requires the development of novel analytical
frameworks, and it has produced results of conceptual and practical
relevance. A coherent and comprehensive review of the vast research
activity concerning epidemic processes is presented, detailing the
successful theoretical approaches as well as making their limits and
assumptions clear. Physicists, mathematicians, epidemiologists,
computer, and social scientists share a common interest in studying
epidemic spreading and rely on similar models for the description of
the diffusion of pathogens, knowledge, and innovation. For this
reason, while focusing on the main results and the paradigmatic models
in infectious disease modeling, the major results concerning
generalized social contagion processes are also presented. Finally,
the research activity at the forefront in the study of epidemic
spreading in coevolving, coupled, and time-varying networks is
reported.
\end{abstract}

\maketitle

\tableofcontents

\section{Introduction}

Since the first mathematical approach to the spread of a disease by
Daniel \citet{bernoulli1760}, epidemic models lie at the core of our
understanding about infectious diseases. As experimenting in-vivo
epidemics is not a viable option, modeling approaches have been the main
resort to compare and test theories, as well as to gauge uncertainties
in intervention strategies. The acclaimed work of \citet{siroriginal},
defining the modern mathematical modeling of infectious diseases, has
evolved along the years in an impressive body of work, whose culmination
is well represented by the monumental summary of \citet{anderson92}. At
the same time, the epidemic modeling metaphor has been introduced to
describe a wide array of different phenomena. The spread of information,
cultural norms and social behavior can be conceptually modeled as a
contagion process. How black-outs spread on a nationwide scale or how
efficiently memes can spread on social networks are all phenomena whose
mathematical description relies on models akin to classic epidemic
models \cite{Vespignani:2012fk}. Although the basic mechanisms of each
phenomenon are different, their effective mathematical description often
defines similar constitutive equations and dynamical behaviors framed in
the general theory of reaction-diffusion processes \cite{vankampen}.  It
is not surprising then that epidemic modeling is a research field that
crosses different disciplines and has developed a wide variety of
approaches ranging from simple explanatory models to very elaborate
stochastic methods and rigorous results \cite{Keeling07book}.

In recent years we are witnessing a second golden age in epidemic
modeling. Indeed, the real-world accuracy of the models used in
epidemiology has been considerably improved by the integration of
large-scale datasets and the explicit simulation of entire populations
down to the scale of single individuals
\cite{Eubank2004,Ferguson2005,Longini2005,Halloran2008,Chao2010,Balcan2009,Merler2011}.
Mathematical models have evolved into microsimulation models that can be
computationally implemented by keeping track of billions of
individuals. These models have gained importance in the public-health
domain, especially in infectious disease epidemiology, by providing
quantitative analyses in support of policy-making processes. Many
researchers are advocating the use of these models as real-time,
predictive tools
\cite{Nishiura2011,Tizzoni2013,Nsoesie2013}. Furthermore, these models
offer a number of interesting and unexpected behaviors, whose
theoretical understanding represents a new challenge, and have
stimulated an intense research activity. In particular, modeling
approaches have expanded into schemes that explicitly include spatial
structures, individual heterogeneity and the multiple time scales at
play during the evolution of an epidemics \cite{Riley2007}.

At the core of all data-driven modeling approaches lies the structure of
human interactions, mobility and contacts patterns that finds its best
representation in the form of networks
\cite{Vespgnani_2009,butts:revisiting,Jackson2010,Newman10,Vespignani:2012fk}.
For a long time, detailed data on those networks was simply unavailable.
The new era of the social web and the data deluge is, however, lifting
the limits scientists have been struggling with for a long time. The
pervasive use of mobile and wifi technologies in our daily life is
changing the way we can measure human interactions and mobility network
patterns for millions of individuals at once. Sensors and tags are able
to produce data at the micro-scale of one-to-one interactions.  Proxy
data derived from the digital traces that individuals leave in their
daily activities (microblogging messages, recommendation systems,
consumer ratings) allow the measurement of a multitude of social
networks relevant to the spreading of information, opinions, habits,
etc.

Although networks have long been acknowledged as a key ingredient of
epidemic modeling, the recent abundance of data is changing our
understanding of a wide range of phenomena and calls for a detailed
theoretical understanding of the interplay between epidemic processes
and networks.  A large body of work has shown that most real-world
networks exhibit dynamic self-organization and are statistically
heterogeneous---typical hallmarks of complex systems
\cite{barabasi02,newman-review,Dorogovtsev:2002,baronchelli13,boccaletti2006cns,caldarelli2007sfn,Newman10,mendesbook,havlinbook,fontourareview}.
Real-world networks of relevance for epidemic spreading are very
different from regular lattices. Networks are hierarchically organized
with a few nodes that may act as hubs and where the vast majority of
nodes have very few interactions. Both social and infrastructure
networks are organized in communities of tightly interconnected
nodes. Although randomness in the connection process of nodes is always
present, organizing principles and correlations in the connectivity
patterns define network structures that are deeply affecting the
evolution and behavior of epidemic and contagion processes. Furthermore,
network's complex features often find their signature in statistical
distributions which are generally heavy-tailed, skewed, and varying over
several orders of magnitude.

The evidence of large-scale fluctuations, clustering and communities
characterizing the connectivity patterns of real-world systems has
prompted the need for mathematical approaches capable to deal with the
inherent complexity of networks.  Unfortunately, the general solution,
handling e.g. the master equation of the system, is hardly achievable
even for very simple dynamical processes. For this reason, an intense
research activity focused on the mathematical and computational modeling
of epidemic and diffusion processes on networks has started across
different disciplines~\cite{dorogovtsev07:_critic_phenom}.  The study of
network evolution and the emergence of macro-level collective behavior
in complex systems follows a conceptual route essentially similar to the
statistical physics approach to non-equilibrium phase transitions
\cite{Henkel}. Hence, statistical physics has been leading the way to
the revamped interest in the study of contagion processes, and more
generally dynamical processes in complex networks. In the last ten
years, an impressive amount of methods and approaches ranging from
mean-field theories to rigorous results have provided new quantitative
insights in the dynamics of contagion processes in complex
networks~\cite{danonreview,keeling05:_networ}.

However, as it is often the case in research areas pursued by different
scientific communities, relevant results are scattered across domains
and published in journals and conference proceedings with completely
different readership. In some cases, relevant advances have been derived
independently by using different jargons as well as different
assumptions and methodologies.  This fragmented landscape does not
advance the field and is, in many cases, leading to the
compartmentalization and duplication of the research effort.  We believe
that a review is timely to contextualize and relate the recent results
on epidemic modeling in complex networks.  Although infectious diseases
will be at the center stage of our presentation, social contagion
phenomena and network dynamics itself are discussed, offering a general
mathematical framework for all social and information contagion
processes that can be cast in the epidemic metaphor.  The final goal is
to provide a coherent presentation of our understanding of epidemic
processes in populations, that can be modeled as complex networks.

After a review of the fundamental results in classical epidemic modeling
and the characterization of complex networks, we discuss the different
methodologies developed in recent years to understand the dynamic of
contagion processes in the case of heterogeneous connectivity
patterns. In particular, in Section IV we specifically spell out the
assumptions inherent to each methodology and the range of applicability
of each approach. In Section V those theoretical approaches are applied
to classic epidemic models such as the susceptible-infected-susceptible
(SIS) and susceptible-infected-removed (SIR) models. In those Sections
particular care is devoted to shed light on the role of the interplay of
the time-scales of the epidemic process and of the network dynamics and 
on the appropriateness of different modeling approximations.  In Sections VI
and VII we focus on various approaches to the mitigation and containment
of epidemic processes and on the analysis of several variations of the
basic epidemic models, aiming at a more realistic description of
contagion processes and contact patterns.  In Section VIII we provide a
summary of recent results concerning time-varying networks. Although
this is an area that is rapidly advancing due to both theoretical and
data gathering efforts, we report on results that are expected to become
foundational.  In Section IX we discuss the generalization of epidemic
processes in complex, multi-species reaction diffusion processes, an
area relevant in the analysis of epidemics in structured populations.
Finally, in Section X, we will review the generalization of epidemic
modeling of social contagion phenomena. The number of specific models
for social contagion is extensive and we therefore confine ourselves to
the most relevant to highlight differences and novel dynamical behaviors
in the evolution of the epidemic process. We conclude with an outlook to
the field and the challenges lying ahead of us.

The upsurge of interest in epidemic modeling in complex networks has led
to an enormous body of work: a query on the Thompson Web of Science
database with the keywords "epidemic" and "networks" returns more than
3600 papers in just the last 15 years. A review of all these papers is
unfortunately hardly feasible. Therefore, we have concentrated our
attention to, what we believe, are the most influential papers. In
providing a unified framework and notation for the various approaches,
we aim at fostering synergies across application domains and provide a
common knowledge platform for future efforts in this exciting research
area.

\section{The mathematical approach to epidemic spreading}

\subsection{Classical models of epidemic spreading}

\label{sec:class-models-epid}

In more than 200 years of its history, the mathematical modeling of
epidemic spreading has evolved into a research area that cuts across
several fields of mathematical biology as well as other disciplines and
is treated in the classic books
by~\citet{anderson92,epidemics,Keeling07book,brauer2010,Diekmann_Heesterbeek_Britton_boek2012,Andersson2000}.
Here, we merely set the notation and present some of the basic elements
and approximations generally used in the modeling of epidemic phenomena,
in order to provide the necessary conceptual toolbox needed in the
following sections.

Epidemic models generally assume that the population can be divided into
different classes or compartments depending on the stage of the disease
\cite{anderson92,epidemics,Keeling07book}, such as susceptibles (denoted
by $S$, those who can contract the infection), infectious ($I$, those
who contracted the infection and are contagious), and recovered ($R$,
those who recovered from the disease). Additional compartments can be
used to signal other possible states of individuals with respect to the
disease, for instance immune individuals. This framework
can be extended to take into account vectors, such as mosquitoes for
malaria, for diseases propagating through contact with an external
carrier.  Epidemic modeling describes the dynamical evolution of the
contagion process within a population.  In order to understand the
evolution of the number of infected individuals in the population as a
function of time we have to define the basic individual-level processes
that govern the transition of individuals from one compartment to
another.

The simplest definition of epidemic dynamics considers the total
population in the system as fixed, consisting of $N$ individuals, and
ignores any other demographic process (migrations, births, deaths,
etc.).  One of the simplest two-state compartmentalizations is the
susceptible-infected-susceptible (SIS) model with only two possible
transitions: The first one, denoted $S\to I$, occurs when a susceptible
individual interacts with an infectious individual and becomes infected.
The second transition, denoted $I\to S$, occurs when the infectious
individual recovers from the disease and returns to the pool of
susceptible individuals. The SIS model assumes that the disease does not
confer immunity and individuals can be infected over and over again,
undergoing a cycle $S \to I \to S$, which, under some conditions, can be
sustained forever.  Another basic model is the classic three-state
susceptible-infected-recovered (SIR) model. In the SIR model, the
transition $I\to S$ of the SIS process is replaced by $I\to R$, which
occurs when an infectious individual recovers from the disease and is
assumed to have acquired a permanent immunity, or is removed (e.g. has
died).  Clearly, the SIR process always stops, when no more infected
individuals are present.

\begin{figure}[t]
  \centering
  \includegraphics[clip=true,width=8.5cm]{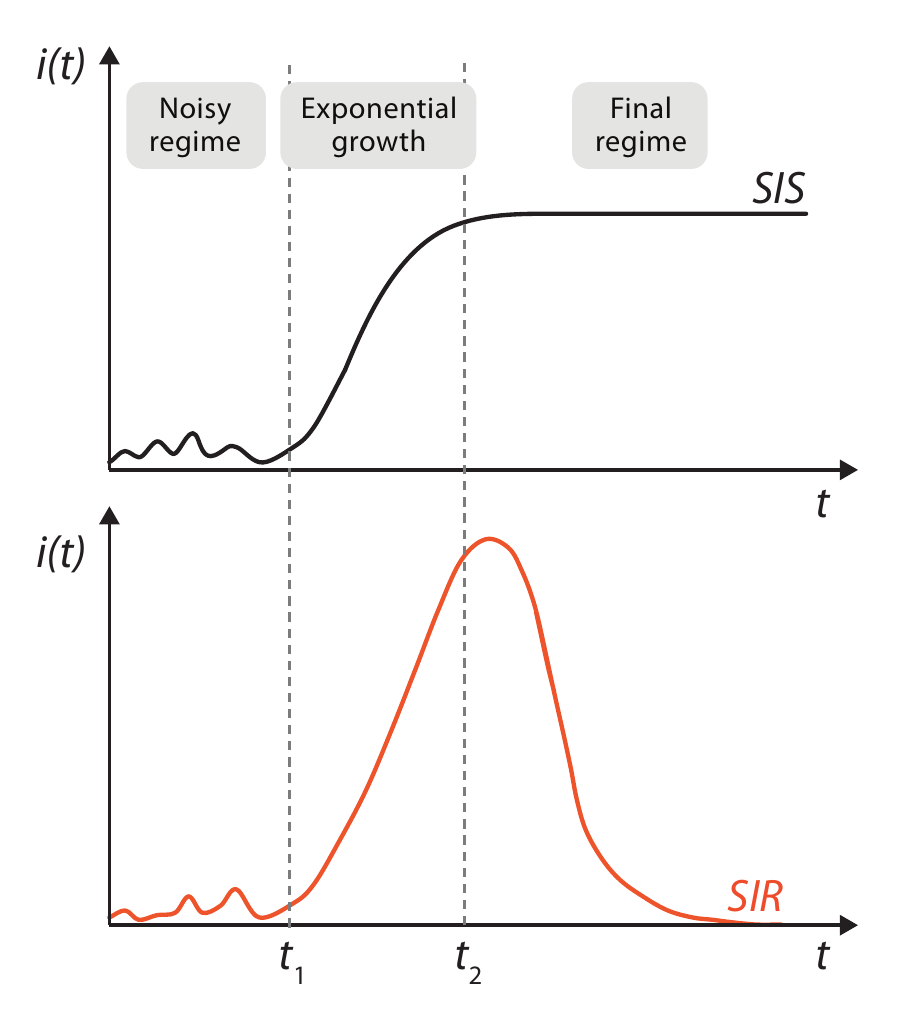}
  \caption{Typical profile of the density $i(t)$ of infected individuals
    versus time in a given epidemic outbreak. In the first regime $t <
    t_1$, the outbreak is subject to strong statistical
    fluctuations. In the second regime, $t_1 < t < t_2$ there is an
    exponential growth characterized by the details of the epidemic
    process. In the final regime ($t > t_2$), the density of infected
    individuals either converges to zero, for SIR-like models, or to a
    constant, possibly zero, for SIS-like models.}
  \label{fig:IIevolution}
\end{figure}

The SIR and SIS models exemplify a basic classification of epidemic
models given in terms of their long time behavior, see
Fig.~\ref{fig:IIevolution}. In the long time regime, the SIS model can
exhibit a stationary state, the \textit{endemic} state, characterized by
a constant (in average) fraction of infected individuals.  In the SIR
model, instead, the number of infected individuals always tends to zero.

In the SIS and SIR models, the infection and recovery processes
completely determine the epidemic evolution. The $I\to R$ and $I\to S$
transitions occur spontaneously after a certain time the individuals
spend fighting the disease or taking medical treatments; the transition
does not depend on any interactions with other individuals in the
population. The $S\to I$ transition instead occurs only because of the
contact/interaction of the susceptible individual with an infectious
one. In this case the interaction pattern among individuals is a
specific feature of the transition and has to be taken into account.
 
For many types of disease, the amount of time spent in the infectious
class is distributed around a well-defined mean value. The distribution
of the "infectious period" and the transition probability can be
generally estimated from clinical data.  However, in a simplistic
modeling scheme, the probability of transition is often assumed
constant.  In this way, a discrete-time formulation defines the recovery
probability $\mu$, that an individual will recover at any time step.
The time an individual will spend on average in the infectious
compartment, the mean infectious period, is then equal to $\mu^{-1}$
time steps. In a continuous-time formulation and assuming a Poisson
process \cite{renewal}, $\mu$ is a rate (probability per unit time) and
the probability that an individual remains infected for a time $\tau$
follows an exponential distribution
$P_\mathrm{inf}(\tau) = \mu e^{-\mu \tau}$, with an average infection
time $\av{\tau} = \mu^{-1}$. The Poisson assumption for the processes of
infection and recovery leads naturally to a Markovian description of
epidemic models \cite{Ross1996}.

The probability of the $S\to I$ transitions is more complicated and it
is dependent on several factors and on the modeling approximations
considered. In the absence of detailed data on human interactions, the
most basic approach considers a homogenous mixing approximation
\cite{anderson92} which assumes that individuals interact randomly with
each other. In this assumption, the larger the number of infectious
individuals among an individual's contacts, the higher the probability
of transmission of the infection. This readily translates to the
definition of the {\it force of infection} $\alpha$, that expresses the
probability, also called the risk, at which one susceptible individual
may contract the infection in a single time step. In the continuous-time
limit we can define $\alpha$ as a rate and assume that
\begin{equation}
  \alpha= \bar{\beta}\frac{N^I}{N},
\end{equation}
where $\bar{\beta}$ 
depends on the specific disease as well as the contact pattern of the
population, and $N^I$ is the number of infected individuals. Thus,
$\alpha$ is proportional to the fraction $\rho^I=N^I/N$ of infected
individuals in the population. In some cases $\bar{\beta}$ is explicitly
split in two terms as $\beta k$, were $\beta$ is now the rate of
infection per effective contact and $k$ is the number of contacts with
other individuals. This form of the force of infection corresponds to
the mass action law \cite{hethcote2000}, a widely used tool in the basic
mean-field description of many dynamical processes in chemistry and
physics.  The force of infection depends only on the density of
infectious individuals and decreases for larger populations, all the other
factors being equal. It is possible however to consider forces of infection of
the type $\alpha=\beta N^I$, where the per capita infection probability
is proportional to the actual number of infected individuals $N^I$, and
assumes that the number of contacts scales proportionally to the size of
the population. Indeed, also intermediate expressions for the force of
infection depending on the size of the population as $N^{-a}$ have been
discussed in the literature \cite{anderson92}.

\begin{figure}[t]
  \centering
  \includegraphics[width=8.5cm]{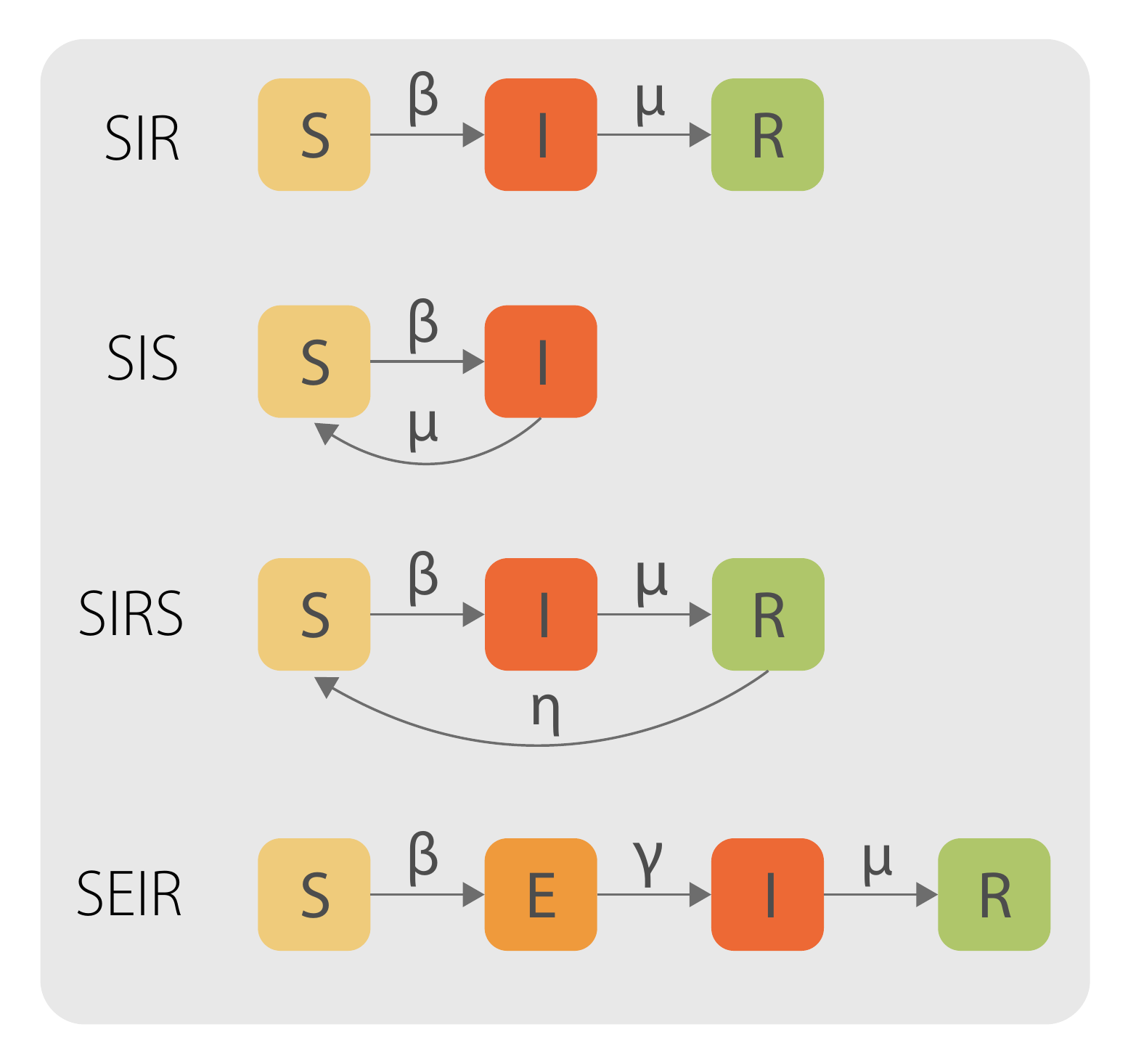}
  \caption{Diagrammatic representation of different epidemic models in
    terms of reaction-diffusion processes.  Boxes stand for different
    compartments, while the arrows represent transitions between
    compartments, happening stochastically according to their respective
    rates.}
  \label{fig:epidiagrams}
\end{figure}

Generalizing the previous approach, an epidemic can be rephrased as a
stochastic \textit{reaction-diffusion process} \cite{vankampen}.
Individuals belonging to the different compartments can be represented
as different kinds of ``particles'' or ``species'', that evolve
according to a given set of mutual interaction rules, representing the
different possible transitions among compartments, and that can be
specified by means of appropriate stoichiometric equations. In the
continous-time limit each reaction (transition) is defined by an
appropriate \textit{reaction rate}. We can therefore adopt the
reaction-diffusion formalism to describe the basic epidemic models, see
Figure~\ref{fig:epidiagrams}.  The SIS model is thus governed by the
reactions
\begin{eqnarray}
  S +I &\overset{\beta}{\rightarrow}& 2 I,\label{eq:sis1} \\
  I &\overset{\mu}{\rightarrow}& S, \nonumber
\end{eqnarray}
where $\beta$ and $\mu$ are transition rates for infection and recovery,
respectively. In this model infection can be sustained forever for
sufficiently large $\beta$ or small $\mu$.  The
Susceptible-Infected-Recovered (SIR) model \cite{siroriginal} is instead
characterized by the three compartments S, I and R, coupled by the
reactions
\begin{eqnarray}\label{eq:sir1}
  S +I &\overset{\beta}{\rightarrow}& 2 I, \\
  I &\overset{\mu}{\rightarrow}& R. \nonumber
\end{eqnarray}
For any value of $\beta$ and $\mu$, the SIR process will always
asymptotically die after affecting a given fraction of the population.

Many more epidemic models can be defined analogously to the SIS and SIR
models.  A useful variant is the SI model, which only considers the
first transition in Eqs.~(\ref{eq:sis1}) and~(\ref{eq:sir1}),
i.e. individuals become infected and never leave this state. While the
SI model is a somewhat strong simplification (valid only in cases where
the time scale of recovery is much larger than the time scale of
infection), it approximates the initial time evolution of both SIS and
SIR dynamics.  More realistic models are defined in order to better
accommodate the biological properties of real diseases.  For instance,
the SIRS (Susceptible-Infected-Removed-Susceptible) model is an epidemic
model incorporating a temporary immunity. It can be defined from the SIR
model by adding a microscopic transition event
\begin{equation}
\label{eq:waning}
  R \overset{\eta}{\rightarrow} S,
\end{equation}
where $\eta$ is the rate at which the immunity of a recovered
individual is lost, rendering him/her susceptible again. The SEIR
model is a variation of the SIR model including the effects of exposed
($E$) individuals, which have been infected by the disease but cannot
yet transmit it. The SEIR model is one of the paradigmatic
models for the spreading of influenza-like illnesses and in the compact
reaction-diffusion notation reads as
\begin{eqnarray}
  S +I &\overset{\beta}{\rightarrow}& E + I,\\
  E &\overset{\gamma}{\rightarrow}& I, \nonumber \\
  I &\overset{\mu}{\rightarrow}& R. \nonumber
\end{eqnarray}
All the above models can be generalized to include demographic effects
(birth and death processes in the population), the age structure of the
population, other relevant compartments (such as asymptomatic infected
individuals), etc.  A more complete and detailed review of epidemic
models and their behavior can be found in
\citet{anderson92,Keeling07book,brauer2010}.

\subsection{Basic results from classical epidemiology}
\label{sec:class-results}

Although epidemic spreading is best described as a stochastic
reaction-diffusion process, the classic understanding of epidemic
dynamics is based on taking the continuous-time limit of difference
equations for the evolution of the average number of individuals in each
compartment.  This deterministic approach relies on the homogeneous
mixing approximation, which assumes that the individuals in the
population are well mixed and interact with each other completely at
random, in such a way that each member in a compartment is treated
similarly and indistinguishably from the others in that same
compartment. This approximation, which is essentially equivalent to the
mean-field approximation commonly used in statistical physics, for both
equilibrium \cite{stanley} and nonequilibrium \cite{Marrobook} systems,
can be shown to be correct in regular lattices with high dimension, but
it is not exact in low dimensions
\cite{havlin_diffusion_reaction}. Under this approximation, full
information about the state of the epidemics is encoded in the total
number $N^\alpha$ of individuals in the compartment $\alpha$ or,
analogously, in the respective densities $\rho^\alpha = N^\alpha / N$,
where $N$ is the population size. The time evolution of the epidemics is
described by deterministic differential equations, which are constructed
applying the law of mass action, stating that the average change in the
population density of each compartment due to interactions is given by
the product of the force of infection times the average population
density \cite{hethcote2000}.

The deterministic equations for the SIR and SIS processes are obtained by applying the law of mass action and read as
\begin{eqnarray}
  \frac{d \rho^I}{dt} &=&  \beta \rho^I \rho^S  - \mu \rho^I \\
  \frac{d \rho^S}{dt} &=& - \beta \rho^I \rho^S + \chi \rho^I ,
\end{eqnarray}
where $\chi = \mu$ for the SIS process and $\chi = 0$ for the SIR model,
and the force of infection is $\alpha=\beta \rho^I$.  These equations
are complemented with the normalization condition,
$\rho^R = 1 - \rho^S - \rho^I$ and $\rho^S = 1 - \rho^I$ for the SIR and
SIS model, respectively.  If we consider the limit $\rho^I \simeq 0$,
generally valid at the early stage of the epidemic, we can linearize the
above equations obtaining for both the SIS and SIR models the simple
equation
\begin{equation}
   \frac{d \rho^I}{dt} \simeq (\beta - \mu) \rho^I.
\end{equation}
whose solution
\begin{equation}
   \rho^I(t) \simeq \rho^I(0) e^{(\beta - \mu)t}
   \label{initial_growth_approx}
\end{equation}
represents the early time evolution. Equation
\ref{initial_growth_approx} illustrates one of the key concepts in the
classical theoretical analysis of epidemic models. The
number of infectious individuals grows exponentially if
\begin{equation}
  \beta - \mu > 0 \quad \Rightarrow \quad R_0 = \frac{\beta}{\mu} >1,
\end{equation}
where we have defined the \textit{basic reproduction number} $R_0$ as
the average number of secondary infections caused by a primary case
introduced in a fully susceptible population \cite{anderson92}.  This
result allows to define the concept of epidemic threshold: only if
$R_0 > 1$ (i.e. if a single infected individual generates on average
more than one secondary infection) an infective agent can cause an
outbreak of a finite relative size (in SIR-like models) or lead to a
steady state with a finite average density of infected individuals,
corresponding to an \textit{endemic} state (in SIS-like models). If
$R_0 < 1$ (i.e. if a single infected individual generates less than one
secondary infection), the relative size of the epidemics is negligibly
small, vanishing in the thermodynamic limit of an infinite
population\footnote{In the present context, since we do not consider
  spatial effects, the thermodynamic limit is simply defined as the
  limit of an infinitely large number of individuals.} (in SIR-like
models) or leading to a unique steady state with all individuals healthy
(in SIS-like models). This concept is very general and the analysis of
different epidemic models \cite{anderson92} reveals in general the
presence of a \textit{threshold behavior}, with a reproduction number
that can be expressed as a function of the rates of the different
transitions describing the epidemic model.

A few remarks are in order here. First, although we have stated that
epidemic processes can be considered as reaction-diffusion systems, the
classic approach completely neglects the diffusion of
individuals. Spatial effects can be introduced by adding diffusive
continuous terms or by considering patch models.
Furthermore, epidemic spreading is governed by an inherently
probabilistic process. Therefore, a correct analysis of epidemic models
should consider explicitly its stochastic nature
\cite{Andersson2000}. Accounting for this stochasticity is particularly
important when dealing with small populations, in which the number of
individuals in each compartment is reduced. For instance, while the
epidemic threshold condition $R_0 > 1$ is a necessary and sufficient
condition for the occurrence of an epidemic outbreak in deterministic
systems, in stochastic systems this is just a necessary
condition. Indeed even for $R_0 > 1$ stochastic fluctuations can lead to
the epidemic extinction when the number of infectious individuals is
small.  Analogously, all the general results derived from deterministic
mean-field equations can be considered representative of real systems
only when the population size is very large (ideally in the
thermodynamic limit) and the fluctuations in the number of individuals
can be considered small.  Indeed, most of the classical results of
mathematical epidemiology have been obtained under these assumptions
\cite{anderson92}.

Another point worth stressing is the Poisson assumption.  Although we
will mostly focus on Poissonian epidemic processes (see Sections
\ref{sec:real-epid-models} and \ref{sec:epid-proc-temporal-nets} for
some remarks on the non-Poissonian case), a different phenomenology,
both more complex and interesting, can be obtained from non
exponentially distributed infection or recovery processes.

Finally, the classic deterministic approach assumes \emph{random and
  homogeneous mixing}, where each member in a compartment is treated
similarly and indistinguishably from the others in that same
compartment.  In reality, however, each individual has his/her own
social contact network over which diseases propagate,
usually differing from that of other members in a group or
compartment. \citet{Diekmann_Heesterbeek_Britton_boek2012} illustrate
the weakness of $R_{0}$ by discussing a line and square lattice topology
and they conclude that network and percolation theory needs to be
consulted to compute the epidemic threshold, leading to a new definition
of the basic reproduction number depending on the topology of the
network. Thus, for example, in the case of a homogeneous contact network
in which every individual is in contact with the same number of
individuals $\avk$, the basic reproduction number takes the form
\begin{equation}  
  R_0 = \avk \frac{\beta}{\mu},
  \label{homogeneousR0}
\end{equation}
The impact of heterogeneous connectivity patterns, reflected by an
underlying network topology, on the epidemic behavior is the focus of
the present review.

\subsection{Connections with other statistical physics models}

\label{sec:2.C}
\begin{figure}[t]
  \includegraphics*[width=8.5cm]{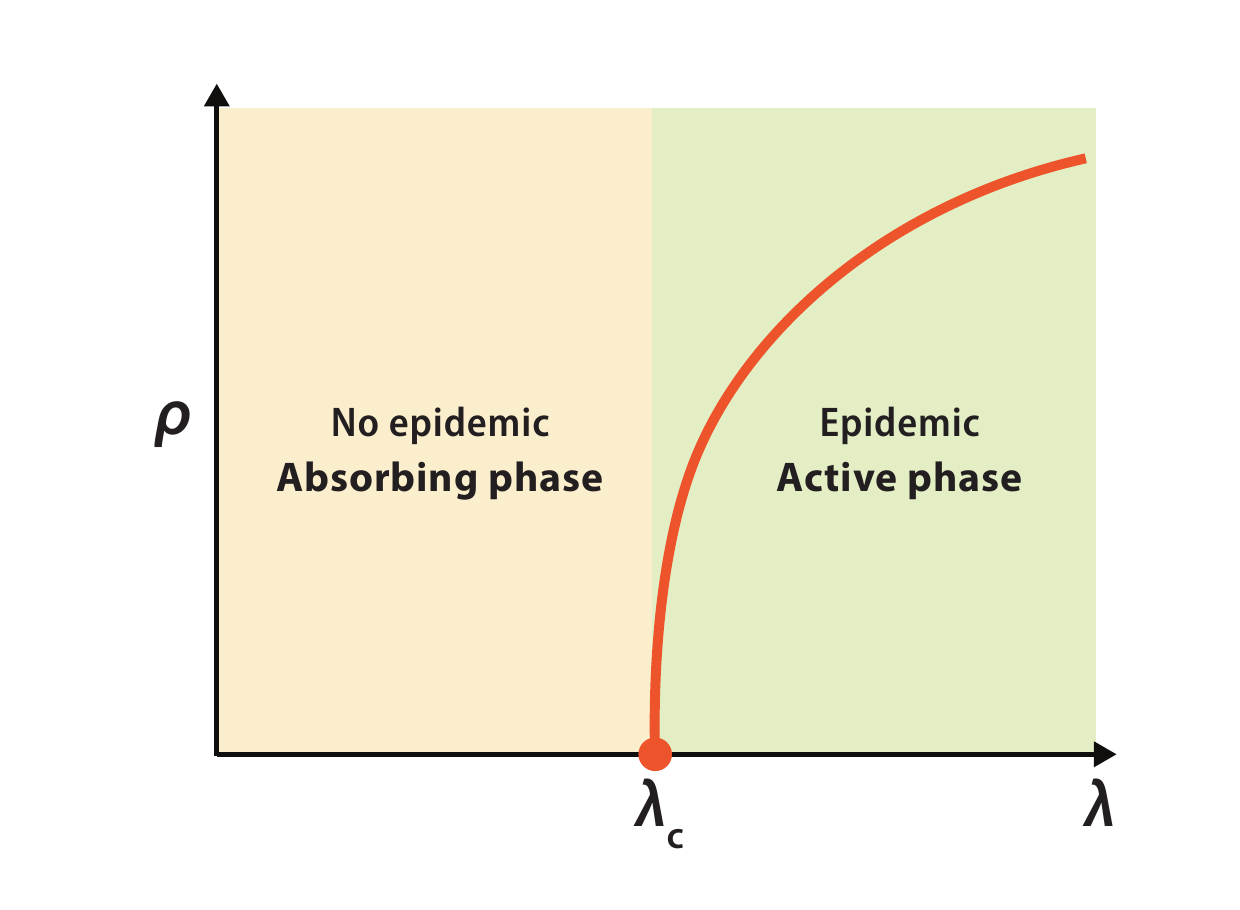}
  \caption{Phase diagram of a typical non-equilibrium absorbing state
    phase transition (SIS-like). 
    Below the critical point $\lambda_c$, the order
    parameter is zero (healthy phase in an epidemics
    interpretation). Above the critical point, the order parameter
    attains a non-zero average value in the long time regime (endemic
    or infected epidemic phase).}
\label{fig:phasetrans}
\end{figure}

The interest that models for epidemic spreading have attracted within
the statistical physics community stems from the close connection
between these models and more standard nonequilibrium problems in
statistical physics~\cite{Marrobook,Henkel}.  In particular, the
epidemic threshold concept is analogous to the concept of phase
transition in non-equilibrium systems.  A phase transition is defined as
an abrupt change in the state (\textit{phase}) of a system,
characterized by qualitatively different properties, and that is
experienced varying a given \textit{control parameter} $\lambda$. The
transition is characterized by an \textit{order parameter} $\rho$
\cite{yeomans}, which takes (in a system of infinite size) a non-zero
value in one phase, and a zero value in another (see
Figure~\ref{fig:phasetrans}).  The phase transition takes place at a
particular value of the control parameter, the so-called
\textit{transition point} $\lambda_c$, in such a way that for
$\lambda>\lambda_c$ we have $\rho >0$, while for
$\lambda \leq \lambda_c$, $\rho =0$. Apart from the determination of the
transition point, the interest in physics lies in the behavior of the
order parameter around $\lambda_c$, which in \textit{continuous, or
  critical phase transitions}\footnote{In \textit{first order
    transtions} the order parameter takes a discontinuous jump at the
  transition point \cite{stanley}.} takes a power law form,
$\rho(\lambda) \sim (\lambda- \lambda_c)^{\beta_{crit}}$, defining the
\textit{critical exponent} $\beta_{crit}$ \cite{yeomans}. 

The SIS dynamics thus belongs to the wide class of
non-equilibrium statistical models possessing \textit{absorbing states},
i.e.  states in which the dynamics becomes trapped with no possibility
to escape. The paradigmatic example of a system with an absorbing state
is the contact process \cite{harris74}, where all nodes of a
lattice or network can be either occupied or empty. Occupied nodes
annihilate at rate $1$; on the other hand, they can reproduce at rate
$\lambda$, generating one offspring that can occupy an empty nearest
neighbor.  The contact process experiences an \textit{absorbing-state
  phase transtion} \cite{Marrobook,Henkel} at a critical point $\lambda_c$ between an active phase, in
which activity lasts forever in the thermodynamic limit, implying a
finite average density of occupied nodes, and an absorbing phase, in
which activity eventually vanishes, corresponding to an empty
system. In
the case of the SIS model, the active phase is given by the infected
state, and the absorbing phase by the state where no individual is
infected, see Figure~\ref{fig:phasetrans}.  The order parameter is
therefore the \textit{prevalence} or density of infected individuals,
and the control parameter is given by the \textit{spreading rate} or {\em effective infection rate}, which
equals $\lambda = \beta/\mu$.
The \textit{epidemic threshold} (critical point) $\lambda_c$ separates
thus the infected from the healthy phase.  While this distinction is
strictly true in the thermodynamic limit, for finite systems the
dynamics for any value of $\lambda$ sooner or later visits the
absorbing-state and remains trapped there.  The absorption event can occur even in
the active phase well above the critical point, because of random
fluctuations, illustrating that the determination of the critical point is a nontrivial task, both for
theoretical approaches and numerical simulations
\cite{Marrobook,Henkel}. It is interesting to note that the dynamics of
the SIS process is essentially identical to that of the contact process
in lattices; indeed, the difference between the SIS and the contact
process lies exclusively in the number of offsprings that an active
individual can generate. While in the contact process one particle
generates always in average one offspring per unit time, an infected
individual in the SIS model can infect all his/her nearest neighbors in the
same time interval. This difference is trivial when the number of
nearest neighbors is fixed, but it can lead to a dramatic difference
when the number of nearest neighbors has large fluctuations (see
Section~\ref{sec:epid-proc-heter}).

The SIR model also exhibits a transition between a phase where the
disease outbreak reaches a finite fraction of the population and a phase
where only a limited number of individuals are affected. This is
strongly reminiscent of the transition occurring in
\textit{percolation}~\cite{stauffer94,Grassberger1983}.  In the simplest
possible setting of (bond) percolation in a lattice, the connections
between nearest neighbors of a lattice or network are erased with
probability $1-p$ and kept with complementary probability $p$.  A
critical value $p_c$ separates a super-critical percolating phase, where
a macroscopic connected cluster spans the whole lattice, from a
sub-critical phase where only connected clusters of finite size
exist. The order parameter describing the transition is the probability
$P_G(p)$ that a randomly chosen site belongs to the spanning cluster.
In the case of networks, the percolating phase corresponds to the
presence of a largest connected component with a size proportional to
the network size (the \textit{giant component}, see
Section~\ref{sec:general-definitions}), while in the sub-critical phase
it has a relative size that vanishes in the termodynamic limit. In the
case of networks, the order parameter is proportional to the
relative size of the giant component.  The mapping between SIR and
bond percolation is made through the assimilation of the size of connected 
components with the size of epidemic outbreaks, with a control parameter 
that depends on the spreading rate
$\lambda=\beta/\mu$. This connection will be further developed and
exploited in Sec.~\ref{sec:4.B}.

Finally, it is worth mentioning first-passage
percolation~\cite{hammersley1965first,kesten2003first} as another
classical problem related to epidemics.  In this model, a nonnegative
value $\tau_{ij}$ is defined on each edge of a graph and interpreted as
the time needed to cross the edge. Given a topology and the distribution
of the times $\tau$, first passage percolation investigates which points
can be reached in a certain time starting from a fixed origin. The SI
model for epidemics can be seen as the limit of first-passage
percolation with all passage times equal.

\section{Network measures and models}

\label{sec:netw-epid-whorps}

Although very common, the homogeneous assumption used in the previous
Section to derive the constitutive deterministic equations of basic
epidemic processes maybe inadequate in several real-world situations
where individuals have large heterogeneity in the contact rate, specific
frozen pattern of interaction or are in contact with only a small part
of the population.  These features may have different relevance
depending on the disease or contagion process considered. However, a
wide range of social and biological contagion processes require
capturing the individuals' contact pattern structure in the mathematical
modeling approaches. This is even more relevant, because most real-world
systems show very complex connectivity patterns dominated by large-scale
heterogeneities described by heavy-tailed statistical distributions.

Network theory \cite{Newman10}
provides a general framework to discuss interactions among individuals
in detail. In this Section, we provide a short summary of the main
definitions and properties of networks, relevant for epidemic spreading,
and a basic introduction to the language of graph theory that is
necessary for a formal analysis of networks properties. Network science
is burgeoning at the moment, and for more extensive accounts of this
field we refer to the books
\cite{mendesbook,caldarelli2007sfn,Dorogobook2010,Newman10,havlinbook,networksciencebook}.

\subsection{General definitions}
\label{sec:general-definitions}
Networks are mathematically described as graphs. A graph is a collection
of points, called \textit{vertices}, (\textit{nodes} in the physics
literature or \textit{actors} in the social sciences). These points are
joined by a set of connections, called \textit{edges}, \textit{links} or
\textit{ties}, in mathematics, physics and social sciences,
respectively. Each edge denotes the presence of a relation or
interaction between the vertices it joins.
Edges can represent a bidirectional interaction between vertices, or
indicate a precise directionality in the interaction.  In the first case
we talk about \textit{undirected} networks, and in the second case,
about \textit{directed} networks or \textit{digraphs}.  From an
epidemiological point of view, the directedness of a network is indeed
relevant since it imposes restrictions on the possible paths of
propagation of the contagion.  A compact way to specify all connections
present in a graph of size $N$ (i.e. with $N$ vertices) is the
$N \times N$ adjacency matrix $A$, with elements $a_{ij}=1$ if an edge
is connecting nodes $i$ and $j$ and zero otherwise.  $A$ is symmetric in
undirected graphs, and asymmetric in directed graphs.

A path $\mathcal{P}_{i_0, i_n}$ connecting vertices $i_0$ and $i_n$ is a
sequence of different edges $\{(i_j, i_{j+1})\}$, $j=0,\ldots, n-1$; the
number of edges traversed, $n$, is the hopcount, also called the length,
of the path. A graph is
\textit{connected} if there exists a path connecting any two vertices in
the graph. A \textit{loop} is a closed path with $i_0 \equiv i_n$.
\begin{figure}[t]
\includegraphics*[width=\columnwidth]{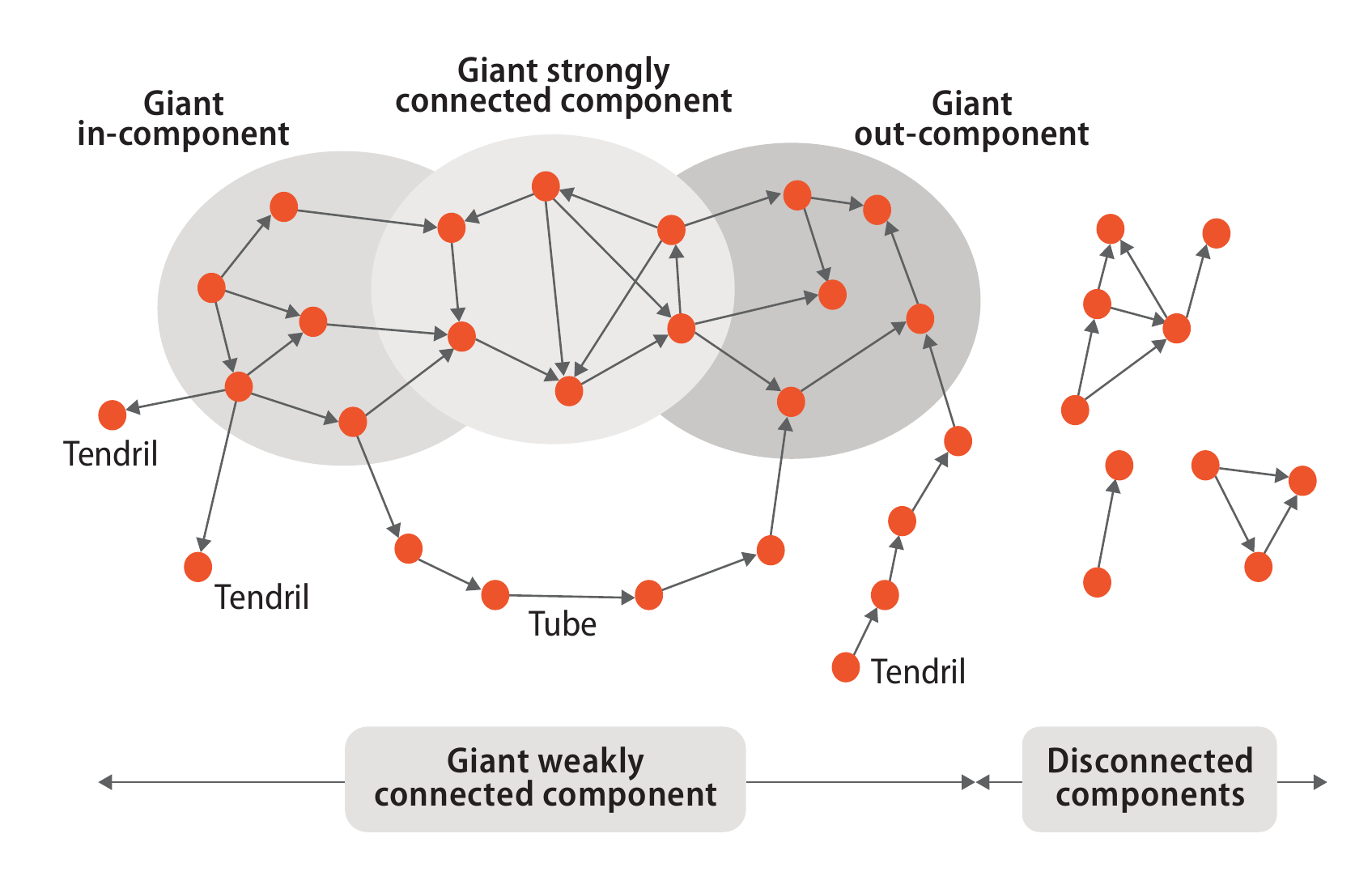}
\caption{Component structure of a directed graph. Figure adapted from
  \citet{dorogodirect01}.}
\label{fig:bowtie}
\end{figure}
A \textit{component} $\mathcal{C}$ of a graph is
\index{component}defined as a connected subgraph.
The \textit{giant component} is the component or subgraph, whose size
scales as the number of vertices in the graph. From an epidemiological
perspective, a disease in the giant component may in principle infect a
macroscopic fraction of the graph, while if the disease starts outside
of the giant component, the total number of infected vertices will be
necessarily limited, representing a fraction that decreases with the
network size.

In the case of directed graphs, the structure of the components is more
complex as the presence of a path from the node $i$ to the node $j$ does
not necessarily guarantee the presence of a corresponding path from $j$
to $i$. In general (see Figure~\ref{fig:bowtie}) the component structure
of a directed network can be decomposed into a giant weakly connected
component (GWCC), corresponding to the giant component of the same graph
in which the edges are considered as undirected, plus a set of smaller
disconnected components. The GWCC is itself composed of several parts
because of the directed nature of its edges: (1) the giant strongly
connected component (GSCC), in which there is a directed path joining
any pair of nodes; (2) the giant in-component (GIN), formed by the nodes
from which it is possible to reach the GSCC by means of a directed path;
(3) the giant out-component (GOUT), formed by the nodes that can be
reached from the GSCC by means of a directed path; (4) the tendrils,
that connect nodes that cannot reach the GSCC or be reached from it and
(5) the tubes, that connect the GIN and GOUT, but do not belong to the
GSCC.

\subsection{Network metrics}
A large number of metrics have been defined to characterize
different aspects of the topology of complex networks.

\subsubsection{Shortest path length and network diameter}
\label{sec:shortest-path-length}
In order to characterize the distance among nodes we introduce the
\textit{shortest path length}, sometimes also referred to as the
\textit{chemical} distance or \textit{geodesical} distance. The
shortest path distance $\ell_{ij}$ between two nodes $i$ and $j$ is
defined as the length of the shortest path (not necessarily unique)
joining $i$ and $j$. The \textit{diameter} of a network is the maximum
value of all the pairwise shortest path lengths, and the average
shortest path length $\av{\ell}$ is the average of the value of
$\ell_{ij}$ over all pairs of vertices in the network.

\subsubsection{Degree and degree distribution}
\label{sec:degr-degr-distr}
The degree $k_i$ of vertex $i$ in an undirected network is the number
of edges emanating from $i$, i.e. $k_i = \sum_j a_{ij}$. In the case
of directed networks, we distinguish between in-degree, $\ki_i$, and
out-degree, $\ko_i$, as the number of edges that end in $i$ or start
from $i$, respectively.  In undirected networks we
define the \textit{degree distribution} $P(k)$ as the probability that
a randomly chosen vertex has degree $k$, or, in finite networks, as
the fraction of vertices in the graph with degree exactly equal to
$k$. In the case of directed networks, there are instead two different
distributions, the out-degree $P_\mathrm{out}(\ko)$ and the in-degree
$P_\mathrm{in}(\ki)$ distributions. The
in-degree and out-degree of a given vertex might not be
independent. Correlations are encoded in the joint probability
distribution $P(\ki, \ko)$ that a randomly chosen vertex has in-degree
$\ki$ and out-degree $\ko$.  It is
useful to consider the moments of the degree distribution, $\av{k^n} =
\sum_k k^n P(k)$. The first moment, the \textit{average degree} $\av{k} = 2L/N$, twice the ratio between the
number $L$ of edges (or links) and the number $N$ of nodes, provides information
about the density of the network. A network is called \textit{sparse}
if its number of edges $L$ grows at most linearly with the network size
$N$; otherwise, it is called \textit{dense}. 
In directed networks, since every edge contributes to one node in-degree
and other node out-degree we have that $\av{\ki} = \av{\ko}$.

\subsubsection{Degree correlations}
\label{sec:degree-correlations}

Two-vertex degree correlations can be conveniently
measured by means of the conditional probabilility $P(k'|k)$ that an
edge departing from a vertex of degree $k$ is connected to a vertex of
degree $k'$ \cite{alexei}. A network is called \textit{uncorrelated}
if this conditional probability is independent of the originating
vertex $k$. In this case, $P(k'|k)$ can be simply estimated as the
ratio between the number of edges pointing to vertices of degree
$k'$, $k' P(k')N /2$, and the total number of edges, $\av{k} N/2$, to yield
$P^\mathrm{un}(k'|k) = \frac{k'P(k')}{\av{k}}$. 
The empirical evaluation of $ P(k'|k)$ turns out to be quite noisy in
real networks, due to finite size effects. A related, simpler,
measure of correlations is the average degree of the
nearest neighbors of vertices of degree $k$, $\bar{k}_{nn}(k)$ which
is formally defined as  \cite{alexei}
\begin{equation}
  \label{eq:knndef}
  \bar{k}_{nn}(k) = \sum_{k'} k' P(k'|k).
\end{equation}
For uncorrelated networks,
$\bar{k}_{nn}^\textrm{un}(k) = \av{k^2}/{\av{k}}$ does not depend on
$k$. Therefore, a varying $\bar{k}_{nn}(k)$ is the signature of degree
correlations. The analysis of empirical networks has suggested a broad
classification of networks in two main classes, according to the nature
of their degree correlations
\cite{assortative}:
\textit{Assortative networks} exhibit an increasing $\bar{k}_{nn}(k)$,
indicative that high degree nodes tend to connect to high degree nodes,
while low degree nodes are preferentially attached to low degree
nodes. \textit{Disassortative networks}, on the other hand, show a
decreasing $\bar{k}_{nn}(k)$ function, suggesting that high degree nodes
connect to low degree nodes, and viceversa. Assortativity by degree can
be characterized by the Pearson correlation coefficient $r$
\cite{assortative}: Uncorrelated networks have
$r=0$, while assortative (disassortative) networks present $r>0$
($r<0$), respectively.

\subsubsection{Clustering coefficient and clustering spectrum}
\label{sec:clust-coeff-clust}
The concept of clustering refers to network transitivity, i.e.
the relative propensity of two nodes to be connected, provided that
they share a common neighbor. The clustering coefficient $C$ is
defined as the ratio between the number of loops of length three in
the network (i.e. triangles), and the number of connected triples (three nodes
connected by two edges). A local measure $c_i$ of clustering
\cite{watts98} can also be defined as the ratio between the actual
number of edges among the neighbors of a vertex $i$, $e_i$, and its
maximum possible value, measuring thus directly the probability that
two neighbors of vertex $i$ are also neighbors of each other. The mean
clustering of the network $\av{c}$ is defined as the average of $c_i$
over all vertices in the network.  The clustering spectrum
$\bar{c}(k)$ is defined as the average clustering coefficient of the
vertices of degree $k$ \cite{alexei02,ravasz_hierarchical_2003},
satisfying $\av{c} =\sum_k P(k) \bar{c}(k)$.

\subsubsection{Centrality and structure in networks}
\label{sec:centrality}
The concept of \textit{centrality} encodes the relative importance of a
node inside a network, a relevant issue in the context of social network
analysis \cite{wass94}. Many different definitions of centrality have
been proposed, based on different indicators of the structural
importance of nodes.  The simplest of them is the degree, referred to as
\textit{degree centrality}.  The higher its degree, the more the node
can be considered influential/central in the network.  Alternative
definitions are based on the shortest paths between vertices. Thus, the
\textit{closeness centrality} $\mathcal{C}_i$ is defined as the inverse
of the average of the shortest path lengths from vertex $i$ to all other
vertices in the network.  With this measure, we consider a vertex
central if it is situated in average at a short distance to all other
vertices in the network.  A very different perspective on centrality is
provided by the \textit{betweenness centrality} $b_i$ of vertex $i$,
defined as number of shortest paths between any two vertices in the
network that pass through vertex $i$. More precisely, if $L_{h,j}$ is
the total number of shortest paths from $h$ to $j$, and $L_{h,i,j}$ is
the number of these shortest paths that pass though vertex $i$, then
$b_i = \sum_{h \neq j} L_{h,i,j}/ L_{h,j}$. Betweeness measures thus
centrality from the perspective of the control of information flowing
between different nodes, assuming this information flows following the
shortest path route \cite{freeman77}. 

Another way to characterize the centrality of nodes resides in the concept
of $K$-coreness. 
The $K$-core of a network is a maximal connected subgraph, such that all
vertices in the subgraph have degree $k \geq K$~\cite{Seidman1983269}. 
The  $K$-core decomposition is an iterative procedure that
classifies the vertices of the network in nested levels of increasing
connectivity (increasing $K$-core). The algorithm runs as follows: One
starts with the complete network, and removes iteratively all
vertices with degree $k=1$, until only vertices with degree $k \geq 2$
are present. The set of removed nodes represents the
$K=1$-\textit{shell}, while the remaining nodes constitute the
$K=2$-core. In the next iteration of the process, all vertices with
degree $k=2$ are removed (the $K=2$-shell), are we are left with the
$K=3$-core. This iterative process is stopped when we arrive at the
maximum $K_S$-core, where one more application of the algorithm
leaves no vertices. At each node is assigned a centrality measure 
equal to its $K$-core index, the deeper the more central.

It is worth remarking that real networks can display higher levels
of architecture that are difficult to capture with a single number.  Many
networks possess a \textit{community structure}, in which
different sets of nodes, called \textit{communities} or
\textit{modules}, have a relatively high density of internal
connections, while they are more loosely connected among them. The
problem of computing the community structure of a given network has
been a very active topic in network science and a large number of
different approaches have been considered (see
\citet{Fortunato201075} for a specific review).

\subsection{Generalizations of simple graphs}
\label{sec:gener-simple-graphs}

The simple concept of graph considered above can be refined at
different levels, adding more and more complexity and detail in order
to better represent the real system under consideration. A first
extension is that of \textit{bipartite} graphs, in which we have $2$
different kinds of nodes, and edges join only two nodes of a different
kind. A classical example are the networks of heterosexual sexual
relationships \cite{liljeros_web_2001}.

Another important generalization consists in the definition of
\textit{weighted networks}, in which a real number $\omega_{ij}$ (the
weight) is associated to the edge between vertices $i$ and
$j$. Weighted networks constitute the natural choice to represent many
systems, including transportation networks (e.g. the airport network),
in which the weight of an edge measures the fraction of people or
goods transported by the edge in a given interval of time, or social
networks, for which weights measure the relative intensity or
frequency of contacts between pairs of vertices. The addition of
weights allows to define a complete new set of topological metrics
\cite{Braunstein03,Barrat16032004,onnela05:_inten,PhysRevE.76.016101,PhysRevE.74.055101}.
Among those, the strength of a node $s_i$, defined as the sum of the
weights of all edges incident to it, i.e. $s_i = \sum_j \omega_{ij}$,
generalizes to weighted networks the concept of degree.

\subsection{Network classes and basic network models}
\label{sec:basic-network-models}
The recent abundance of data and measurements of real-world networks
has highlighted the existence of different classes of networks,
characterized by a large variability in basic metrics and statistical
properties.  This classification in its turn has fueled an intense
theoretical research effort devoted to the study of different network
generation models. The usefulness of these models in the present
context is that 
they serve as generators of synthetic networks, with controlled
topological properties, in which the behavior of dynamical processes
such as epidemics can be studied in detail.  In the following we will
survey some of the main network classes and models that are used for
exploring the properties of epidemic processes.

\subsubsection{Random homogenous networks}
The first theoretical model of random networks is the \textit{classical
  random graph} model \cite{solomonoff51,gilbert59,erdos59}. In its
simplest formulation, the graph $G_{p}(N)$ is constructed from a set of
$N$ nodes in which each one of the $N(N-1)/2$ possible links is present
with probability $p$.  The degree distribution is given by a binomial
form, which, in the limit of constant average degree (i.e.
$p = \av{k}/ N$) and large $N$ can be approximated by a Poisson
distribution $ P(k) = e^{-\av{k}} \frac{\av{k}^k}{k!}$.  The clustering
coefficient is simply given by $\av{c}=p$, and the average shortest path
length is $\av{\ell} \simeq \log N / \log \av{k}$ \cite{Dorogobook2010}.
This model is therefore adequate in the case of networks governed only
by stochasticity, although $G_{p}(N)$ tends to a regular graph for large
$N$ and constant $p$.  The degree distribution is peaked around the
average value, thus denoting a statistical homogeneity of the nodes.
Interestingly, the model features for $\av{k} > 1$ the small diameter
observed in most real-world networks.  However, any other structural
properties, including the generally high clustering coefficient observed
in real world networks, cannot be reproduced by this model.

\subsubsection{Small-world networks}
The \textit{small-world model} of \citet{watts98} represents a first
attempt to obtain a network with small diameter $\av{\ell}$ and large
clustering coefficient.  This model considers an ordered lattice, 
such as a ring of $N$
vertices, each one of which symmetrically connected to its $2 m$
nearest neighbors. This initial configuration has large clustering
coefficient 
and large average shortest path length.
Starting from it, a fraction $p$ of edges in the network are rewired, by
visiting all $m$ clock-wise edges of each vertex and reconnecting them,
with probability $p$, to a randomly chosen node.  In another version of
the model \cite{Monasson-1999}, a fraction $p$ of edges are added
between randomly chosen pairs of vertices.  The overall effect of the
rewiring processes is to add long-range shortcuts, that, even for a
small value of $p \sim N^{-1}$, greatly reduce the average shortest path
length, while preserving a large clustering for not very large values of
$p$.  This model, although better suited for social networks with high
clustering coefficient, has a degree distribution and centrality
measures decaying exponentially fast away from the average value.  The
small-world model thus generates homogeneous networks where the average
of each metric is a typical value shared, with little variations, by all
nodes of the network.

\subsubsection{Heavy-tailed networks}
Empirical evidence from different research areas has shown that many
real-world networks exhibit levels of heterogeneity not anticipated
until few years ago. The statistical distributions characterizing
heterogeneous networks are generally skewed, and varying over several
orders of magnitude. Thus, real-world networks are structured in a
hierarchy of nodes with a few nodes having very large connectivity (the
hubs), while the vast majority of nodes have much smaller degrees. More
precisely, in contrast with regular lattices and homogeneous graphs
characterized by a typical degree $k$ close to the average $\av{k}$,
heterogeneous networks exhibit heavy-tailed degree distributions often
approximated by a power-law behavior of the form $P(k)\sim k^{-\gamma}$,
which implies a non-negligible probability of finding vertices with very
large degree.  The degree exponent $\gamma$ of many real-world networks
takes a value between $2$ and $3$. In such cases networks are called
\textit{scale-free}, since the second moment of the degree distribution
diverges in the infinite network size limit ($N\to\infty$).  It is
understood that in real-world networks the finite size $N$ and the
presence of biological, cognitive and physical constraints impose an
upper limit to the second degree moment. However, the second moment of
the distribution is in many case overwhelmingly large, reflecting
enormous connectivity fluctuations.  The presence of large-scale
fluctuations associated with heavy-tailed distributions is often true
not only for the degree of nodes but it is also observed for the
intensity carried by the connecting links, transport flows, and other
basic quantities.

Several variations of the classical random graph model have been
proposed in order to generate networks with a power-law degree
distribution.  One variation, the so-called \textit{configuration model}
\cite{benderoriginal,molloy95}, considers a random network with a fixed
degree distribution, instead of the fixed average degree of classical
random graphs. Its construction is as follows: To each of the vertices,
we assign a degree $k_i$, given by a random number selected from the
probability distribution $P(k)$, subject to the conditions
$m \leq k_i \leq N$, where $m$ is the desired minimum degree, and such
that $\sum_i k_i$ is an even number. The actual graph is constructed by
randomly connecting the nodes with $\sum_i k_i /2 $ edges, preserving
the degree originally assigned.
In finite networks, an \textit{average} maximum degree or degree
\textit{cut-off} $k_m$, known as the \textit{natural cut-off} of the
network~\cite{mariancutofss} is often observed, which is a function of the network size
of the form $k_m(N) \sim N^{1/(\gamma-1)}$~\cite{Cohen00}.
The original \textit{configuration model} leads for power-law
distributions with $\gamma \leq 3$ to the formation of networks with
multiple and self-connections. The additional prescription that
multiple and self-connections are removed leads to the generation of
disassortative correlations~\cite{maslovcorr,PhysRevE.68.026112}.
These correlations are avoided in the \textit{uncorrelated
  configuration model}~\cite{Catanzaro05} by imposing a hard
\textit{structural cut-off} $k_m \sim N^{1/2}$.

A different modeling paradigm, namely the class of growing network
models, is based on the empirical observation that many real networks do
not have a constant number of vertices and edges, but are instead
growing entities, in which nodes and links are added over time.  The
first undirected model of this kind is the Barab\'asi-Albert (BA) model
\cite{Barabasi:1999}, based on the assumption that newly added edges
will tend in general to be connected to nodes chosen via some
\textit{preferential attachment} rule. The simplest of these
preferential rules is a degree-biased rule, in which the probability to
add a connection to a vertex $i$ is some function $F(k_i)$ of its
degree. The \citet{Barabasi:1999} model, assuming the simplest, linear,
form for the preferential attachment function, is defined as follows:
(\textit{i}) The network starts with a small nucleus of $m_0$ connected
vertices; every time step a new node is added, with $m$ ($m \leq m_0)$
edges which are connected to old vertices in the network. (\textit{ii})
New edges are connected to the $i$-th node in the network with
probability equal to $F(k_i) = k_i / \sum_j k_j$.  In the long time
limit, the network thus generated has a degree distribution
$P(k) \sim k^{-3}$ \cite{Barabasi:1999,mendes99}.  The original growing
network model has been subject to an impressive number of variations and
extensions towards realistic growing dynamics and to accommodate for
different exponents of the degree distribution and other properties such
as high clustering and tunable degree-degree
correlations~\cite{Newman10}.

\subsection{Static versus dynamic networks}
\label{sec:static-vs-dynamic}

So far, we
have assumed that the topology
defining the network is \textit{static}: the set of nodes and links do not change over time. 
However, many other real
networks are far from static, their links being created, destroyed and
rewired at some intrinsic time scales. In some of these dynamical
networks, such as the Internet \cite{romuvespibook}, the time
scale of the network evolution is quite slow. A static network provides a good approximation, when the
properties of dynamical processes evolve at a much faster time
scale than topological changes.  The opposite limit defines the so-called \textit{annealed
  networks}
\cite{gil05:_optim_disor,stauffer_annealed2005,PhysRevE.76.046111,Boguna09},
which describe the case when the evolution of the network is much
faster than the dynamical processes. In this limit, the
dynamical process unfolds on a network that is rapidly rewiring so that
 the dynamics effectively occurs on an average network in which each
connection is possible according to a specific probability that
depends on the degree distribution $P(k)$ and the two-node degree
correlations $P(k'|k)$.  An annealed network is thus described by a
mean-field version of the adjacency matrix that will be presented in
Section~\ref{sec:theoreticalmethods}.

The two above limits are relevant in the definition of the
approximations and the limits of applicability of the most commonly used
theoretical approaches to epidemic spreading in networks. There are,
however, several other instances of networks, such as in social systems,
where the connectivity pattern varies over time scales comparable to
those of the dynamical processes on top of it and it is crucial to take
explicitly into account the concurrent dynamics of the spreading process
and the connectivity pattern.  The effect on epidemic spreading of the
dynamical nature of such \textit{temporal} \cite{Holme:2011fk} networks
is discussed in Section~\ref{sec:epid-proc-temporal-nets}.

Finally, 
co-evolution of the network and the dynamical process occurs
when the topological structure of a network \textit{reacts}
dynamically to the evolution of a dynamical process taking place on
top of it. Indeed, individual social activity can be altered by the
presence of an epidemic outbreak (e.g. avoiding contacts that amount
to link deletion), thus affecting the topology of the underlying social
network, which in turn feeds back nontrivially on the spreading
dynamics. The coupling of topology with disease evolution in such
\textit{coevolving} networks is discussed in Section~\ref{sec:6.C}.

\section{Theoretical approaches for epidemic modeling on networks}

\label{sec:theoreticalmethods}
A continuous-time epidemic process with constant transition rates
between compartments on any graph can be described by Markov theory.
Let us consider a network
defined by its adjacency matrix $A$ and a general epidemic process with
$q$ compartments. The state of node $i$ at time $t$ is specified by a
random variable $X_{i}\left( t\right) \in\{0,1, \ldots, q-1\}$, where
$X_{i}\left( t\right) = \alpha$ means that node $i$ belongs to
compartment $\alpha$ at time $t$.  We assume that all transitions
between compartments are given by independent Poisson processes with
given rates.  Under these conditions, the evolution of the epidemic
process can be described in terms of a Markov chain
\cite{vankampen,PVM_PAComplexNetsCUP}. In a network with $N$ nodes, the
total number of states equals $q^{N}$, all possible combinations in
which all $N$ nodes can take a value from $0$ to $q-1$.  The elements of
the $q^{N} \times q^{N}$ infinitesimal generator $Q$ of the
continuous-time Markov chain are explicitly computed for $q=2$ in
\citet{PVM_ToN_VirusSpread,PVM_EpsilonSIS_PRE2012,Simon_Taylor_Kiss_MathBiol2011},
while the general case is treated in \citet{PVM_GEMF}. Once the
infinitesimal generator $Q$ and the initial infection probabilities are
known, the state probabilities
$\Pr\left[ X_{1}\left( t\right) =x_{1},\ldots,X_{N}\left( t\right)
  =x_{N}\right] $
at time $t$, for each $x_{j}=0, 1, \ldots, q-1$, can be computed using
ordinary matrix operations, from which all desired information can be
deduced in principle.

Although the Markov approach is exact, its use has been limited to a
few exact results in the case of the SIS model.  Indeed, using an
exact Markov approach is impervious for a number of reasons. First,
the linear set of $q^{N}\times q^{N}$ equations to be solved limits
the analysis to very small graphs.  Second, the structure of the
infinitesimal generator $Q$ is rather complex, which prevents from
gaining general insights, although it is possible
\citep{PVM_EpsilonSIS_PRE2012} to deduce a recursion relation between
the $Q$ matrix in a graph with $N$ and $N+1$ nodes. Third, in most
cases, we are interested in the steady-state (or stationary) behavior
or in the final size of the epidemic. The peculiar property of the
exact continuous-time Markov process is the appearance of an absorbing
state, which is equal to the overall-healthy state ($x_{j}=0$ for each
node $j$) in which the activity (virus, information spreading etc.)
has disappeared from the network. Mathematically, an absorbing state
means that the $Q$ matrix has a row of zero elements, the Markov chain
is reducible and the steady-state is equal to this overall-healthy
state for finite $N$. These complications mean that only a
time-dependent analysis, focusing on metastable states, may answer
questions of practical interest. 

More in general, few exact results have been derived for epidemic 
spreading in networks.  For this reason, the derivation of explicit 
results on the behavior of epidemic
spreading processes in networks mostly relies on mean-field
theoretical approaches of different kind.  In the following we review these
approaches, and discuss the different approximations and assumptions on which they are based.
The detailed applications of these approaches to the paradigmatic cases of the SIS
and SIR models will be presented in
Section~\ref{sec:epid-proc-heter}.

\subsection{Individual-based  mean-field approach}
\label{sec:quenched-mean-field}

Individual-based mean-field theory (IBMF) represents a drastic 
simplification of  the exact description presented above.  The basic
idea \cite{Wang03,Chakrabarti_2008,PVM_ToN_VirusSpread,Gomez10} is to
write down evolution equations for the probability $\rho^\alpha_i$
that the node $i$ belongs to the compartment $\alpha$, for any node $i$,
assuming that the dynamic state of every node is statistically independent 
of the state of its nearest neighbors.  
The mean-field equations can be obtained, under
this assumption, by applying an extended version of the law of mass
action, i.e. assuming that the probability that node $i$ is in state
$\alpha$ and its neighbor node $j$ in state $\alpha'$ is
$\rho_i^\alpha \rho_j^{\alpha'}$. More systematically, they can be
obtained directly from the governing equations derived from the
$q^N$-state Markov chain, assuming that the expected values 
of variables pairs factorize: $E[X_i X_j] = E[X_i] E[X_j]$. 
This method is akin to the classic assumption of the
mean-field theory, while keeping the full topological structure of the
network encoded in all the entries of the adjacency matrix $a_{ij}$,
that it is considered to be static or \textit{quenched}, using the
language of mean-field theory in statistical mechanics.

The solutions of IBMF theories depend in general on the spectral
properties of the adjacency matrix, and in particular on the value of
its largest eigenvalue $\Lambda_1$.  Their predictions are generally in
agreement with numerical simulation results obtained for static
networks. As well-known from the theory of critical phenomena, the
agreement tends to decrease, when the densities $\rho_i^\alpha \to 0$ and the
independence assumption breaks down.

Individual-based mean-field approximations can be extended by using
pair-approximation approaches \cite{PhysRevA.45.8358}, in which the expectation $E[X_i X_j]$  are
considered as relevant dynamical quantities, for which the
evolution equations are written. In order to provide these equations
in closed form, the three-point correlations functions $E[X_i X_j
X_m]$ are factorized as a function of the single and two points
correlation functions. By the same token it is possible to derive
exact equations for the correlation functions up to $n$
points~\citet{PVM_upperbound_SIS_epidemic_threshold}.  An
approximation is, however, always required to close the set of equations
by expressing $n+1$-points correlations as functions of correlations
of lower order. As the order $n$ grows, these approximations are
characterized in general by increasing levels of accuracy.

Although the IBMF method can be generalized to time-dependent adjacency matrices and adaptive models, explicit solutions
have been obtained mainly for the SIS models on static
networks.

\subsection{Degree-based mean-field approach}
\label{sec:heter-mean-field-1}

Degree-based mean field (DBMF) theory was the first theoretical
approach proposed for the analysis of general dynamical processes on
complex networks, and its popularity is due to its applicability to a
wide range of dynamical processes on
networks~\cite{dorogovtsev07:_critic_phenom,barratbook}.  The DBMF
approximation for dynamical processes on networks starts with the
assumption that all nodes of degree $k$ are statistically
equivalent. This assumption implies that, instead of working with
quantities $\Phi_i$ specifying the state of vertex $i$ (as in IBMF
theory), the relevant variables $\Phi_k$ are specifying the state of
all vertices with degree $k$, the \textit{degree class}
$k$~\cite{marian1}. The assumption also implies that any given vertex
of degree $k$, is connected with the same probability $P(k'|k)$ to any
node of degree $k'$. The approach is a convenient complexity reduction
technique that consists in a drastic reduction in the number of
degrees of freedom of the system.

DBMF theory for epidemic models focuses on   
the partial densities of individuals of degree $k$ in the compartment $\alpha$,
$\rho^\alpha_k(t)$, or, in other words, the probability that an
individual in the population with degree $k$ is in the compartment
$\alpha$. These variables are not independent, but fulfill the
condition $\sum_\alpha \rho^\alpha_k(t) =1$.  The total fraction of individuals in the compartment $\alpha$ is
$\rho^\alpha(t) = \sum_k P(k) \rho_k^\alpha(t)$.  The explicit rate equations
for the quantities $\rho^\alpha_k(t)$ are obtained by using the law of mass action 
and assuming the independence of the expectation
values  (see Section~\ref{sec:class-results}).

The DBMF theory implicitly contains an approximation that is not
always clearly stated.  The statistical equivalence within degree classes
considers the network itself in a mean-field perspective, in which the
adjacency matrix $a_{ij}$ is completely destroyed, only
the degree and the two-vertex correlations of each node being preserved.
This is equivalent to replacing the adjacency matrix in the IBMF
theory by its ensemble average $\bar{a}_{ij}$, 
expressing the probability that vertices $i$
and $j$ are connected (\textit{annealed network approximation}), 
taking the form \cite{dorogovtsev07:_critic_phenom,Boguna09}
\begin{equation}
  \bar{a}_{ij} = \frac{k_j P(k_i|k_j)}{N P(k_i)}.
  \label{eq:annealedadjacencymatrix}
\end{equation}
In the case of uncorrelated networks, the simple form $\bar{a}_{ij} =
k_i k_j /(N \av{k})$ is obtained.  

The solutions obtained from DBMF theories depend in general on the
statistical topological properties of the underlying networks, and in
the case of uncorrelated networks, on the moments of its degree
distribution. Although the DBMF theory is obviously a strong
approximation in the case of dynamical processes occurring on static
networks, it appears to be a suitable approximation to capture the
behavior of epidemics mediated by interaction patterns changing on a
time scale much faster than the timescales of the spreading
process. In this limit, we can consider the epidemic process to spread
on a network that is constantly rewired, while preserving the given
functional form for $P(k)$ and $P(k'|k)$.  This process amounts to a
contagion process spreading on an effective mean-field network
specified by the \textit{annealed network approximation}. Furthermore,
the DBMF provides a good description of a wide range of dynamical
processes that include complex compartment transitions, multiple
occupancy of nodes and time-varying connectivity patterns.

\subsection{Generating function approach}
\label{sec:mapping-percolation}
For the SIR model and similar models without steady-state, the long time
(static) properties of the epidemic outbreak can be mapped into a
suitable bond percolation problem (see Section~\ref{sec:2.C}). In this
framework, the probability $p$ that a link exists is related to the
probability of transmission of the disease from an infected node to a
connected susceptible one.

The problem of percolation in networks
\cite{molloy95,Cohen00,Callaway2000} can be elegantly tackled with
generating functions ~\cite{Wilf:2006:GEN:1204575}.  
Let us consider the case of bond percolation, in which
edges in a network are removed with probability $1-p$ and kept with
probability $p$ (see Section~\ref{sec:2.C}). Let us define $u$ as the
probability that a randomly chosen edge \textit{does not} lead to a
vertex connected to the (possibly existing) giant component. A randomly
chosen edge is not connected to the giant component if either it has
been removed, or if it leads to a vertex of degree $k$, whose remaining
$k-1$ edges either do not exist or do not lead to the giant component,
i.e.:
\begin{equation}
  u = 1-p + \sum_{k} \frac{k P(k)}{\av{k}} (1-p+pu)^{k-1}.
  \label{eq:perco1}
\end{equation}
This equation is valid for degree uncorrelated networks which have no
loops\footnote{The formalism can be extended to degree correlated
  networks, see Section~\ref{sec:effects-degr-corr} and
  \citet{PhysRevE.78.051105}.}, in which a randomly chosen edge points
to a vertex of degree $k$ with probability $k P(k)/\av{k}$, see
Section~\ref{sec:degree-correlations}. The probability $1-P_G$ that a 
randomly chosen vertex does not belong to the giant component, 
is proportional to the probability that it has degree $k$, and all 
of its outgoing edges either have been removed or do not
lead to the giant component, i.e.
\begin{equation}
  P_G(p) = 1 - \sum_k P(k) (1-p+up)^k.
  \label{eq:perco2}
\end{equation}
Eqs~(\ref{eq:perco1}) and~(\ref{eq:perco2}) can be conveniently
written in terms of the degree distribution generating function
\cite{Wilf:2006:GEN:1204575} $G_0(z) = \sum_k P(k) z^k$ and the excess
degree generating function $G_1(z) = \sum_k (k+1)P(k+1) z^k/\av{k}$,
taking the form
\begin{eqnarray}
  u & = & 1 - p + G_1(1-p+pu) \label{eq:perco3}\\
  P_G(p) & = & 1 - G_0(1-p+pu). 
\end{eqnarray}
The condition for the existence of a giant component translates into
the condition for the existence of a nonzero solution of
Eq.~(\ref{eq:perco3}), which is \cite{Callaway2000}
\begin{equation}
  p > p_c = \frac{G_0'(1)}{G_0''(1)}= \frac{\avk}{\av{k^2}-\avk}.
  \label{eq:percothreshold}
\end{equation}
In the vicinity of the critical point, the expansion of the generating
functions around the nonzero solution yields the scaling behavior of the
order parameter, $P_G(p) \sim (p-p_c)^{\beta_{perc}}$, with
$\beta_{perc}=1$ in the case of homogeneous networks.  In the case of
heterogeneous networks with degree distribution $P(k)\sim k^{-\gamma}$,
we surprisingly find that the percolation threshold tends to zero for
$\gamma<3$ in the limit of an infinite network size, $N\to\infty$
~\cite{Cohen02}.  The critical exponent $\beta_{perc}$ assumes in this
class of networks the following values~\cite{Cohen02}
\begin{equation}
  \beta_{perc} = \left\{\begin{array}{ll}
    1/(3-\gamma) & \mathrm{for} \; \gamma < 3\\
    1/(\gamma-3) & \mathrm{for} \; 3< \gamma \leq  4\\
    1 & \mathrm{for} \; \gamma \geq 4\\
  \end{array} \right. . \label{eq:SIRbetaexponent}
\end{equation}
For the case $\gamma=3$, a stretched exponential form $P_G(p) \sim
e^{1/p}$ is expected, based on the mapping to the SIR model, see
Sec.~\ref{sec:heter-mean-field}.

The above expressions are very general, and can be used also to study
immunization strategies and other containment measures in the case of
SIR-like models. See also~\citet{Karrer2014,Hamilton2014} for very recent
further improvements on these results.

\section{Epidemic processes in heterogeneous networks}
\label{sec:epid-proc-heter}

\subsection{Susceptible-Infected-Susceptible model}
\label{sec:susc-infect-susc}

An impressive research effort has been devoted to understanding
the effects of complex network topologies on the SIS model.  The SIS
dynamics involves only two-state variables and may reach a stationary
state, making it ideal for the application of several theoretical
approaches.  For this reason, there are a large number of results
concerning the SIS model, obtained with approaches ranging from
approximate mean-field theories to exact methods. In the following, we
will follow a historical perspective that starts with the basic and
easily generalizable mean-field approaches and moves then to recent
exact results that put our understanding of the SIS model in complex
networks on firm theoretical ground.

\subsubsection{Degree-based mean-field theory}
\label{sec:heter-mean-field-2}

The first approach to the study of the SIS model in complex networks
\cite{pv01a} used a degree-based mean-field (DBMF) theory (commonly
referred in the physics literature as the heterogeneous mean-field
approach), whose general methodology can be extended to a wealth of
dynamical processes in networks~\cite{barratbook}.  In the DBMF
approach, the SIS model is described in terms of the probability
$\rho^I_k(t)$ that a node of degree $k$ is infected at time $t$,
assuming the statistical equivalence of all nodes of degree $k$.  The
SIS dynamical equation for $\rho^I_k(t)$ is derived by applying the law
of mass action,

\begin{equation}
  \frac{d \rho^I_k(t)  }{dt} 
  = - \rho^I_k(t)+ \lambda k [1- \rho^I_k(t)]\sum_{k'}  P(k'|k)
  \rho^I_{k'}(t), 
  \label{eq:HMFSISequation}
\end{equation}
where, without loss of generality, we  have
rescaled time by $\mu^{-1}$, so that the recovery rate is unitary and the infection rate is equivalent to the spreading rate $\lambda=\beta/\mu$.
The first term accounts for the recovery of nodes
of degree $k$, proportional to the probability $\rho^I_k(t)$ that a
node of degree $k$ is infected. 
The second term accounts for the infection of new nodes, and is proportional
to the probability that a node of degree $k$ is susceptible,
$1-\rho^I_k(t)$, times the probability $P(k'|k)$ that this node is
connected to a node of degree $k'$,
multiplied by the probability $\rho^I_{k'}(t)$ that this
last node is infected, times the rate of infection $\lambda$. This
factor is summed over all the possible values of $k'$. The extra
factor $k$ takes into account all the possible edges through which the
disease can arrive at a node of degree $k$. 

The set of Eqs.~(\ref{eq:HMFSISequation}) for the DBMF approximation to
the SIS model cannot be solved in a closed form for general
degree correlations. The value of the epidemic threshold can however
be obtained by means of a linear stability analysis~\cite{marian1}.
Performing an expansion of
Eq.~(\ref{eq:HMFSISequation}) at first order in  $\rho^I_k(t)$ leads to 
\begin{equation}
   \frac{d \rho^I_k(t)  }{dt} \simeq \sum_{k} J_{k k'}  \rho^I_{k'}(t),
\end{equation}
where the Jacobian matrix element is 
$J_{kk'} = - \delta_{kk'} + \lambda k P(k'|k)$ and
where $ \delta_{ij}$ is the Kronecker delta symbol. 
A null steady state, corresponding to the healthy phase, is stable
when the largest eigenvalue of the Jacobian is negative. The endemic
phase will thus take place when  $-1 + \lambda \Lambda_M >0$,
where $ \Lambda_M$ is the largest eigenvalue of the 
\textit{connectivity matrix} \cite{marian1}, whose elements are
\begin{equation}
  C_{kk'} = k P(k'|k).
  \label{eq:SISconnectivitymatrix}
\end{equation}
From Perron-Frobenius Theorem \cite{Gantmacher}, since $C$ is
non-negative, and assuming that it is irreducible, its largest
eigenvalue is real and positive.  Therefore, the endemic state occurs
for
\begin{equation}
  \lambda > \lambda_c^\mathrm{DBMF} = \frac{1}{\Lambda_M}.
  \label{eq:HMSSISthreshold}
\end{equation}

In the case of uncorrelated networks, in which $ P(k'|k) = k' P(k')/\av{k}$,
it is possible to obtain an explicit solution of  the DBMF equations by writing 
\begin{equation}
  \frac{d \rho^I_k(t)  }{dt} = - \rho^I_k(t)+ \lambda k [1-
  \rho^I_k(t)] \Theta,
\end{equation}
where
\begin{equation}
  \Theta = 
  \sum_{k'} \frac{k'
    P(k')}{\av{k}}  \rho^I_{k'}(t)
  \label{eq:HMFSISDeftheta}
\end{equation}
The latter expression gives the probability to find an infected node
following a randomly chosen edge.
In the steady state, imposing the stationarity
condition $ \frac{d \rho^I_k(t) }{dt} = 0$, we obtain
\begin{equation}
   \rho^I_{k} = \frac{ \lambda k \Theta (\lambda)}{1 +  \lambda k
     \Theta (\lambda)},
   \label{eq:HMFSISStationary}
\end{equation}
where $\Theta$ is now a constant that depends on the spreading rate $\lambda$.
The set of Eqs.~(\ref{eq:HMFSISStationary}) shows that the
higher the degree of a node, the higher its infection probability, indicating that strongly
 inhomogeneous connectivity patterns impact the epidemic spreading. The factor $\Theta(\lambda)$
can be computed self-consistently, introducing
(\ref{eq:HMFSISStationary}) into the definition
Eq.~(\ref{eq:HMFSISDeftheta}), to obtain
\begin{equation}
  \Theta (\lambda) = \frac{1}{\av{k}}  \sum_{k} k P(k)  \frac{ \lambda
    k \Theta (\lambda)}{1 +  \lambda k 
     \Theta (\lambda)}.
   \label{eq:DefTheta}
\end{equation}
The self-consistent equation (\ref{eq:DefTheta}) admits a non-zero
solution, corresponding to the endemic state, only when the following
threshold condition for uncorrelated networks is fulfilled~\cite{pv01a}
\begin{equation}
   \lambda > \lambda_c^\mathrm{DBMF, unc} = \frac{\av{k}}{\av{k^2}}.
  \label{eq:HMSSISthresholdunc}
\end{equation}
The uncorrelated threshold can also be obtained from the general
expression Eq.~(\ref{eq:HMSSISthreshold}) by noticing that the
elements of the connectivity matrix reduce to $C_{kk'}
= k k'P(k')/\av{k}$, which has a unique non-zero eigenvector with
eigenvalue $\av{k^2}/\av{k}$. For a fully homogeneous (regular) network with
$\av{k^2} = \av{k}^2$, Eq.~\eqref{eq:HMSSISthresholdunc} recovers the
result $\lambda_c^\mathrm{DBMF} = 1 / \av{k}$, as expected from the
simple arguments from Section~\ref{sec:class-results} (see
Eq.~\eqref{homogeneousR0}).

Eq.~(\ref{eq:HMSSISthresholdunc}) implies that, in networks with a
power-law degree distribution with exponent $2<\gamma \leq3$, for
which $\av{k^2} \to \infty$ in the limit of a network of infinite
size, the epidemic threshold tends asymptotically to
\textit{zero}. This was one of the first results pointing out the crucial effect of degree heterogeneities on epidemic
spreading.  The critical behavior of the prevalence in the vicinity of
the epidemic threshold can be obtained by solving
Eq.~(\ref{eq:DefTheta}) for $\Theta$ in the continuous degree
approximation and introducing the result into the definition
$\rho^I(\lambda) = \sum_k P(k) \rho^I_k$.  From these manipulations,
one obtains \cite{Pastor01b} $\rho^I(\lambda) \sim (\lambda -
\lambda_c^\mathrm{DBMF})^{\beta^\mathrm{DBMF}_\mathrm{SIS}}$, with the
critical exponent
\begin{equation}
   \beta^\mathrm{DBMF}_\mathrm{SIS} = \left\{\begin{array}{ll}
     1/(3-\gamma) & \mathrm{for} \; \gamma < 3\\
     1/(\gamma-3) & \mathrm{for} \; 3< \gamma \leq  4 \\
     1 & \mathrm{for} \; \gamma \geq 4\\
   \end{array} \right. . \label{eq:DBMFSISbetaexponent}
\end{equation}
For the case $\gamma=3$, a prevalence following a stretched exponential
form is obtained, namely $\rho^I(\lambda) \sim e^{-1/(m \lambda)}$
\cite{pv01a}. Noticeably, this exponents take the exact same form as
those observed for the percolation problem,
Eq.~(\ref{eq:SIRbetaexponent}).  It is interesting to note that for
$2<\gamma \leq3$ the exponent governing the prevalence behavior close to
the threshold is larger than one.  As noted in~\citet{pv01a} this
implies that, while the vanishing threshold makes the spreading of
pathogens more easy, the very slow growth of the epidemic activity for
increasing spreading rates makes epidemic in these networks less
threatening.

\subsubsection{Individual-based mean-field theory}
\label{sec:quenched-mean-field-1}

As introduced in Section~\ref{sec:theoreticalmethods}, the state of
the system in the SIS model is fully defined by a set of Bernoulli
random variables $X_{i}\left( t\right) \in\{0,1\}$: $X_{i}\left(
  t\right) =0$ for a healthy, susceptible node and $X_{i}\left(
  t\right) =1$ for an infected node. It is possible to
construct a $2^N$ Markov chain
\cite{PVM_ToN_VirusSpread,PVM_EpsilonSIS_PRE2012,Simon_Taylor_Kiss_MathBiol2011},
specifying exactly the time evolution of the SIS model. While exact,
as mentioned above, the Markov chain approach complicates
analytical calculations. A simpler route to derive rigorous
results on the SIS model is to use the property of a Bernoulli random variable $X_{i}$ that the expectation
$E\left[ X_{i}\right]$ is equal to the probability that
node $i$ is infected, i.e. $E\left[ X_{i}\right] =\Pr\left[
  X_{i}=1\right] \equiv \rho^I_i(t)$.  This allows to write the
exact equations for the expectation of being infected for each
node $i$ of the SIS model
\citep{PVM_upperbound_SIS_epidemic_threshold,PVM_PAComplexNetsCUP},
\begin{equation}
\frac{dE\left[  X_{i}\left(  t\right)  \right]  }{dt}  =E\left[  -\mu
X_{i}\left(  t\right)  +\left(  1-X_{i}\left(  t\right)  \right)  \beta
\sum_{j=1}^{N}a_{ij}X_{j}\left(  t\right)  \right] 
\label{governing_eq_SIS}
\end{equation}
Eq.~(\ref{governing_eq_SIS}) holds also for asymmetric adjacency
matrices, i.e. for both directed and undirected networks and for
time-varying networks where the adjacency matrix $A(t)$ depends on
time $t$~\citep{Guo2013}.  The SIS governing equation
(\ref{governing_eq_SIS}) states that the change over time of the
probability of infection $E\left[ X_{i}\left( t\right) \right]
=\Pr\left[ X_{i}\left( t\right) =1\right] $ of node $i$ equals the
average of two competing random variables: (a) if the node $i$ is
infected ($X_{i} = 1$), then $\frac{dE\left[ X_{i}\right] }{dt}$ decreases
with rate equal to the curing rate $\mu$ and (b) if the node is
healthy ($X_{i} = 0$), it can be infected with infection rate $\beta$
from each infected neighbor. The total number of infected neighbors of
node $i$ is $\sum_{j=1}^{N}a_{ij}X_{j}$.

For a static network, Eq.~(\ref{governing_eq_SIS}) reduces to
\cite{Sharkey2011,PhysRevE.87.042815,PVM_PAComplexNetsCUP}
\begin{eqnarray}
\frac{d \rho^I_i(t)  }{dt}&=&- 
 \rho^I_i(t)  +\lambda\sum_{j=1}^{N}a_{ij} \rho^I_j(t)
  \nonumber\\
&&-\lambda\sum_{j=1}^{N}a_{ij}E\left[
X_{i}\left(t \right)  X_{j}\left(t \right)  \right],
\label{dE[X_i]_met_joint_probabilities}
\end{eqnarray}
where $t$ has been rescaled by $1/\mu$ and $\lambda=\beta/\mu$.

The above equations do not lend themselves to an explicit solution because the equation
for $\rho^I_i(t)$ depends on the two-node expectation $E\left[
  X_{i}\left( t\right) X_{j}\left( t\right) \right]$. Its exact
computation requires the knowledge of the joint
probability distribution $\Pr\left[ X_{i}=1, X_{j}=1\right]$ for the
state of nodes $i$ and $j$. In order to derive a closed set of $N$ 
dynamical equations, the
Individual-Based Mean-Field (IBMF) approximation is usually made [also termed Quenched
Mean-Field (QMF) or N-Intertwined Mean-Field Approximation (NIMFA)], 
which assumes that
the states of neighboring nodes are statistically~\textit{independent}, i.e.
\begin{equation}
E\left[ X_{i}\left( t\right) X_{j}\left( t\right) \right]
\equiv E\left[ X_{i}\left( t\right) \right] E\left[ X_{j}\left(
    t\right) \right] = \rho^I_i(t) \rho^I_j (t)
\label{independence_assumption_NIMFA} 
\end{equation}
Under this approximation the dynamical 
equations~(\ref{dE[X_i]_met_joint_probabilities}) for the SIS model
become~\cite{hethcote1984gonorrhea,Wang03,Chakrabarti_2008,PVM_ToN_VirusSpread}
\begin{equation}
\frac{d \rho^I_i(t)  }{dt}=-
 \rho^I_i(t)  +\lambda [1-\rho^I_i(t)] \sum_{j=1}^{N}a_{ij} 
\rho^I_j(t).
\label{eq:IBMFSISequations}
\end{equation}
The physical interpretation is immediate: the change in the probability
$\rho^I_i$ has a destruction term, equal to the probability that node $i$ is
infected times the rate of recovery $\mu=1$, and a creation term, equal
to the probability that node $i$ is susceptible, times the total probability
that any of its nearest neighbors is infected, times the effective
transmission rate $\lambda=\beta/\mu$. Again, time has been rescaled in
Eq.~(\ref{eq:IBMFSISequations}). 
Noticeably, Eq.~(\ref{eq:IBMFSISequations}) can be derived using other
approaches. For example, \citet{Gomez10} propose a discrete time
equation taking additionally into account the possibility of reinfection
in a single time step of length $\Delta t$. The equation thus obtained
leads to Eq.~(\ref{eq:IBMFSISequations}) in the continuous time limit
$\Delta t \to 0$.

To obtain a prediction of the threshold, we can apply a linear 
stability analysis on
Eq.~(\ref{eq:IBMFSISequations}).  Indeed, linearizing
Eq.~(\ref{eq:IBMFSISequations}) leads to the Jacobian matrix, with
elements $J_{ij} = -\delta_{ij} + \lambda a_{ij}$.
An endemic state occurs when the largest
eigenvalue of $J$ is positive. This condition translates in the epidemic
threshold
\begin{equation}
  \lambda \geq \lambda^\mathrm{IBMF}_c, \quad \lambda^\mathrm{IBMF}_c = 
  \frac{1}{\Lambda_1}, 
  \label{eq:IBMFthreshold}
\end{equation}
where $\Lambda_1$ is the largest eigenvalue of the adjacency matrix
\cite{Wang03,Chakrabarti_2008,PVM_ToN_VirusSpread}.

In networks with a power-law degree
distribution, $P(k) \sim k^{-\gamma}$, eq. (\ref{eq:IBMFthreshold})
can be combined with $\Lambda_1 \sim
\max\{\sqrt{k_\mathrm{max}}, \AV \}$ \cite{Chung03}, where
$k_\mathrm{max}$ is the maximum degree in the network, to produce an
  expression for the scaling of the epidemic threshold \cite{Castellano2010,Castellano2012}
\begin{equation}
  \lambda^\mathrm{IBMF}_c \simeq \left \{ 
    \begin{array}{lr}
      1/\sqrt{k_\mathrm{max}} & ~~~~~~~ \ \ \gamma > 5/2 \\
        \av{k}/\av{k^2} & ~~~~~~~ 2< \gamma < 5/2
\end{array} \right. .
\label{together}
\end{equation}
The relevance of this result is the prediction, in the thermodynamic
limit, of a vanishing epidemic threshold for \textit{every} network for
which the maximum degree is a growing function of the network size,
which is essentially the case for all random, non-regular networks.
Although the expression for the epidemic threshold obtained from the
IBMF theory is not exact, (see~\citet{Givan2011} for a detailed
assessment of the independence assumption), it provides a relatively
good accuracy when compared with the results of extensive numerical
simulations, see Section~\ref{sec:numerical-results}.

It is worth bridging the IBMF approach with the DBMF approach presented in the
previous section.  
As stated in Section~\ref{sec:theoreticalmethods}, the DBMF approach is
based on the assumption of the statistical equivalence of all nodes
with the same degree $k$, actually defining the spreading process on
an effective mean-field graph,  whose adjacency matrix is given by the
annealed form $\bar{a}_{ij} = k_j P(k_i|k_j)/(N P(k_i))$. 
This elucidates the connection between the IBMF and DBMF approaches.
The latter can be simply derived by substituting the annealed adjacency
matrix in the Eqs.~(\ref{eq:IBMFSISequations}).
By performing a degree-based average
$\rho^I_k = \sum_{i \in k} \rho^I_i/ (NP(k))$, 
the
equations~(\ref{eq:HMFSISequation}) are thus recovered from the IBMF
approach. Hence, DBMF is equivalent to IBMF with the additional
approximation that the detailed topological network structure is
replaced by its annealed version.

Within the framework of IBMF theory, it is also possible to derive 
the behavior of the prevalence $\rho^I$ in the stationary state
just above the epidemic
threshold~\cite{PVM_epidemic_phase_transition2011,Goltsev12}
\begin{equation}
  \rho^I  \left(  \lambda \right)  \simeq \frac{1}{
    N}\frac{\sum_{j=1}^{N}\left(  x_{1}\right)  _{j}}{\sum_{j=1}^{N}\left(
      x_{1}\right)  _{j}^{3}} \frac{\lambda -\lambda_c}{\lambda_c} 
\label{y_infty_general_behavior_just_above_epidemic_threshold}%
\end{equation}
 where $\vec{x}_1$ is the principal eigenvector (PEV) corresponding
to the largest eigenvalue of the adjacency matrix. 
The complete expansion of the prevalence in the stationary state around the epidemic threshold is derived in \citet{PVM_viral_conductance}.

Based on Eq.~(\ref{y_infty_general_behavior_just_above_epidemic_threshold}),
the validity of the IBMF prediction for the epidemic threshold has been
recently questioned~\cite{Goltsev12} according to the following
argument.  For $\lambda_c^\mathrm{IBMF}$ to be the true epidemic
threshold, the stationary state above it must be endemic, with a
finite fraction of the network infected. This requires that for
$N\to\infty$ the prefactor
\begin{equation}
  \mathcal{A} = \frac{1}{%
    N}\frac{\sum_{j=1}^{N}\left(  x_{1}\right)  _{j}}{\sum_{j=1}^{N}\left(
      x_{1}\right)  _{j}^{3}}
\end{equation}
in Eq.~(\ref{y_infty_general_behavior_just_above_epidemic_threshold})
must tend to a constant of $\mathcal{O}(1)$.  Whether $\mathcal{A}$ is
constant or not depends on the localization of the PEV,
i.e. whether its weight is evenly distributed (delocalized) on all
nodes of the network, or localized in a few nodes.  Goltsev \textit{et
  al.} apply this idea to the analysis of power-law distributed
networks, arguing by means of analytical calculations and numerical
experiments (see also \citet{2014arXiv1401.5093M}) that, for $\gamma
\leq 5/2$, the PEV is delocalized, while it is localized for $\gamma >
5/2$.  This would imply that, while $\lambda_c^\mathrm{IBMF}$ always
marks a transition to an active state, this one is endemic only for
$\gamma<5/2$, corresponding to a delocalized PEV; for $\gamma>5/2$,
instead, a localized PEV indicates that the transition at
$\lambda_c^\mathrm{IBMF}$ is not to an endemic state, but to a
subendemic state, in which activity is restricted to the neighborhood
of the hubs with largest degree.  Support to this argument (which is
mean-field in nature, based on
Eq.~(\ref{y_infty_general_behavior_just_above_epidemic_threshold}))
is provided in ~\citet{Lee2013}, who
characterize the sub-endemic state as a Griffiths phase 
(see also~\citet{PhysRevLett.111.068701}).

\subsubsection{Extensions of degree-based and individual-based mean-field
approaches}
\label{sec:extens-mean}

Several extensions of the degree-based and individual-based mean-field
theories have been proposed, taking into account the role of dynamical correlations, 
which are neglected in both approaches. 

A natural way to include the effect of correlations is to consider
additional variables representing the state of pairs, triples etc. of
neighboring nodes.  \citet{Eames2002} introduced an extended
degree-based approach where the evolution of the average number
$\av{I^k}$ of nodes of degree $k$ in the infected state depends on the
number $\av{S^k I^l}$ of connections between susceptibles of degree $k$
with infected nodes of degree $l$. The dynamics
can be written in terms of the properties of triples, such as 
$\av{S^k S^l I^m}$ and so on so forth.  If averages for triples are
approximated with averages for pairs and single nodes, the dynamical
equations are reduced to a set of $O(k_{max}^2)$ nonlinear ordinary
differential equations.  This procedure can be iterated, but
the increased accuracy is counteracted by a rapid growth in the number of
equations.

Similarly, \citet{Gleeson11}, building on the results
of~\citet{Marceau2010}, proposed a general theory for binary-state
dynamics in networks. This approach takes into account explicitly
the dynamical correlations between adjacent nodes (see
also~\citet{Lindquist2011} for a similar approach).  The theory is based
on a set of master equations for the quantities $s_{k,m}(t)$ and
$i_{k,m}(t)$ which, in the context of the SIS model, are defined as the
fraction of nodes of degree $k$ which are susceptible (resp. infected)
at time $t$ and are connected to $m \leq k$ infected neighbors. By means
of combinatorial arguments, these quantities can be related to the
prevalence $\rho^I_k$ of nodes of degree $k$, allowing the determination
of the prevalence and epidemic threshold. This theoretical approach
provides a good description of the time evolution of the prevalence
\cite{PhysRevX.3.021004} and good estimates of the epidemic threshold
for random regular lattices \cite{Gleeson11}. Gleeson's approach
presents again the drawback that the estimation of the threshold in more
complex networks requires the numerical solution of large sets of
coupled equations, which hinders the analysis of large network sizes.

Another degree-based approach, proposed by \citet{PhysRevLett.111.068701}, 
takes into account long distance
correlations by considering explicitly the possibility of reinfection
between nodes $i$ and $j$, separated by a topological distance
$\ell_{ij}$ possibly larger than one.
For this purpose, the original SIS dynamics is replaced by a modified
description valid over coarse-grained time scales. In such longer
temporal intervals, a given infected node $i$
can propagate the infection to any other node $j$ at distance $\ell_{ij}$
in the network, via a sequence of microscopic infection events of 
intermediate, nearest neighbors nodes. 
The infection rate $\beta$ is then replaced by an
effective rate $\bar{\beta}(\ell_{ij}, \beta)$.
On the coarse-grained time scale also the recovery rate $\mu$ of node $i$ is
replaced by an effective rate  $\bar{\mu}(k_i, \beta)$.
Both parameters
$\bar{\beta}(\ell_{ij}, \beta)$ and $\bar{\mu}(k_i, \beta)$ can be
estimated from the properties of the network and the SIS
model. Writing down a mean-field theory for such extension of the SIS
model
, upper bounds for the epidemic threshold $\lambda_c$
of the original SIS model are deduced, which are in good agreement with
numerical simulations, see Section~\ref{sec:numerical-results}.

For individual-based approaches, the consideration of dynamical
correlations can be introduced in a systematic way, by the analogue of a
cluster expansion \cite{PhysRevA.45.8358}. The exact SIS
Eqs.~(\ref{dE[X_i]_met_joint_probabilities}) are,
as discussed above, not closed, due to the presence of the term
involving dynamical correlations between pairs of adjacent nodes. One
way to proceed 
consists in complementing Eq.~(\ref{dE[X_i]_met_joint_probabilities})
with an equation for the evolution of the pair correlations $E\left[
  X_{i}\left( t\right) X_{k}\left( t\right) \right]$. The $\binom{N}{2}$ 
governing equations for
$\frac{dE\left[ X_{i}X_{j}\right] }{dt}$ for $i\neq j$ take the
form \cite{PVM_secondorder_SISmeanfield_PRE2012}
\begin{eqnarray}
\frac{dE\left[  X_{i}X_{j}\right]  }{dt}  
&  =-2\mu E[X_{i}X_{j}]+\beta\sum_{k=1}^{N}a_{ik}E[X_{j}X_{k}]\nonumber\\
&  \hspace{0.5cm}+\beta
\sum_{k=1}^{N}a_{jk}E[X_{i}X_{k}]\nonumber\\
&  \hspace{0.5cm}-\beta\sum_{k=1}^{N}(a_{ik}+a_{jk})E[X_{i}X_{j}X_{k}]
\label{second_order_pair_correlation_E[XiXj]}%
\end{eqnarray}
while for $i=j$, obviously Eq.~(\ref{governing_eq_SIS}) holds. 
Equations~(\ref{governing_eq_SIS}) 
and~(\ref{second_order_pair_correlation_E[XiXj]}) are still an exact 
description of the dynamics involving now the terms
$E\left[ X_{i} X_{j} X_{k} \right] $, that in turn need to be determined,
via $\binom{N}{3}$ differential equations involving joint fourth order
expectations and so on. In summary,
the approach leads to a set of $\sum_{k=1}^{N}\binom{N} {k}=2^{N}-1$ 
exact equations describing the evolution of the SIS process
(to be complemented with the conservation of probability)
that form a hierarchy: the equations for the evolution of correlations of
order $n$ depending on those of order $n+1$. 

To allow computations in practice, this hierarchy must be limited to some small $n$ by imposing a
closure condition for the set of equations. The simplest closure
condition, $ E[X_{i}X_{j}] = E[X_{i}]E[X_{j}]$, leads to the IBMF
approximation. Higher order closures include dynamical correlations in a
more detailed way, thus providing a more accurate description of the
system dynamics.  The assumption of different closure relations leads to
different degrees of tractability of the ensuing equations. Some of
those can be proved to be exact for simple networks
\cite{2013arXiv1307.7737K}.  For example, focusing on general closure
forms, \citet{PVM_secondorder_SISmeanfield_PRE2012} propose the
expression $ E[X_{i}X_{j}X_k] = E[X_{i}X_j]E[X_{k}]$. Analogously,
\citet{0295-5075-103-4-48003}, applying standard techniques from pair
approximations in statistical physics, propose the closure
\begin{equation}
   E[X_{i}X_{j}X_k] = \frac{ E[X_{i}X_{j}]  E[X_{j}X_k]}{E[X_{j}]}.
\label{eq:pairapproxSilvio}
\end{equation}
The particular interest of the closure (\ref{eq:pairapproxSilvio}) is that it allows deriving an explicit 
expression for the epidemic threshold in terms of the largest
eigenvalue of the new Jacobian matrix of the dynamical equations
\cite{0295-5075-103-4-48003}:
\begin{equation}
  J_{ij} = - \left(1+ \frac{\lambda^2 k_i}{2 \lambda+2}\right)
  \delta_{ij} + \frac{\lambda(2+\lambda)}{2 \lambda+2}a_{ij}.
  \label{eq:jacobianSilvio}
\end{equation}

A completely different approach to determine the epidemic threshold for the
SIS model has been proposed by~\citet{Parshani2010}. The idea is to
map the SIS dynamics with fixed infection time, to a percolation process,
mirroring the approach successfully used for the 
SIR model (see Sec.~\ref{sec:mapping-percolation-SIR}). 
In SIS dynamics, however, the mapping is approximate and one
has to take into account the reinfection
probability $\pi$, i.e. the probability that an infected node reinfects 
the node from which it originally received the disease. 
By estimating $\pi$ and using it in a modified percolation approach, values 
of the epidemic threshold are derived, in good agreement with numerical 
simulations, also for heavy-tailed degree distributions.

\subsubsection{Exact results}
\label{sec:exact-results}
Although the above mean-field approaches provide a general theoretical
picture of the behavior of the SIS model in networks, 
a few  exact results exist that  provide rigorous
bounds for the threshold and the dynamical behavior of the model.
A first exact result concerning the  lower bound of the epidemic
threshold \cite{van_mieghem_non-markovian_2013} can be achieved by revisiting
Eq.~(\ref{dE[X_i]_met_joint_probabilities}).
Since $0\leq\sum_{k=1}^{N}a_{ki}X_{i}\left( t\right) X_{k}\left(
  t\right) $, it is possible to write the inequality:
 \begin{equation}
\frac{d \rho^I_i(t)  }{dt}\leq- \rho^I_i(t)
 +\lambda \sum_{k=1}^{N}a_{ki}
\rho^I_k(t)
\label{eq:sisboundeq}
\end{equation}
Denoting the vector 
$W=\left( \rho^I_{1},\rho^I_{2},\cdots,\rho^I_{N}\right) $, the solution of
the inequalities (\ref{eq:sisboundeq}) is
\begin{equation}
W\left(t \right)  \leq e^{\left(  \lambda A-I\right) t } W\left(0\right).
 \label{upper_bound_W(t)}
\end{equation}
The exponential factor is dominated by the fastest growing mode, 
which is $\lambda\Lambda_{1}-1$,
where $\Lambda_{1}$ is the largest eigenvalue of the non-negative
matrix $A$, which is real and positive, by the Perron-Frobenius
Theorem~\cite{Gantmacher}.
When $\lambda\Lambda_{1}-1\leq0$, then $W_{i}=\rho^I_i(t)$ decreases
exponentially in $t$ towards zero and the epidemic dies out fast, so that 
\begin{equation}
\lambda_{c}\geq\frac{1}{\Lambda_{1}}.
\label{lowerbound_SIS_epidemic_threshold}%
\end{equation}
Interestingly, this lower bound coincides with the IBMF result.

\citet{Ganesh05} have proven that the average time $E\left[
T\right]$ for the SIS Markov process to hit the absorbing state, when
the effective infection rate $\lambda<\frac{1}{\Lambda_{1}}$, obeys
\begin{equation}
E\left[  T\right]  \leq\frac{\log N+1}{1- \lambda \Lambda_{1}}
\label{mean_time_to_absorption_below_tau_c}
\end{equation}
from which Eq.~(\ref{lowerbound_SIS_epidemic_threshold}) is deduced.

Above the epidemic threshold instead, the activity must be endemic,
so that the average time to absorption is
$E\left[ T\right] = O(e^{cN})$ for some constant $c>0$. 
\citet{Chatterjee_Durret2009} proved that in graphs with power-law degree
distribution $E\left[ T\right] > O(e^{bN^{1-\delta}})$ for any
$\delta>0$. This result pointed to a vanishing threshold in the
large $N$ limit, but still left the possibility open for nonendemic
long-lived metastable states, as those predicted by~\citet{Goltsev12,Lee2013}.
This possibility has been recently ruled
out by the work of \citet{Mountford2013},
showing that for any $\lambda>0$ and large $N$, the time to absorption
on a power law graph grows exponentially in $N$, implying
that there is endemic activity for any $\lambda>0$.

For the complete graph, the exact average survival time has been
determined using the Markov theory \cite{XXXXXX}. In
particular, for the complete graph, the average survival time for all
$\lambda$ and $N$ is
\begin{equation}
E[T] = \sum_{j=1}^{N}\sum_{r=0}^{j-1}\frac{(N-j+r)!}{j(N-j)!}\lambda^{r} \label{ET_KN}
\end{equation}
whose asymptotic for large $N$ is
\[
E\left[  T\right]  \sim\frac{1}{\mu}\frac{\frac{\lambda}{\lambda_{c}}\sqrt{2\pi}%
}{\left(  \frac{\lambda}{\lambda_{c}}-1\right)  ^{2}}\frac{\exp\left(  N\left\{
\log\frac{\lambda}{\lambda_{c}}+\frac{\lambda_{c}}{\lambda}-1\right\}  \right)  }{\sqrt
{N}}
\]
for an effective infection rate $\lambda=\frac{\beta}{\mu}$ above the
epidemic threshold $\lambda_{c}$. Since an infection can survive the
longest in the complete graph, the maximum average lifetime (or
survival) time of an SIS epidemic in any network with $N$ nodes is not
larger than (\ref{ET_KN}), or than
$E[T]=O\left( e^{N\ln\frac{\lambda}{\lambda_{c}}}\right)$.

For power-law graphs,  \citet{Chatterjee_Durret2009} provide exact bounds
for the exponent $\beta_{SIS}$ governing the singular behavior 
$\rho^I \sim \lambda^{\beta_{SIS}}$ of the activity at the 
transition, namely $\gamma-1 \leq \beta_{SIS} \leq 2 \gamma-3$.
This implies that the mean-field value $\beta_{SIS}=1$ does not hold
for any $\gamma>2$, as well as the failure of the DBMF
prediction, Eq.~(\ref{eq:DBMFSISbetaexponent}).

For a few special classes of simple graphs such as the complete graph
and the star, the $2^N$-state Markov chain can be reduced to a much
smaller number of states, enabling an exact solution
\cite{PVM_EpsilonSIS_PRE2012,PVM_MSIS_star_PRE2012,PhysRevE.87.042815,
  PVM_decay_SIS2014}.  More results can be classified as
\emph{asymptotic} exact results, where the network size
$N\rightarrow\infty$.  An overview of asymptotic exact results is given
by \citet{Durrett_PNAS2010}.

\subsubsection{Numerical simulations of the SIS model on networks}
\label{sec:numerical-results}

As presented above, the different approximations of the SIS process on networks yield different results
for the numerical value of the epidemic threshold. This is
particularly important in the case of networks with a heavy-tailed
degree distribution $P(k) \sim k^{-\gamma}$, where the two main
approximations, IBMF and DBMF, lead to the same result for $\gamma <
5/2$, but to noticeable differences for $\gamma > 5/2$, especially in
the case $\gamma > 3$.  In this region, while DBMF predicts a finite
threshold, IBMF indicates a vanishing one, albeit at a relatively
small rate with the system size.

Computational efforts have been mostly devoted to the numerical
determination of the epidemic threshold of the SIS model on
power-law distributed networks, in order to assess the validity of the
different theoretical approaches. For a detailed study on graphs of
small size see~\cite{PVM_comparisonSIS_meanfield_PRE2012}.

The standard numerical procedure to study absorbing phase transitions,
such as the epidemic transition of SIS, is based on
the determination of the average of the order parameter (in this case
the density of infected nodes), restricted only to surviving runs
\cite{Marrobook}, i.e., runs which have not reached the absorbing
state up to a given time $t$. Such a technique is not efficient,
because close to the threshold long time surviving configurations are very rare 
and an exceedingly large number of realizations of the
process are needed in order to get substantial statistics. This
problem is particularly severe for a large network size, for which
very large simulation times are required, due to the presence
of a long initial transient. 
These issues make the standard procedure impractical and have not
led to reliable conclusions until recently.

\begin{figure}[t]
\includegraphics*[width=\columnwidth]{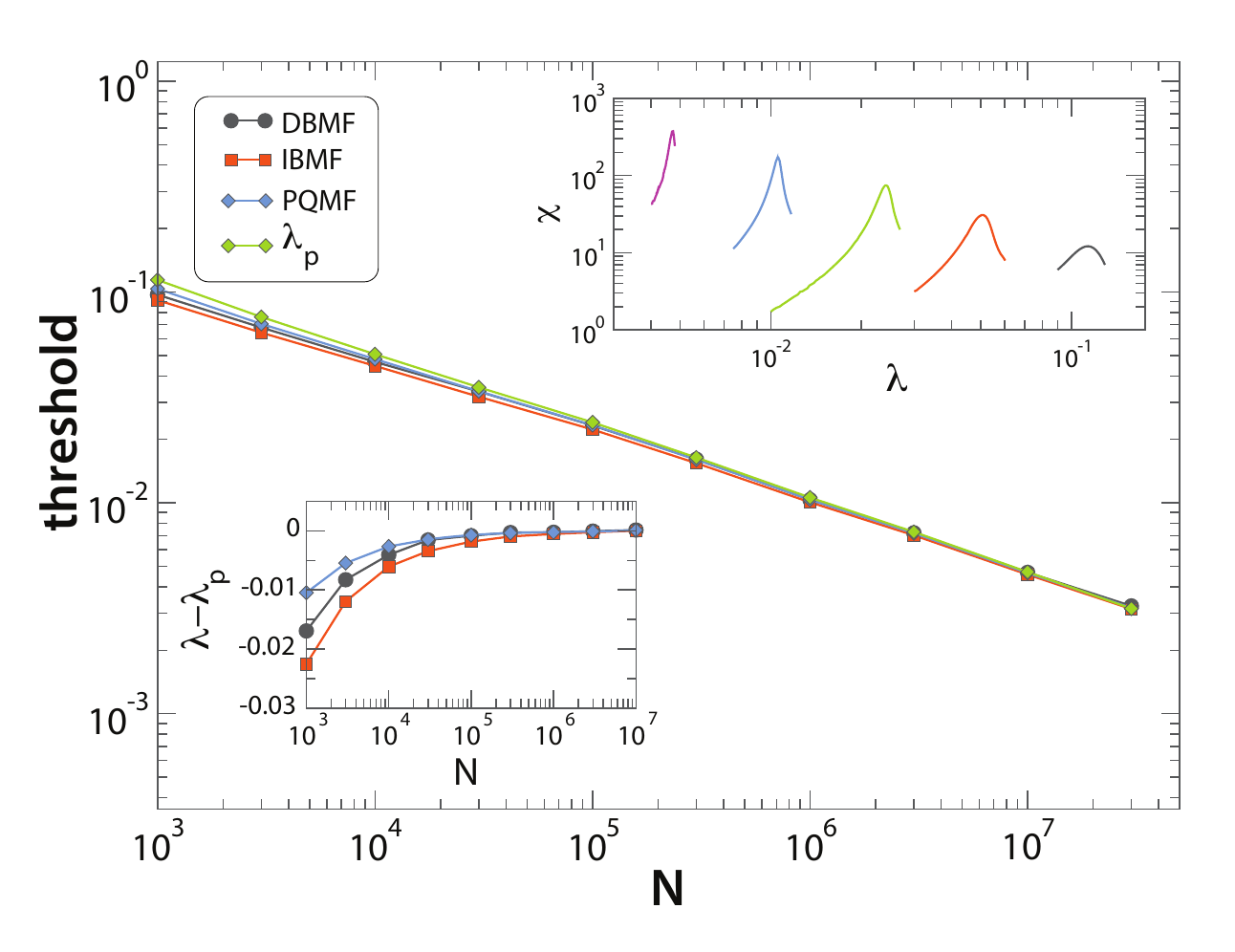}
\caption{Numerical thresholds for the SIS model as a function of the
  network size $N$ in scale-free networks with degree exponent
  $\gamma=2.25$, computed using the QS method, compared with different
  theoretical predictions. Upper inset shows the behavior of the
  susceptibility as a function of the spreading rate for different
  values of $N=10^3, 10^4, 10^5, 10^6, 10^7$, from right to left. Lower
  inset shows the difference between the different theoretical
  thresholds and the peaks of the susceptibility. Figure adapted from
  \citet{0295-5075-103-4-48003}.}
\label{fig:IBMFSis}
\end{figure}

\begin{figure}[t]
\includegraphics*[width=\columnwidth]{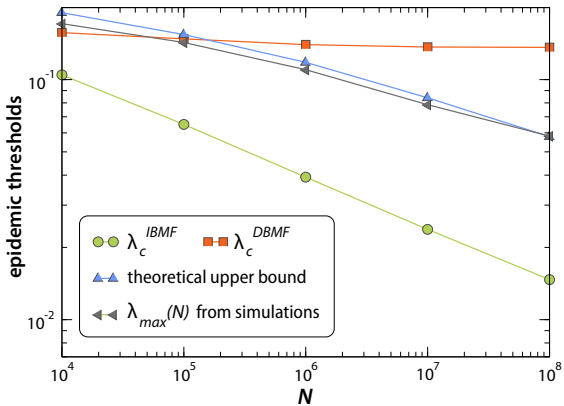}
\caption{Numerical thresholds for the SIS model as a function of the
  network size $N$ in power-law distributed networks with degree
  exponent $\gamma=3.5$, computed from the average lifetime method
  proposed by \citet{PhysRevLett.111.068701}. Numerical data are
  compared with different theoretical approaches as well as with the the
  upper bound obtained from the DBMF theory with long range dynamical
  correlations, developed by \citet{PhysRevLett.111.068701}. Figure
  adapted from \citet{PhysRevLett.111.068701}}
\label{fig:LifetimeSis}
\end{figure}

In order to overcome the restrictions of the surviving runs method,
\citet{Ferreira12, 0295-5075-103-4-48003} use the quasi-stationary state
(QS) method~\cite{DeOliveira05,FFCR11}, based on the idea of
constraining the system in an active state.  This procedure is
implemented by replacing the absorbing state, every time the system
tries to visit it, with an active configuration randomly taken from the
history of the simulation (see also~\citet{PVM_EpsilonSIS_PRE2012} for
an implementation of the same idea by means of an external field).  With
this technique, the threshold is estimated by studying the
susceptibility \cite{Ferreira12}, defined as
\begin{equation}
  \label{eq:susceptdef}
  \chi = N \frac{\fluc{\rho^I}-\av{\rho^I}^2}{\av{\rho^I}}.
\end{equation}
When plotted as a function of $\lambda$ in a system of size $N$, the
susceptibility $\chi$ exhibits a maximum at a value $\lambda_p(N)$,
corresponding to a transition rounded by finite size effects.  In the
thermodynamic limit, the position of the peak tends to the critical
point as
$\lambda_p(N) - \lambda_c(\infty) \sim
N^{-1/\bar{\nu}}$~\cite{Binder2010}.
Large scale simulations performed using the QS method~\cite{Ferreira12,
  0295-5075-103-4-48003}, see Figure~\ref{fig:IBMFSis}, show that, for
$\gamma < 5/2$, the IBMF and a pair approximation at the individual level
(PQMF) are almost exact, coinciding asymptotically with the DBMF result
in this range of degree exponents.  For $5/2 < \gamma<3$, on the other
hand, the IBMF result provides the correct scaling of the threshold with
network size.
For the crucial case $\gamma>3$, where IBMF and DBMF provide radically
different predictions, the results are not as conclusive. A new
numerical approach has been proposed to explore this region
\cite{PhysRevLett.111.068701}, based on the study of the
\textit{lifetime} of individual realizations of the SIS process
starting with a single infected node.  Each realization is
characterized by duration $T$ and coverage $C$, where the latter is
the fraction of distinct nodes ever infected during the realization.
In the thermodynamic limit, realizations can be either finite
(i.e. having a finite lifetime and, therefore, vanishing coverage) or
endemic (i.e. having an infinite lifetime and coverage equal to 1.)
The average lifetime $E[ T ]$ of finite realizations plays
the role of a susceptibility, exhibiting a peak at the transition,
whose position can then be used to estimate the threshold.  The
nontrivial problem to determine whether, in a finite system, a
realization is endemic or not, can be overcome by declaring endemic
all realizations for which the coverage reaches a predefined value
(e.g.  $C=0.5$). Numerical simulations performed with this method
indicate that the extended DBMF approach by
\citet{PhysRevLett.111.068701} provides a very good fit to the
numerical threshold for $\gamma>3$, see Figure~\ref{fig:LifetimeSis},
with a scaling with network size that is essentially given by the IBMF
expression Eq.~(\ref{together}).

\subsubsection{Finite size effects and the epidemic threshold}

As we have seen in the previous sections, the connectivity pattern of
the network enters explicitly in the determination of the epidemic
threshold that generally depends on the moments of the degree
distribution and/or the maximum degree of the network.  This finding has
particular relevance in networks with heavy-tailed degree distributions,
where the probability of nodes with very large degree is appreciable. In
the limit of infinite size networks, the epidemic threshold may be
vanishing, thus prompting to the disruption of the classical epidemic
framework where the disease can spread only for adequate
transmissibility of the pathogen. While mathematically compelling, the
argument of a vanishing threshold has been soon recognized as not
realistic in real-world
networks~\cite{may2001infection,pastor2002fs}. Even if the connectivity
pattern of a network is well approximated by a heavy-tailed distribution
in a given range of degree values, any real-world network is composed by
a finite number of nodes $N$. For instance, the finite size of
scale-free networks is generally related to the presence of a natural
maximum degree $k_\mathrm{max}\sim N^{1/(\gamma-1)}$, as reported in
Section~\ref{sec:basic-network-models}, that translates into a finite
effective epidemic threshold. Although the finite size of the network is 
often a determinant element in the estimation of the epidemic threshold, for
instance in the analysis of numerical simulations (see Section
\ref{sec:numerical-results}), there are many other limitations to the
maximum degree of the network. These limits are often imposed by
spatio-temporal constraints, such as maximum occupancy in spatial
locations and the finite time each individual can interact with other
individuals. As well, intrinsic cognitive and biological constraints may
be at work in real-world systems. One example is provided by the
so-called Dunbar's number that limits humans' degree to
between 100 and 200 individuals, a size apparently imposed by the finite
neocortical processing capacity of the
brain~\cite{dunbar1998social}. Interestingly, Dunbar's number has been
observed in a wide range of human activities, including communication on
modern information technologies, making it a relevant limit in the case
of many information diffusion
processes~\cite{Goncalves2011Dunbar,miritello2013time}.

In view of these inherent limitations, it is often convenient to assume
that even in the case of heavy-tailed networks the degree distribution
is characterized by the analytic form
$P(k)\simeq k^{-\gamma}\exp{(-k/k_{c})}$, where $k_c$ is a
characteristic degree size. The exponential cut-off makes it extremely
unlikely to observe nodes with degree much larger than $k_c$,
effectively introducing an intrinsic limit to the connectivity capacity
of nodes~\cite{pastor2002fs}. Within the DBMF approach this leads,
For large $k_c$ and $2 < \gamma < 3$, to
$\lambda_c^{DBMF,unc} \simeq \left( k_c/m \right)^{\gamma-3}$
where $m$ is the minimum degree of the network,
which can be generalized for other values of $\gamma$ and which shows
the effect of the degree limitations imposed by the intrinsic
biological, social and cognitive constraints in real-world
networks. Similar finite size effects and considerations also apply to
the epidemic threshold obtained with the IBMF theory and other
approaches.

It is important to stress however that the presence of an epidemic
threshold because of finite size effects and other connectivity
limitations should not be considered as an argument to neglect the
network heterogeneity. It is indeed possible to show with simple
calculations~\cite{pastor2002fs} that simple homogenous approaches can
overestimate the actual epidemic threshold in heterogeneous networks by
one or more orders of magnitude.

\subsection{Susceptible-Infected-Removed model}

\label{sec:4.B}
The SIR model is a cornerstone  in infectious disease modeling.  It
applies to the wide range of diseases that do provide immunity to the
host and it is also a widely used modeling scheme in knowledge and
information diffusion (see Sec.~\ref{sec:7.A}).  Theoretically, the SIR
model represents a different challenge with respect to the SIS model
because it does not allow for a stationary state.  The two most used
routes to a general analysis of the SIR model have been initially the
DBMF theory and the mapping of static properties to the percolation
model. Here, we start with a presentation of the
DBMF approach, focusing then on other degree-based, individual-based and
alternative methods which have been completing the understanding of the
SIR dynamics in networks in recent years. We end the subsection with an
overview of the exact results on static properties which can be obtained
by mapping SIR to bond percolation.

\subsubsection{Degree-based mean-field approach}
\label{sec:heter-mean-field}
The DBMF approach can be easily adapted to provide 
insight into the dynamical and statical properties of the SIR model.
In the DBMF approximation, we can define as a function of time
three different partial densities, namely of infected, susceptible 
and recovered nodes of degree $k$, denoted by the variables 
$\rho_k^I(t)$, $\rho_k^S(t)$ and $\rho_k^R(t)$, respectively. 

The order parameter (prevalence) of the SIR model, defined as the number
of removed individuals at the end of the epidemics, is then given by
$\rho^R_\infty = \lim_{t\to\infty} \sum_k P(k) \rho_k^R(t)$.
In describing the time evolution of these densities, one can follow
the analogy with the SIS model, to obtain the set of
equations~\cite{refId0,lloyd01}
\begin{eqnarray}
\label{eq:SIR_HMF}
  \frac{d \rho_k^I(t)}{d t} &=& - \rho_t^I(t) + \lambda k \rho_k^S(t)
  \Gamma_k(t), \\ \nonumber
  \frac{d \rho_k^R(t)}{d t} &=& \rho_k^I(t),
\end{eqnarray}
complemented with the normalization condition $ \rho_k^S(t) = 1 -
\rho_k^I(t)- \rho_k^R(t)$, where
\begin{equation}
  \Gamma_k(t) = \sum_{k'} P(k'|k) \rho_{k'}^I(t). 
\end{equation}
The value of the epidemic threshold in the case of general
correlations can be obtained as in the SIS case, by performing a
linear stability analysis. The same result follows,
with the epidemic threshold given by the inverse of the largest
eigenvalue $\Lambda_M$ of the connectivity matrix, 
Eq.~(\ref{eq:SISconnectivitymatrix}). As for SIS, 
in the case of uncorrelated networks the epidemic threshold is
given by $\lambda_c = \av{k} / \av{k^2}$~\cite{refId0,lloyd01}.
For uncorrelated networks,
within the same DBMF approximation, it is also possible
to integrate the rate equations over time, 
starting form a small seed, thus obtaining the full temporal evolution of
the spreading process. The solution depends on a differential equation
for an auxiliary function $\phi(t)$, which cannot be solved
analytically in general. However, in the infinite time limit, it is
possible to determine the dependence of the final prevalence
$\rho^R_\infty$ on $\lambda$
\begin{equation}
  \rho_\infty^R = \sum_k P(k) (1- e^{-\lambda k \phi_\infty}),
\end{equation}
where
\begin{equation}
  \phi_\infty =  1 - \frac{1}{\av{k}}  - \sum_k
 \frac{k P(k)}{\av{k}} e^{-\lambda k \phi_\infty}.
\label{eq4B2:1}
\end{equation}
The solution of Eq.~(\ref{eq4B2:1}) leads again to the epidemic
threshold $\lambda_c = \av{k} / \av{k^2}$, a result that again recovers
the naive expectation for regular networks, see
Eq.~\eqref{homogeneousR0}, $\lambda_c = 1 / \av{k}$.  For a power-law
degree distribution, $P(k) \sim k^{-\gamma}$, a detailed analysis
\cite{refId0} leads to a prevalence, in the vicinity of the epidemic
threshold, of the form $\rho_\infty^R \sim (\lambda -
\lambda_c)^{\beta_\mathrm{SIR}}$, with exponent $\beta_\mathrm{SIR}$
coinciding with the value for bond percolation,
Eq.~(\ref{eq:SIRbetaexponent}).  The above results are exact for
annealed networks, when the topology changes (preserving $P(k)$) at a
very fast rate \cite{volz_epidemic_2009}. Instead, when considering it
as an approach to static networks, the DBMF can be improved taking
into account that, in the SIR process, a vertex cannot propagate the
disease to the neighbor who originally infected it, because the latter
is necessarily not susceptible. This effect can be included in the
DBMF equations by discounting, from the number of edges pointing from
infected individuals of degree $k'$ to vertices of degree $k$, the
edge from which the original infection arrived to the vertices of
degree $k'$. In this way, the Eqs.~(\ref{eq:SIR_HMF}) are recovered
but now the $\Gamma_k(t)$ function takes the form \cite{marianproc}
\begin{equation}
  \Gamma_k(t) = \sum_{k'} \frac{k' -1}{k'} P(k'|k) \rho_{k'}^I(t). 
\end{equation}
The value of the epidemic threshold in this case is given by
$\lambda_c=1 / \tilde{\Lambda}_M$, where  $\tilde{\Lambda}_M$ is the
largest eigenvalue of the new connectivity matrix
\begin{equation}
  \tilde{C}_{k k'} = \frac{k(k'-1)}{k'} P(k'|k).
  \label{sirconnectmatrix}
\end{equation}
In the case of uncorrelated networks, 
the largest eigenvalue of the matrix $\tilde{C}_{k k'}$ is
$\tilde{\Lambda}_M = \av{k^2}/ \av{k}-1$ 
(the corresponding eigenvector has components $\tilde{v}_k = k$)
so that the epidemic threshold is
\begin{equation}
  \lambda_c = \frac{\av{k}}{\av{k^2}-\av{k}}.
\label{eq:4B2:thres}
\end{equation}
As shown below, Eq.~(\ref{eq:4B2:thres}) is
an approximation of the exact result~(\ref{Tlambda}).
However, this modified DBMF approach captures the correct 
qualitative behavior, discriminating between vanishing threshold,
for scale-free networks, and finite threshold, for $\gamma>3$.

The DBMF approach allows also to tackle the scaling of
the \textit{time evolution} of the epidemic outbreak. This is particularly important in the
context of models like SIR that do not have a stationary state. 
For the sake of simplicity let us initially focus on the SI model
\cite{anderson92}, representing a disease in which infected
individuals never recover and keep propagating the disease
forever. The SI model can be considered the limit
of the SIR model in which the recovery rate $\mu$ is set to zero. While
this simplification leads to a trivial asymptotic state in which the whole
population becomes eventually infected, it is nevertheless interesting
due to its simplicity, which allows to obtain explicit results for 
the initial time evolution of epidemic outbreaks. 
The DBMF analysis of the SI model proceeds from the analogue of
Eq.~(\ref{eq:SIR_HMF}), valid for generic networks
\cite{sievolution,Barthelemy2005275}
\begin{equation}
  \frac{d \rho_k^I(t)}{d t} = \beta k [1-\rho_k^I(t)] \Gamma_k(t),
\label{eq:siequation}
\end{equation}
with 
\begin{eqnarray}
\label{eq:sigammafunc}
\Gamma_k(t) &=& \sum_{k'} P(k'|k) \rho_{k'}^I(0) \\ \nonumber
& + & \sum_{k'} \frac{k'-1}{k'}  P(k'|k) [\rho_{k'}^I(t) - \rho_{k'}^I(0)].
\end{eqnarray}
The first term in Eq.~(\ref{eq:sigammafunc}) accounts for a very small
initial seed of infected individuals, with initial partial density
$\rho_{k}^I(0)$, which can infect all their neighbors. The second term
represents the contribution of individuals infected during the
outbreak, which can infect all their neighbors, with the exception of
those who transmitted the disease. 
Linear stability analysis shows that the time evolution at very short times
(when the partial densities of infected individuals are very small) follows an
exponential growth, $\rho^I(t) \sim e^{t/\tau}$, where the
characteristic time is given by $\tau = (\beta \tilde{\Lambda}_M)^{-1}$,
where again $\tilde{\Lambda}_M$ is the largest eigenvalue of the
connectivity matrix in Eq.~(\ref{sirconnectmatrix}). 
In the case of uncorrelated networks this implies
\cite{sievolution,Barthelemy2005275}
\begin{equation}
  \label{tausi}
  \tau = \frac{\av{k}}{\beta[\av{k^2}-\av{k}]}.
\end{equation}
The solution for the SI model can be extended to the case of the general
SIR model by allowing a nonzero healing rate, which leads to the general
time scale of the initial growth~\cite{Barthelemy2005275}
\begin{equation}
  \tau = \frac{\av{k}}{\beta \av{k^2}-(\mu+\beta)\av{k}}.
\end{equation}

These results readily indicate that the growth time scale of an
epidemic outbreak is inversely proportional to the second moment of
the degree distribution $\av{k^2}$; when this quantity diverges, as in
the case of scale-free networks, not only the threshold tends to
vanish, but also the time until the establishment of
the infection becomes very small (vanishing in the thermodynamic limit).
Computer simulations allow to obtain a detailed picture of the
mechanism of spreading of a disease in a scale-free network
\cite{sievolution,Barthelemy2005275}: Initially, the infection reaches
the hubs and from them it quickly invades the rest of the network via
a cascade through progressively smaller degree classes. The dynamical
structure of the spreading is therefore characterized by a
hierarchical cascade from hubs to intermediate $k$, and finally to
small $k$ classes.

\subsubsection{Individual and pair-based mean-field approaches}

As in the SIS case, a systematic way to attack the SIR model is based on
the full master equation for the exact evolution of probabilities of
microscopic states, and the derivation, starting from it, of deterministic
evolution equations for dynamical quantities.  In this framework,
\citet{Sharkey2008} considers SIR with Poissonian infection and recovery
processes and derives from the master equation the $2N$ equations for
the probabilities for the state of individuals
\begin{eqnarray}
\label{SIR_individual}
\frac{d \rho^S_i(t)}{dt} & = & - \beta \sum_{j} a_{ij} \av{S_i I_j} \\ \nonumber
\frac{d \rho^I_i(t)}{dt} & = &  \beta \sum_{j} a_{ij} \av{S_i I_j} - \mu \rho^I_i
\end{eqnarray}
where $S_i$ and $I_j$ are Bernoulli variables equal to $1$ when the
node is susceptible (infected, respectively) and $0$ otherwise,
$\rho^S_i = \av{S_i}$ is the probability that node $i$ is in state
S, $\rho^I_i = \av{I_i}$, is the analogue for state I and $\av{S_i I_j}$ is
the joint probability of state $S_i I_j$.
In order to close the equations~(\ref{SIR_individual}), the
simplest possibility is to assume that the state of neighbors is
independent (individual-based mean-field approximation).
Alternatively, one can derive from the master equation the evolution
of the probabilities of pairs of neighbors, which
depend in turn on the state of triples of neighboring nodes. The closure
of the hierarchy at this level (pair-based mean-field) requires
the approximation of probabilities for triples with moments of lower order.
There are several possible ways to implement the closure and the best 
choice is not a trivial problem. 
The validity of the different approximation schemes is investigated 
in ~\citet{Sharkey2011}, who shows
that replacing $\av{S_i I_j} = \av{S_i} \av{I_j}$ is equivalent to writing down
an equation for the evolution of $\av{S_i I_j}$ containing unphysical terms 
(i.e. terms assuming that a node is at the same time susceptible and infected).
The consequences of these unphysical terms are relevant: from the
individual-based mean-field approach one can derive an expression for the
SIR epidemic threshold equal to what is found for the SIS 
case~\cite{Youssef2011,Prakash2012}:
$\lambda_c = 1/\Lambda_1$, where $\Lambda_1$ is the largest eigenvalue of
the adjacency matrix. This result, however, is even qualitatively 
not correct, as it predicts a vanishing threshold for power-law distributed
networks with $\gamma>3$, at odds with exact results (see below) and numerical
simulations~\cite{Castellano2010}.
The pair-based approach instead, complemented with the closure in 
Eq.~(\ref{eq:pairapproxSilvio}), is proved to be an exact description of 
the stochastic system for a tree  topology~\cite{Sharkey2013}. 
In the case of networks with loops
it is possible to find a precise connection between
the detailed loop structure and the closures that leave the description 
exact~\cite{2013arXiv1307.7737K}.
From these individual and pair-based approaches, by summing over 
all nodes, the equations for the probabilities
of the global quantities $\rho^I$ and $\rho^S$ can be obtained, thus 
providing a microscopic foundation of equations obtained at population level 
by means of the mass action principle.
Eq.~(\ref{SIR_individual}) and similar pair-based
approaches can be written also for heterogeneous infection and recovery
rates~\cite{Sharkey2008}. Hence, the approaches apply in full generality
also to directed and weighted networks.

\subsubsection{Other approaches}

Due to its great relevance, the time evolution of the SIR dynamics has
been tackled with many other approaches.

The extended degree-based approach of~\citet{Eames2002} (see 
Sec.~\ref{sec:susc-infect-susc}) can be applied also to the SIR model,
providing a set of closed ODEs that can be integrated numerically or
used to derive an expression for the basic reproductive ratio $R_0$.
Also the other extended degree-based approach of ~\citet{Lindquist2011} can
be applied to SIR, by categorizing each node by its disease state (i.e., S,
I, R), as well as by the number of neighbors in each disease state.
In this way, an excellent agreement with numerical simulations for
both the temporal evolution and the final outbreak size is found.
The threshold condition derived analytically turns out to be equal 
to the exact one obtained using percolation theory, Eq.~(\ref{Tlambda})
in Sec.~\ref{sec:mapping-percolation-SIR}.

An alternative approach by ~\citet{Volz2008} describes the Poissonian SIR
epidemics at the global population level.  Based on the probability generating function for
the degree distribution, it describes the evolution of the
infection using only 3 coupled nonlinear ordinary differential
equations. The solution of these equations is in excellent agreement
with numerical simulations~\cite{Lindquist2011}; it is shown to be
exact in the thermodynamic limit~\cite{Decreusefond2012,Janson2013} 
and it allows to derive the exact expression, Eq.~(\ref{Tlambda}), 
for the epidemic threshold, in the case of static uncorrelated networks.
In this case, the approach of~\citet{Volz2008}
can be  shown~\cite{House2011} to be a specific case of the extended
degree-based theory of~\citet{Eames2002}.  
Volz's approach can be made more physically transparent and
simpler, reducing to a single evolution equation~\cite{Miller2011}.
The basic idea of this improved approach is to focus on the state of a 
random partner instead of a random individual.
From this starting point, a fully general theoretical framework 
(edge-based compartmental modelling) can be developed, allowing 
to deal with many different scenarios, including static 
and dynamic networks, both undirected and 
directed~\cite{Miller2012,10.1371/journal.pone.0069162,Valdez2012b}. 
For other approaches to SIR dynamics based on
the probability generating function,
see~\citet{marder_dynamics_2007,noel_time_2009,Noel2012}.

A derivation of a condition for the possibility of a
global spreading event starting from a single seed in SIR-like models on
generic networks is presented in \citet{Dodds2011} and generalized 
in~\citet{Payne2011}.  
The approach is based on the state of "node-edge" 
pairs and relates the possibility of spreading to the condition that the
largest eigenvalue of a "gain ratio" matrix (encoding information on
both the topology and the spreading process) is larger than 1.

Finally, a new, substantial step forward in the understanding of the SIR
model is the recent application of the 
message-passing approach to SIR dynamics~\cite{Karrer2010}. 
This approach provides an exact description of the dynamics on trees, 
via a closed set of integro-differential equations,
allowing the calculation of the probabilities to be in state $S$, $I$ or
$R$ 
for any node and any time.
When loops are present, the method gives instead a rigorous bound on the
size of disease outbreaks. 
On generic (possibly directed) trees the approach of ~\citet{Karrer2010}
has been shown~\cite{Wilkinson2014} to coincide for Poissonian
infections with the pair-based moment-closure presented
by~\citet{Sharkey2013}.
Remarkably, the message-passing approach allows dealing with
fully generic (non-Poissonian) infection and recovery processes.

\subsubsection{Mapping the SIR model to a percolation process}
\label{sec:mapping-percolation-SIR}
The connection between the static properties of the SIR model 
and bond percolation (see Section~\ref{sec:mapping-percolation})
was recognized long
ago~\cite{Ludwig1975,Grassberger1983,Andersson2000}.  
In the context of epidemics
on complex networks, the mapping has been studied in detail
by~\citet{newman02}.  Considering a SIR model with uniform infection
time $\tau$, i.e. where infected nodes become removed at time $\tau$
after infection\footnote{Notice that this does not coincide exactly
with the definition given in Section~\ref{sec:class-models-epid}},
and infection rate $\beta$, the \textit{transmissibility}
$T$ is defined as the probability that the infection will be transmitted from an
infected node to a connected susceptible neighbor before recovery
takes place. For continuous-time dynamics the transmissibility can be
computed as \cite{newman02}
\begin{equation}
T =  1 - \lim_{\delta t \to 0} (1-\beta \delta t)^{\tau/\delta t} = 1-e^{-\tau \beta}. 
\end{equation}

The set of removed nodes generated by an SIR
epidemic outbreak originated from a single node is nothing else than
the cluster of the bond percolation problem (with occupation
probability $T$) to which the initial node belongs.  The correspondence is
exact: all late-time static properties of the SIR model can be derived
as direct translations of the geometric properties of the percolation
problem. For tree-like networks the exact epidemic threshold is given 
by Eq.~(\ref{eq:percothreshold}), so that
\begin{equation}
T_c = \frac{\avk}{\av{k^2}-\avk} \Rightarrow
\beta_c = \frac{1}{\tau}\ln\frac{\av{k^2}-\avk}{\av{k^2}-2\avk},
\label{ExactTc}
\end{equation}
The behavior of the outbreak size close to the epidemic
threshold, ruled by the equivalent percolating giant component, is
given in terms of the exponents in Eq.~(\ref{eq:SIRbetaexponent}).
Expression (\ref{ExactTc}) confirms for the SIR model that the epidemic
threshold has a qualitatively different behavior for scale-free
networks ($\gamma<3$) and for scale-rich ones ($\gamma>3$). In the
former case the second moment of the degree distribution diverges,
so that the threshold vanishes: scale-free networks are extremely
vulnerable to disease spreading. 

The above results can be considered
exact only for a tree (completely loopless) structure.
In other networks, the presence of loops and multiple spreading paths
leads in general to correlations, which may invalidate the 
results obtained for trees. However, for random networks which are 
locally tree-like the presence of long loops (infinitely long in the 
thermodynamic limit) is not sufficient to perturb the validity of
the results obtained using the tree
ansatz~\cite{dorogovtsev07:_critic_phenom}. A different conclusion
holds instead in networks with short loops (finite clustering)
as discussed in Sec.~\ref{sec:effects-clust-coeff}.

The derivation of Eq.~(\ref{ExactTc}) is based on a uniform infection time.
More realistically, we assume that 
infection times $\tau_i$ and rates $\beta_{ij}$ vary between individuals. This implies that the transmissibility $T_{ij}$
depends on the specific edge $(i,j)$.
One possible approach, that reduces to the solution of
the homogeneous case~\citep{newman02}, is to neglect
fluctuations, and replace $T_{ij}$ by its mean value
\begin{equation}
\langle T_{ij} \rangle = 1- \int d\tau \int d\beta e^{-\beta \tau} 
Q(\beta) P(\tau),
\label{Taveraged}
\end{equation}
where $Q$ and $P$ are the distributions of $\beta_{ij}$ and $\tau_i$,
respectively. The case of
nondegenerate $\tau_i$ includes the usual definition of the SIR model
with constant recovery rate $\mu$ for which recovery times are
distributed exponentially with average $\langle \tau_i \rangle =
1/\mu$. In such a case, performing the integral in
Eq.~(\ref{Taveraged}) and setting $\beta \langle \tau_i \rangle =
\beta/\mu = \lambda$, yields $\langle T_{ij} \rangle =
\lambda/(1+\lambda)$, implying
\begin{equation}
\lambda_c = \frac{\avk}{\av{k^2}-2 \avk}.
\label{Tlambda}
\end{equation}
This approximation leads to the exact
epidemic threshold, the mean outbreak size below it and the final size
above it, but fails in other respects~\citep{Kenah2007} (see 
also~\citet{Trapman2007}).
The discrepancy is due to
correlations~\cite{Karrer2010}: ``if an individual recovers quickly,
then the probability of transmission of the disease to any of its
neighbors is small; if it takes a long time to recover the probability
is correspondingly larger."  Newman's approximation is not exact also
when the $\tau_i$ are degenerate and the $\beta_{ij}$ vary~\cite{Miller2007}.

The correct way to take into account the heterogeneous transmissibility
maps the disease spreading to a bond percolation process,
involving now a semi-directed network (epidemic percolation
network)~\citep{Miller2007,Kenah2007}, see
Section~\ref{sec:general-definitions}. 
The mapping works as follows.
For each pair of connected nodes $i$ and $j$ in the contact network,
place a directed edge from $i$ to $j$ with probability $1-e^{-\beta_{ij}\tau_i}$
and a directed edge from $j$ to $i$ with probability $1-e^{-\beta_{ji}\tau_j}$.
Tools from percolation theory on directed
networks~\cite{Boguna2005}, see Section~\ref{sec:directed-networks},
allow to characterize exactly the long time features
of the epidemic process. 
In particular the epidemic transition is
associated with the formation of a giant strongly connected component
(GSCC) in the directed network.  
If such a component exists, then an infection originating in
one of its nodes or in the giant in-component (GIN) will spread to all
nodes in the GSCC and in the giant out-component (GOUT), giving rise
to a macroscopic outbreak. It is crucial to recognize that the GIN and
GOUT components play completely different roles: nodes in GOUT are
necessarily part of macroscopic outbreaks but cannot originate them.
The opposite is true for nodes in GIN. As a consequence the
probability that an epidemic occurs (given by the size of
GIN $\cup$ GSCC) and the size of the epidemic (equal to the
size of GSCC $\cup$ GOUT) do not
coincide~\citep{Meyers2006,Miller2007}.
The mapping to percolation on semi-directed networks is valid for any
type of contact network underlying the SIR epidemics. For trees and locally
tree-like networks it is again possible to apply the machinery of
probability generating functions to derive explicit results for the
related percolation properties.

Other discrepancies of the mapping to percolation approach to the SIR
model are reported in~\citet{Lagorio2009755}.

\section{Strategies to prevent or maximize spreading}

\subsection{Efficient immunization protocols}
\label{sec:effic-immun-prot}

The fact that epidemic processes in heavy-tailed networks have a
vanishing threshold in the thermodynamic limit, or a very
small one in large but finite networks (see
Sec.~\ref{sec:epid-proc-heter}), prompted the study of immunization
strategies leveraging on the network structure in order to protect the
population from the spread of a disease.  Immunization strategies are
defined by specific rules for the identification of the individuals
that shall be made immune, taking into account (local or non-local)
information on the network connectivity pattern. Immunized nodes are
in practice removed from the network, together with all the links
incident to them, and each strategy is assessed by the
effects of immunizing a variable fraction $g$ of nodes in the network.
The application of immunization does not only protect directly
immunized individuals, but can also lead, for a sufficiently large
fraction $g$, to an increase of the epidemic threshold up to an
effective value $\lambda_c(g) > \lambda_c(g=0)$, precluding the global
propagation of the disease.  This effect is called \textit{herd
  immunity}.  The main objective in this context is to determine the
new epidemic threshold, as a function of the fraction of immunized
individuals.  Indeed, for a sufficiently large value of
$g$, any strategy for selecting immunized nodes will lead to an
increased threshold. We define the \textit{immunization threshold}
$g_c(\lambda)$, for a fixed value of $\lambda$ such that, for values
of $g > g_c(\lambda)$ the average prevalence is zero, while for $g \leq
g_c(\lambda)$ the average prevalence is finite.

The simplest immunization protocol, using essentially no information at
all, is the random immunization, in which a number $g N$ of nodes is
randomly chosen and made immune.  While random immunization in the SIS
model (under the DBMF approximation) can depress the prevalence of the
infection, it does so too slowly to increase the epidemic threshold
substantially. Indeed, from Eq.~(\ref{eq:HMFSISequation}), an epidemics
in a randomly immunized network is equivalent to a standard SIS process
in which the spreading rate is rescaled as $\lambda \to \lambda(1-g)$,
i.e. multiplied by the probability that a given node is not immunized,
so that the immunization threshold becomes \cite{PhysRevE.65.036104}
\begin{equation}
  g_c(\lambda) = 1- \frac{\av{k}}{\lambda \av{k^2}}.
\end{equation}
For heterogeneous networks, for which $\av{k^2}$ diverges and any
value of $\lambda$, $g_c(\lambda)$ tends to
$1$ in the limit $N\to\infty$, indicating that almost the whole network must be immunized to
suppress the disease.

This example shows that an effective level of protection in
heavy-tailed networks must be achieved by means of \textit{optimized}
immunization strategies \cite{anderson92}, taking into account the
network heterogeneity. Large degree
nodes (the hubs leading to the large degree distribution variance)
are potentially the largest spreaders. Intuitively, an optimized
strategy should be targeting those hubs rather than small
degree vertices. Inspired by this observation, the targeted
immunization protocol proposed by \citet{PhysRevE.65.036104} considers
the immunization of the $g N$ nodes with largest degree. A simple DBMF
analysis leads to an immunization threshold given, for the SIS model, by
the implicit equation \cite{PhysRevE.65.036104}
\begin{equation}
  \frac{\av{k^2}_{g_c}}{\av{k}_{g_c}} = \frac{1}{\lambda},
\label{eq:immuno1}
\end{equation}
where $\av{k^n}_{g}$ is the $n$th moment of the degree distribution
$P_g(k)$ of the network resulting after the deletion of the $gN$ nodes
of highest degree, which takes the form \cite{havlin01}
\begin{equation}
  P_g(k) = \sum_{k' \geq k}^{k_c} P(k') \binom{k'}{k}(1-g)^k g^{k'-k}.
\end{equation}
Eq.~(\ref{eq:immuno1}) can be readily solved in the case of scale-free
networks. For a degree exponent $\gamma=3$, the immunization
threshold reads $g_c(\lambda) \simeq \exp[-2/(m \lambda)]$, where $m$ is the
minimum degree in the network. This result highlights the
convenience of targeted immunization, with an immunization threshold
that is exponentially small over a large range of the
spreading rate $\lambda$. A similar effect can be obtained with a
\textit{proportional} immunization strategy \cite{PhysRevE.65.036104}
(see also \citet{aidsbar} for a similar approach involving the cure of
infected individuals with a rate proportional to their degree), in
which nodes of degree $k$ are immunized with probability $g_k$, which
is some increasing function of $k$.  In this case, the infection is
eradicated when $g_k \geq 1 - 1/(\lambda k)$, leading to an
immunization threshold \cite{PhysRevE.65.036104}
\begin{equation}
  g_c(\lambda) = \sum_{k > \lambda^{-1}} \left(1-\frac{1}{k \lambda}
  \right) P(k),
\end{equation}
which takes the form $g_c(\lambda) \simeq (m \lambda)^2/3$ for
scale-free networks with $\gamma=3$.

Other approaches to immunization stress that not only the behavior
close to the critical point should be taken into account, but also the
entire prevalence curve (the so-called viral conductance)
\cite{Rob_VC_networking2009,Mina_Caterina_VC2011,PVM_viral_conductance}. Additionally,
strategies involving possible different interventions on different
nodes have been analyzed within a game-theoretic formalism
\cite{PVM_heterogeneous_virusspread,PVM_Jasmina_Game_protection_INFOCOM2009,PVM_Gourdin_Networkprotection_DRCN2011}).

The previously discussed immunization protocols are based on a global
knowledge of the network properties (the whole degree sequence must be
known to target selectively the nodes to be immunized). Actually, the more a global knowledge  
of the network is available, the more effective is the immunization strategy.
For instance, one of the
most effective targeted immunization strategies is based on the
betweenness centrality (see Sec.~\ref{sec:centrality}), which combines
the  bias towards high degree nodes and the inhibition of the
most probable paths for infection transmission
\cite{PhysRevE.65.056109}. This approach can be even improved by
taking into account the order in which nodes are immunized in a
sequential scheme in which the betweenness centrality is recomputed
after the removal of every single node, and swapping the order of
immunization in different immunization sequences, seeking to minimize
a properly defined size for the connected component of susceptible individuals. This approach has been proved to
be highly efficient in the case of the SIR model
\cite{PhysRevE.84.061911}.  Improved immunization performance in
the SIR model has been found with an ``equal graph
partitioning'' strategy \cite{PhysRevLett.101.058701} which seeks to
fragment the network into connected components of approximately the
same size, a task that can be achieved by a much smaller number of
immunized nodes, compared with a targeted immunization scheme.

The information that makes targeted strategies
very effective, also makes them hardly feasible in real-world situations,
where the network structure is only partially known. In
order to overcome this drawback, several local immunization strategies
have been considered. A most ingenious one is the
\textit{acquaintance} strategy proposed by \citet{Cohen03}, and
applied to the SIR model. In this protocol, a number $gN$ of
individuals is chosen at random and each one is asked to point to one
of his/her nearest neighbors. Those nearest neighbors, instead of the
nodes, are selected for immunization. Given that a randomly chosen
edge points with high probability to a large degree node, this
protocol realizes in practice a preferential immunization of the hubs, 
that results to be very effective in hampering epidemics.
An analogous result can be obtained by means of a random
walk immunization strategy
\cite{0295-5075-68-6-908,1009-1963-15-12-003}, in which a random
walker diffuses in the network and immunizes every node that it
visits, until a given degree of immunization is reached. Given that a
random walk visits a node of degree $k_i$ with probability
proportional to $k_i$ \cite{PhysRevLett.92.118701}, this protocol
leads to the same effectiveness as the acquaintance immunization.

The acquaintance immunization protocol can be improved by
allowing for the consideration of additional information, always at
the local level. For example, allowing for each node to have knowledge
on the number of connections of its nearest neighbors, a large
efficiency is attained by immunizing the neighboring nodes with the
largest degree \cite{0295-5075-68-6-908}. As more information is
available, one can consider the immunization of the nodes with highest
degree found within short paths of length $\ell$ starting from a
randomly selected node \cite{gomez2006}. The random walk immunization
strategy, on the other hand, can be improved by allowing a bias
favoring the exploration of high degree nodes during the random walk
process \cite{PhysRevE.74.056105}. Variations of the acquaintance
immunization scheme have also been used for weighted networks. 
The acquaintance immunization for weighted networks is outperformed by a
strategy in which the immunized neighbors are selected among those
with large edge weights \cite{Deijfen201157}.

A different approach to immunization, the \textit{high-risk
immunization strategy}, applied by \citet{nian2010} to the SIRS
model, considers a dynamical formulation, in which nodes
in contact with one or more infected individuals are immunized with a
given probability. Again, by immunizing only a small fraction of the
network, a notable reduction of prevalence and increase of the
epidemic threshold can be achieved.

Finally, for the SIR model, the mapping to percolation
suggests which nodes to target in a vaccination campaign, depending on
whether the probability of an outbreak or its size are to be
minimized~\citep{Kenah2011}.  A targeted vaccination of
nodes in the GSCC implies both a reduction of the probability of a
major epidemics and of its size.

\subsection{Relevant spreaders and activation mechanisms}
\label{sec:5.A}
Although the problem of immunization is central in the study of
epidemics because of its practical implications, the attention of the
research community has been recently attracted by the somewhat related
theme of discovering which nodes are most influential/effective in the
spreading process. For instance, what node should be chosen as initial
seed in a SIR epidemic, in order to maximize the total number of
nodes eventually reached by the outbreak?  This is a very natural
question to be posed~\citep{kitsak2010}, in particular when the
propagation process does not involve a disease to be contained but
rather a positive meme (such as a crucial piece of information, see
Section~\ref{sec:7.A}) whose spreading is instead to be maximized.

The traditional common wisdom, derived from early studies on the
immunization problem~\citep{PhysRevE.65.036104}, was that nodes with
the highest degree play the role of superspreaders in networks. This
view has been challenged by \citet{kitsak2010} who pointed out that
the $K$-core index  (see Section~\ref{sec:centrality}) is a much
better predictor of the final outbreak size in the SIR model 
spreading on several real
networks where (as opposed to uncorrelated networks) the set of nodes
with large degrees does not coincide with high $K$.
The intuitive reason is that the most densely connected core gets
easily infected by an outbreak initiated by one of its
vertices, finally transmitting the infection to a large portion of
the entire network.  High degree nodes which are not part of the core may
spread the activity to a large number of neighbors but the infection
hardly extends further.

These findings have stimulated a flurry of activity aimed at
understanding which of several possible topological centrality
measures (degree, betweenness,
$K$-core index, closeness and many others) are more correlated with
spreading influence in various types of networks and contagion dynamics
~\citep{Chen2012,Li2012,Chung2012,Hou2012,daSilva2012,Bauer2012,Zeng2013,
  Liu2013,Chen2013,HebertDufresne2012}.  These studies consider 
different issues and features of the interplay between the network 
and the spreading process, and such a large variability does not allow 
to reach firm conclusions.  
Various quantities are used to evaluate the spreading
effectiveness: in some cases only top influential spreaders are
considered, in others complete rankings of all nodes are compared.
Moreover, the consideration of different real networks in different
papers does not help in comparing approaches and in particular to
disentangle the effects of specific topological features such as
degree heterogeneity, clustering, or assortativity.  Finally not all
studies take properly into account the fact that results may be largely
different depending on which part of the epidemic phase-diagram is considered:
the absorbing phase, the transition regime or the phase where activity
is widespread.  As a consequence, a clear picture that uniquely determines
the best centrality measure that identifies superspreaders for
different epidemic models and different networks has yet to emerge.

The $K$-core decomposition is in many cases a good predictor of
spreading efficiency.  Nevertheless an interesting
finding~\citep{Klemm2012,HebertDufresne2012} is that the removal of a
node with high $K$-core index has a limited effect as multiple paths exist 
among the nodes in the central cores. 
Thus in general efficient spreaders are not necessarily also good
targets for immunization protocols.  An extension of the $K$-core
decomposition to weighted networks with application to a SIR epidemics
on weighted networks (see Sec.~\ref{sec:weighted-networks}) has also
been proposed~\citep{Garas2012}.

Similar to the problem of finding efficient spreaders is the
identification of nodes which are infected earlier than the others, 
thus playing the role of ``sensors'' for epidemic 
outbreaks~\citep{Christakis2010,Garcia-Herranz14}.  The strategy
of considering friends of randomly chosen nodes allows to select,
without any knowledge of the global network structure, individuals
with high degree, high betweenness, small clustering and high $K$-core
index, which are actually reached early by epidemic outbreaks.
This effect lies at the basis of the acquaintance immunization strategy
\cite{Cohen03} discussed above.

Another problem, conceptually close to the search for superspreaders,
is the identification of what topological features trigger global 
epidemics, i.e. what network subsets determine the position
of the epidemic threshold~\citep{Castellano2012}.  For SIS, the
epidemic threshold scales, within the IBMF approximation, as the
inverse of the largest eigenvalue of the adjacency matrix $\Lambda_1$
(see Section~\ref{sec:quenched-mean-field-1}).  Applying the scaling
form of $\Lambda_1$ for large uncorrelated scale-free
networks~\citep{Chung03}, the scaling of the threshold with network
size is given by Eq.~\eqref{together}.  This result can be interpreted
as follows \cite{Castellano2012}: For $\gamma>5/2$,  the node
with the largest degree (hub) together with its direct neighbors
forms a self-sustained nucleus of activity above $\lambda_c$ which
propagates to the rest of the system.  For $\gamma<5/2$ instead, the
threshold position is dictated by the set of most densely
interconnected nodes, as identificated by the $K$-core of largest
index. Topological correlations may alter the picture. For SIR
dynamics instead, the largest hub is not able to trigger the
transition and the position of the threshold is always dictated by the
max $K$-core.

All investigations described so far attempt to relate dynamical
properties of the spreading process to purely geometric features of
the contact pattern.  Taking a more general
perspective,~\citet{Klemm2012} define a "dynamical influence''
centrality measure, which incorporates not only topological but also
dynamical information. The dynamical influence is the leading left
eigenvector of a characteristic matrix that encodes the interplay
between topology and dynamics. When applied to SIR and SIS epidemic
models, the characteristic matrix coincides with the adjacency matrix.
The ``dynamical influence'' predicts well which nodes are active
around the transition, while it is outperformed by other centrality
measures far from the threshold \cite{Klemm2012}.

A growing activity has also recently been concerned with the inverse
problem of inferring statistically, from the configuration of the
epidemics at a given time, which of the nodes was the initial seed
originating the 
outbreak~\citep{Comin2011,Pinto2012,Lokhov2013,Altarelli2013,Brockmann2013}.

Finally, the problem of finding efficient
spreaders is not limited to disease epidemics models; it is possibly
even more important for complex contagion phenomena (such as rumor
spreading or the diffusion of innovations), see Section~\ref{sec:7.A}.

\section{Modeling realistic epidemics}

\subsection{Realistic models}
\label{sec:real-epid-models}

The simple SIS and SIR models considered so far can be generalized to
provide a more realistic description of the disease progression by
introducing additional compartments (see
Sec.~\ref{sec:class-models-epid}) and/or by allowing additional
transitions between the different compartments. These variations, that
can be studied analytically or most often numerically, may alter the
basic phenomenology of the epidemic process.  In this section, we
survey some of those models and refer the reader to the work of
\citet{Masuda200664} for more complicated models that include
pathogens' competition and game-theoretical inspired
\cite{gametheorywebb} contagion processes.

\subsubsection{Non-Markovian epidemics on networks}
\label{sec:non-mark-react}
The modeling framework presented in the previous sections is mostly
based on the Poisson approximation \cite{tijms2003first} for both the
transmission and recovery processes. The Poisson approximation assumes
that the probabilities per unit time of transmitting the disease through
a given edge, or recovering for a given infected node, are constant, and
equal to $\beta$ and $\mu$, respectively. Equivalently, the total time
$\tau_i$ that a given node $i$ remains infected is a random variable
with an exponential distribution, $P_i(\tau_i) = \mu e^{-\tau_i \mu}$,
and that the time $\tau_a$ for an infection to propagate from an
infected to a susceptible node along a given edge (the inter-event time)
is also exponentially distributed,
$P_a(\tau_a) = \beta e^{-\tau_a \beta}$.  A notable variation assumes
that all infected nodes remain infective for a fixed time $\tau$. The
SIR model can be analyzed exactly in this setting by means of the
generating function approach (see
Sec.~\ref{sec:mapping-percolation-SIR}).

From a practical point of view, the Poisson assumption leads to an
increased mathematical tractability. Indeed, since the rates of
transmission and recovery are constant, they do not depend on the
previous history of the individual, and thus lead to memoryless,
Markovian processes
\cite{Ross1996,tijms2003first,vankampen,PVM_PAComplexNetsCUP}.
While the Poisson approximation may be justified when only the average
rates are known \cite{burstylambiotte2013}, it is at odds with empirical
evidence for the time duration of the infective period in most diseases
\cite{BLYTHE01011988}, whose distribution usually features a peak
centered on the average value but exhibits strongly non-exponential
tails. Furthermore, the interest in non exponential transmission
processes has been also fueled by the recent evidence on the patterns of
social and communication contacts between individuals, which have been
observed to be ruled by heavy-tailed distributions of inter-event times
(see Sec.~\ref{sec:epid-proc-temporal-nets}).

The framework of non-Poissonian infection and recovery
processes can be set up as follows, for either the SIS and SIR models
\cite{boguna_simulating_2013}: Infected individuals remain infective
for a period of time $\tau_i$, after which they recover, that follows
the (non exponential) $P_i(\tau_i)$ distribution. For the sake of
simplicity, it is assumed that this distribution is the same for all
nodes. Infection events take place along \textit{active} links,
connecting an infected to a susceptible node. Active links transmit
the disease at times following the inter-event distribution
$P_a(\tau_a)$, i.e. a susceptible individual connected to an infected
node becomes infected at a time $\tau_a$, measured from the instant
the link became active. If a susceptible node is connected to more
than one infected node, it becomes infected at the time of the first
active link transmitting the disease. The complexity of this
non-Markovian process is now evident: the infection of a node does not
only depend on the number of neighbors, but also on the time at which
each connection became active.

Numerical results on non-Poissonian epidemics in networks are relatively scarce. Simple event-driven
approaches rely on a time ordered sequence of events (tickets), that
represent actions to be taken (recovery or infection) at given fixed
times, which are computed from the inter-event distributions
$P_i(\tau_i)$ and $P_a(\tau_a)$. These approaches are quite demanding,
so only small system sizes can be considered. For example,
\citet{van_mieghem_non-markovian_2013} report results for the SIS
model with Poissonian recovery, with rate $\mu$, while infection
happens with a non-exponential distribution following the
Weibull form, $P(\tau_a) \sim (x/b)^{\alpha-1}
e^{-(x/b)^{\alpha}}$. In this case, strong variations in the value of the prevalence
and of the epidemic threshold are found when varying the parameter
$\alpha$. A promising approach is provided by the general
simulation framework proposed by \citet{boguna_simulating_2013}, based
on the extension of the the Gillespie algorithm for Poissonian
processes \cite{gillespie_exact_1977}. This algorithm allows
the simulation of much larger network sizes.

The consideration of non-Poissonian infection or recovery processes
does not lend itself easily to analytical
approaches \cite{burstylambiotte2013}.
Some simple forms for the distribution of
infectious periods, such as the Erlang distribution, which can be
described as the convolution of identical Poisson processes
\cite{renewal}, can be tackled analytically by postulating an extended
epidemic model with different infective phases and Poissonian
transitions among them \cite{Lloyd07052001,Lloyd200159}. However,
general non-Poissonian forms lead to convoluted sets of
integro-differential equations \cite{Keeling03011997}.
As a consequence there are not many analytical results for
non-Poissonian transitions in complex networks.
We can mention the
results of \citet{min_suppression_2013} which consider the SIR process
on a network in which infection events follow an inter-event
distribution $P_a(\tau_a)$. Assuming that infected nodes remain in that
state for a fixed amount of time $\tau_i$, it is possible to compute
\cite{min_suppression_2013} the disease transmissibility as
\begin{equation}
  T(\tau_i) = 1 - \int_{\tau_i}^\infty \psi(\Delta) d \Delta,
  \label{transmissibility_T_tau_i}
\end{equation}
where $\psi(\Delta) = \int_{\Delta}^\infty P_a(\tau_a) d\tau_a /
\int_{0}^\infty P_a(\tau_a) d\tau_a$ is the probability distribution of
the time between infection (assumed uniform) and the next activation
event.  Eq. (\ref{transmissibility_T_tau_i}) assumes that the dynamics of
infections follows a stationary renewal process \cite{renewal,PVM_PAComplexNetsCUP}. Applying
the generating function approach (see Sec.~\ref{sec:4.B}), the epidemic
threshold is obtained, as a function of $\tau_i$, from the implicit
equation
\begin{equation}
  T({\tau_i}_c) = \frac{\av{k}}{\av{k^2} - \av{k}}.
\end{equation}
For a power-law distribution $P_a(\tau_a) \sim \tau_a^{-\alpha}$,
 it is found that ${\tau_i}_c$ diverges as
$\alpha \to 2$, implying that only diseases without recovery are able
to spread through the network \cite{min_suppression_2013}.
An important step forward in the treatment of generic nonexponentially
distributed recovery and transmission times in the SIR model is
the application of a message-passing method,
as reported by \citet{Karrer2010}. This approach leads to an exact
description in terms of integro-differential equations for trees and
locally tree-like networks, and to exact bounds for non-tree-like
networks, in good agrement with simulations.

Finally, \citet{PVM_nonMarkovianSIS_NIMFA_2013} propose an extension of
the SIS IBMF theory for non-exponential distributions of infection or
healing times. Using renewal theory, their main result is the 
observation that the functional
form of the prevalence in the metastable state is the same as in the
Poissonian SIS model, when the spreading rate $\lambda = \beta/\mu$ is
replaced by the average number of infection attempts during a recovery
time. The theory by \citet{PVM_nonMarkovianSIS_NIMFA_2013} also allows
to estimate the epidemic threshold in non-Markovian SIS
epidemics.

\subsubsection{The SIRS model}
\label{sec:sirs-model}
The behavior of the SIRS model on complex networks
has been analytically considered by
\citet{ForestFireSatorras09} at the DBMF level. Within this approximation,
the steady-state solution of the SIRS model can be exactly mapped to that
of the SIS model, via the identification of the densities of infected
individuals
\begin{equation}
  \rho_\mathrm{SIRS}(\eta, \lambda) = \frac{\eta}{\eta+1}
  \rho_\mathrm{SIS}(\lambda),
\end{equation}
where $\eta$ is the immunity decay rate.
Therefore, within DBMF, all the critical properties of the SIRS model are the same
as the SIS model, the only effect of $\eta$ being
a rescaling of the density of infected individuals.

Numerically, the SIRS model was studied by \citet{abramson01} on
small-world Watts-Strogatz networks (see
Sec.~\ref{sec:basic-network-models}) within a discrete time
deterministic framework, in which infected individuals remain infective
for a fixed time $\tau_I$, after which they recover, while recovered
individuals remain in this state for a fixed time $\tau_R$.  For large
values of the Watts-Strogatz model rewiring probability $p$, a periodic
steady state is observed, in which the state of all nodes stays
synchronized~\cite{abramson01}.  The level of synchronization increases
with the average degree and also with $p$, after a threshold $p_c$
depending on $\avk$ for fixed network size.

The SIRS model can be also interpreted in terms of a disease that
causes death ($I \to R$), leading to an empty node that can be later
occupied by the birth of a new, susceptible individual ($R \to S$).
Within this interpretation, \citet{liu04:_spread} consider a generalized
SIRS model, allowing additionally for simple recovery ($I \to S$ with
rate $\gamma$) and
death of susceptible individuals due to other causes ($S\to R$ with
rate $\alpha$). Applying a DBMF formalism, they recover again a
threshold inversely proportional to the second moment of the degree
distribution, modulated by the diverse parameters in the model, in
agreement with the SIS result.

\subsubsection{The SEIR model}
\label{sec:seir-model}
The SEIR model is generally used to model influenza-like-illness and
other respiratory infections.  In the context of networks, this model
has been used by \citet{doi:10.1142/S0218127405012776} to study
numerically the evolution of the Severe Acute Respiratory Syndrome
(SARS) in different social settings, using both deterministic and
stochastic versions of the model, in which different reaction rates
were adjusted using empirical spreading data of the disease.
The edge-based compartmental modelling approach can be adapted
to deal with multiple infectious stages, including SEIR as a
particular case~\cite{10.1371/journal.pone.0069162}.

Exposed individuals can also play a role in more complex epidemiological
models. Thus, for example, the SEIRS model can be used to mimic the
eventual waning of the immunization of recovered individuals, which
implies one additional transition rule, Eq.~(\ref{eq:waning}).
The properties of the SEIRS model in Watts-Strogatz small-worlds
networks (see Sec.~\ref{sec:basic-network-models}) have been described
by \citet{5365379}. A variation of the SEIRS model without the
recovered compartment, or in other words, in the limit of the reaction rate
$\eta \to \infty$ (SEIS),  which coincides with a two-stage variation of the
classical contact process \cite{1999} has been analyzed in
heterogeneous networks by \citet{Masuda200664}. Application of DBMF
theory recovers the mapping to the simple SIS model obtained in the
case of the SIRS epidemics.

\subsection{Realistic static networks}
The analytical and numerical results presented so far for the
paradigmatic SIS and SIR models have focused mainly on random
undirected uncorrelated networks, which are only
characterized by their degree distribution, assuming that the rest of
the properties are essentially random. However, real networks are far
from being completely random. Beyond the degree distribution, a
multitude of other topological properties, such as clustering, degree
correlations, weight structure, etc. (see
Sec.~\ref{sec:general-definitions}), are needed to
characterize them.

\subsubsection{Degree correlations}
\label{sec:effects-degr-corr}

Most theoretical results on epidemic spreading in networks, especially
at the DBMF level, are obtained imposing a lack of correlations at the
degree level, that is, assuming that the probability that a vertex of
degree $k$ is connected to a vertex of degree $k'$ is given by
$P(k'|k) = k' P(k') / \av{k}$ \cite{Dorogovtsev:2002}. However, most
natural networks show different levels of correlations, which can have
an impact on dynamical processes running on top of them.

From a theoretical point of view, the specific effect of degree
correlations, as measured by the different observables detailed in
Sec.~\ref{sec:degree-correlations}, is difficult to assess. However,
some specific results are available. At the level of DBMF theory (see
Sec.~\ref{sec:heter-mean-field-2}) it has been shown that for
scale-free networks with $\gamma<3$, no sort of degree correlations
is able to alter the vanishing of the epidemic threshold in the
thermodynamic limit \cite{marian3,marianproc}. From a numerical point
of view, however, the precise determination of the effects of degree
correlations on the position of the epidemic threshold and the shape
of the prevalence function is problematic.
Indeed, it is generally not possible to
ascertain if the changes in the epidemic process are due to the
presence of correlations or other
topological properties generally related to correlations, such as
local clustering. Initial simulations on network models
\cite{structured,sander} claimed that disassortative degree
correlations could induce a finite threshold in the SIS model in
scale-free networks. However, those claims were based on networks with
an underlying finite-dimensional structure \cite{structurednets}, and
most probably the finite threshold observed was due to this
effect.

For the SIS model, the main IBMF result,
Eq.~(\ref{eq:IBMFthreshold}), stating that the epidemic threshold is the
inverse of the largest eigenvalue of the adjacency matrix $\Lambda_1$,
remains unaltered. The presence of correlations has only the
effect of changing the largest eigenvalue. In this respect,
\citet{PVM_assortativity_EJB2010} showed that increasing the degree
assortativity, by means of an appropriately defined degree preserving
rewiring scheme, increases the largest eigenvalue of the adjacency
matrix, thus reducing the effective IBMF epidemic threshold, in a
network of fixed size $N$. On the other hand, the induction of degree
disassortativity reduces the largest eigenvalue, with a corresponding
increase of the effective IBMF threshold. This observation is
confirmed by \citet{Goltsev12} who estimate, by means
of the power iteration method, the
largest eigenvalue of the adjacency matrix as
\begin{equation}
  \Lambda_1 \simeq \frac{\av{k^2}}{\av{k}} + \frac{\av{k} \sigma^2
    r}{\av{k^2}},
\end{equation}
where $\sigma$ is a positive function of the moments of the degree
distribution and $r$ is the Pearson correlation coefficient (see
Sec.~\ref{sec:degr-degr-distr}). Thus assortativity with $r>0$
(resp. disassortativity with $r<0$) is associated with an increase
(resp. decrease) of the largest eigenvalue. Other properties of the
largest eigenvalue in general networks with any kind of correlations,
such as the bound
 \[
\max\left(\sqrt{\av{k^2}},\sqrt{k_{max}}\right)\leq\Lambda_1\leq k_{max},
 \]
 are derived in \citet{PVM_graphspectra}.

Regarding the SIR model, the mapping to percolation (see
Sec.~\ref{sec:4.B}) allows to obtain more precise information. Assortative
correlations can induce a vanishing threshold in networks with
finite second moment of the degree distribution~\cite{morenopercolation}.
The more general treatment by \citet{PhysRevE.78.051105}, considering the
\textit{branching matrix} $B_{k, k'} = (k'-1)P(k'|k)$
\cite{marianproc}, allows to explicitly check the effects of degree
correlations on the epidemic threshold. Indeed disassortative
correlations increase the threshold from its uncorrelated value, while
assortative correlations decrease
it \cite{PhysRevE.78.051105,Miller2009}. These results, valid for the
SIR model, can also be extended to the SEIR model~\cite{Kenah2011}.
While no explicit
expression for the threshold can be obtained, it is possible to work
out upper and lower bounds, in terms of the transmissibility $T$, that
read as
\begin{equation}
  \frac{1}{\max_k B(k)} \leq T_c \leq \frac{\av{k(k-1)}}{\sum_k
      k(k-1)B(k)P(k)},
\end{equation}
where $B(k) = \sum_{k'} B_{k, k'}$~\cite{PhysRevE.78.051105}.  With
respect to the behavior of the outbreak size close to the epidemic
threshold, degree correlations are irrelevant, in the sense that the
critical exponents are not changed, when the following conditions are
fulfilled \cite{PhysRevE.78.051105}: (i) The largest eigenvalue of the
branching matrix is finite if $\av{k^2}$ is finite, and infinite if
$\av{k^2} \to \infty$; (ii) the second largest eigenvalue of $B_{k, k'}$
is finite; (iii) the eigenvector associated to the largest eigenvalue
has nonzero components in the limit $k \to \infty$. On the other hand,
if any one of these conditions is not fulfilled (large assortativity
leads to the failure of condition (ii), while strong disassortativity
affects condition (iii)), degree correlations become relevant and they
lead to new critical exponents.  At the DBMF level the results of
\citet{marian3} for the SIS model extend to the SIR case, implying again
the inability of degree correlations to alter the vanishing of the
epidemic threshold in the thermodynamic limit for $\gamma<3$. This
result has been confirmed numerically by means of the direct numerical
solution of the DBMF equations of the SIR model on scale-free networks
with weak assortative correlations \cite{PhysRevE.68.035103}.  The main
effect of these correlations is to induce a smaller overall prevalence
and a larger average lifetime of epidemic outbreaks.

\subsubsection{Effects of clustering}
\label{sec:effects-clust-coeff}

While a priori entangled with degree correlations and other
topological observables, the effect of clustering on epidemic
spreading has been the subject of a large interest, due to the fact
that social networks, the basic substrate for human epidemic
spreading, are generally highly clustered. Initial work in this area
\cite{keeling99}, based on a simple mean-field approximation (and thus
valid in principle for homogeneous networks) already pointed out the
effects of clustering (measured as the clustering coefficient $C$,
see Sec.~\ref{sec:clust-coeff-clust}) on the SIR dynamics.
A noticeable departure from the standard mean-field
results in the absence of clustering is observed,
and in particular a decrease of the outbreak size when increasing $C$.
In the case of the Watts-Strogatz model (see
Sec.~\ref{sec:basic-network-models}), the paradigm of a network with large
clustering, exact analytical results, confirmed by numerical
simulations, were obtained by \citet{moore00} for any value of the
rewiring probability $p$. 
Another analytical approach was proposed by
\citet{PhysRevE.68.026121}, who considered a network model based on a
one-mode projection of a bipartite network (see
Sec.~\ref{sec:gener-simple-graphs}) and applied the usual mapping to
percolation.  Apart from confirming the observation by
\citet{keeling99} that epidemic outbreaks are a decreasing function of
$C$, it was interestingly observed that, at odds with the behavior of
networks with no clustering, for large $C$ the outbreak size saturates
to a constant value when increasing the transmissibility even for
moderate values of $T$, suggesting that ``in clustered networks
epidemics will reach most of the people who are reachable even for
transmissibilities that are only slightly above the epidemic
threshold'' \cite{PhysRevE.68.026121}.  Along the same line,
\citet{Miller2009}, considering a model of random networks with
assortative correlations and tunable clustering, was able to show
that, for a SIR dynamics with uniform transmissibility $T$, clustering
hinders epidemic spreading by increasing the threshold and reducing
prevalence of epidemic outbreaks.

A more general approach, valid for any network,
confirms the previous observations~\cite{PhysRevLett.97.088701}. In
this approach, the generating function calculation scheme includes the
concept of edge multiplicity $m_{ij}$, defined as the number of
triangles in which the edge connecting nodes $i$ and $j$ participate.
In the limit of weak clustering, corresponding to constant $m_{ij}= m_0$,
the clustering spectrum (see
Sec.~\ref{sec:clust-coeff-clust}) follows the scaling
$\bar{c}(k) \sim k^{-1}$, which is
essentially decoupled from two-vertex degree correlations. The epidemic threshold depends on $m_0$ and is shifted with
respect to the unclustered result; however, for
scale-free networks, this shift is not able to restore a finite
threshold in the thermodynamic limit. For strong clustering, with a
clustering spectrum decaying more slowly than $k^{-1}$, numerical
simulations in a model with tunable clustering coefficient
\cite{PhysRevE.72.036133} confirm the inability of clustering to
restore a finite threshold in scale-free networks.  Other numerical and
anaytical works \cite{Miller2009,Miller06122009} have confirmed these
results in different clustered network models.

Within the context of IBMF theory for the SIS model, it is possible to
find bounds for the largest eigenvalue of the adjacency matrix as a
function of the clustering (measured by the number of triangles in the
network), indicating that SIS epidemic threshold decreases
with increasing clustering coefficient  \cite{PVM_graphspectra}.

\subsubsection{Weighted networks}
\label{sec:weighted-networks}

If we want to take into account that not all contacts in a
social network are equally facilitating contagion (e.g. due to
the different relative frequency of physical contacts associated to
different edges), we must consider weighted networks, where
a weight $\omega_{ij} \ge 0$ is assigned to the edge between connected
nodes $i$ and $j$ (see Sec.~\ref{sec:gener-simple-graphs}).
The models for epidemic spreading are generalized assuming the rate of
disease transmission between two vertices equal to some function
of the weight of the link joining them.
The simplest possibility occurs when the probability of infection
transmission along an edge is directly proportional to the edge weight.

The IBMF theory for the SIS model is readily applied,
just replacing in Eq.~(\ref{eq:IBMFSISequations}) the adjacency matrix
$a_{ij}$ by the matrix $\Omega_{ij} = \omega_{ij} a_{ij}$. The IBMF
threshold is just the inverse of the largest eigenvalue of $\Omega$
\cite{4610111}.  \citet{Peng2010549} consider a generalized SIS model
defined by the matrix $\beta_{ij}$, whose terms are the probabilities
that node $i$ is infected by node $j$ through an edge joining them.
Defining the \textit{parametrized adjacency matrix}
$M_{ij} = \beta_{ij} + (1-\mu_i) \delta_{ij}$, where $\mu_i$ is the
recovery probability of node $i$, \citet{Peng2010549} (see also
\citet{PVM_heterogeneous_virusspread}) show that endemic
states occur when the largest eigenvalue (in absolute value) of the
parametrized adjacency matrix is larger than one.

The DBMF approach to the SIS process on weighted networks is simplified
by the introduction of additional assumptions, such as a functional
dependence of the weights of edges on the degree of the nodes at their
endpoints \cite{dynam_in_weigh_networ}.  \citet{karsai:036116}
consider the SIS process in a network with local spreading rate, at
the DBMF level, $\lambda_{k k'} \sim (k k')^{-\sigma}$, with $\sigma$
in the range $[0,1]$.
The resulting equations are found to
depend on the effective degree exponent $\gamma' =
(\gamma-\sigma)/(1-\sigma)$. For $\gamma'<3$, a null threshold in the
thermodynamic limit is obtained, while for $\gamma'>3$, the threshold
is finite. \citet{karsai:036116} discuss additionally a finite-size
scaling theory, relating the average prevalence with the network size,
which is checked against numerical simulations. The strict correlation
between weights and degrees is relaxed in other works, such as
\citet{PhysRevE.85.056106}, where a purely edge-based mean-field
approach for weighted homogeneous networks for the SIS model is
proposed. By means of this approach, and  focusing on bounded and
power-law weight distributions, \citet{PhysRevE.85.056106} show that the more homogeneous
the weight distribution, the higher is the epidemic prevalence.

Other approaches to the SIS model include a pair-based mean-field
approach~\cite{2012arXiv1208.6036R} for networks with
random and fixed deterministic weight distributions. The main result
is the observation that a weight distribution leads to the
concentration of infectiousness on fewer target links (or individuals)
which causes an increase in the epidemic threshold in both kinds of
networks considered.

\citet{0256-307X-22-2-068} report numerical results for the behavior
of the SI model on the growing weighted network model proposed by
\citet{barrat04:_weigh}
with a local spreading rate of the form
$\lambda_{ij} \sim (\omega_{ij})^\alpha$. The main
results obtained concern the slowing down of the disease spread in
weighted networks with respect to their unweighted counterparts, which
is stronger for larger weight dispersion. Interesting, they also
report a decay in the velocity of spread, after a sharp peak, taking a
slow power law form, at odds with the exponential form obtained in
nonweighted networks \cite{Barthelemy2005275}.

In the case of the SIR model \citet{Chu2011471} present a DBMF analysis
in the case of weights correlated with the degree. The analysis is
based on a trasmission rate $\lambda_{k'k}$ from vertices of degree
$k'$ to vertices of degree $k$, taking the form $\lambda_{k k'} =
\lambda k \omega_{k k'}/s_k$ (where $s_k$ is the strength of a $k$
node) and on an infectivity of nodes $\phi(k)$, denoting the rate at
which a node of degree $k$ transmits the disease. Writing down rate
equations for the usual relevant DBMF quantities for the SIR model, and
assuming $ \omega_{k k'} \sim (k k')^\sigma$ and $\phi(k) \sim
k^\alpha$, \citet{Chu2011471} find the threshold
\begin{equation}
  \lambda_c = \frac{\av{k^{\sigma+1}}}{\av{k^{\alpha+\sigma+1}}}.
\end{equation}
By means of numerical simulations, \citet{Chu2011471} report additionally
that the size of epidemic outbreaks increases with the exponent $\alpha$,
while it decreases with increasing $\sigma$.
An analysis of the SIR model in terms of pair approximations for IBMF
theory is presented by \citet{2012arXiv1208.6036R}, reaching analogous
results as those obtained for the SIS model within the same formalism.

It is also noteworthy the numerical work of \citet{Eames200970} on the
SIR model in a realistic social network constructed from actual survey
data on social encounters recorded from a peer-group sample of 49
people. 
The results of \citet{Eames200970} highlight the strong correlations
between infection risk and node degree and weight, in correspondence
with the observations at the DBMF level. Additional simulations
considering different immunization strategies (see
Sec.~\ref{sec:effic-immun-prot}) indicate that, for this particular
realistic network, targeting for total degree or total weight provides
approximately the same efficiency levels.

Concerning other models,
\citet{brittonweight2011} have discussed an epidemic model in
a weighted network in which the weights attached to nodes of degree
$k$ are random variables with probability
distributions $q(\omega|k)$, in a construction akin to a weighted
configuration model (see Sec.~\ref{sec:basic-network-models}). In this
kind of network, \citet{brittonweight2011} observe, by means of an
analysis based on branching theory, that both the epidemic threshold
and the outbreak probability are affected by the correlations between
the degree of a node and the weights attached to it. This observation
is confirmed by numerical simulations of their weighted
network model fitted to empirical data from different network
examples, showing that the epidemic threshold is different in the
original network with respect to a network with reshuffled weights.
On the other hand, \citet{Deijfen201157} analyzes immunization of
weighted networks with random and degree dependent weights, observing,
in agreement with~\citet{Eames200970}, that targeting the largest weights
outperforms other immunization strategies.

In the framework of epidemic models on weighted networks it is possible
to include also the contact process (CP) on networks. In this model
each infected node may transmit the disease to at most one neighbor
for each time step.
This can be intepreted in continuum time as a SIS-like model with a
spreading rate $\lambda_{k k'} = 1/k$ for any edge departing from a node
of degree $k$. This modification has the effect of reducing the
importance of degree fluctuations in the spreading dynamics: the
threshold is finite for any value of the exponent
$\gamma$~\cite{Castellano2006,Olinky2004}. The same conclusion can
be drawn also for a model where multiple neighbors can be infected
simultaneously, but up to a fixed maximum value of neighbors
(and not for any $k$ as in SIS)~\cite{Joo2004}.

\subsubsection{Directed networks}
\label{sec:directed-networks}

Directed networks are useful to represent specific types of epidemic
transmission in which there is an intrinsic directionality in the
propagation. An example is given by diseases communicated by means of
blood transfusions or needle sharing. The study of epidemic processes in
directed networks is difficult due to the component structure of this
kind of networks (see Sec.~\ref{sec:general-definitions}). Indeed, the
position of a node in a specific network component can restrict or
enhance its spreading capabilities with respect to other
positions. Thus, in order to be able to generate a macroscopic outbreak,
a seed of infection should be located on the GIN or GSCC components;
seeds on the GOUT or the tendrils will in general produce small
outbreaks, irrespective of the spreading rate. In this sense, the
distribution of outbreak sizes starting from a randomly chosen vertex is
proportional to the distribution of outcomponents.

In the case of the SIR model, the mapping to percolation allows to apply
the generating function formalism developed for percolation in random
directed networks
\cite{Newman2001,PhysRevE.66.015104}.  For purely directed networks
(i.e. in which all edges have assigned a directionality), computations
depend on the joint probability $P(\ki, \ko)$, see
Section~\ref{sec:degr-degr-distr}, that a randomly chosen node has
in-degree $\ki$ and out-degree $\ko$, which in general exhibits
correlations between the two values.
In the absence of correlations among the degrees of
neighbors~\footnote{Notice that these are correlations among two
connected vertices, while correlations between $\ki$ and $\ko$ are
for the {\em same } node.}, under the tree-like assumption, the
critical transmissibility is
\begin{equation}
\label{eq:SIRdirectthres}
  T_c = \frac{\av{\ki}}{\av{ \ki \ko}},
\end{equation}
where averages are taken over the distribution $P(\ki, \ko)$
\cite{Newman2001}.
The same result can be obtained by means of more intuitive arguments
\cite{PhysRevE.66.015104}. Eq.~(\ref{eq:SIRdirectthres}) highlights
the important role of correlations between the in-degree and
out-degree in directed networks. Its full discussion is, however, not
easy, since one cannot impose arbitrary forms to $P(\ki,
\ko)$ given the explicit constraint $\av{\ki} =
\av{\ko}$. \citet{PhysRevE.66.015104} discuss the effects of
scale-free degree distributions with exponents $\gamma_\mathrm{in}$
and $\gamma_\mathrm{out}$ for in-degree and out-degree, respectively,
and given correlations $P(\ki, \ko)$
With this distribution, epidemics in the GWCC behave as in an undirected network
with effective degree distribution $P(k) = \sum_{\ki=0}^kP(\ki,
k-\ki)$, while the $\beta_{SIR}$ exponent characterizing the size of
supercritical outbreaks takes the form of
Eq.~(\ref{eq:SIRbetaexponent}), with an effective $\gamma^* =
\gamma_\mathrm{out} + (\gamma_\mathrm{in} -
\gamma_\mathrm{out})/(\gamma_\mathrm{in}-1)$~\cite{PhysRevE.66.015104}.

More generally, it is possible to
consider semi-directed networks, in which edges maybe directed or
undirected \cite{Meyers2006}.
The network specification is then given in terms of the probability $P(\ki, \ko,
k)$ that a vertex has $\ki$ incoming edges, $\ko$ outgoing edges and
$k$ bidirectional edges.
The presence of undirected links implies the existence of short loops
of length $2$, and thus the violation of the tree-like assumption.
Considering the possibility of different
transmissibilities $T_u$ and $T_d$ for undirected and directed edges,
respectively, \citet{Meyers2006} find expressions for the
critical values of one of them, keeping the other fixed. The rather
involved expressions simplify when imposing that the in-degree, out-degree and
undirected degree of each vertex are uncorrelated.
In particular, when these quantities obey Poisson distributions,
the epidemic threshold is given by
\cite{Meyers2006}
\begin{equation}
  T_{uc} \av{k}_u + T_{dc} \av{k}_d =1,
\end{equation}
where $\av{k}_u$ and $\av{k}_d$ are the undirected and directed average
degrees, respectively.  The analysis of these results allows the
identification of the key epidemiological difference between directed
and undirected networks: while in undirected networks the probability of
an outbreak and the expected fraction of the population affected (if
there is one) are equal, they differ in directed networks: depending on
the topology any of the two can be larger~\cite{Meyers2006}.

The generic case of semi-directed networks with arbitrary one-point
and two-point correlations is treated in~\citet{Boguna2005}.
The temporal evolution of epidemic outbreaks is considered
using the edge-based compartmental modelling
in~\citet{10.1371/journal.pone.0069162}.

Epidemic processes on purely directed networks can be tackled by an
extension of the standard DBMF. The key point is the consideration of
new degree classes which are defined in terms of the pair of in-degree
and out-degree values, $(\ki, \ko)$.  This implies that the dynamical
quantities characterizing the processes also depend on these two
values, $\rho^{\alpha}_{\ki, \ko}$, see
Sec.~\ref{sec:heter-mean-field-1}
and~\ref{sec:heter-mean-field}. Equations for the SIS and SIR models
(Eqs.~(\ref{eq:HMFSISequation}) and Eq.~(\ref{eq:SIR_HMF})) translate
directly with just one caveat: degree-degree two-vertex correlations (see
Sec.~\ref{sec:degree-correlations}) in purely directed networks
translate into the conditional probability $P^\mathrm{out}({\ki}',
{\ko}'|\ki, \ko)$ that an outgoing edge from a vertex $(\ki, \ko)$ is
connected to a vertex $({\ki}', {\ko}')$. Lack of two-point
degree-degree correlations implies
\begin{equation}
  P^\mathrm{out}({\ki}', {\ko}'|\ki, \ko) = \frac{{\ki}'
    P({\ki}',{\ko}')}{\av{\ko}}.
\end{equation}
\citet{Boguna2005} developed this DBMF formalism for the SIR model,
finding a threshold that, in the general case, is a function of the
largest eigenvalue of the extended connectivity matrix
${\ki}' P({\ki}', {\ko}'|\ki, \ko)$, and that, without degree-degree
correlations, reduces to Eq.~(\ref{eq:SIRdirectthres}).

In the case of the SIS model, the IBMF result is the same as in
undirected networks, since directionality (i.e. the asymmetry of the
adjacency matrix) does not explicitly enter into the theory. See also
the generalization of the IBMF theory presented by \citet{Peng2010549}
(Sec.~\ref{sec:weighted-networks}). The value of the largest eigenvalue
has been numerically studied in synthetic semi-directed networks with
directionality $\xi$, defined as the fraction of directed edges
\cite{PhysRevE.88.062802}. The main result obtained is the increase of
the epidemic threshold lower bound when increasing directionality $\xi$,
implying that directed networks hinder the propagation of epidemic
processes.  At the DBMF level, an extension analogous to the one
considered for the SIR model leads to a threshold with the same
functional form, Eq.~(\ref{eq:SIRdirectthres}), in degree-degree
uncorrelated networks~\cite{tanimoto_epidemic_2011}.

\subsubsection{Bipartite networks}

Bipartite networks (see Sec.~\ref{sec:gener-simple-graphs}) represent
the natural substrate to understand the spreading of sexually
transmitted diseases, in which two kinds of individuals (males and
females) are present and the disease can only be transmitted between
individuals of different kinds\footnote{We neglect here homosexual
  contacts.}. In other contexts, bipartite networks can be used to
represent vector-borne diseases, such as malaria, in which the
transmission can only take place between the vectors and the hosts
\cite{10.1371/journal.pone.0013796}, or the spreading of diseases in
hospitals, in which the different kinds of nodes account for (isolated)
patients and caregivers \cite{Ancel-Meyers:2003fe}.

Dealing with the SIR dynamics,
\citet{newman02} considers a variation of the mapping to percolation,
for a model on bipartite networks characterized by the partial degree
distributions $P_m(k)$ and $P_f(k)$, finding that the epidemic
threshold takes the form of a hyperbola in the space defined by the
male and female transmissibilities, $T_m$ and $T_f$,
\begin{equation}
  T_m T_f = \frac{\av{k}_m\av{k}_f}{\av{k(k-1)}_m \av{k(k-1)}_f},
\label{SIR_bipartite}
\end{equation}
where the moments $\av{k}_\alpha$ and $\av{k(k-1)}_\alpha$ are
computed for the degree distribution $P_\alpha(k)$.

In the case of the SIS model on bipartite networks,
\citet{Gomez-Gardenes05022008}
find analogous results at the DBMF level, with threshold
on the hyperbola defined by the male and female spreading rates,
$\lambda_m$ and $\lambda_f$, of the form
\begin{equation}
  \lambda_m \lambda_f = \frac{\av{k}_m\av{k}_f}{\av{k^2}_m
    \av{k^2}_f},
\label{SIS_bipartite}
\end{equation}
see \citet{Wen2012967} for further results with the DBMF
formalism. The general behavior of the SIS model on multipartite
networks, allowing for more than two different classes of nodes, is
discussed by \citet{2013arXiv1306.6812S}.

Expressing in Eq.~(\ref{SIR_bipartite})
the transmissibility in terms of the spreading rate,
$T_i = \lambda_i/(\lambda_i+1)$ (see Sec.~\ref{sec:4.B}) and comparing
with Eq.~(\ref{SIS_bipartite}), an interesting observation emerges
~\cite{Hernandez2013}. In the SIR case, when $\lambda_f$ diverges the
threshold value for $\lambda_m$ goes to a finite value. Hence the
possibility of an endemic outbreak is completely ruled out by reducing
the spreading rate of a {\em single} type of nodes. In the SIS case
instead, the asymptotic value is $\lambda_m=0$ and as a consequence reducing only
one spreading rate may not be sufficient to guarantee no endemic spreading.
This last conclusion, however, turns out to be an artifact of the DBMF approach
~\cite{Hernandez2013}: also for SIS dynamics a finite asymptotic threshold
is found in a theoretical approach based on a pair approximation, confirmed
by numerical simulations.
The previous conclusions hold when the topology-dependent factors appearing
on the right-hand-sides of Eqs.~(\ref{SIR_bipartite})
and~(\ref{SIS_bipartite}) are finite. However it is enough that
one of the restricted degree distributions has a diverging second
moment to have an epidemics spreading over the whole network, no
matter how small are the spreading rates $\lambda_i$.

\subsubsection{Effect of other topological features}

Many works have dealt with networks endowed with a modular (community)
structure, i.e., subdivided in groups with a relative high density of
connections within groups and a smaller density of inter-group links,
see Section~\ref{sec:centrality}.
SIS dynamics has been studied by~\citet{Liu2005} on a generalization of
the classical random graph model with probability $p$ ($q$) of
intra-(inter-) community links. The epidemic threshold is found to
decrease with $p/q$; this effect, however, cannot be attributed to the
community structure only, because of the concurrent change of the degree
distribution, which gets broader.  Other studies have decoupled the two
effects, by comparing spreading dynamics on modular networks and on
randomized networks with the same $P(k)$, obtained by suitable
reshuffling~\cite{maslov02}. They support instead the opposite view that
the community structure of a network tends to hinder epidemic spreading.
Using IBMF, \cite{PVM_SIS_communityNetworks2014} express the epidemic
threshold explicitly in terms of the sizes and spreading rates in the
clusters.

For the SI dynamics, the modular structure makes the
growth of the infection slower: prevalence at fixed time is reduced in
networks with community structure~\cite{Huang2007}.  The
interpretation is that the presence of communities tends to confine
the outbreak around the initial seed and hinders the transmission to
other communities. This effect is further enhanced in weighted social
networks~\cite{Onnela:2007} by the correlation between topology and
weights~\cite{Granovetter1973}: the ties bridging between strongly connected
communities are typically weak and this greatly delays the propagation among
different communities~\cite{Onnela:2007,PhysRevE.83.025102}.
Investigations on the SIRS model with fixed
infection and recovery times have focused on the oscillations of the
number of infected nodes in the stationary
state~\cite{Yan2007,Zhao2007}.  Both for topologies with scale-free
and non scale-free degree distributions it turns out that the modular
structure reduces the synchronization.
Also for SIR dynamics
modularity is found to make spreading more difficult: the final value
of $\rho^R$ is smaller for stronger community structure~\cite{Wu2008}.
More convincingly, ~\citet{Salathe2010}, show, both for empirical and
synthetic networks that community structure has a major hindering effect
on spreading: the final value of $\rho^R$ and the height of the peak of
$\rho^I$ decrease with the modularity.
Moreover, they show that in
networks with strong community structure targeting vaccination
interventions at individuals bridging communities is more effective
than simply targeting highly connected individuals.

It is also worth to mention the observation that SIS-like processes on
complex networks may give rise to the nontrivial scenario of Griffiths
phases~\cite{Vojta2006}, regions of the phase-space where the only
stationary state is the absorbing one, which is however reached via
anomalously long nonuniversal relaxation~\cite{Munoz2010}.  This
behavior arises because of rare-regions effects, which can be due either
to quenched local fluctuations in the spreading rates or to subtle
purely topological heterogeneities~\cite{Juhasz2012,PhysRevE.86.026117}.
Such rare-region effects have been discussed in the case of the SIS
model on loopless (tree) weighted networks
\cite{Buono13,PhysRevE.87.042132,PhysRevE.88.032109}, where they have
been related to the localization properties of the largest eigenvalue of
the adjacency matrix \cite{PhysRevE.88.032109}.

\subsubsection{Epidemics in adaptive networks}

\label{sec:6.C}
Previous sections have focused on the evolution of epidemics on static
networks or on annealed topologies where connections are rewired on a
time scale much smaller than the characteristic time scale of the
infection process.
For real human disease epidemics, however, the assumption
that the structure of contacts does not depend on the progression of
the contagion is often unrealistic: In the presence of infectious spreading,
human behavior tends to change spontaneously, influencing the
spreading process itself in a nontrivial feedback loop.
The modifications induced by this coupling may be distinguished
depending on several features~\cite{Funk2010}: the source of information
about the contagion, the type of information considered and the type
of behavioral change induced.
The source of information about the spreading process may be local
(individuals decide depending on the state of their direct contacts) or
global (info on the state of the whole system is publicly available).
Different types of information may influence the behavioural choice:
in prevalence-based models decisions are taken based on the observation
of the epidemic state of others; in belief-based models matters the awareness or the risk perception which may be (at least partially)
independent from the actual disease dynamics and often behaves in
turn as a spreading
process~\cite{Granell2013,Galvani2013,Perra2011,Salathe2008,Bagnoli2007,Funk2009}.
Finally, the behavioral change can be of different types: affecting
the state of the individual (for example via voluntary vaccination)
or the structure of contacts (eliminating existing connections or creating
new ones).
Many models incorporating these features have been investigated in mathematical epidemiology, generally assuming well-mixed
populations~\cite{Funk2010}.
Here we focus on epidemic spreading on adaptive
(or coevolving) contact networks, where the topology of the interaction
pattern changes in response to the contagion.
The coevolution between structure and dynamics is a common theme in many
contexts, from game theory to opinion dynamics~\cite{Gross2008,Nardini2008}.

The first investigation of an adaptive topology for SIS dynamics~\cite{Gross2006} includes
the possibility for individuals to protect themselves by
avoiding contacts with infected people. Infected individuals are allowed at each time step to infect each
of their susceptible contacts with probability $p$ or recover with
probability $r$ (usual SIS dynamics); in addition, susceptibles can
decide (with probability $w$) to sever a link with an infected and
reconnect to a randomly chosen susceptible.
The possibility of rewiring links drastically changes the
phase-diagram of the model.
\begin{figure}[t]
\includegraphics*[width=\columnwidth]{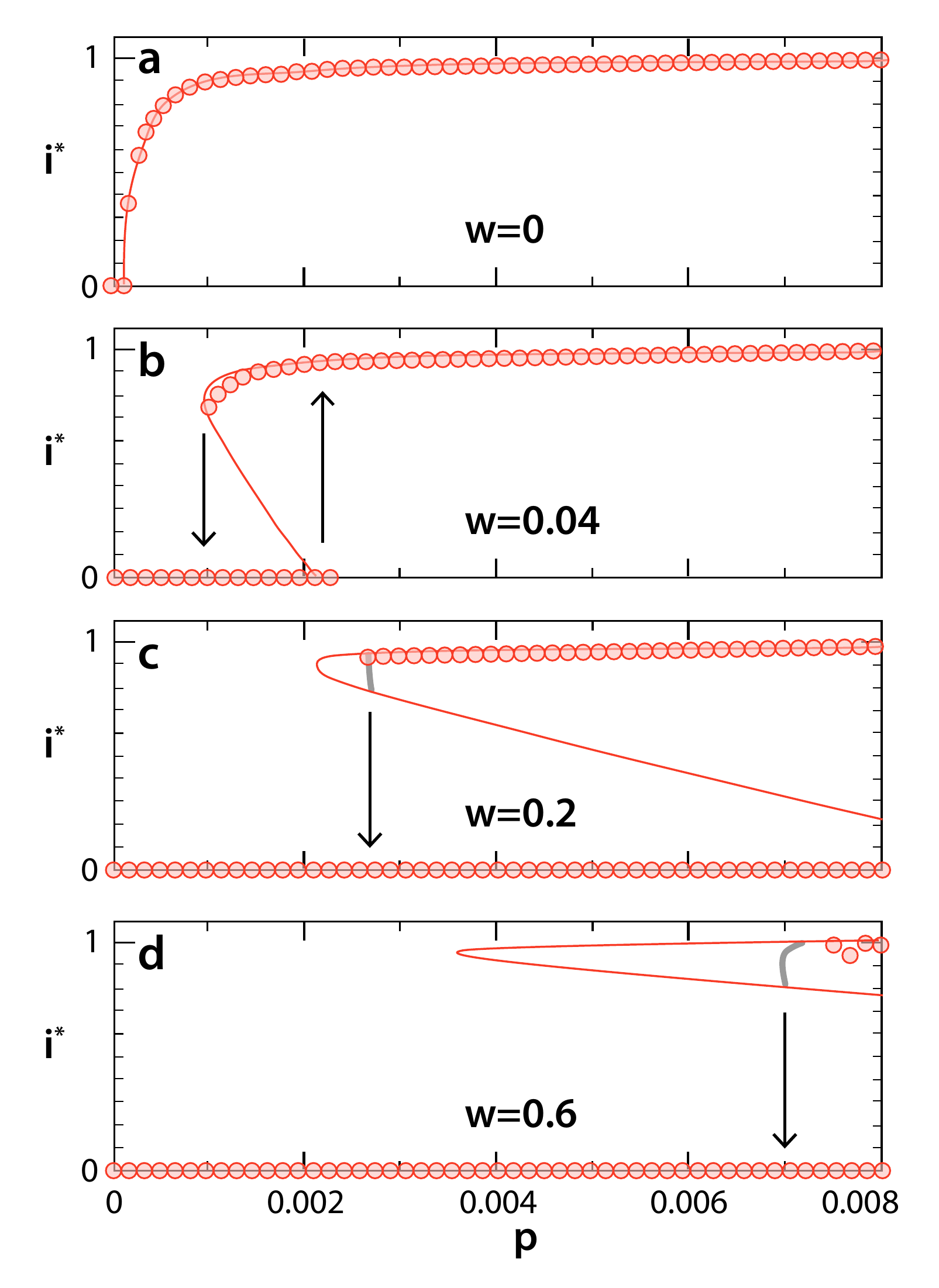}
\caption{Density of the infected nodes $i^*$ as a function of the
  infection probability $p$ for different values of the rewiring rate
  $w$. In each diagram thin lines are computed using a homogeneous
  mean-field approach while circles are the results of numerical
  simulations.  Without rewiring only a single continuous transition
  occurs for $p_c \approx 0.0001$ (a). By contrast, rewiring causes a
  number of discontinuous transitions, bistability, and hysteresis loops
  (indicated by arrows) in (b), (c), (d).  Figure adapted from
  ~\citet{Gross2006}.}
\label{fig:3Gross}
\end{figure}
The threshold $p_c$, below which the system always converges to the
absorbing healthy state, is much larger than in the case of no
coevolution ($w=0$): rewiring hinders the disease propagation. More
interestingly, above this threshold a bistability region appears (see
Fig.~\ref{fig:3Gross}) with associated discontinuous transitions and
hysteresis.  In this region both the healthy and the endemic state are
stable and the fate of the system depends on the initial condition.  If
$p$ is further increased above a second threshold, bistability ends and
the endemic state is the only attractor of the dynamics.  The
coevolution has also strong effects on the topology of the contact
network, leading to the formation of two loosely connected clusters of
infecteds and susceptibles, with a general broadening of the degree
distribution and buildup of assortative correlations.  The rich
phase-diagram is recovered by a simple homogeneous mean-field approach
which complements the equation for the prevalence with two additional
equations for the density of links of $I$-$I$ and $S$-$I$ type.  A
bifurcation analysis predicts also the existence of a very narrow region
with oscillatory dynamics.  A more detailed approach to the same
dynamics~\cite{Marceau2010} takes into account explicitly the degree of
nodes, writing equations for the evolution of the probabilities $S_{kl}$
($I_{kl}$) that nodes in state $S$ ($I$) have degree $k$ and $l$
infected neighbors. The numerical integration of the equations is in
excellent agreement with numerical simulations both with respect to the
transient evolution and to the stationary state. Different initial
topologies (degree-regular, Poisson, power-law distributed) with the
same average connectivity may lead to radically different stationary
states: either full widespread contagion or rapid disease extinction.

The qualitative picture emerging from the model of \citet{Gross2006}
is found also for the adaptive SIRS model~\cite{Shaw2008} and for the
SIS dynamics where a susceptible individual rewires to any randomly
chosen other vertex (not necessarily susceptible)~\cite{Zanette2008}.
The possibility that also infected individuals decide to rewire their
connections is discussed in~\citet{Risau-Gusman2009}.
In the SIS model, also the interplay of the adaptive
topology and vaccination has been investigated~\cite{Shaw2010}.
It turns out that the vaccination frequency needed to significantly lower
the disease prevalence is much smaller in adaptive networks than in
static ones.

The effect of the very same type of adaptive rewiring introduced for SIS
has been studied also for SIR dynamics~\cite{Lagorio2011}.
In this case the effects of the coevolution are less strong, as the
time needed to reach the stationary (absorbing) state is short
(logarithmic in the system size $N$) and the global topology is only
weakly perturbed in this short interval.
The phase-diagram remains qualitatively the same of the nonadaptive case
with a single epidemic
transition separating a healthy state from an endemic one.
The mapping to percolation (see Sec.~\ref{sec:4.B}) is useful also here.
Coevolution leads to an effective transmissibility $T$ which decreases
with the rewiring probability $w$. One can then identify a critical value
$w_c$ above which the adaptive behavior is sufficient to completely
suppress the epidemics.

The assumptions that disconnected links are immediately rewired and
that the target vertices of the reconnection step are randomly
selected in the whole network are highly implausible in real world
situations. Attempts to go beyond these limitations include the
consideration of different rates for breaking and establishing
links~\cite{VanSegbroek2010,Guo2013} and ``intermittent'' social
distancing strategies, such that a link is cut and recreated (between
the same vertices) after a fixed time interval~\cite{Valdez2012} or
with a certain rate after both endpoints have healed~\cite{Tunc2013}.
The latter strategies are intended to mimic what happens with friends
or working partners, with which connections are reestablished after
the disease.  The overarching structure of the network
remains static and there is no real coevolution (no new links are
formed). As a consequence the phase-diagram of epidemic models remains
the same found on static networks, with only an increase in the
epidemic threshold due to social distancing.

\subsection{Competing pathogens}

Another generalization of the basic modeling scheme considers the evolution
of multiple epidemic processes in competition in the same network,
a scenario with clear relevance for realistic situations.
The crucial concept here is cross-immunity, i.e. the possibility that
being infected by one pathogen confers partial or total immunity against
the others.

\citet{Newman2005} considers two SIR epidemic processes occuring
one after the other in the same static network, in conditions
of total cross-immunity: The second
pathogen can affect only survivors of the first, i.e. in the "residual"
network obtained once the nodes recovered when the first epidemics
ends are removed. The mapping of SIR static properties to bond percolation
allows to understand this case.
If the first pathogen is characterized by a transmissibility above
a certain value (coexistence threshold) the residual network has
no giant component and the second pathogen cannot spread globally,
even if it has a huge transmissibility.
Global spreading of both pathogens can occur only for values of the
transmissibility of the first infection in an interval between the epidemic
and the coexistence thresholds.
A generalization to the case of partial cross-immunity is discussed by \citet{Funk2010b}.
The case of competing SIR infections spreading concurrently is
investigated in~\citet{Newmancompeting2011}, again in the case of complete
cross-immunity: Infection by one pathogen confers immunity for both.
Nontrivial effects occur when both transmissibilities are above the
threshold for single spreading (otherwise one of the pathogens does not
spread globally and there is no real interference).
If one of the pathogens has a transmissibility significantly
larger than the other, it spreads fast and the second spreads afterwards
in the residual network, much as in the case of subsequent infections.
If the growth rates are very similar the final outcome shows strong dependence
on stochastic fluctuations in the early stages of growth, with
very strong finite size effects.
An alternative approach, based on the edge-based compartmental modelling
allows to investigate theoretically also the dynamics of two competing
infectious diseases~\cite{Miller13}.
\citet{Poletto2013} consider cross-immune pathogens in competition within a
metapopulation framework (see Sec.\ref{metapop}).
The dominance of the strains depends in this case also on the mobility
of hosts across different subpopulations.

Mutual cross-immunity for two competing SIS dynamics is considered
by~\citet{Trpevski2010} (see also~\citet{Ahn2006}), while the domination
time of two competing SIS viruses is analysed in
\cite{PVM_SIS_competing_virus_PRE2014}.
Depending on the network topology, for some values of the parameters it
is possible to find a steady state where the two processes coexist, each
having a finite prevalence.

Another nontrivial and relevant example of interacting epidemics is the case of
coinfection processes, where the opposite of cross-immunity holds: The
second pathogen can spread only to individuals that have been already
infected by the first.
~\citet{Newmaninteracting2013} report a first theoretical and numerical
investigation of this type of dynamics on complex networks.

\section{Epidemic processes in temporal networks}

\label{sec:epid-proc-temporal-nets}

\begin{figure}[t]
\includegraphics*[width=\columnwidth]{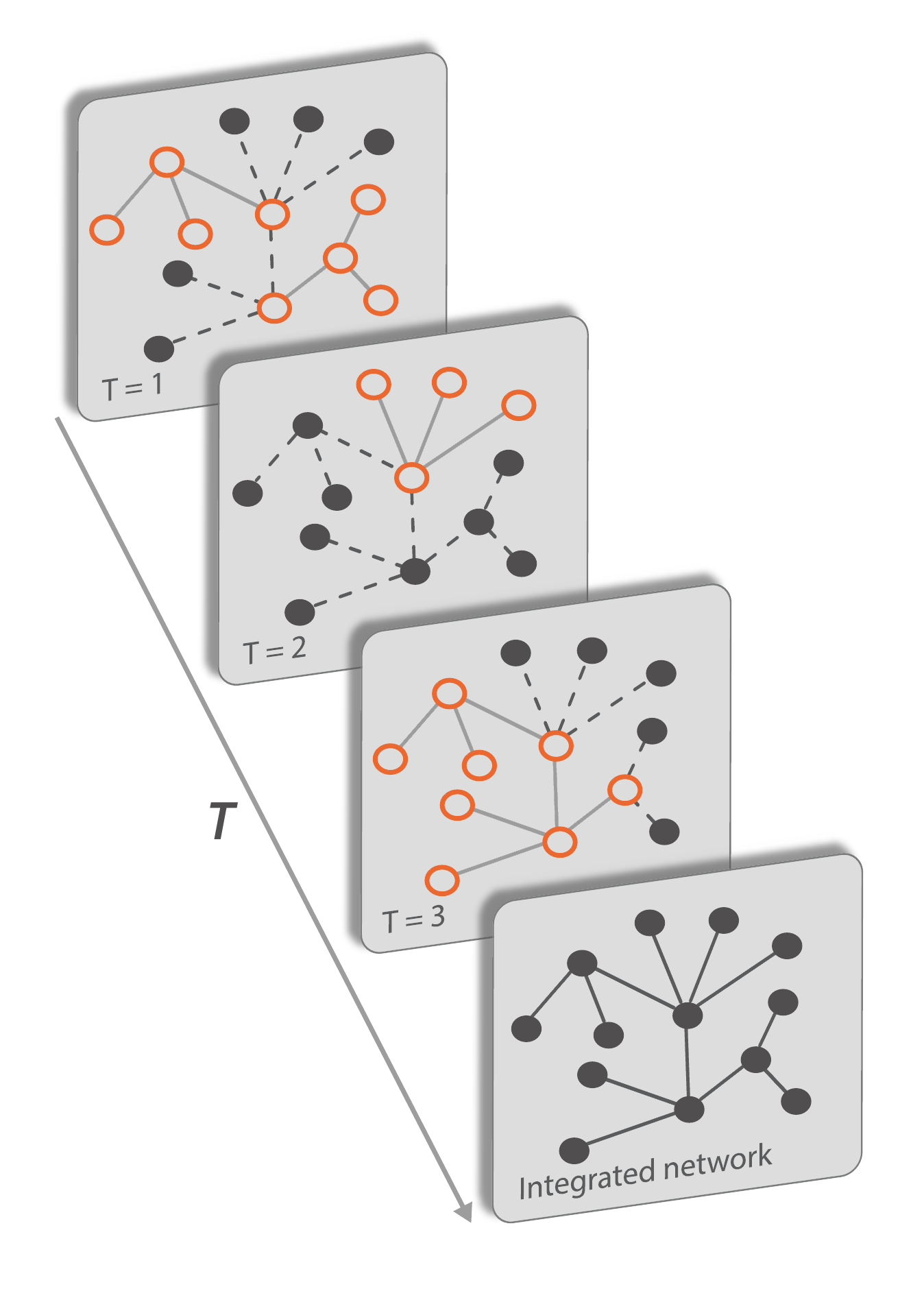}
\caption{A temporal (or time-varying) network can be represented as a
  set of nodes that, at every instant of time, are connected by a
  different set of edges. A integrated network over a time window $T$ is
  constructed by considering that nodes $i$ and $j$ are connected by an
  edge if they were ever connected at any time $t \leq T$.  Figure
  adapted from \citet{2012arXiv1203.5351P}}
\label{fig:temporal_net}
\end{figure}

The majority of the results presented so far considered the spreading of
epidemic process in the limit of extreme time scale separation between
the network and the contagion process dynamics (see however
Sec.~\ref{sec:6.C} for a discussion of adaptive networks, whose topology
changes in reaction to a disease).  In \textit{static} networks, the
epidemic spreads on a network that is virtually frozen on the time scale
of the contagion process. On the opposite limit, the DBMF theory
considers an effective mean-field network where nodes are effectively
rewired on a time-scale much faster than the contagion process.
However, in the case of many real-world networks those assumptions are
rather simplistic approximations of the real interplay between time
scales.  For instance, in social networks, no individual is in contact
with \textit{all} his/her friends \textit{simultaneously} all the
time. On the contrary, contacts are changing in time, often on a time
scale that is comparable with the one of the spreading process.  Real
contact networks are thus essentially dynamic, with connections
appearing, disappearing and being rewired with different characteristic
time scales, and are better represented in terms of a \textit{temporal}
or time-varying network \cite{Holme:2011fk,temporalnetworksbook}, see
Fig.~\ref{fig:temporal_net}.

Temporal networks are defined in terms of a \textit{contact sequence},
representing the set of edges present at a given time $t$.  By
aggregating the instantaneous contact sequence at all times $t<T$, a
static network projection can be constructed, see
Fig.~\ref{fig:temporal_net}.  In this aggregated network, the edge
between nodes $i$ and $j$ is present if it ever appeared at any time
$t<T$.  A more informative static representation is a weighted network,
in which the weight associated to each edge is proportional to the total
number of contacts (or the total amount of time the contact was active)
between each pair of individuals.  These static network projections,
however, do not account for the nontrivial dynamics of the temporal
network and are thus often inappropriate when considering dynamical
processes unfolding on time-varying connectivity patterns.

\begin{figure}[t]
\includegraphics*[width=\columnwidth]{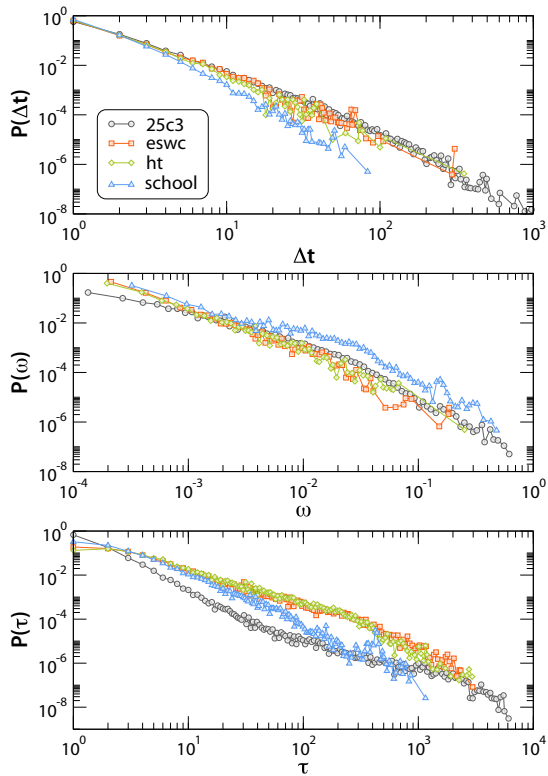}
\caption{Statistical properties of a temporal face-to-face contact
  network \cite{10.1371/journal.pone.0011596}. The probability
  distributions of the lenght of conversations $\Delta t$, total time
  spent in conversation between pairs of individuals $\omega$, and the
  gap $\tau$ between conversation with different individuals, all show a
  long-tailed form, compatible with a power law. Figure adapted from
  \citet{PhysRevE.85.056115}. }
\label{fig:Sociobursts}
\end{figure}

Recent technological advances allow gathering large amounts of
data on social temporal networks, such as mobile phone communications
\cite{Onnela:2007} and face-to-face
interactions~\cite{10.1371/journal.pone.0011596}.  From the analysis
of these datasets, social interactions are
characterized by temporally heterogeneous contact patterns.  Indeed it
is more the norm than the exception to find that the temporal behavior
of social interactions is characterized by heavy-tail and skewed
statistical distributions. For instance, the probability distributions
of the length of contacts between pairs of individuals, of times between
consecutive interactions involving the same individual, etc., all
follow a heavy tailed form (see Fig.~\ref{fig:Sociobursts})
\cite{Onnela:2007,Hui:2005,PhysRevE.71.046119,Tang:2010,10.1371/journal.pone.0011596,PVM_Twitter_lognormal}.
These properties contrast with the Poissonian behavior expected in
purely random interactions, thus catalyzing the recent interest in the
study of the \textit{burstiness} of human behavior
\cite{Oliveira:2005fk}.

The time-varying connectivity pattern of networks affects epidemic
processes in a number of different ways.  First, the presence of a
temporal ordering in the connections of the network limits the possible
paths of propagation of the epidemic process. In particular, not all the
edges of the eventually aggregated network projection are available for
the propagation of a disease.  Starting on a given node, only the nodes
that belong to its \textit{set of influence} \cite{PhysRevE.71.046119},
defined as the nodes that can be reached through paths that respect time
ordering, may propagate the disease.  Furthermore, the Poissonian
approximation for the transmission rate of infectious individuals is not
correct because the time between consecutive nodes' contacts is
generally power-law distributed. However, this non-Poissonian behavior
is different from the one presented in Sec.~\ref{sec:non-mark-react},
where we considered fixed networks in which a disease takes, to
propagate from an infected individual to a susceptible one along a fixed
link, a time $\tau_a$ that is not exponentially distributed. Here we
have the situation in which the very link that can propagate the disease
appears at instants of time that are separated by an inter-event time
$\tau_l$, that can be distributed non-exponentially. Finally, the
relation between the intrinsic time scales of the temporal network and
those of the dynamics plays a substantial role.  Thus, for slow dynamics
with a very large relative time scale, it can be a good approximation to
consider as a substrate the weighted integrated network. If the dynamics
is fast, with a small relative time scale, comparable to that of the
temporal network, then the substrate must be the actual contact sequence
defining the temporal network.

Among the effects that a non-Poissonian temporal network induces on
epidemic spreading, one of the most remarkable is a substantial slowing
down of the spread velocity.  This observation was first made by using
an SI model~\cite{PhysRevLett.98.158702} (see also
\citet{min_spreading_2011}) in the context of the spreading of email
worms among email users.  Empirical data show that the time between
consecutive email activities is heavy-tailed and well approximated by
the form $P(\tau) \sim \tau^{-1-\beta}$.  The generation time $\tau$,
defined as the time between the infection of the primary individual and
the infection of a secondary individual is given by the residual waiting
time distribution, assuming a stationary process,
\cite{renewal}
$g(\tau) = \int_\tau^\infty P(\tau')d\tau' / \av{\tau} \sim
\tau^{-\beta}$,
where it is assumed that the time at which emails are received is
uniformly random. The average number of new infections at time $t$,
$n(t)$ is estimated as $n(t) = \sum_{d=1}^D Z_g \hat{g}_d(t)$, where
$Z_d$ is the average number of users at a distance $d$ (at $d$ email
steps) from the first infected user, $D$ is the maximum possible value
of $d$, and $\hat{g}_d(t)$ is the convolution of order $d$ of
$g(\tau)$. Assuming that the integrated network of email contacts is
sparse, \citet{min_spreading_2011} find that $n(t) \sim t^{-\beta}$,
independently of the integrated network structure. This result implies
that the disease spreads much more slowly than in a regular static
network, where an exponential increase of infected individuals is
observed. The slowing down in temporal networks has been empirically
measured in different systems
\cite{PhysRevLett.98.158702,PhysRevE.83.025102,dynnetkaski2011,Stehle:2011nx},
and also reported in other dynamical processes, such as diffusion
\cite{PhysRevE.85.056115,perra_random_2012,hoffmann_generalized_2012} or
synchronization \cite{albert2011sync}. The situation is however not
completely clear, since other works suggest instead a dynamic
acceleration \cite{2013arXiv1309.0701J}. These temporal effects are,
moreover, entangled with topological ones, as shown by
\citet{Rocha:2010} analyzing the SI and SIR models in empirical
spatio-temporal networks.  Temporal correlations accelerate epidemic
outbreaks, especially in the initial phase of the epidemics, while the
network heterogeneity tends to slow them down.

The time-varying structure of temporal networks is also able to alter
the value of the epidemic threshold, as analytically shown for the SIS
and SIR processes in \textit{activity driven} network
models~\cite{2012arXiv1203.5351P}. The activity-driven network class of
models \cite{2012arXiv1203.5351P,PhysRevE.87.062807} is based on the
concept of \textit{activity potential}, defined as the probability per
unit time that an individual engages in a social activity. Empirical
evidence shows that the activity potential varies considerably from
individual to individual and the dynamics of the networks is encoded in
the function $F(a)$ that characterizes the probability for a node to
have an activity potential $a$.  The activity driven network model
considers $N$ nodes whose activity $a_i$ is assigned randomly according
to the distribution $F(a)$. During each time step the node $i$ is
considered active with probability $a_i$. Active nodes generate $m$
links (engage in $m$ social interactions) that are connected to $m$
individuals chosen uniformly at random. Finally, time is updated
$t \to t+1$. The model output is a sequence of graphs, depending on the
distribution $F(a)$, which is updated at every time step $t$. An
integrated network at time $T$ can be constructed by considering the
union of the sequence of graphs, see Fig.~\ref{fig:temporal_net}. This
integrated network has a degree distribution which depends on the
activity distribution as
$P_T(k) \simeq \frac{1}{T} F\left(\frac{k}{T} - \av{a} \right)$
\cite{PhysRevE.87.062807}, where $\av{a}$ is the average activity and
for simplicity we take $m=1$. The empirically observed power-law
activity distributions, $F(a)$, can thus explain the long tails in the
degree distribution of social networks \cite{2012arXiv1203.5351P}.
\begin{figure}[t]
\includegraphics*[width=\columnwidth]{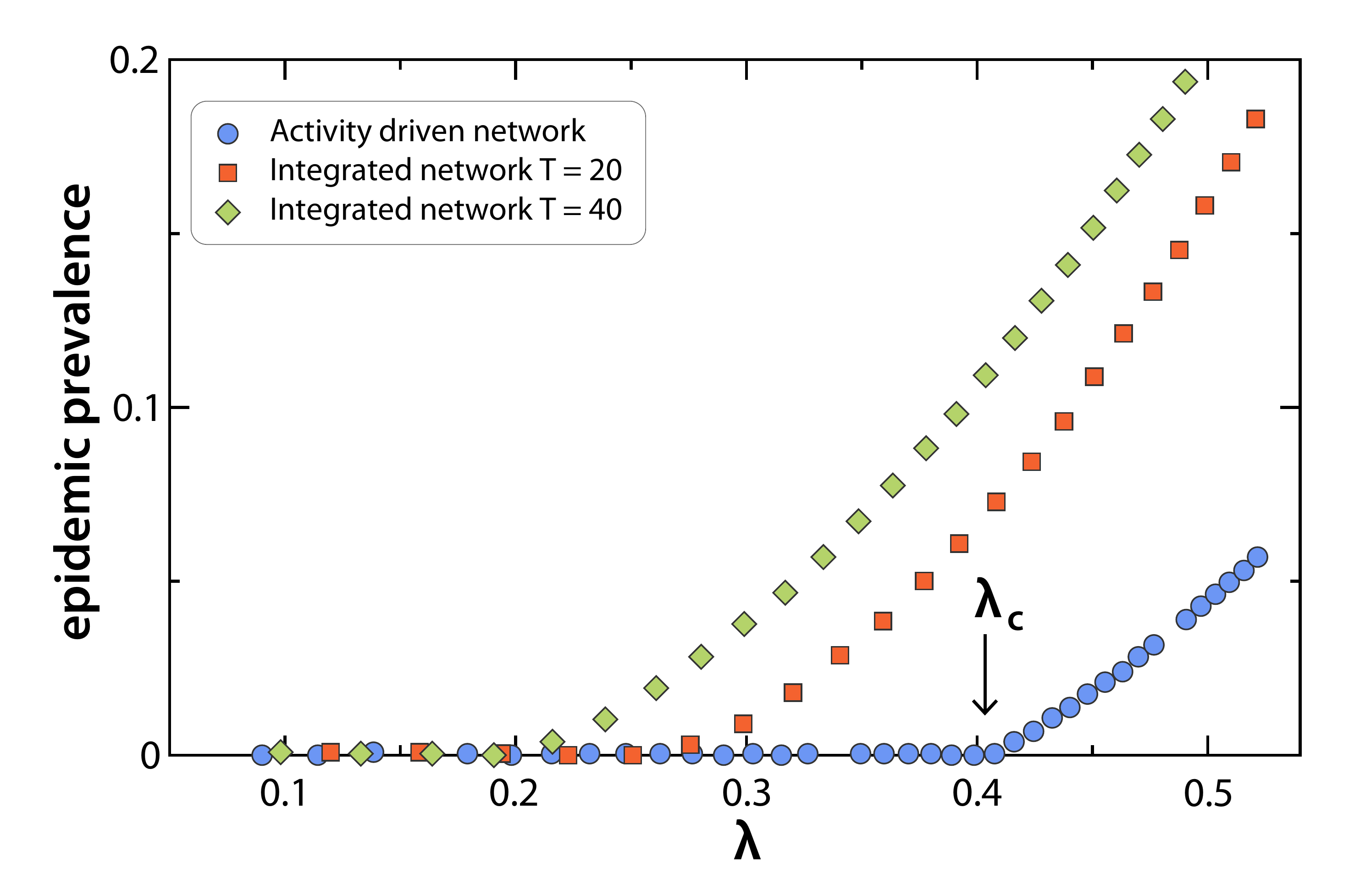}
\caption{Prevalence of the SIS model on the temporal network defined by
  the activity driven model, as a function of the basic transmission
  probability $\lambda$. The threshold observed for the dynamics on the
  temporal network coincides with the theoretical prediction
  Eq.~\eqref{eq:SISActivityThreshold}. Simulations on integrated
  networks show instead a threshold that becomes smaller when increasing
  the integration time $T$. Figure adapted
  from~\citet{2012arXiv1203.5351P}}
\label{fig:sistemporalnets}
\end{figure}
\citet{2012arXiv1203.5351P} consider the behavior of the  SIS
model in activity driven networks, writing dynamical mean field
equations for the infected individuals in the class of activity rate
$a$, at time $t$, namely $I_{a}(t)$. The discrete time dynamical
evolution considers concurrently the dynamics of the network and the
epidemic model, yielding:
\begin{eqnarray}
\label{pp1activity}
I^{t+1}_{a} &=&  \lambda m (N_{a}-I_{a}^{t})a \int d a'  \frac{I_{a'}^{t}}{N} + \\ \nonumber
&+&\lambda  m(N_{a}-I_{a}^{t})\int d a'  \frac{I_{a'}^{t}a'}{N},
\end{eqnarray}
where $N_a = F(a)N$ is the total number of individuals with
activity $a$ and where the recovery probability $\mu=1$.
In Eq.~(\ref{pp1activity}), the first term on the right
side takes into account the probability that a susceptible of class
$a$ is active and acquires the infection getting a connection from any
other infected individual (summing over all different classes), while
the last term takes into account the probability that a susceptible,
independently of his activity, gets a connection from any infected
active individual. A linear stability analysis of
Eq.~(\ref{pp1activity}) leads to an epidemic threshold
\begin{equation}
  \lambda_c = \frac{1}{m(\av{a} + \sqrt{\av{a^2}})},
\label{eq:SISActivityThreshold}
\end{equation}
which is independent of the integration time. The same epidemic threshold is obtained for the SIR model,
applying mean-field approximations \cite{2013arXiv1309.7031L} and a
mapping to percolation \cite{2013arXiv1312.5259S}.
This result highlights the crucial fact that scale-free integrated networks can
lead to a vanishing threshold for epidemics with a very large time
scale, while epidemics with a short time scale, comparable to the one
of the contact sequence, can be associated with a finite,
non-vanishing threshold, see Fig.~\ref{fig:sistemporalnets}. This
observation has been confirmed in studies of other temporal network
models \cite{Rocha:2013}.

Finally, a very recent avenue of research in this area has been the
identification of effective immunization protocols for temporal
networks \cite{Lee:2010fk}. The idea here is to define a
\textit{training window} $\Delta T$, such that information is gathered
from the contact sequence at times $t< \Delta T$. A set of individuals
to be immunized is chosen, and effectively vaccinated at time $\Delta
T$. The effects of the immunization are then observed 
for $t > \Delta T$.  \citet{Lee:2010fk}
explore two local strategies, inspired by the acquittance immunization
protocol for static networks \cite{Cohen03}: In the \textit{Recent}
strategy, a randomly chosen individual is asked at time $\Delta T$ for
its last contact; this last contact is immunized. In the
\textit{Weight} strategy, a randomly chosen individual at time $\Delta T$
is asked for its most frequently contacted peer, up to time $\Delta
T$; this most frequent contact is immunized. By means of numerical
simulations \citet{Lee:2010fk} observe that both protocols offer, for
a limited amount of local information, a reasonable level of
protection against the disease propagation. An interesting issue is the question about the amount of information (the
length $\Delta T$ of the training window) sufficient to achieve an
optimal level of immunization for a fixed fraction of immunized
individuals. \citet{2013arXiv1305.2357S} find a
saturation effect in the level of immunization for training windows of
about a $20\%$ - $40\%$ of the total length of the contact sequence,
for several immunization protocols, indicating that a limited amount
of information is actually enough to optimally immunize a temporal
network. In the case of the activity driven networks, analytical expressions for several immunization
strategies can be obtained \cite{2013arXiv1309.7031L}.

\section{Reaction-diffusion processes and metapopulation models}
\label{metapop}

\begin{figure*}[t]
  \centering
  \includegraphics*[width=\textwidth]{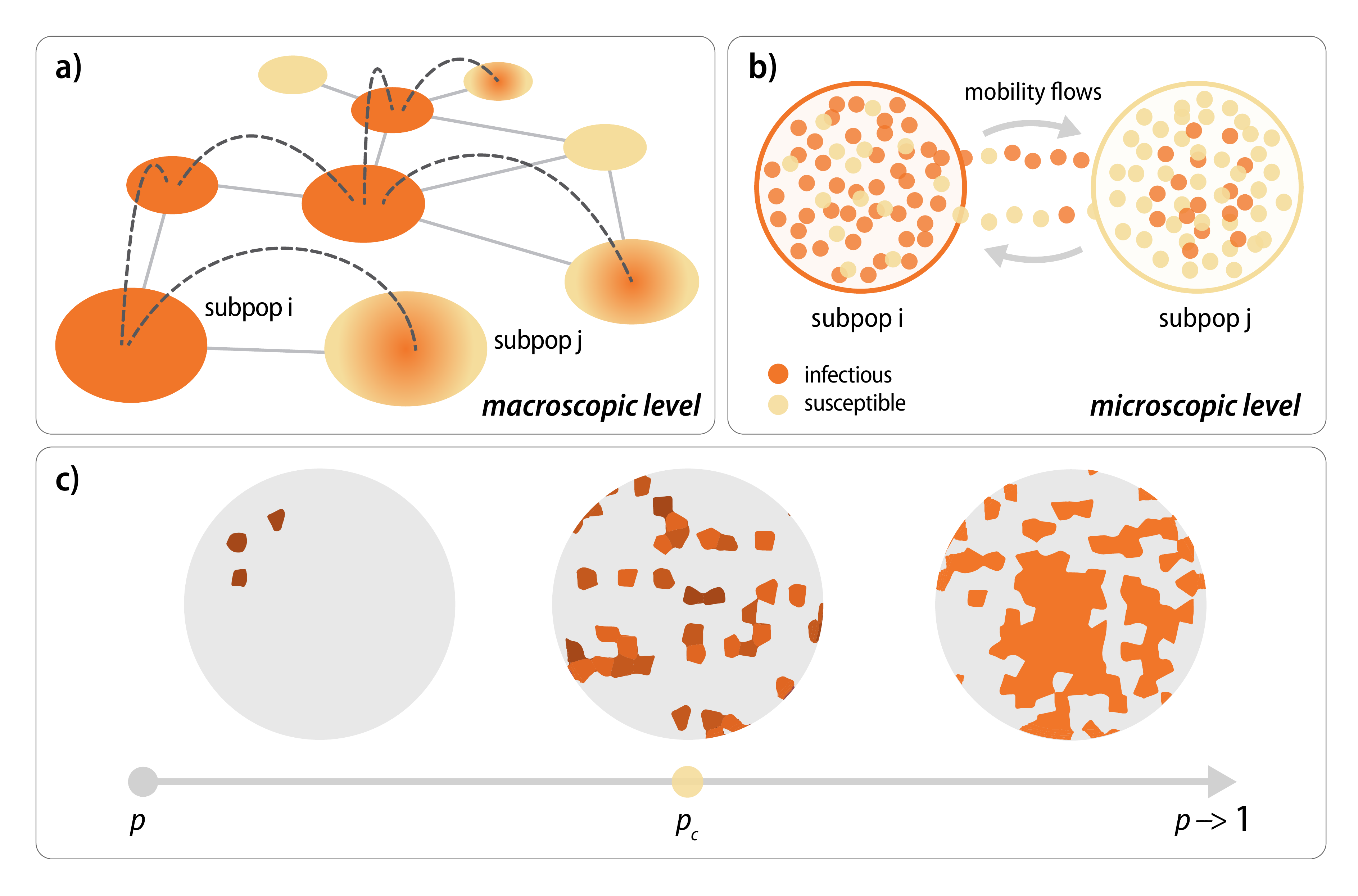}
  \caption{{\bf a} Schematic illustration of the simplified modeling
    framework based on the particle-network scheme. At the macroscopic
    level the system is composed of a heterogeneous network of
    subpopulations. The contagion process in one subpopulation (marked
    in red) can spread to other subpopulations because of particles
    diffusing across subpopulations. {\bf b} At the microscopic level,
    each subpopulation contains a population of individuals. The
    dynamical process, for instance a contagion phenomenon, is
    described by a simple compartmentalization (compartments are
    indicated by different colored dots in the picture). Within each
    subpopulation, individuals can mix homogeneously or according to a
    subnetwork and can diffuse with probability $p$ from one subpopulation to
    another following the edges of the network. {\bf c} A critical
    value $p_c$ of the individuals/particles diffusion identifies a
    phase transition between a regime in which the contagion affects a
    large fraction of the system and one in which only a small
    fraction is affected (see the discussion in the text).}
  \label{fig:metapop}
\end{figure*}

So far we have reviewed results concerning spreading and contagion
processes in which each node of the network corresponds to a single
individual of the population. A different framework emerges if
we consider nodes as entities where multiple individuals/particles can
be located and eventually wander by moving along the links connecting
the nodes.  Examples of such systems are provided by mechanistic
epidemic models where particles represent people moving between
different locations or by the routing of information packets in
technological networks
\cite{Keeling:2002,Sattenspiel:1995,gallos2004absence,Wattsresurgent:2005}. 
More in general, models of social behavior and human mobility are often framed
as reaction-diffusion processes where each node $i$ is allowed to host
any nonnegative integer number of particles $\mathcal{N}(i)$, so that
the total particle population of the system is $\mathcal{N}=\sum_i
\mathcal{N}(i)$. This particle-network framework considers that each
particle diffuses along the edges connecting nodes with a diffusion
coefficient that depends on the node degree and/or other node
attributes. Within each node particles may react according to
different schemes characterizing the interaction dynamics of the
system. A simple sketch of the particle network framework is
represented in Figure~\ref{fig:metapop}.

In order to have an analytic description of reaction-diffusion systems
in networks one has to allow the possibility of heterogeneous
connectivity patterns among nodes. 
A first analytical approach to these systems considers the extension
of the degree-based mean-field approach to reaction-diffusion systems
in networks with arbitrary degree distribution.  For the sake of
simplicity, let us first consider the DBMF approach to the case of a
simple system in which non interacting particles (individuals) diffuse
on a network with arbitrary topology. A convenient representation of
the system is therefore provided by quantities defined in terms of the
degree $k$
\begin{equation}
  \mathcal{N}_k=\frac{1}{N_k}\sum_{i \in{\mathcal V}(k)}\mathcal{ N}(i)\,, 
\end{equation}
where $N_k = NP(k)$ is the number of nodes with degree $k$ and the
sum runs over the set of nodes ${\mathcal V}(k)$ having degree equal to $k$.  The degree
block variable $\mathcal{N}_k$ represents the average number of
particles in nodes with degree $k$. The use of the DBMF approach
amounts to the assumption that nodes with degree $k$, and thus the
particles in those nodes, are statistically equivalent. In this
approximation the dynamics of particles randomly diffusing on the
network is given by a mean-field dynamical equation expressing the
variation in time of the particle subpopulation $\mathcal{N}_k(t)$ in
each degree block $k$. This can be easily written as:
\begin{equation}
  \frac{d\mathcal{ N}_k}{d t}= -d_k\mathcal{ N}_k(t) +
  k\sum_{k'}P(k'|k)d_{k'k}\mathcal{N}_{k'}(t). 
\end{equation}
The first term of the equation considers that only a fraction $d_k$ of
particles moves out of the node per unit time. The second term
instead accounts for the particles diffusing from the neighbors into
the node of degree $k$. This term is proportional to the number of
links $k$ times the average number of particles coming from each
neighbor. This is equal to the average over all possible degrees $k'$ of the
fraction of particles moving on that edge,
$d_{k'k}\mathcal{N}_{k'}(t)$, according to the conditional probability
$P(k'|k)$ that an edge belonging to a node of degree $k$ is pointing
to a node of degree $k'$.  Here the term $d_{k'k}$ is the diffusion
rate along the edges connecting nodes of degree $k$ and $k'$.  The
rate at which individuals leave a subpopulation with degree $k$ is
then given by $d_k=k\sum_{k'}P(k'|k)d_{kk'}$. In the simplest case of
homogeneous diffusion each particle diffuses with rate $r$ from the
node in which it is and thus the diffusion per link $d_{k'k}=r/k'$.
On uncorrelated networks $P(k'|k)=k'P(k')/{\langle k\rangle}$ and
hence one easily gets, in the stationary state $d \mathcal{N}_k/dt=0$
the solution~\cite{v.07:_react,PhysRevLett.92.118701}
\begin{equation}
\mathcal{N}_k= \frac{k}{\langle k\rangle}\frac{\mathcal{N}}{N}.
\label{eq:Wk2}
\end{equation}

The above equation explicitly brings the diffusion of particles in the
description of the system and points out the importance of network
topology in reaction-diffusion processes. This expression indicates
that the larger the degree of a node, the larger the probability to be
visited by the diffusing particles.

\subsection{SIS model in metapopulation networks}

The above approach can be generalized to reacting particles with
different states by adding a reaction term to the above
equations~\cite{v.07:_react}. We now describe a generalization to this
setting of the standard SIS model in discrete time, with probability
per unit time $\beta$ of infection and probability $\mu$ of recovery.
We consider $\mathcal{N}$ individuals diffusing in a heterogeneous
network with $N$ nodes and degree distribution $P(k)$. Each node $i$
of the network has a number $I(i)$ of infectious and $S(i)$ of 
susceptible individuals, respectively. The occupation numbers $I(i)$
and $S(i)$ can have any integer value, including $I(i)=S(i)=0$, that is,
void nodes with no individuals. This modeling scheme describes
spatially structured interacting subpopulations, such as city
locations, urban areas, or defined geographical
regions~\citep{Hanski:2004,grenfell1997meta} and is usually referred
to as \textit{metapopulation approach}.  Each node of the network
represents a subpopulation and the compartment dynamics accounts for
the possibility that individuals in the same location may get into
contact and change their state according to the infection
dynamics. The interaction among subpopulations is the result of the
movement of individuals from one subpopulation to the other.  We have
thus to associate to each individual's class a diffusion probability
$p_I$ and $p_S$ that indicates the probability for any individual to
leave its node and move to a neighboring node of the network.  In
general the diffusion probabilities are heterogeneous and can be node
dependent; however for the sake of simplicity we assume that
individuals diffuse with probability $p_I=p_S=1$ along any of the
links departing from the node in which they are. This implies that at
each time step an individual sitting on a node with degree $k$ will
diffuse into one of its nearest neighbors with probability $1/k$. In
order to write the dynamical equations of the system we define the
following quantities:
\begin{equation}
I_k=\frac{1}{N_k}\sum_{i \in{\mathcal V}(k)} I(i); 
\quad S_k=\frac{1}{N_k}\sum_{i \in{\mathcal V}(k)} S(i),
\end{equation}
where the sums $\sum_{i \in{\mathcal V}(k)}$ are performed over nodes of degree $k$.
These two quantities express the average number of
susceptible and infectious individuals in nodes with degree $k$.
Clearly, $\mathcal{N}_k=I_k+S_k$ is the average number of
individuals in nodes with degree $k$. 
 These
quantities allow to write the discrete time equation describing
the time evolution of $I_k(t)$ for each class of degree $k$ as
\begin{equation}
  I_k(t+1)=k\sum_{k'}P(k|k')\frac{1}{k'}\left[(1-\mu)I_{k'}(t)+\beta
    \Gamma_{k'}(t)\right]\,
  \label{eq:sismetapop}
\end{equation}
where $\Gamma_{k'}(t)$ is an interaction kernel, function of $I_{k'}$
and $S_{k'}$.  The equation is obtained by considering that at each
time step the particles present on a node of degree $k$ first react
and then diffuse away from the node with probability $1$. The value of
$I_k(t+1)$ is obtained by summing the contribution of all particles
diffusing to nodes of degree $k$ from their neighbors of any degree
$k'$, including the new particles generated by the reaction term
$\Gamma_{k'}$.  In the case of uncorrelated networks,
Eq.~\eqref{eq:sismetapop} reduces to
\begin{equation}
  I_k(t+1)=\frac{k}{\langle k\rangle} \left[(1-\mu)\bar{I}(t)+
    \beta \Gamma \right],
\end{equation}
where $\bar{I}(t)=\sum_k P(k) I_k$ is the average number of infected
individuals per node in the network and $\Gamma=\sum_k P(k)\Gamma_k$.
Analogously the equation describing the dynamics of susceptible
individuals is
\begin{equation}
  S_k(t+1)=\frac{k}{\langle k\rangle} \left[ \bar{S}(t) + 
    \mu\bar{I}(t) - \beta \Gamma \right],
\end{equation}
where $\bar{S}(t)=\sum_k P(k) S_k$.

In order to explicitly solve these equations we have to specify the
type of interaction among individuals.  In the usual case of a
mass-action law for the force of infection, we have $\Gamma_k=I_k
S_k/\mathcal{N}_k$.  This implies that each particle has a finite
number of contacts with other individuals.
Considering the stationary state $t\to\infty$, and by using some
simple algebra, we can find that an endemic state $\bar{I}>0$ occurs
only if $\beta/\mu > 1$, thus recovering the classic epidemic
threshold in homogeneous systems~\cite{v.07:_react}.

A very different result is obtained if we consider the case in which
each susceptible individual may react with all the infectious
individuals in the same node. In this case $\Gamma_k=I_k S_k$,
i.e. all individuals are in contact with the same probability
(absorbed in the factor $\beta$), independently of the total
population present in each node. This law, referred to as pseudo
mass-action law, is sometimes used to model animal diseases as well as
mobile phone malwares.  In this case, an active stationary solution
$\bar{I}>0$ occurs if \cite{v.07:_react}
\begin{equation}
  \bar{\mathcal{N}}\geq \bar{\mathcal{N}}_c \equiv
  \frac{\langle k \rangle}{\langle k^2 \rangle} \frac{\mu}{\beta},
\end{equation}
where $\bar{\mathcal{N}}=\sum P(k)\mathcal{N}_k=\mathcal{N}/N$ is the
average number of individuals per node.  
This result implies that a stationary state with infectious
individuals is possible only if the particle density average
$\bar{\mathcal{N}}$ is larger than a specific critical threshold. However the network
topological fluctuations affect the critical value. In particular, in
heavy-tailed networks with $\langle k^2 \rangle \to \infty$ we have
that $\bar{\mathcal{N}}_c \to 0$, i.e. topological fluctuations induce a vanishing of the threshold in the limit of an infinite 
network.

The different behavior obtained in the two types of processes can be
understood qualitatively by the following argument~\cite{v.07:_react}.
In a process governed by the mass action law the epidemic activity in
each node is rescaled by the local population $\mathcal{N}_i$ and it is
therefore the same in all nodes. In this case, the generation of
infected individuals is homogeneous across the network and an epidemic
active state depends only on the balance between $\beta$ and $\mu$,
whose values must poise the system above the critical threshold. In
contagion processes determined by the pseudo-mass action law, whatever
the parameters $\beta$ and $\mu$, there exists a local density of
individuals able to sustain the generation of infected individuals to
keep the system in the active state. In this case topological
fluctuations induce density fluctuations in the network as the
diffusion process brings individuals to each node proportionally to
the degree $k$, Eq.~\eqref{eq:Wk2}.  Whatever the average number of
individuals per node in the thermodynamic limit, there is always a
node (with a virtually infinite degree) with enough individuals to
keep alive the contagion process, leading to the disappearance of the
phase transition.

Although the above results are obtained by a discrete formulation that
generally suits well simulation schemes in which reactions and
diffusion  are executed sequentially, the continuum
formalism of the above models has been derived
in~\citet{Saldana:2008} (see also \citet{baronchelli08:_boson_react}).  In the continuum derivation the same
phenomenology is obtained although the results concerning the critical
value in pseudo mass reaction-like processes scales as the maximum
degree in the network: $\bar{\mathcal{N}}_c\sim k_\mathrm{max}^{-1}$.

It is worth stressing that in most contagion processes, the mobility
of individuals is generally extremely heterogeneous and not simply
mimicked by constant diffusion probabilities as those used in the
previous simple example. The interaction among subpopulations is the
result of the movement of individuals from one subpopulation to the
other.  For instance, it is clear that one of the key issues in the
modeling of contagion phenomena in human populations is the accurate
description of the commuting patterns or traveling of people. In many
instances even complicated mechanistic patterns can be accounted for
by effective couplings expressed as a force of infection generated by
the infectious individuals in subpopulation $j$ on the individuals in
subpopulation $i$. More realistic descriptions are provided by
approaches which include explicitly the detailed rate of
traveling/commuting obtained from data or from an empirical fit to
gravity law models ~\cite{Viboud:2006}). For analytical studies,
simplified approaches use the Markovian assumption in which at each
time step the movement of individuals is given according to a matrix
$d_{ij}$ that expresses the rate at which an individual in the
subpopulation $i$ is traveling to the subpopulation $j$.  This
approach is extensively used in large populations where the
traffic $w_{ij}$ between subpopulations is known, stating that
$d_{ij}\sim w_{ij}/\mathcal{N}_j$.  Several modeling approaches to the
large scale spreading of infectious disease~\cite{Baroyan:1969,
  Rvachev:1985,Flahault:1991,Grais:2004,
  Hufnagel:2004,colizza06:_predic,Colizza:2007, Balcan2009} use this
mobility process based on real data about transportation networks.  A
detailed description of different mobility and diffusion schemes can
be found in~\citet{colizza07:_epidem_model_published}.

\subsection{SIR model in metapopulation networks and the global
  invasion threshold}

In the analysis of contagion processes in metapopulation networks, the
diffusion parameters that mimic the mobility rate of
individuals/particles in the system may cause severe changes to the
phase diagram by inducing a novel type of
critical threshold. To see these effects we consider SIR-like models
with no stationary state possible.  If we assume a diffusion 
probability $p$ for each
individual and that the single population reproductive number of the
SIR model is $R_0>1$, we can easily identify two different limits. If
$p=0$ any epidemic occurring in a given subpopulation will remain
confined; no individual can travel to a different subpopulation and
spread the infection across the system. In the limit $p \to 1$
we have that individuals are constantly wandering from one subpopulation
to the others and the system is in practice equivalent to a well mixed
unique population. In this case, since $R_0>1$, the epidemic will
spread across the entire system. A transition point between these two
regimes is therefore occurring at a threshold value $p_c$ of the
diffusion rate, identifying a global invasion threshold that depends
on the mobility as well as the parameters of the contagion process
(see Fig.~\ref{fig:metapop}). In other words, in a model such as the
SIR model, the epidemic within each subpopulation generates a finite
fraction of infectious individuals in a finite amount of time, and
even if $R_0>1$ the diffusion rate must be large enough to ensure the
diffusion of infected individuals to other subpopulations before the
local epidemic outbreak dies out. It is worth remarking that this does not apply 
in models with endemic states such as the SIS model. In this case the disease produces
infectious individuals indefinitely in time and sooner or later the epidemic will be exported 
to other subpopulations.

The invasion threshold is encoded in a new quantity $R_*$
characterizing the disease invasion of the metapopulation
system. $R_*$ denotes the number of subpopulations that become
infected from a single initially infected subpopulation; i.e. the
analogue of the reproductive number $R_0$ at the subpopulation level.
It defines the critical values of parameters that allow the contagion
process to spread across a macroscopic fraction of subpopulations.
Interestingly, this effect cannot be captured by a continuous
description that would allow any fraction $p \bar{I}$ of diffusing
infected individual to inoculate the virus in a subpopulation not yet
infected.  In certain conditions this fraction $p \bar{I}$, that is a
mean-field average value, may be a number smaller than 1.  This is a
common feature of continuous approximations that allow the infection
to persist and diffuse via ``nano-individuals'' that are not capturing
the discrete nature of the real systems. The discrete nature of
individuals and the stochastic nature of the diffusion can therefore
have a crucial role in the problem of resurgent epidemics, extinction
and eradication~\citep{Ball:1997,Cross:2005,Wattsresurgent:2005,Vazquez:2007,Cross:2007}.

In order to provide an analytical estimate of the invasion threshold,
we consider a metapopulation network with arbitrary degree
distribution $P(k)$, where each node of degree $k$ has a stationary
population $\mathcal{N}_k$. By using a Levins-type
approach~\cite{colizza07:_invas_thres} it is possible to characterize the invasion
dynamics by looking at the tree-like branching process describing the contagion process
at the subpopulation level~\cite{Levins:1970}.  Let us define
$D^0_k$ as the number of \emph{diseased} subpopulations of degree $k$
at generation $0$, i.e. those which are experiencing an outbreak at
the beginning of the process. Each infected subpopulation will
seed---during the course of the outbreak---the infection in
neighboring subpopulations, defining the set $D^1_k$ of infected
subpopulations at generation 1, and so on.  This corresponds to a
basic branching process where the number of infected subpopulations
of degree $k$ at the $n-$th generation is denoted as $D^n_k$.
We can write the iterative equation relating $D^n_k$ and $D^{n-1}_k$ as
\begin{eqnarray}
  D_k^n & =& \sum_{k'}D_{k'}^{n-1} (k'-1) 
  P(k|k')
  \nonumber\\
  &&
~~~~~~ \times \left(1-\frac{D_k^{n-1}}{N_k}\right)\left[1-\left(\frac{1}{R_0}\right)^{\lambda_{k'k}}\right] .
\label{poptree-het}
\end{eqnarray}
In this expression we assume that each infected subpopulation of
degree $k'$ at the $(n-1)-$th generation may seed the infection in a
number of subpopulations of degree $k$ according to the number of
neighboring subpopulations $(k'-1)$ that discount the neighboring population from which the infection was originally transmitted.
The right term takes into account  the probability $P(k|k')$ that each
of the $k'-1$ neighboring populations has degree $k$, the probability
that the seeded population is not infected, and the probability to
observe an outbreak in the seeded population.  This last probability
stems from the probability of extinction $P_{ext}=1/R_0$ of an
epidemic seeded with a single infectious
individual~\cite{Bailey_book}, when one considers a seed of size
$\lambda_{kk'}$ given by the number of infected individuals that move
into a connected subpopulation of degree $k'$ during the duration of
the local outbreak in the subpopulation of degree $k$.

The quantity $\lambda_{kk'}$ can be explicitly calculated by
considering that in the case of a macroscopic outbreak in a closed
population, the total number of infected individuals during the
outbreak evolution will be equal to $\bar{\alpha} \mathcal{N}_{k}$ where
$\bar{\alpha}$ depends on the specific disease model and parameter values
used. Each infected individual stays in the infectious state for a
time $\mu^{-1}$ equal to the inverse of the recovery rate, during
which it can travel to the neighboring subpopulation of degree $k'$
with rate $p$. Here, for the sake of simplicity we consider that the
mobility coefficient $p$ is the same for all individuals. Under this
condition the number of infected individuals that may move into a
connected subpopulation of degree $k'$ during the duration of the
local outbreak in the subpopulation of degree $k$ is given by
\begin{equation}
  \lambda_{kk'}= p \frac{\bar{\mathcal{N}}\bar{\alpha}\mu^{-1}}{\langle k\rangle},
\label{eq:Nk}
\end{equation}
where we have considered that each individual will diffuse with the
same probability in any of the $k$ available connections and that
$\mathcal{N}_k$ is given by Eq.~\eqref{eq:Wk2}.

In order to provide an explicit solution to the above iterative
equation we consider in the following that $R_0-1\ll 1$, thus
assuming that the system is very close to the epidemic
threshold.  In this limit we can approximate the outbreak probability
as $1-R_0^{-\lambda_{k'k}} \simeq \lambda_{k'k}(R_0-1)$.  In addition,
we assume that at the early stage of the epidemic $D_k^{n-1}/N_k \ll
1$, and we consider the case of uncorrelated networks, obtaining
\begin{equation} 
  D_k^n =(R_0-1)\frac{kP(k)}{\langle k\rangle^2}
  \frac{p\bar{\mathcal{N}}\bar{\alpha}}{\mu} \sum_{k'}D_{k'}^{n-1} (k'-1).
\end{equation}
By defining $\Theta^n= \sum_{k'}D_{k'}^{n} (k'-1)$, the 
last expression can be conveniently written in the iterative form
\begin{equation} 
\Theta^n =(R_0-1)\frac{\langle k^2 \rangle-\langle k\rangle}{\langle k \rangle^2}
\frac{p\bar{\mathcal{N}}\bar{\alpha}}{\mu} \Theta^{n-1},
\end{equation}
that allows a growing epidemic only if 
\begin{equation} 
  R_*=(R_0-1)\frac{\langle k^2\rangle-\langle k\rangle}{\langle k\rangle^2}
  \frac{p\bar{\mathcal{N}}\bar{\alpha}}{\mu} >1,
\end{equation}
defining the  {\em global invasion threshold} of the metapopulation system.

The explicit form of the threshold condition
can be used to find the minimum mobility rate ensuring that on
average each subpopulation can
seed more than one neighboring subpopulation. 
The constant $\bar{\alpha}$ is larger than zero for any
$R_0>1$, and in the SIR case for $R_0$ close to 1 it can be approximated 
by $\bar{\alpha} \simeq 2(R_0-1)/R_0^2$~\cite{Bailey_book}, yielding
a critical mobility
value $p_c$ below which the epidemics cannot invade the metapopulation
system given by the equation
\begin{equation} 
p_c\bar{\mathcal{N}}\geq \frac{\langle k\rangle^2}{\langle k^2\rangle-\langle k\rangle}
\frac{\mu R_0^2}{2(R_0-1)^2}.
\label{eq:glth1}
\end{equation}

\begin{figure}[t]
  \centering
  \includegraphics*[width=\columnwidth]{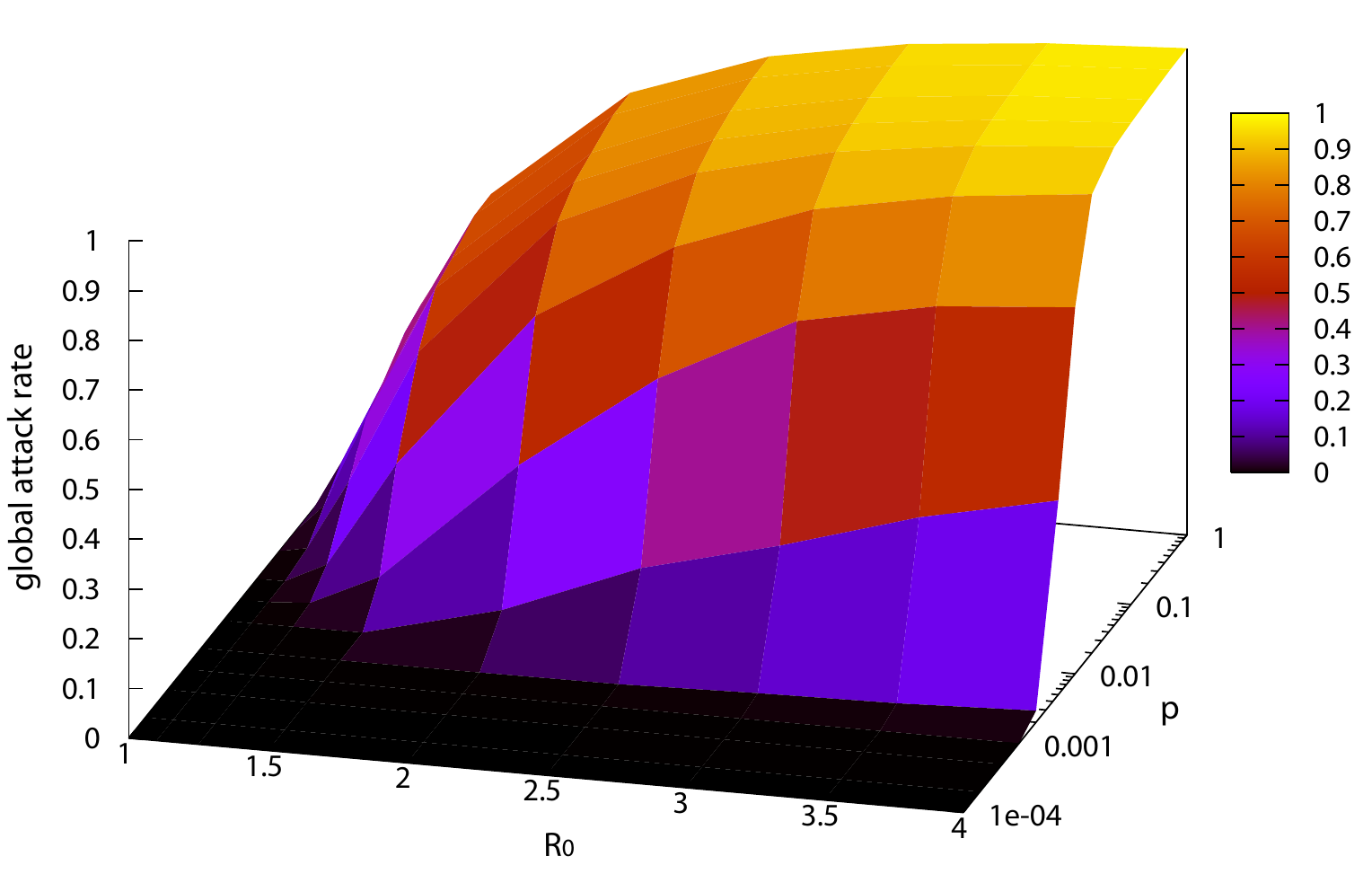}
  \caption{ Global threshold in a heterogeneous metapopulation
    system. The left panel shows a 3D surface representing the value of
    the final epidemic size in the metapopulation system as a function
    of the local threshold $R_0$ and of the diffusion probability
    $p$. If $R_0$ approaches the threshold, larger values of the
    diffusion probability $p$ need to be considered in order to observe
    a global outbreak in the metapopulation system. Figure adapted from
    Colizza \& Vespignani, 2007.}
  \label{fig:3Dinv}
\end{figure}

In Fig.~\ref{fig:3Dinv} we show the total number of infected
individuals across all subpopulations, also called the global attack
rate, as a function of both $R_0$ and $p$, as obtained from extensive
Monte Carlo simulations in an uncorrelated metapopulation network with
$P(k)\sim k^{-2.1}$, $N=10^5$, $\bar{\mathcal{N}}=10^3$ and $\mu=0.2$.
The global attack rate surface in the $p$-$R_0$ space shows that the smaller
the value of $R_0$, the higher the mobility $p$ in order for the contagion
process to successfully invade a finite fraction of the
subpopulations.

The invasion threshold $R_*>1$ implicitly defines the critical
mobility rate of individuals and is an indicator as important as the
basic reproductive number $R_0>1$ in assessing the behavior of
contagion processes in structured populations. It shifts the attention
from the local outbreak to a global perspective where the
interconnectivity and mobility among subpopulations is extremely
important in possibly hampering the spreading process. The presence of
the factor $\langle k \rangle^2/\langle k^{2}\rangle$ in the explicit
expression of the threshold points out that also at the global level
the heterogeneity of the network plays a very important role.  In
other words, the topological fluctuations favor the subpopulation
invasion and suppress the phase transition in the infinite size limit.

While the analysis we have presented here is extremely simplified, in
the last years several studies have provided insight
on metapopulation spreading fully considering the
stochastic and discrete nature of the process in various realistic
contexts: heterogenous schemes for the diffusion of
individuals~\cite{colizza07:_epidem_model_published,Ni:2009,Ben-Zion:2010,Bart:2008};
heterogeneous populations~\cite{Apolloni:2013,Poletto:2012};
non-markovian recurrent mobility patterns mimicking commuting among
geographical regions~\cite{Balcan:2011,Belik:2011,Balcan:2012} and the
introduction of individual behavioral responses to the presence of
disease~\cite{Meloni:2011,Nicolaides:2013}. Indeed one of the
interesting applications of the particle-network framework and the
study of reaction-diffusion processes in metapopulation networks
consists in providing analytic rationales for data driven epidemic
models. 

\subsection{Agent Based Models and Network Epidemiology}

In recent years, mathematical and computational approaches to the study
of epidemics have been increasingly relevant in providing quantitative
forecast and scenario analysis of real infectious disease
outbreaks~\cite{Lofgren2014}. For this reason, epidemic models have
evolved into large-scale microsimulations, data-driven approaches that
can provide information at very detailed spatial resolutions. An example
is provided by agent based, spatially structured models that consider
the discrete nature of individuals and their mobility and are generally
including the stochasticity of interactions and mobility of
individuals. These models, are based on the construction of synthetic
populations characterizing each individual in the population and its
mobility pattern, often down to the level of households, schools and
workplaces~\cite{Hufnagel:2004,Eubank2004,Longini2005,Ferguson2005,Colizza:2007,Chao2010}.
The synthetic population construction is a data hungry process and the
resulting model is in most of the cases non-transparent to an analytical
understanding. For this reason, the analysis of these models relies on
computational microsimulations of the epidemic evolution that keep track
of each single individual in the population. The resulting ensemble of
possible epidemic evolutions is then leveraged to provide the usual
quantitative indicators such as median, mean, and reference ranges for
epidemic observables, such as newly generated cases, seeding events,
time of arrival of the infection. The statistical information generated
by the computational approaches is then exploited with different
visualization techniques that reference the data geographically. At
first sight this modeling approach seems unrelated to network
epidemiology.  In reality, most of the data driven computational
approaches are relying on the construction of synthetic populations and
interaction patterns that are effectively encoded as multiscale networks
of individuals and locations~\cite{Marathe2013}.

\begin{figure*}[t]
  \centering
  \includegraphics*[width=\textwidth]{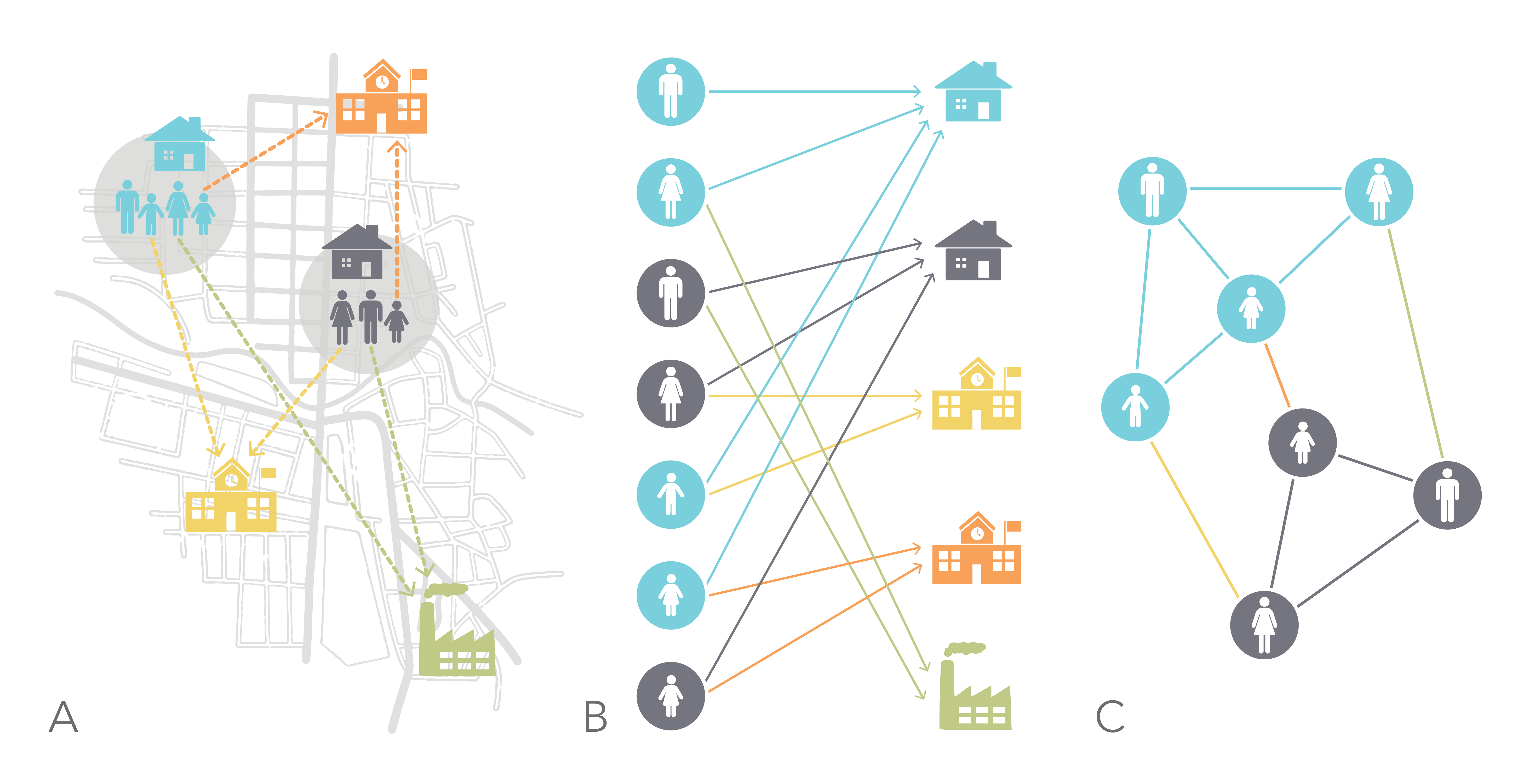}
  \caption{ Schematic illustration of the construction of a synthetic
    population and the resulting contact network.  {\bf a} At the
    macroscopic level, a synthetic population and its movements are
    constructed from census and demographic data. {\bf b} A bipartite
    network associating individuals to locations, and eventually
    weighting the links with the time spent in the location, is derived
    from the synthetic population.  {\bf c} The unipartite projection of
    the bipartite network provides a contact network for the contagion
    process. Different transmission rates and weights on the network
    depends on the location and type of interactions. }
  \label{fig:synthpop}
\end{figure*}
An example of the underlying network structure of data driven epidemic
models is provided by the GLobal Epidemic and Mobility (GLEAM) model
that integrates census and mobility data in a fully stochastic
meta-population network model that allows for the detailed simulation of
the spread of influenza-like illnesses around the
globe~\cite{Broeck2011}.  This model uses real demographic and mobility
data. The world population is divided into geographic census areas that
are defined around transportation hubs and connected by mobility
fluxes. Within each subpopulation, the disease spreads between
individuals. Individuals can move from one subpopulation to another
along the mobility network according to high quality transportation
data, thus simulating the global spreading pattern of epidemic
outbreaks.  
At the finer scale of urban areas, synthetic population
constructions are even more refined and consider a classification of
location such as house, schools, offices etc.  The movement and time
spent in each location can be used to generate individuals-location
bipartite networks whose unipartite projection defines the
individual-level, synthetic interaction network that governs the
epidemic
spreading~\cite{Eubank2004,Halloran2008,Merler2011,Fumanelli2012}.  Also
in this case, although the model underlying the computational approach
is a network model, each individual is annotated with the residence
place, age, as well as many other possible demographic information, that
can be exploited in the analysis of the epidemic outbreak (see
Fig.~\ref{fig:synthpop}).

Data driven computational approaches can generate results at
unprecedented level of detail, and have been used successfully in the
analysis and forecast of real epidemics
\cite{Hufnagel:2004,Balcan2009,Balcan:2009BMC,Merler2011},
and policy making scenario
analysis~\cite{Eubank2004,Longini2005,Ferguson2005,Colizza:2007,Brockmann2013}. Similar
approaches are becoming more and more popular in the simulation of
generalized contagion processes and social
behavior~\cite{Marathe2013}. Although realistic and detailed,
computational approaches often provides non-intuitive results and the
key mechanisms underlying the epidemic evolution are difficult to
identify because of the amount of details integrated in the models. In
such cases, the analytic understanding of the basic models presented in
this review can be the key to the systematic investigation of the impact
of the various complex features of real systems on the basic properties
of epidemic outbreaks.  For instance, the simple calculation of the
invasion threshold explains why travel restrictions appear to be highly
ineffective in containing epidemics in large-scale data driven
simulation: the complexity and heterogeneity of the present time human
mobility network favor considerably the global spreading of infectious
diseases. Only unfeasible mobility restrictions reducing the global
travel fluxes by $90\%$ or more would be
effective~\cite{Cooper:2006,Hollingsworth:2006,colizza07:_epidem_model_published,Bajardi:2011}. The
understanding of the behavior of reaction-diffusion processes in complex
networks is therefore a crucial undertaking if we want to answer many
basic questions about the reliability and predictive power of data
driven computational models.

\section{Generalizing epidemic models as social contagion processes}

\label{sec:7.A}
Infectious diseases certainly represent the central focus of epidemic
modeling because of the relevance they played, and continue to play in
present days, in human history.  The contagion metaphor however applies
in several other domains and in particular in the social context: the
diffusion of information~\citep{Bikhchandani1992}, the propagation of
rumors, the adoption of innovations or
behaviors~\citep{Bass1969,Rogers2010}, are all phenomena for which the
state of an individual is strongly influenced by the interaction with
peers. Mediated by the network of social contacts, these
interactions can give rise to epidemic-like outbreaks: fads, information
cascades, memes going viral online, etc. The term social (or complex)
contagion generally denotes these type of phenomena.  New communication
technologies, online social media, the abundance of digital fingerprints
that we, as individuals, disseminate in our daily life, provide an
unprecedented wealth of data about social contagion phenomena, calling
for theoretical approaches to measure, interpret, model and predict
them.  Simple models for disease epidemics are the natural paradigm for
this endeavour and have been applied to social spreading
phenomena~\citep{Goffman1964,Goffman1966,Bettencourt2006}.  Some
specific features of social contagion, however, are qualitatively
different from pathogen spreading: the transmission of information
involves intentional acts by the sender and the receiver, it is often
beneficial for both participants (as opposed to disease spreading), and
it is influenced by psychological and cognitive factors.  This leads to
the introduction of new ingredients in the models, from which the name
{\it complex contagion} derives.  In this Section we will discuss recent
developments in this modeling effort, which we divide in two broad
categories depending on whether the spreading process (threshold models)
or the recovery process (rumor spreading models) of the disease epidemic
propagation is changed. In the light of the modeling efforts, a review
of papers analyzing empirical data follows next.

As the topics presented here encompass a vast spectrum of disciplines,
including physics, computer science, mathematics, and social sciences,
the usual caveat about the impossibility of an exhaustive review of
all the literature is to be particularly stressed. Our limited
goal  is to try to outline the most important contributions in a
unitary framework.  This endeavor is made even more difficult by the
fact that the propagation of social contagion is also close to other
processes such as failure cascades (in network routing protocols or
mechanical failure~\citep{Motter2002}) or the adoption of strategies
in game-theoretic context~\citep{Easley2010} that are beyond the scope of this review.

\subsection{Threshold models}

\begin{figure}[t]
\includegraphics*[width=\columnwidth]{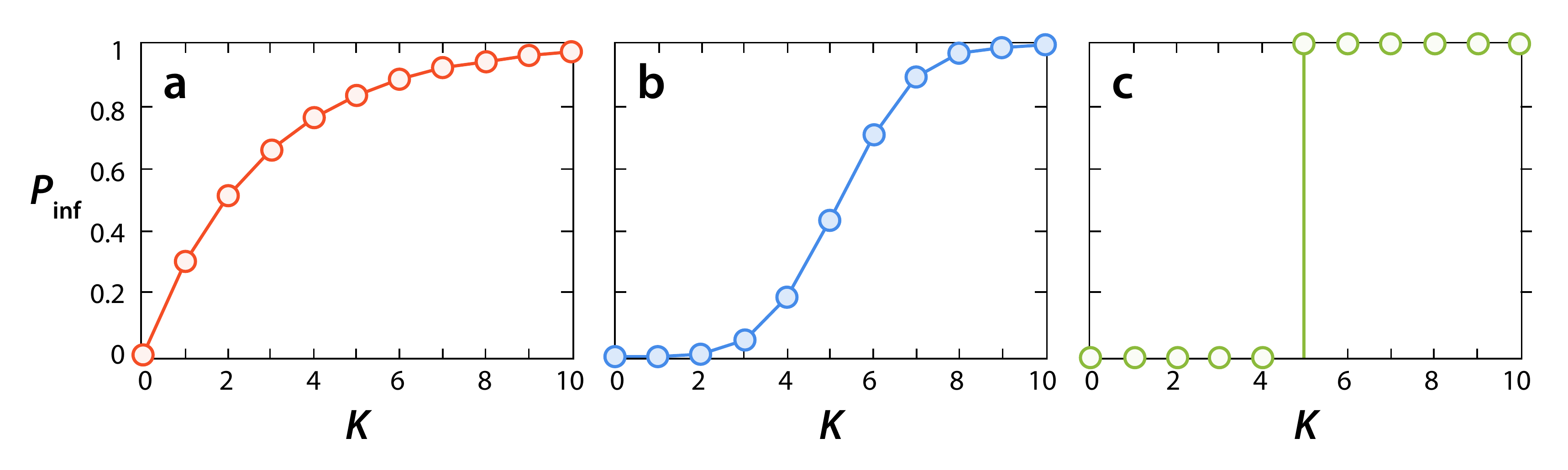}
\caption{Probability $P_{inf}$ of infection for a susceptible individual
  after $K$ contacts with infected individuals.  (a) Independent
  interaction (e.g., SIR-type) model.  (b) Stochastic threshold
  model. (c) Deterministic threshold model.  Adapted
  from~\citet{Dodds2004}}
\label{fig:Dodds04} 
\end{figure}

For disease epidemics it is customary to assume that a susceptible
individual has a constant probability to receive the infection from a
peer upon every exposure, independently of whether other infected
individuals are simultaneously in contact or other exposures have
occurred in the past.  While generally reasonable for the transmission
of pathogens (though exceptions may occur~\citep{Joh2009}) this
hypothesis is clearly unrealistic in most situations where a social
meme is spreading: a piece of information is more credible if arriving
from different sources; the push to adopt a technological innovation
is stronger if neighboring nodes in the social network have already
adopted it.  These considerations lead naturally to the introduction
of ``threshold models'' for spreading phenomena, where the effect of
multiple exposures changes from low to high as a function of their
number.  Fig.~\ref{fig:Dodds04} displays the probability of infection
(adoption) $P_{inf}$ after $K$ attempts in the different scenarios. In
the case of SIR (left panel) each attempt has a fixed probability $p$
of success and $P_{inf} = 1-(1-p)^K$.

\begin{figure}[t]
\includegraphics*[width=\columnwidth]{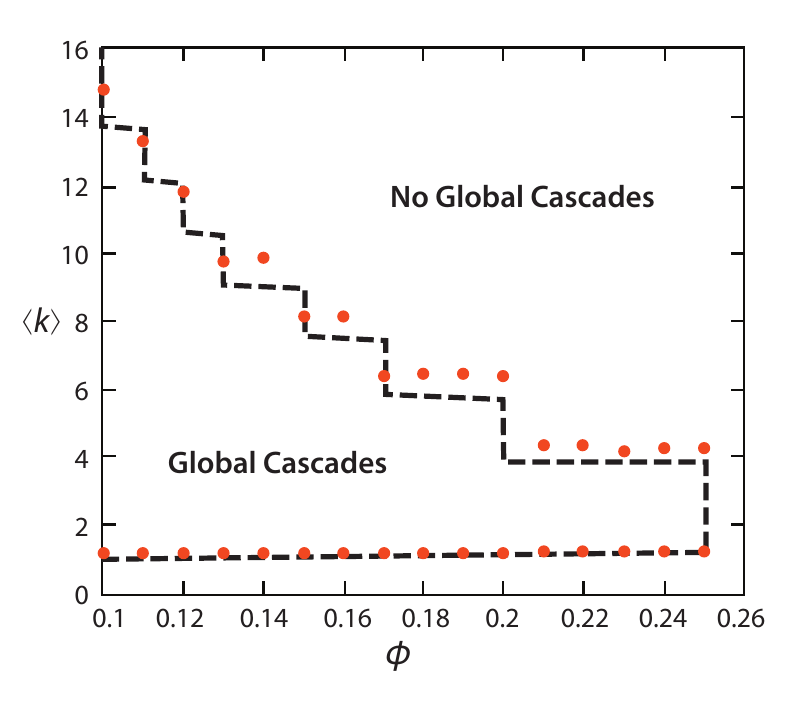}
\caption{Phase-diagram of Watts' threshold model. 
The dashed line encloses the region of the $(\phi, \av{k})$ plane in which 
the condition for the existence of global cascades is satisfied 
for a uniform random graph with uniform threshold $\phi$. 
The solid circles outline the region in which global cascades occur for
the same parameter settings in the full dynamical model for $N = 10000$ 
(averaged over $100$ random single-node perturbations).
Adapted from~\citet{Watts2002}.}
\label{fig:Watts02} 
\end{figure}

Threshold models have a long tradition in the social and economical
sciences~\citep{Granovetter1978,Morris2000}.  In the context of
spreading phenomena on complex networks, a seminal role has been
played by the model introduced by \citet{Watts2002}.  Each individual
can be in one of two states ($S$ and $I$) and is endowed with a quenched,
randomly chosen ``threshold'' value $\phi_i$.  In an elementary step
an individual agent in state $S$ observes the current state of its
neighbors, and adopts state $I$ if at least a threshold fraction
$\phi_i$ of its neighbors are in state $I$; else it remains in state $S$. 
\footnote{This is the definition for {\em relative} threshold models.  
  In many cases {\em  absolute} thresholds are considered
  ~\citep{Granovetter1978,Kempe2003,Galstyan2007,Centola2007a,
    Kimura2009,Karimi2013}. For strongly heterogeneous networks the
  different definitions may lead to important changes.}.
No transition from $I$ back to $S$ is possible.
Initially all nodes except for a
small fraction are in state $S$.  Out of these initiators a {\em
  cascade} of transitions to the $I$ state is generated.  The nontrivial
question concerns whether the cascade remains local, i.e. restricted
to a finite number of individuals, or it involves a finite fraction of
the whole population.  Given an initial seed, the spreading can occur
only if at least one of its neighbors has a threshold such that
$\phi_i \le 1/k_i$.  A cascade is possible only if a cluster of these
``vulnerable'' vertices is connected to the initiator.  For global
cascades to be possible it is then conjectured that the subnetwork of
vulnerable vertices must percolate throughout the network. The
condition for global cascades can then be derived applying on locally
tree-like networks the machinery of generating functions for
branching processes.  In the
simple case of a uniform threshold $\phi$ and an Erd\H{o}s-R\'enyi pattern
of interactions the phase diagram as a function of the threshold
$\phi$ and of the average degree $\av{k}$ is reported in
Fig.~\ref{fig:Watts02}.
For fixed $\phi$, global cascades occur only for intermediate values of
the mean connectivity $1<\av{k}<1/\phi$.  The transition occurring for
small $\av{k}$ is trivial and is not due to the spreading dynamics: the
average cascade size is finite for $\av{k}<1$ because the network itself
is composed of small disconnected components: the transition is
percolative with power-law distributed cascade size.  For large
$\av{k}>1/\phi$ instead, the propagation is limited by the local
stability of nodes. As the transition is approached increasing $\av{k}$
the distribution of cascade size is bimodal, with an exponential tail at
small cascade size and global cascades increasingly larger but more
rare, until they disappear altogether, implying a discontinuous (i.e.,
first-order) phase transition in the size of successful cascades.
Heterogeneous thresholds reduce the system stability, increasing the
range of parameters where global cascades occur. Degree heterogeneity
has instead the opposite effect.

The critical value of the threshold $\phi_c=1/\av{k}$, separating global
cascades for $\phi<\phi_c$ from localized spreading for $\phi>\phi_c$ 
highlights the peculiar features of threshold dynamics~\citep{Centola2007}.
Adding new links to the network makes $\av{k}$ grow, thus reducing
$\phi_c$ and making system-wide spreading more difficult; the
opposite of what occurs for SIR epidemics.
Notice indeed that the dependence of the threshold on the average
degree is the same (for homogeneous networks) in both the threshold
model and in SIR dynamics, but in the latter case the global spreading
occurs {\em above} the threshold
(for $\lambda>1/\av{k}$), while in the former case global cascades
are possible {\em below} the threshold ($\phi<1/\av{k})$.
By the same token,
link rewiring which destroys clustering of a network is seen to
reduce the average cascade size for the threshold model. Instead of the
{\em strength of the weak ties}~\citep{Granovetter1973} here the
{\em weakness of long ties}~\citep{Centola2007a} is at work.

Watts' model can be seen as a particular instance of a more general
model~\citep{Dodds2004}, which includes also independent interaction
models (SIR, SIRS) as particular cases.
The model incorporates individual memory, variable magnitude
of exposure (dose amount) and heterogeneity in the susceptibility
of individuals.
At each contact with an infected neighbor a susceptible receives with 
probability $p$ a random dose $d(t)$ (distributed according
to $f(d)$).
A susceptible individual $i$ accumulates the doses $d_i(t)$ over a time $T$
and it becomes infected if at some time the accumulated dose 
$D_i(t)=\sum_{t'=t-T+1}^t d_i(t')$ is larger than a threshold $d_i^*$ 
(random for each node with distribution $g(d^*)$).
Recovery is possible with probability $r$ provided the dose $D_i(t)$ 
falls below $d_i^*$.
The probability that a susceptible individual who encounters 
$K \le T$ infected individuals in $T$ time steps becomes infected is
therefore
\be
P_{inf}(K) = \sum_{k=1}^K \binom {K}{k} p^k (1-p)^{K-k} P_k
\ee
where
\be
P_k = \int_0^{\infty} dd^* g(d^*) P\left(\sum_{i=1}^k d_i \ge d^*\right) 
\ee
is the average fraction of individuals infected after receiving
$k$ positive doses in $T$ time steps.
When all doses $d_i$ are identical, all members of the population have the
same threshold $d^*$, and $p<1$,
then the model reduces to the standard SIR
(see Fig.~\ref{fig:Dodds04}a).
In other cases it is a deterministic or stochastic threshold model,
depending on whether thresholds vary (see Fig.~\ref{fig:Dodds04}b)
or are all identical (see Fig.~\ref{fig:Dodds04}c).

Adding a probability $\rho$ that a recovered individual becomes
susceptible again leads to a SIRS-like dynamics. Setting $r=1$ and
$\rho=1$ gives a SIS-like model, for which the stationary fraction of
active nodes as a function of $p$ is the order parameter.  Three
qualitatively different shapes of the phase-diagram are found,
depending only on $T$ and $P_1$ and $P_2$, the probabilities that an
individual will become infected as a result of one and two exposures,
respectively.  If $P_1>P_2/2$ there is a standard epidemic transition
between an absorbing healthy phase and an active infected one. The
phenomenology is the same of SIS, indicating that successive exposures
are effectively independent.  The two other possible behaviors both
exhibit a discontinuous phase transition for finite $p$, differing
in the sensitivity with respect to the size of the initial seed.

By means of an analytical approach for locally tree-like networks,
\citet{Gleeson2007} extended Watts' approach to consider
a finite fraction of initiators $p^{in}$. It turns out that this change
may have dramatic effects on the location of the transitions as a 
function of $\av{k}$ and even make the transition for small $\av{k}$
discontinuous.
\citet{Singh2013} have shown that for any $\phi<1$ there is a critical 
value $p^{in}_{c}(\phi)$ such that for $p>p^{in}_c(\phi)$ the cascades are
global.
Further work along the same lines has generalized
the analytical treatment to modular 
networks~\citep{Gleeson2008}, degree-correlated
networks~\citep{Gleeson2008,Dodds2009} and to networks with tunable
clustering~\citep{Hackett2011}. In the latter case, it turns out that
for large and small values of $\av{k}$ clustering reduces the
size of cascades, while the converse occurs for intermediate
values of the average degree.

Watts' threshold model has been extended in many directions, 
to take into account other potentially relevant effects that may
influence the spreading process.
Interaction patterns described by layered
networks are found to increase the cascade size~\citep{Brummitt2012}
while the consideration of temporal networks~\citep{Holme:2011fk}
with the associated bursty activity of individuals may either
facilitate~\citep{Takaguchi2012} or hinder~\citep{Karimi2013}
the spreading process.
Watts' model on a basic two-community network is considered
in~\citet{Galstyan2007}.
Finally it is worth mentioning the work of~\citet{Lorenz2009} 
which propose a very general classification of models for cascades,
including, among many others, standard epidemic models and Watts' model
as particular cases.

A large interest in threshold models has also be spurred by the
goal of identifying influential spreaders, i.e.
the starting nodes which maximize the size of cascades, a
topic of interest also for traditional epidemic models
(see Section~\ref{sec:5.A}).
\citet{Kempe2003} show that the problem of finding
the set of initiator nodes such that the total size of the cascade
is maximal~\citep{Domingos2001} is NP-hard, both for linear
threshold models and for an independent cascade model,
which is essentially an inhomogeneous SIR. 
Moreover, they provide a greedy hill-climbing algorithm
that provides an efficient approximation
to the NP-hard solution, outperforming random choice as well as
choices based on degree centrality and distance centrality, when
tested on some empirical networks.
\citet{Kempe2003}'s method is computationally costly. An improvement
which makes it much faster is provided by~\citet{Kimura2009}. 

\subsection{Rumor spreading}

Models for rumor spreading are variants of the SIR model for disease
epidemics in which the recovery process does not occur spontaneously,
but rather is a consequence of interactions.  The basic idea behind this
modification is that it is worth propagating a rumor as long as it is
novel for the recipient: If the spreader finds that the recipient
already knows the rumor he/she might lose interest in spreading it any
further. The formalization of this process is due
to~\citet{Daley1964,Daley1965}; individuals can be in one of three
possible states\footnote{For consistency, we use the same symbols of the
  SIR model.}: ignorant (S, equivalent to susceptible in SIR), spreader
(I, equivalent to infected) and stifler (R, equivalent to removed).  The
possible events, and the corresponding rates are: \be \left\{
    \begin{array}{rcl}
       S + I & \xrightarrow{\beta} & 2 I \\
       R + I & \xrightarrow{\alpha} & 2 R \\
       2 I & \xrightarrow{\alpha} & 2 R 
    \end{array} 
  \right. .
\label{DK}
\ee

In a slightly distinct version, introduced by~\citet{Maki1973},
the third process is different: when a spreader
contacts another agent and finds it in state I, only the former
turns into a stifler, the latter remaining unchanged, i.e. the third
process is
\be
       2 I \xrightarrow{\alpha} R + I .
\label{MT}
\ee

As for the SIR model, starting from a single informed individual the
rumor propagates through the network with an increase in the number of
spreaders. Asymptotically all spreaders turn into stiflers and in the
final absorbing state there are only ignorants or stiflers.  The
``reliability'', i.e. the fraction $r_{\infty}$ of stiflers in this
asymptotic state, quantifies whether the rumor remains localized
($r_{\infty} \to 0$ for system size $N \to \infty$) or spreads
macroscopically.  The solution of both versions of the model on the
complete graph~\citep{Sudbury1985,barratbook} gives the whole temporal
evolution of the reliability, yielding $r_{\infty}$ as the solution of
\be r_{\infty} = 1 - e^{-(1+\beta/\alpha) r_{\infty}}
\label{MT_r}
\ee
As a consequence, $r_{\infty}$ is positive for any $\beta/\alpha>0$. 
i.e. the rumor spreads macroscopically for any value of the
spreading parameters, at odds with what happens for the SIR dynamics,
which has a finite threshold for homogeneous networks.

Since models for disease epidemics are strongly affected by complex
topologies, it is natural to ask what happens for rumor dynamics.
When the Maki-Thompson model is simulated on scale-free networks it turns
out that heterogeneity hinders the propagation dynamics by reducing
the final reliability $r_{\infty}$, still without introducing a finite
threshold~\citep{Liu2003,Moreno2004a,Moreno2004b}.  
Why this happens is easily understood: large hubs are rapidly reached by
the rumor, but then they easily turn into stiflers, thus preventing the
further spreading of the rumor to their many other neighbors. This is
confirmed by the observation that the density of ignorants of degree $k$
at the end of the process decays exponentially with
$k$~\citep{Moreno2004a}.  Degree-based mean-field
approaches~\citep{Nekovee2007,Zhou2007} are in good agreement with the
numerical findings.  The phenomenology of rumor spreading is markedly
different from the behavior of the SIR model and this is due to the
healing mechanism involving two individuals, present in both
Maki-Thompson and Daley-Kendall dynamics. If spontaneous recovery is
also allowed with rate $\mu$, justified as the effect of forgetting, it
turns out that the model behaves exactly as SIR: macroscopic spreading
occurs only above a threshold inversely proportional to the second
moment $\av{k^2}$, which then vanishes in the large network size limit
for scale-free networks~\citep{Nekovee2007}. Again the interpretation of
this outcome is not difficult: the forgetting term is linear in the
density of spreaders and thus dominates for small densities, since the
healing terms, due to the processes in Eqs.~(\ref{DK}) and~(\ref{MT}),
are quadratic.

When the pattern of interactions among individuals is given by the
Watts-Strogatz topology, rumor dynamics gives rise to a nontrivial
phenomenon: a phase-transition occurring at a critical value of the
rewiring probability $p$~\citep{Zanette2001}: For large values of $p$
the network is essentially random and the rumor reaches a finite
fraction of the vertices. For small values of $p$ the spreading occurs
only in a finite neighborhood of the initiator, so that the density of
stiflers vanishes with the system size. In other transitions occurring
on the Watts-Strogatz network, the critical point scales to zero with
the system size $N$, a consequence of the fact that the geometric
crossover between a one-dimensional lattice and a small-world structure
scales as $1/N$~\citep{watts98}.  Strikingly instead, the threshold
$p_c$ for macroscopic rumor spreading converges to a finite value as the
system size grows.  This indicates that the transition cannot be
explained only in geometrical terms; some nontrivial interplay between
topology and dynamics is at work.  Interestingly, the transition at
finite $p_c$ persists also when an annealed Watts-Strogatz network is
considered.

Recently, some activity has been devoted to the investigation of the
role of influential spreaders in rumor spreading, in analogy to what has
been done for disease epidemics (Sec.~\ref{sec:5.A}).
\citet{Borge-Holthoefer2012a} have looked for the role of nodes with
large $K$-core index for the Maki-Thompson dynamics on several empirical
networks.  It turns out that the final density of stiflers does not
depend on the $K$-core value of the initiator. Nodes with high $K$-core
index are not good spreaders; they are reached fast by the rumor and
short-circuit its further spreading. An empirical investigation of
cascades on the Twitter social network~\citep{Borge-Holthoefer2012b}
points out instead that privileged spreaders (identified by large degree
$k$ or large $K$) do exist in real world spreading phenomena, in
patent contrast with the predictions of rumor spreading models. To
reconcile theoretical predictions and empirical observations it is
necessary to amend Maki-Thompson dynamics. Two possible modifications
are proposed in~\citet{Borge-Holthoefer2012c}. In one case individuals
are not always active and do not spread further twits reaching them
while inactive.  In the second an ignorant contacted by a spreader turns
into a spreader only with probability $p$, while with probability
(1-$p$) it turns {\em directly} into a stifler. Both modified rumor
spreading models are able to reproduce qualitatively the empirical
findings, provided (for the first) that the probability to be active is
proportional to the node degree or (for the second) that the probability
$p$ to actually spread is very small (of the order of $10^{-3}$).

\subsection{Empirical studies}

Empirical data for a large number of spreading processes in the
real world have been analyzed in terms of epidemic-like phenomena.
Here we outline some of the most important contributions.

\citet{Leskovec2007a} analyze an instance of viral
marketing, in the form of the email recommendation network for products
of a large retailer. There are large variations depending on the type
of goods recommended, its price and the community of customers targeted,
but in general recommendations turn out not to be
very effective and cascades of purchases are not very extended.
The key factor, different from disease epidemics, is that
the ``infection probability'' quickly saturates to a low value
with the number of recommendations received. Moreover, as an
individual sends more and more recommendations the success per
recommendation declines (high degree individuals are not so influent).
Overall, viral marketing is very different from epidemic-like spreading.

A case where cascades are large and the spreading is a real collective
phenomenon is the propagation of chain letters on the
internet. \citet{Liben-Nowell2008} found tree-like dissemination patterns,
very deep but not large.  A simple epidemic-like model,
with an individual having a probability to forward
the message to a fraction of his/her contacts, gives instead wide and
shallow trees.  More realistic propagations are obtained introducing
two additional ingredients, asynchronous response times and
"back-response''~\citep{Liben-Nowell2008}.

Cascading behavior in large blog graphs is actively
investigated~\citep{Gruhl2004,Adar2005}.
\citet{Leskovec2007} found that in this case cascades 
tend to be wide, not deep, with a size distribution
following a power law with slope -2.
The shape of cascades is often star-like. 
A single-parameter generative model (essentially a SIS-like
model in the absorbing phase)
is in good agreement with empirical observations regarding
frequent cascades shapes and size distributions.

Also the behavior of individuals is subject to social
influence and thus giving rise to collective spreading.
Obesity, smoking habits and even 
happiness~\citep{Christakis2007,Christakis2008,Fowler2008}
have been claimed to spread as epidemics in social networks
(see however~\citet{Shalizi2011} for a criticism of these results).
In a nice empirical investigation \citet{Centola2010}
analyzed an artificially structured online community, devised
to check whether spreading is favored by random unclustered
structures (as in the ``strength of weak ties'' 
hypothesis~\citep{Granovetter1973}) or by clustered ones with
larger diameter~\citep{Centola2007a}. 
The latter structures turn out to favor spreading, the more so for
increasing degree. At the individual level, the presence of 2 or 3
neighbors adopting a behavior leads to an increase in the probability of
doing the same. For 4 and more neighbors the probability remains
instead constant.

For a long time empirical investigations of spreading phenomena
suffered of the drawback that the network mediating the propagation
was unknown and its properties had to be in some way guessed from how
the spreading process itself unfolds.  Online social networks, (such
as Facebook and Twitter) are an ideal tool to bypass this problem as
they provide both the topology of existing connections and the actual
path followed by the spreading process on top of the contact
graph~\citep{Lerman10icwsm}.  In one of such social networks (Digg),
\citet{VerSteeg2011} find that while the network of contacts has a
scale-free degree distribution, the size of cascades is lognormally
distributed, with essentially all propagations limited to a fraction
smaller than $1\%$ of the whole network. Within the framework of a
SIR model this would imply that the spreading parameter of each
cascade is fine-tuned around the transition point.  Two additional
ingredients help to reconcile the empirical findings with models: on
the one hand Digg contact network has a high clustering and this
feature leads to a reduction of outbreak size; on the other hand, as
in~\citet{Centola2010}, the probability to transmit the spreading
quickly saturates as a function of the number of active
neighbors. 
Another empirical investigation of Digg~\citep{Doerr2012} (see also
\citet{PVM_lognormal_Digg_EJPb2011}) finds that links between friends
in the social network contribute surprisingly little to the
propagation of information.

Another critical element of the spreading of memes in modern
online social networks is the competition among a large number
of them.~\citet{Weng2013} have analyzed Twitter, finding a very
broad variability of the lifetime and popularity of spreading
memes. A minimalistic model, based on the heterogeneous structure
of Twitter graphs of followers and on ``limited attention'', i.e.
the survival of memes in agents' memory for only a finite time due
to competition with others,
is sufficient to reproduce the empirical findings. Surprisingly,
it is not necessary to assume a variability in the intrinsic appeal
of memes to explain the very heterogeneous persistence and popularity
of individual memes.

Another information spreading experiment was performed by
\citet{Iribarren2009}, in which subscribers to an online newsletter in
11 European countries were offered a reward to recommend it via
email. The recommendations were tracked at every step by means of
viral propagation and it was thus possible to reconstruct the
recommendation cascades originated by 7154 initiators. The topology
of the observed cascades is essentially tree-like, in agreement with
the results of \citet{Liben-Nowell2008}, and of very small size,
suggesting again a behavior at or below a possible critical point. The
heterogeneity of the viral spreading process was quantified by looking
at the distribution of time elapsed between receiving an invitation
email, and forwarding it to another individuals. This distribution can
be fitted to a long-tailed log-normal form. On the other hand, the
average number of informed individuals forwarding the message at time
$t$ was also found to decay slowly (with a log-normal shape), in
contrast with the exponential decay expected in epidemics below the
threshold. Similar results were reported for the retweet time of
Twitter messages, see \citet{PVM_Twitter_lognormal}.

\section{Outlook}

In the last years the whole field of epidemic modeling in networks has
enormously progressed in the understanding of the interplay between
network properties and contagion processes. We hope to have fairly
portrayed the major advances and achieved clarity of presentation on
the various theoretical and numerical approaches in a field that has literally exploded. However the results and
understanding achieved so far have opened the door to new questions and
problems, often stimulated by the availability of new data. For this
reason, the research activity has not slowed its pace and there is
still a number of major challenges.

As shown in the previous sections, we are just moving the first
steps to access to the mathematical and statistical laws that
characterize the co-evolution mechanisms between the network evolution
and the dynamical process. This is a key element in most social
networks, where it is almost impossible to disentangle the agents
cognitive processes shaping the network evolution and their
perception/awareness of the contagion processes. 

Indeed, the adaptive
behavior of individuals in response to the dynamical processes they are
involved in represents a serious theoretical challenge dealing with the
feedback among different and competing dynamical processes. 
For instance, some activity has already been devoted to coupled 
behavior-disease models and to the competition among different 
contagion processes in networks, as reviewed in the previous sections, 
but much more work is needed to build a comprehensive picture.
The final goal is not only to understand epidemic processes, and
predict their behavior, but also to control their dynamics. The
development of strategies for favoring or hindering contagion
processes is crucial in a wide range of applications that span from
the optimization of disease containment and eradication to viral
marketing. Also in this case, much more work is needed 
investigating how co-evolution and feedback mechanisms between the
network evolution and the spreading dynamics affect our influence and
ability to control epidemic processes.
 
Networks show also a large number of interdependencies of various
nature: physical interdependency when energy, material or people flow
from one infrastructure to another; cyber interdependency when
information is transmitted or exchanged; geographic interdependency
signaling the co-location of infrastructural elements; logical
interdependency due to financial, political coordination, etc.
Interdependence is obviously a major issue also in diffusion and spreading processes. One simple example is
provided by the spreading of information in communication networks
that induces an alteration of the physical proximity contact pattern of
individuals or of the flows and traffic of mobility infrastructure. This
has triggered interest in the understanding of contagion processes in
coupled interdependent networks~\cite{interdependent12} More broadly,
the community is becoming aware that, especially in the area of modern
social networks populating the information technology ecosystem,
epidemic spreading may occur on different interacting networks that
however affect each other. This is obviously the case of information
processes where different type of social communication networks (phone,
real-world, digital) coexist and contribute to the spreading
process.  This evidence has led recently to the introduction of
multilayer or multiplex networks \cite{Kivela2013,Boccaletti2014}. Multiplex networks
are defined by a set of $N$ nodes and a set of $L$ layers, which
represent ``dimensions'' or ``aspects'' that characterize a node. A
node can belong to any subset of layers, and edges represent
interactions between nodes belonging to the same layer.  We can
consider that a vertex is connected to itself across the different
layers, or allow for inter-layer connections between nodes in
different layers. Every layer is represented thus by a network, and
the whole multiplex by a set of interconnected networks.  The
analysis of epidemic processes in these networks shows very
interesting and peculiar behaviors. Several studies have focused  on physical-information layered 
networks and studied the epidemic dynamics on the different layers as a function of the 
inter-layer coupling and the epidemic threshold values on each layer~\cite{Marceau11,Yagan2013,Buono2014} 
For the SIR model it is also observed that depending on the average degree of
inter-layer connections~\cite{Dickison2012} a
global endemic state may arise in the interconnected system even if no
epidemics can survive in each network
separately~\cite{Saumell-Mendiola2012,DarabiSahneh_Scogio_interdependentnets2013}.  SIS
dynamics on multiple coupled layers is also analyzed
by~\citet{Cozzo2013} and by
\citet{DarabiSahneh_Scogio_interdependentnets2013} in a generalized
mean-field framework. However, epidemic behavior on multiplex networks
is still largely unexplored for more complex models, complex contagion
phenomena and in data-driven settings.

The ever increasing computational power is also favoring very detailed models that simulate large-scale
population networks, including geographic and demographic
attributes on an individual by individual basis.  These models
can generate information at unprecedented level of detail and guide
researchers in identifying typical non-linear behavior and
critical points that often challenge our intuition. These results
call for a theoretical understanding and a systematic classification
of the models' dynamical behaviors, thus adding transparency to the
numerical results.  Results raise new general questions such as:
What are the fundamental limits in the predictability of epidemics on
networks? How does our understanding depend on the level of data
aggregation and detail? What is the impact of the knowledge on the
state and initial conditions of the network on our understanding of
its dynamical behavior?  These are all major conceptual and technical
challenges that require the involvement of a vast research community
and a truly interdisciplinary approach, rooted in the combination of
large-scale data mining techniques, computational methods and
analytical techniques.

The study of epidemic spreading is a vibrant research area that is
finding more and more applications in a wide range of domains. The need
of quantitative and mathematical tools able to provide understanding
in areas ranging from infectious diseases to viral marketing is
fostering the intense research activity at the forefront in the
investigation of epidemic
spreading in networks. We hope that the present review will be a
valuable reference for all researchers that will engage in this field.

\section*{Acknowledgments}

R.P.-S. acknowledges financial support from the Spanish MINECO, under
projects Nos. FIS2010-21781-C02- 01 and FIS2013-47282-C2-2, EC
FET-Proactive Project MULTIPLEX (Grant No. 317532), and ICREA Academia,
funded by the Generalitat de Catalunya. P.V.M. was partially funded by
the EU CONGAS project (no. 288021).  A.V. has been partially funded by
the DTRA-1-0910039, NSF CMMI-1125095, MIDAS-National Institute of
General Medical Sciences U54GM111274 awards.  The views and conclusions
contained in this document are those of the authors and should not be
interpreted as representing the official policies, either expressed or
implied, of the funding agencies or the U.S. Government. We thank Nicole
Samay for help with the diagrams and figures.


\begin{thebibliography}{458}%
\makeatletter
\providecommand \@ifxundefined [1]{%
 \@ifx{#1\undefined}
}%
\providecommand \@ifnum [1]{%
 \ifnum #1\expandafter \@firstoftwo
 \else \expandafter \@secondoftwo
 \fi
}%
\providecommand \@ifx [1]{%
 \ifx #1\expandafter \@firstoftwo
 \else \expandafter \@secondoftwo
 \fi
}%
\providecommand \natexlab [1]{#1}%
\providecommand \enquote  [1]{``#1''}%
\providecommand \bibnamefont  [1]{#1}%
\providecommand \bibfnamefont [1]{#1}%
\providecommand \citenamefont [1]{#1}%
\providecommand \href@noop [0]{\@secondoftwo}%
\providecommand \href [0]{\begingroup \@sanitize@url \@href}%
\providecommand \@href[1]{\@@startlink{#1}\@@href}%
\providecommand \@@href[1]{\endgroup#1\@@endlink}%
\providecommand \@sanitize@url [0]{\catcode `\\12\catcode `\$12\catcode
  `\&12\catcode `\#12\catcode `\^12\catcode `\_12\catcode `\%12\relax}%
\providecommand \@@startlink[1]{}%
\providecommand \@@endlink[0]{}%
\providecommand \url  [0]{\begingroup\@sanitize@url \@url }%
\providecommand \@url [1]{\endgroup\@href {#1}{\urlprefix }}%
\providecommand \urlprefix  [0]{URL }%
\providecommand \Eprint [0]{\href }%
\providecommand \doibase [0]{http://dx.doi.org/}%
\providecommand \selectlanguage [0]{\@gobble}%
\providecommand \bibinfo  [0]{\@secondoftwo}%
\providecommand \bibfield  [0]{\@secondoftwo}%
\providecommand \translation [1]{[#1]}%
\providecommand \BibitemOpen [0]{}%
\providecommand \bibitemStop [0]{}%
\providecommand \bibitemNoStop [0]{.\EOS\space}%
\providecommand \EOS [0]{\spacefactor3000\relax}%
\providecommand \BibitemShut  [1]{\csname bibitem#1\endcsname}%
\let\auto@bib@innerbib\@empty
\bibitem [{\citenamefont {Abramson}\ and\ \citenamefont
  {Kuperman}(2001)}]{abramson01}%
  \BibitemOpen
  \bibfield  {author} {\bibinfo {author} {\bibnamefont {Abramson},
  \bibfnamefont {G.}}, \ and\ \bibinfo {author} {\bibfnamefont
  {M.}~\bibnamefont {Kuperman}}} (\bibinfo {year} {2001}),\ \href@noop {}
  {\bibfield  {journal} {\bibinfo  {journal} {Phys. Rev. Lett.}\ }\textbf
  {\bibinfo {volume} {86}},\ \bibinfo {pages} {2909}}\BibitemShut {NoStop}%
\bibitem [{\citenamefont {Adar}\ and\ \citenamefont {Adamic}(2005)}]{Adar2005}%
  \BibitemOpen
  \bibfield  {author} {\bibinfo {author} {\bibnamefont {Adar}, \bibfnamefont
  {E.}}, \ and\ \bibinfo {author} {\bibfnamefont {L.}~\bibnamefont {Adamic}}}
  (\bibinfo {year} {2005}),\ in\ \href
  {http://ieeexplore.ieee.org/xpls/abs\_all.jsp?arnumber=1517844} {\emph
  {\bibinfo {booktitle} {Proceedings of the 2005 IEEE/WIC/ACM International
  Conference on Web Intelligence}}},\ \bibinfo {series and number} {WI '05}\
  (\bibinfo  {publisher} {IEEE Computer Society},\ \bibinfo {address}
  {Washington, DC, USA})\ pp.\ \bibinfo {pages} {207--214}\BibitemShut
  {NoStop}%
\bibitem [{\citenamefont {Ahn}\ \emph {et~al.}(2006)\citenamefont {Ahn},
  \citenamefont {Jeong}, \citenamefont {Masuda},\ and\ \citenamefont
  {Noh}}]{Ahn2006}%
  \BibitemOpen
  \bibfield  {author} {\bibinfo {author} {\bibnamefont {Ahn}, \bibfnamefont
  {Y.-Y.}}, \bibinfo {author} {\bibfnamefont {H.}~\bibnamefont {Jeong}},
  \bibinfo {author} {\bibfnamefont {N.}~\bibnamefont {Masuda}}, \ and\ \bibinfo
  {author} {\bibfnamefont {J.~D.}\ \bibnamefont {Noh}}} (\bibinfo {year}
  {2006}),\ \href {\doibase 10.1103/PhysRevE.74.066113} {\bibfield  {journal}
  {\bibinfo  {journal} {Phys. Rev. E}\ }\textbf {\bibinfo {volume} {74}},\
  \bibinfo {pages} {066113}}\BibitemShut {NoStop}%
\bibitem [{\citenamefont {Ahnert}\ \emph {et~al.}(2007)\citenamefont {Ahnert},
  \citenamefont {Garlaschelli}, \citenamefont {Fink},\ and\ \citenamefont
  {Caldarelli}}]{PhysRevE.76.016101}%
  \BibitemOpen
  \bibfield  {author} {\bibinfo {author} {\bibnamefont {Ahnert}, \bibfnamefont
  {S.~E.}}, \bibinfo {author} {\bibfnamefont {D.}~\bibnamefont {Garlaschelli}},
  \bibinfo {author} {\bibfnamefont {T.~M.~A.}\ \bibnamefont {Fink}}, \ and\
  \bibinfo {author} {\bibfnamefont {G.}~\bibnamefont {Caldarelli}}} (\bibinfo
  {year} {2007}),\ \href {\doibase 10.1103/PhysRevE.76.016101} {\bibfield
  {journal} {\bibinfo  {journal} {Phys. Rev. E}\ }\textbf {\bibinfo {volume}
  {76}},\ \bibinfo {pages} {016101}}\BibitemShut {NoStop}%
\bibitem [{\citenamefont {Albert}\ and\ \citenamefont
  {Barab{\'a}si}(2002)}]{barabasi02}%
  \BibitemOpen
  \bibfield  {author} {\bibinfo {author} {\bibnamefont {Albert}, \bibfnamefont
  {R.}}, \ and\ \bibinfo {author} {\bibfnamefont {A.-L.}\ \bibnamefont
  {Barab{\'a}si}}} (\bibinfo {year} {2002}),\ \href@noop {} {\bibfield
  {journal} {\bibinfo  {journal} {Rev. Mod. Phys.}\ }\textbf {\bibinfo {volume}
  {74}},\ \bibinfo {pages} {47}}\BibitemShut {NoStop}%
\bibitem [{\citenamefont {Altarelli}\ \emph {et~al.}(2014)\citenamefont
  {Altarelli}, \citenamefont {Braunstein}, \citenamefont {Dall'Asta},
  \citenamefont {Lage-Castellanos},\ and\ \citenamefont
  {Zecchina}}]{Altarelli2013}%
  \BibitemOpen
  \bibfield  {author} {\bibinfo {author} {\bibnamefont {Altarelli},
  \bibfnamefont {F.}}, \bibinfo {author} {\bibfnamefont {A.}~\bibnamefont
  {Braunstein}}, \bibinfo {author} {\bibfnamefont {L.}~\bibnamefont
  {Dall'Asta}}, \bibinfo {author} {\bibfnamefont {A.}~\bibnamefont
  {Lage-Castellanos}}, \ and\ \bibinfo {author} {\bibfnamefont
  {R.}~\bibnamefont {Zecchina}}} (\bibinfo {year} {2014}),\ \href@noop {}
  {\bibfield  {journal} {\bibinfo  {journal} {Phys. Rev. Lett.}\ }\textbf
  {\bibinfo {volume} {112}},\ \bibinfo {pages} {118701}}\BibitemShut {NoStop}%
\bibitem [{\citenamefont {Ancel}\ \emph {et~al.}(2003)\citenamefont {Ancel},
  \citenamefont {Newman}, \citenamefont {Martin},\ and\ \citenamefont
  {Schrag}}]{Ancel-Meyers:2003fe}%
  \BibitemOpen
  \bibfield  {author} {\bibinfo {author} {\bibnamefont {Ancel}, \bibfnamefont
  {L.~W.}}, \bibinfo {author} {\bibfnamefont {M.~E.}\ \bibnamefont {Newman}},
  \bibinfo {author} {\bibfnamefont {M.}~\bibnamefont {Martin}}, \ and\ \bibinfo
  {author} {\bibfnamefont {S.}~\bibnamefont {Schrag}}} (\bibinfo {year}
  {2003}),\ \href {http://europepmc.org/abstract/MED/12603991} {\bibfield
  {journal} {\bibinfo  {journal} {Emerging Infectious Diseases}\ }\textbf
  {\bibinfo {volume} {9}},\ \bibinfo {pages} {204}}\BibitemShut {NoStop}%
\bibitem [{\citenamefont {Anderson}\ and\ \citenamefont
  {May}(1992)}]{anderson92}%
  \BibitemOpen
  \bibfield  {author} {\bibinfo {author} {\bibnamefont {Anderson},
  \bibfnamefont {R.~M.}}, \ and\ \bibinfo {author} {\bibfnamefont {R.~M.}\
  \bibnamefont {May}}} (\bibinfo {year} {1992}),\ \href@noop {} {\emph
  {\bibinfo {title} {Infectious diseases in humans}}}\ (\bibinfo  {publisher}
  {Oxford University Press},\ \bibinfo {address} {Oxford})\BibitemShut
  {NoStop}%
\bibitem [{\citenamefont {Andersson}\ and\ \citenamefont
  {Britton}(2000)}]{Andersson2000}%
  \BibitemOpen
  \bibfield  {author} {\bibinfo {author} {\bibnamefont {Andersson},
  \bibfnamefont {H.}}, \ and\ \bibinfo {author} {\bibfnamefont
  {T.}~\bibnamefont {Britton}}} (\bibinfo {year} {2000}),\ \href@noop {} {\emph
  {\bibinfo {title} {{Stochastic epidemic models and their statistical
  analysis}}}},\ \bibinfo {series} {Lecture Notes in Statistics}, Vol.\
  \bibinfo {volume} {151}\ (\bibinfo  {publisher} {Springer US},\ \bibinfo
  {address} {New York})\BibitemShut {NoStop}%
\bibitem [{\citenamefont {Apolloni}\ \emph {et~al.}(2013)\citenamefont
  {Apolloni}, \citenamefont {Poletto},\ and\ \citenamefont
  {Colizza}}]{Apolloni:2013}%
  \BibitemOpen
  \bibfield  {author} {\bibinfo {author} {\bibnamefont {Apolloni},
  \bibfnamefont {A.}}, \bibinfo {author} {\bibfnamefont {C.}~\bibnamefont
  {Poletto}}, \ and\ \bibinfo {author} {\bibfnamefont {V.}~\bibnamefont
  {Colizza}}} (\bibinfo {year} {2013}),\ \href {\doibase
  10.1186/1471-2334-13-176} {\bibfield  {journal} {\bibinfo  {journal} {BMC
  Infectious Diseases}\ }\textbf {\bibinfo {volume} {13}},\ \bibinfo {pages}
  {176}}\BibitemShut {NoStop}%
\bibitem [{\citenamefont {Bagnoli}\ \emph {et~al.}(2007)\citenamefont
  {Bagnoli}, \citenamefont {Li\`o},\ and\ \citenamefont
  {Sguanci}}]{Bagnoli2007}%
  \BibitemOpen
  \bibfield  {author} {\bibinfo {author} {\bibnamefont {Bagnoli}, \bibfnamefont
  {F.}}, \bibinfo {author} {\bibfnamefont {P.}~\bibnamefont {Li\`o}}, \ and\
  \bibinfo {author} {\bibfnamefont {L.}~\bibnamefont {Sguanci}}} (\bibinfo
  {year} {2007}),\ \href {\doibase 10.1103/PhysRevE.76.061904} {\bibfield
  {journal} {\bibinfo  {journal} {Phys. Rev. E}\ }\textbf {\bibinfo {volume}
  {76}},\ \bibinfo {pages} {061904}}\BibitemShut {NoStop}%
\bibitem [{\citenamefont {Bailey}(1975)}]{Bailey_book}%
  \BibitemOpen
  \bibfield  {author} {\bibinfo {author} {\bibnamefont {Bailey}, \bibfnamefont
  {N.~T.~J.}}} (\bibinfo {year} {1975}),\ \href@noop {} {\emph {\bibinfo
  {title} {The Mathematical Theory of Infectious Diseases and its
  Applications}}},\ \bibinfo {edition} {2nd}\ ed.\ (\bibinfo  {publisher}
  {Charlin Griffin \& Company},\ \bibinfo {address} {London})\BibitemShut
  {NoStop}%
\bibitem [{\citenamefont {Bajardi}\ \emph {et~al.}(2011)\citenamefont
  {Bajardi}, \citenamefont {Poletto}, \citenamefont {Ramasco}, \citenamefont
  {Tizzoni}, \citenamefont {Colizza},\ and\ \citenamefont
  {Vespignani}}]{Bajardi:2011}%
  \BibitemOpen
  \bibfield  {author} {\bibinfo {author} {\bibnamefont {Bajardi}, \bibfnamefont
  {P.}}, \bibinfo {author} {\bibfnamefont {C.}~\bibnamefont {Poletto}},
  \bibinfo {author} {\bibfnamefont {J.}~\bibnamefont {Ramasco}}, \bibinfo
  {author} {\bibfnamefont {M.}~\bibnamefont {Tizzoni}}, \bibinfo {author}
  {\bibfnamefont {V.}~\bibnamefont {Colizza}}, \ and\ \bibinfo {author}
  {\bibfnamefont {A.}~\bibnamefont {Vespignani}}} (\bibinfo {year} {2011}),\
  \href {\doibase 10.1371/journal.pone.0016591} {\bibfield  {journal} {\bibinfo
   {journal} {PLoS One}\ }\textbf {\bibinfo {volume} {6}},\ \bibinfo {pages}
  {e16591}}\BibitemShut {NoStop}%
\bibitem [{\citenamefont {Balcan}\ \emph
  {et~al.}(2009{\natexlab{a}})\citenamefont {Balcan}, \citenamefont {Colizza},
  \citenamefont {Goncalves}, \citenamefont {Hu}, \citenamefont {Ramasco},\ and\
  \citenamefont {Vespignani}}]{Balcan2009}%
  \BibitemOpen
  \bibfield  {author} {\bibinfo {author} {\bibnamefont {Balcan}, \bibfnamefont
  {D.}}, \bibinfo {author} {\bibfnamefont {V.}~\bibnamefont {Colizza}},
  \bibinfo {author} {\bibfnamefont {B.}~\bibnamefont {Goncalves}}, \bibinfo
  {author} {\bibfnamefont {H.}~\bibnamefont {Hu}}, \bibinfo {author}
  {\bibfnamefont {J.}~\bibnamefont {Ramasco}}, \ and\ \bibinfo {author}
  {\bibfnamefont {A.}~\bibnamefont {Vespignani}}} (\bibinfo {year}
  {2009}{\natexlab{a}}),\ \href@noop {} {\bibfield  {journal} {\bibinfo
  {journal} {Proceedings of the National Academy of Sciences}\ }\textbf
  {\bibinfo {volume} {106}},\ \bibinfo {pages} {21484}}\BibitemShut {NoStop}%
\bibitem [{\citenamefont {Balcan}\ \emph
  {et~al.}(2009{\natexlab{b}})\citenamefont {Balcan}, \citenamefont {Hu},
  \citenamefont {Goncalves}, \citenamefont {Bajardi}, \citenamefont {Poletto},
  \citenamefont {Ramasco}, \citenamefont {Paolotti}, \citenamefont {Perra},
  \citenamefont {Tizzoni}, \citenamefont {Van Den~Broeck}, \citenamefont
  {Colizza},\ and\ \citenamefont {Vespignani}}]{Balcan:2009BMC}%
  \BibitemOpen
  \bibfield  {author} {\bibinfo {author} {\bibnamefont {Balcan}, \bibfnamefont
  {D.}}, \bibinfo {author} {\bibfnamefont {H.}~\bibnamefont {Hu}}, \bibinfo
  {author} {\bibfnamefont {B.}~\bibnamefont {Goncalves}}, \bibinfo {author}
  {\bibfnamefont {P.}~\bibnamefont {Bajardi}}, \bibinfo {author} {\bibfnamefont
  {C.}~\bibnamefont {Poletto}}, \bibinfo {author} {\bibfnamefont
  {J.}~\bibnamefont {Ramasco}}, \bibinfo {author} {\bibfnamefont
  {D.}~\bibnamefont {Paolotti}}, \bibinfo {author} {\bibfnamefont
  {N.}~\bibnamefont {Perra}}, \bibinfo {author} {\bibfnamefont
  {M.}~\bibnamefont {Tizzoni}}, \bibinfo {author} {\bibfnamefont
  {W.}~\bibnamefont {Van Den~Broeck}}, \bibinfo {author} {\bibfnamefont
  {V.}~\bibnamefont {Colizza}}, \ and\ \bibinfo {author} {\bibfnamefont
  {A.}~\bibnamefont {Vespignani}}} (\bibinfo {year} {2009}{\natexlab{b}}),\
  \href {\doibase 10.1186/1741-7015-7-45} {\bibfield  {journal} {\bibinfo
  {journal} {BMC Med}\ }\textbf {\bibinfo {volume} {7}},\ \bibinfo {pages}
  {45}}\BibitemShut {NoStop}%
\bibitem [{\citenamefont {Balcan}\ and\ \citenamefont
  {Vespignani}(2011)}]{Balcan:2011}%
  \BibitemOpen
  \bibfield  {author} {\bibinfo {author} {\bibnamefont {Balcan}, \bibfnamefont
  {D.}}, \ and\ \bibinfo {author} {\bibfnamefont {A.}~\bibnamefont
  {Vespignani}}} (\bibinfo {year} {2011}),\ \href {\doibase 10.1038/nphys1944}
  {\bibfield  {journal} {\bibinfo  {journal} {Nat Phys}\ }\textbf {\bibinfo
  {volume} {7}},\ \bibinfo {pages} {581}}\BibitemShut {NoStop}%
\bibitem [{\citenamefont {Balcan}\ and\ \citenamefont
  {Vespignani}(2012)}]{Balcan:2012}%
  \BibitemOpen
  \bibfield  {author} {\bibinfo {author} {\bibnamefont {Balcan}, \bibfnamefont
  {D.}}, \ and\ \bibinfo {author} {\bibfnamefont {A.}~\bibnamefont
  {Vespignani}}} (\bibinfo {year} {2012}),\ \href {\doibase
  http://dx.doi.org/10.1016/j.jtbi.2011.10.010} {\bibfield  {journal} {\bibinfo
   {journal} {Journal of Theoretical Biology}\ }\textbf {\bibinfo {volume}
  {293}},\ \bibinfo {pages} {87 }}\BibitemShut {NoStop}%
\bibitem [{\citenamefont {Ball}\ \emph {et~al.}(1997)\citenamefont {Ball},
  \citenamefont {Mollison},\ and\ \citenamefont {Scalia-Tomba}}]{Ball:1997}%
  \BibitemOpen
  \bibfield  {author} {\bibinfo {author} {\bibnamefont {Ball}, \bibfnamefont
  {F.}}, \bibinfo {author} {\bibfnamefont {D.}~\bibnamefont {Mollison}}, \ and\
  \bibinfo {author} {\bibfnamefont {G.}~\bibnamefont {Scalia-Tomba}}} (\bibinfo
  {year} {1997}),\ \href@noop {} {\bibfield  {journal} {\bibinfo  {journal}
  {Ann Appl Probab}\ }\textbf {\bibinfo {volume} {7}},\ \bibinfo {pages}
  {46}}\BibitemShut {NoStop}%
\bibitem [{\citenamefont {Bancal}\ and\ \citenamefont
  {Pastor-Satorras}(2010)}]{ForestFireSatorras09}%
  \BibitemOpen
  \bibfield  {author} {\bibinfo {author} {\bibnamefont {Bancal}, \bibfnamefont
  {J.-D.}}, \ and\ \bibinfo {author} {\bibfnamefont {R.}~\bibnamefont
  {Pastor-Satorras}}} (\bibinfo {year} {2010}),\ \href@noop {} {\bibfield
  {journal} {\bibinfo  {journal} {Eur. Phys. J. B}\ }\textbf {\bibinfo {volume}
  {76}},\ \bibinfo {pages} {109}}\BibitemShut {NoStop}%
\bibitem [{\citenamefont {Barab{\'a}si}\ and\ \citenamefont
  {Albert}(1999)}]{Barabasi:1999}%
  \BibitemOpen
  \bibfield  {author} {\bibinfo {author} {\bibnamefont {Barab{\'a}si},
  \bibfnamefont {A.-L.}}, \ and\ \bibinfo {author} {\bibfnamefont
  {R.}~\bibnamefont {Albert}}} (\bibinfo {year} {1999}),\ \href@noop {}
  {\bibfield  {journal} {\bibinfo  {journal} {Science}\ }\textbf {\bibinfo
  {volume} {286}},\ \bibinfo {pages} {509}}\BibitemShut {NoStop}%
\bibitem [{\citenamefont {BarabasiLab}(2014)}]{networksciencebook}%
  \BibitemOpen
  \bibfield  {author} {\bibinfo {author} {\bibnamefont {BarabasiLab},}}
  (\bibinfo {year} {2014}),\ \href@noop {} {\enquote {\bibinfo {title} {Network
  science book},}\ }\bibinfo {howpublished}
  {\url{http://barabasilab.neu.edu/networksciencebook/}}\BibitemShut {NoStop}%
\bibitem [{\citenamefont {Baronchelli}\ \emph {et~al.}(2008)\citenamefont
  {Baronchelli}, \citenamefont {Catanzaro},\ and\ \citenamefont
  {Pastor-Satorras}}]{baronchelli08:_boson_react}%
  \BibitemOpen
  \bibfield  {author} {\bibinfo {author} {\bibnamefont {Baronchelli},
  \bibfnamefont {A.}}, \bibinfo {author} {\bibfnamefont {M.}~\bibnamefont
  {Catanzaro}}, \ and\ \bibinfo {author} {\bibfnamefont {R.}~\bibnamefont
  {Pastor-Satorras}}} (\bibinfo {year} {2008}),\ \href@noop {} {\bibfield
  {journal} {\bibinfo  {journal} {Phys. Rev. E}\ }\textbf {\bibinfo {volume}
  {78}},\ \bibinfo {pages} {016111}}\BibitemShut {NoStop}%
\bibitem [{\citenamefont {Baronchelli}\ \emph {et~al.}(2013)\citenamefont
  {Baronchelli}, \citenamefont {{Ferrer-i-Cancho}}, \citenamefont
  {Pastor-Satorras}, \citenamefont {Chater},\ and\ \citenamefont
  {Christiansen}}]{baronchelli13}%
  \BibitemOpen
  \bibfield  {author} {\bibinfo {author} {\bibnamefont {Baronchelli},
  \bibfnamefont {A.}}, \bibinfo {author} {\bibfnamefont {R.}~\bibnamefont
  {{Ferrer-i-Cancho}}}, \bibinfo {author} {\bibfnamefont {R.}~\bibnamefont
  {Pastor-Satorras}}, \bibinfo {author} {\bibfnamefont {N.}~\bibnamefont
  {Chater}}, \ and\ \bibinfo {author} {\bibfnamefont {M.~H.}\ \bibnamefont
  {Christiansen}}} (\bibinfo {year} {2013}),\ \href@noop {} {\bibfield
  {journal} {\bibinfo  {journal} {Trends in Cognitive Sciences}\ }\textbf
  {\bibinfo {volume} {17}},\ \bibinfo {pages} {348}}\BibitemShut {NoStop}%
\bibitem [{\citenamefont {Baronchelli}\ and\ \citenamefont
  {Pastor-Satorras}(2010)}]{dynam_in_weigh_networ}%
  \BibitemOpen
  \bibfield  {author} {\bibinfo {author} {\bibnamefont {Baronchelli},
  \bibfnamefont {A.}}, \ and\ \bibinfo {author} {\bibfnamefont
  {R.}~\bibnamefont {Pastor-Satorras}}} (\bibinfo {year} {2010}),\ \href@noop
  {} {\bibfield  {journal} {\bibinfo  {journal} {Phys. Rev. E}\ }\textbf
  {\bibinfo {volume} {82}},\ \bibinfo {pages} {011111}}\BibitemShut {NoStop}%
\bibitem [{\citenamefont {Baroyan}\ \emph {et~al.}(1969)\citenamefont
  {Baroyan}, \citenamefont {Genchikov}, \citenamefont {Rvachev},\ and\
  \citenamefont {Shashkov}}]{Baroyan:1969}%
  \BibitemOpen
  \bibfield  {author} {\bibinfo {author} {\bibnamefont {Baroyan}, \bibfnamefont
  {O.~V.}}, \bibinfo {author} {\bibfnamefont {L.~A.}\ \bibnamefont
  {Genchikov}}, \bibinfo {author} {\bibfnamefont {L.~A.}\ \bibnamefont
  {Rvachev}}, \ and\ \bibinfo {author} {\bibfnamefont {V.~A.}\ \bibnamefont
  {Shashkov}}} (\bibinfo {year} {1969}),\ \href@noop {} {\bibfield  {journal}
  {\bibinfo  {journal} {Bull. Int. Epidemiol. Assoc.}\ }\textbf {\bibinfo
  {volume} {18}},\ \bibinfo {pages} {22}}\BibitemShut {NoStop}%
\bibitem [{\citenamefont {Barrat}\ \emph
  {et~al.}(2004{\natexlab{a}})\citenamefont {Barrat}, \citenamefont
  {Barth{\'e}lemy}, \citenamefont {Pastor-Satorras},\ and\ \citenamefont
  {Vespignani}}]{Barrat16032004}%
  \BibitemOpen
  \bibfield  {author} {\bibinfo {author} {\bibnamefont {Barrat}, \bibfnamefont
  {A.}}, \bibinfo {author} {\bibfnamefont {M.}~\bibnamefont {Barth{\'e}lemy}},
  \bibinfo {author} {\bibfnamefont {R.}~\bibnamefont {Pastor-Satorras}}, \ and\
  \bibinfo {author} {\bibfnamefont {A.}~\bibnamefont {Vespignani}}} (\bibinfo
  {year} {2004}{\natexlab{a}}),\ \href@noop {} {\bibfield  {journal} {\bibinfo
  {journal} {Proc. Natl. Acad. Sci. USA}\ }\textbf {\bibinfo {volume} {101}},\
  \bibinfo {pages} {3747}}\BibitemShut {NoStop}%
\bibitem [{\citenamefont {Barrat}\ \emph
  {et~al.}(2004{\natexlab{b}})\citenamefont {Barrat}, \citenamefont
  {Barth{\'e}lemy},\ and\ \citenamefont {Vespignani}}]{barrat04:_weigh}%
  \BibitemOpen
  \bibfield  {author} {\bibinfo {author} {\bibnamefont {Barrat}, \bibfnamefont
  {A.}}, \bibinfo {author} {\bibfnamefont {M.}~\bibnamefont {Barth{\'e}lemy}},
  \ and\ \bibinfo {author} {\bibfnamefont {A.}~\bibnamefont {Vespignani}}}
  (\bibinfo {year} {2004}{\natexlab{b}}),\ \href@noop {} {\bibfield  {journal}
  {\bibinfo  {journal} {Phys. Rev. Lett.}\ }\textbf {\bibinfo {volume} {92}},\
  \bibinfo {pages} {228701}}\BibitemShut {NoStop}%
\bibitem [{\citenamefont {Barrat}\ \emph {et~al.}(2008)\citenamefont {Barrat},
  \citenamefont {Barth\'{e}lemy},\ and\ \citenamefont
  {Vespignani}}]{barratbook}%
  \BibitemOpen
  \bibfield  {author} {\bibinfo {author} {\bibnamefont {Barrat}, \bibfnamefont
  {A.}}, \bibinfo {author} {\bibfnamefont {M.}~\bibnamefont {Barth\'{e}lemy}},
  \ and\ \bibinfo {author} {\bibfnamefont {A.}~\bibnamefont {Vespignani}}}
  (\bibinfo {year} {2008}),\ \href@noop {} {\emph {\bibinfo {title} {Dynamical
  Processes on Complex Networks}}}\ (\bibinfo  {publisher} {Cambridge
  University Press},\ \bibinfo {address} {Cambridge})\BibitemShut {NoStop}%
\bibitem [{\citenamefont {Barth{\'e}lemy}\ \emph {et~al.}(2004)\citenamefont
  {Barth{\'e}lemy}, \citenamefont {Barrat}, \citenamefont {Pastor-Satorras},\
  and\ \citenamefont {Vespignani}}]{sievolution}%
  \BibitemOpen
  \bibfield  {author} {\bibinfo {author} {\bibnamefont {Barth{\'e}lemy},
  \bibfnamefont {M.}}, \bibinfo {author} {\bibfnamefont {A.}~\bibnamefont
  {Barrat}}, \bibinfo {author} {\bibfnamefont {R.}~\bibnamefont
  {Pastor-Satorras}}, \ and\ \bibinfo {author} {\bibfnamefont {A.}~\bibnamefont
  {Vespignani}}} (\bibinfo {year} {2004}),\ \href@noop {} {\bibfield  {journal}
  {\bibinfo  {journal} {Phys. Rev. Lett.}\ }\textbf {\bibinfo {volume} {92}},\
  \bibinfo {pages} {178701}}\BibitemShut {NoStop}%
\bibitem [{\citenamefont {Barthelemy}\ \emph {et~al.}(2005)\citenamefont
  {Barthelemy}, \citenamefont {Barrat}, \citenamefont {Pastor-Satorras},\ and\
  \citenamefont {Vespignani}}]{Barthelemy2005275}%
  \BibitemOpen
  \bibfield  {author} {\bibinfo {author} {\bibnamefont {Barthelemy},
  \bibfnamefont {M.}}, \bibinfo {author} {\bibfnamefont {A.}~\bibnamefont
  {Barrat}}, \bibinfo {author} {\bibfnamefont {R.}~\bibnamefont
  {Pastor-Satorras}}, \ and\ \bibinfo {author} {\bibfnamefont {A.}~\bibnamefont
  {Vespignani}}} (\bibinfo {year} {2005}),\ \href@noop {} {\bibfield  {journal}
  {\bibinfo  {journal} {Journal of Theoretical Biology}\ }\textbf {\bibinfo
  {volume} {235}},\ \bibinfo {pages} {275 }}\BibitemShut {NoStop}%
\bibitem [{\citenamefont {Bass}(1969)}]{Bass1969}%
  \BibitemOpen
  \bibfield  {author} {\bibinfo {author} {\bibnamefont {Bass}, \bibfnamefont
  {F.~M.}}} (\bibinfo {year} {1969}),\ \href
  {http://ideas.repec.org/a/inm/ormnsc/v15y1969i5p215-227.html} {\bibfield
  {journal} {\bibinfo  {journal} {Management Science}\ }\textbf {\bibinfo
  {volume} {15}},\ \bibinfo {pages} {215}}\BibitemShut {NoStop}%
\bibitem [{\citenamefont {Bauch}\ and\ \citenamefont
  {Galvani}(2013)}]{Galvani2013}%
  \BibitemOpen
  \bibfield  {author} {\bibinfo {author} {\bibnamefont {Bauch}, \bibfnamefont
  {C.~T.}}, \ and\ \bibinfo {author} {\bibfnamefont {A.~P.}\ \bibnamefont
  {Galvani}}} (\bibinfo {year} {2013}),\ \href {\doibase
  10.1126/science.1244492} {\bibfield  {journal} {\bibinfo  {journal}
  {Science}\ }\textbf {\bibinfo {volume} {342}},\ \bibinfo {pages}
  {47}}\BibitemShut {NoStop}%
\bibitem [{\citenamefont {Bauer}\ and\ \citenamefont
  {Lizier}(2012)}]{Bauer2012}%
  \BibitemOpen
  \bibfield  {author} {\bibinfo {author} {\bibnamefont {Bauer}, \bibfnamefont
  {F.}}, \ and\ \bibinfo {author} {\bibfnamefont {J.~T.}\ \bibnamefont
  {Lizier}}} (\bibinfo {year} {2012}),\ \href {\doibase
  10.1209/0295-5075/99/68007} {\bibfield  {journal} {\bibinfo  {journal}
  {Europhysics Letters}\ }\textbf {\bibinfo {volume} {99}},\ \bibinfo {pages}
  {68007}}\BibitemShut {NoStop}%
\bibitem [{\citenamefont {Belik}\ \emph {et~al.}(2011)\citenamefont {Belik},
  \citenamefont {Geisel},\ and\ \citenamefont {Brockmann}}]{Belik:2011}%
  \BibitemOpen
  \bibfield  {author} {\bibinfo {author} {\bibnamefont {Belik}, \bibfnamefont
  {V.}}, \bibinfo {author} {\bibfnamefont {T.}~\bibnamefont {Geisel}}, \ and\
  \bibinfo {author} {\bibfnamefont {D.}~\bibnamefont {Brockmann}}} (\bibinfo
  {year} {2011}),\ \href@noop {} {\bibfield  {journal} {\bibinfo  {journal}
  {Phys Rev X}\ }\textbf {\bibinfo {volume} {1}},\ \bibinfo {pages}
  {011001}}\BibitemShut {NoStop}%
\bibitem [{\citenamefont {{ben-Avraham}}\ and\ \citenamefont
  {Havlin}(2005)}]{havlin_diffusion_reaction}%
  \BibitemOpen
  \bibfield  {author} {\bibinfo {author} {\bibnamefont {{ben-Avraham}},
  \bibfnamefont {D.}}, \ and\ \bibinfo {author} {\bibfnamefont
  {S.}~\bibnamefont {Havlin}}} (\bibinfo {year} {2005}),\ \href@noop {} {\emph
  {\bibinfo {title} {{Diffusion and Reactions in Fractals and Disordered
  Systems}}}}\ (\bibinfo  {publisher} {Cambridge University Press},\ \bibinfo
  {address} {Cambridge, U.K.})\BibitemShut {NoStop}%
\bibitem [{\citenamefont {{ben-Avraham}}\ and\ \citenamefont
  {K\"ohler}(1992)}]{PhysRevA.45.8358}%
  \BibitemOpen
  \bibfield  {author} {\bibinfo {author} {\bibnamefont {{ben-Avraham}},
  \bibfnamefont {D.}}, \ and\ \bibinfo {author} {\bibfnamefont
  {J.}~\bibnamefont {K\"ohler}}} (\bibinfo {year} {1992}),\ \href {\doibase
  10.1103/PhysRevA.45.8358} {\bibfield  {journal} {\bibinfo  {journal} {Phys.
  Rev. A}\ }\textbf {\bibinfo {volume} {45}},\ \bibinfo {pages}
  {8358}}\BibitemShut {NoStop}%
\bibitem [{\citenamefont {Ben-Zion}\ \emph {et~al.}(2010)\citenamefont
  {Ben-Zion}, \citenamefont {Cohen},\ and\ \citenamefont
  {Shnerb}}]{Ben-Zion:2010}%
  \BibitemOpen
  \bibfield  {author} {\bibinfo {author} {\bibnamefont {Ben-Zion},
  \bibfnamefont {Y.}}, \bibinfo {author} {\bibfnamefont {Y.}~\bibnamefont
  {Cohen}}, \ and\ \bibinfo {author} {\bibfnamefont {N.~M.}\ \bibnamefont
  {Shnerb}}} (\bibinfo {year} {2010}),\ \href {\doibase
  http://dx.doi.org/10.1016/j.jtbi.2010.01.029} {\bibfield  {journal} {\bibinfo
   {journal} {Journal of Theoretical Biology}\ }\textbf {\bibinfo {volume}
  {264}},\ \bibinfo {pages} {197 }}\BibitemShut {NoStop}%
\bibitem [{\citenamefont {Bender}\ and\ \citenamefont
  {Canfield}(1978)}]{benderoriginal}%
  \BibitemOpen
  \bibfield  {author} {\bibinfo {author} {\bibnamefont {Bender}, \bibfnamefont
  {E.~A.}}, \ and\ \bibinfo {author} {\bibfnamefont {E.~R.}\ \bibnamefont
  {Canfield}}} (\bibinfo {year} {1978}),\ \href@noop {} {\bibfield  {journal}
  {\bibinfo  {journal} {Journal of Combinatorial Theory A}\ }\textbf {\bibinfo
  {volume} {24}},\ \bibinfo {pages} {296}}\BibitemShut {NoStop}%
\bibitem [{\citenamefont {Bernoulli}(1760)}]{bernoulli1760}%
  \BibitemOpen
  \bibfield  {author} {\bibinfo {author} {\bibnamefont {Bernoulli},
  \bibfnamefont {D.}}} (\bibinfo {year} {1760}),\ \href@noop {} {\bibinfo
  {journal} {Mem. Math. Phys. Acad. Roy. Sci., Paris}\ }\BibitemShut {NoStop}%
\bibitem [{\citenamefont {Bettencourt}\ \emph {et~al.}(2006)\citenamefont
  {Bettencourt}, \citenamefont {Cintr\'{o}n-Arias}, \citenamefont {Kaiser},\
  and\ \citenamefont {Castillo-Ch\'{a}vez}}]{Bettencourt2006}%
  \BibitemOpen
\bibfield  {journal} {  }\bibfield  {author} {\bibinfo {author} {\bibnamefont
  {Bettencourt}, \bibfnamefont {L.~M.}}, \bibinfo {author} {\bibfnamefont
  {A.}~\bibnamefont {Cintr\'{o}n-Arias}}, \bibinfo {author} {\bibfnamefont
  {D.~a.~I.}\ \bibnamefont {Kaiser}}, \ and\ \bibinfo {author} {\bibfnamefont
  {C.}~\bibnamefont {Castillo-Ch\'{a}vez}}} (\bibinfo {year} {2006}),\ \href
  {\doibase 10.1016/j.physa.2005.08.083} {\bibinfo  {journal} {Physica A}\ ,\
  \bibinfo {pages} {513}}\BibitemShut {NoStop}%
\bibitem [{\citenamefont {Bikhchandani}\ \emph {et~al.}(1992)\citenamefont
  {Bikhchandani}, \citenamefont {Hirshleifer},\ and\ \citenamefont
  {Welch}}]{Bikhchandani1992}%
  \BibitemOpen
\bibfield  {journal} {  }\bibfield  {author} {\bibinfo {author} {\bibnamefont
  {Bikhchandani}, \bibfnamefont {S.}}, \bibinfo {author} {\bibfnamefont
  {D.}~\bibnamefont {Hirshleifer}}, \ and\ \bibinfo {author} {\bibfnamefont
  {I.}~\bibnamefont {Welch}}} (\bibinfo {year} {1992}),\ \href@noop {}
  {\bibinfo  {journal} {Journal of Political Economy}\ ,\ \bibinfo {pages}
  {992}}\BibitemShut {NoStop}%
\bibitem [{\citenamefont {Binder}\ and\ \citenamefont
  {Heermann}(2010)}]{Binder2010}%
  \BibitemOpen
\bibfield  {journal} {  }\bibfield  {author} {\bibinfo {author} {\bibnamefont
  {Binder}, \bibfnamefont {K.}}, \ and\ \bibinfo {author} {\bibfnamefont
  {D.~W.}\ \bibnamefont {Heermann}}} (\bibinfo {year} {2010}),\ \href@noop {}
  {\emph {\bibinfo {title} {Monte Carlo Simulation in Statistical Physics}}},\
  \bibinfo {edition} {5th}\ ed.\ (\bibinfo  {publisher} {Springer-Verlag},\
  \bibinfo {address} {Berlin})\BibitemShut {NoStop}%
\bibitem [{\citenamefont {Bisanzio}\ \emph {et~al.}(2010)\citenamefont
  {Bisanzio}, \citenamefont {Bertolotti}, \citenamefont {Tomassone},
  \citenamefont {Amore}, \citenamefont {Ragagli}, \citenamefont {Mannelli},
  \citenamefont {Giacobini},\ and\ \citenamefont
  {Provero}}]{10.1371/journal.pone.0013796}%
  \BibitemOpen
  \bibfield  {author} {\bibinfo {author} {\bibnamefont {Bisanzio},
  \bibfnamefont {D.}}, \bibinfo {author} {\bibfnamefont {L.}~\bibnamefont
  {Bertolotti}}, \bibinfo {author} {\bibfnamefont {L.}~\bibnamefont
  {Tomassone}}, \bibinfo {author} {\bibfnamefont {G.}~\bibnamefont {Amore}},
  \bibinfo {author} {\bibfnamefont {C.}~\bibnamefont {Ragagli}}, \bibinfo
  {author} {\bibfnamefont {A.}~\bibnamefont {Mannelli}}, \bibinfo {author}
  {\bibfnamefont {M.}~\bibnamefont {Giacobini}}, \ and\ \bibinfo {author}
  {\bibfnamefont {P.}~\bibnamefont {Provero}}} (\bibinfo {year} {2010}),\ \href
  {\doibase 10.1371/journal.pone.0013796} {\bibfield  {journal} {\bibinfo
  {journal} {PLoS ONE}\ }\textbf {\bibinfo {volume} {5}},\ \bibinfo {pages}
  {e13796}}\BibitemShut {NoStop}%
\bibitem [{\citenamefont {Blythe}\ and\ \citenamefont
  {Anderson}(1988)}]{BLYTHE01011988}%
  \BibitemOpen
  \bibfield  {author} {\bibinfo {author} {\bibnamefont {Blythe}, \bibfnamefont
  {S.~P.}}, \ and\ \bibinfo {author} {\bibfnamefont {R.~M.}\ \bibnamefont
  {Anderson}}} (\bibinfo {year} {1988}),\ \href@noop {} {\bibfield  {journal}
  {\bibinfo  {journal} {Mathematical Medicine and Biology}\ }\textbf {\bibinfo
  {volume} {5}},\ \bibinfo {pages} {181}}\BibitemShut {NoStop}%
\bibitem [{\citenamefont {Boccaletti}\ \emph {et~al.}(2014)\citenamefont
  {Boccaletti}, \citenamefont {Bianconi}, \citenamefont {Criado}, \citenamefont
  {del Genio}, \citenamefont {Gomez-Gardenes}, \citenamefont {Romance},
  \citenamefont {Sendina-Nadal}, \citenamefont {Wang},\ and\ \citenamefont
  {Zanin}}]{Boccaletti2014}%
  \BibitemOpen
  \bibfield  {author} {\bibinfo {author} {\bibnamefont {Boccaletti},
  \bibfnamefont {S.}}, \bibinfo {author} {\bibfnamefont {G.}~\bibnamefont
  {Bianconi}}, \bibinfo {author} {\bibfnamefont {R.}~\bibnamefont {Criado}},
  \bibinfo {author} {\bibfnamefont {C.}~\bibnamefont {del Genio}}, \bibinfo
  {author} {\bibfnamefont {J.}~\bibnamefont {Gomez-Gardenes}}, \bibinfo
  {author} {\bibfnamefont {M.}~\bibnamefont {Romance}}, \bibinfo {author}
  {\bibfnamefont {I.}~\bibnamefont {Sendina-Nadal}}, \bibinfo {author}
  {\bibfnamefont {Z.}~\bibnamefont {Wang}}, \ and\ \bibinfo {author}
  {\bibfnamefont {M.}~\bibnamefont {Zanin}}} (\bibinfo {year} {2014}),\ \href
  {\doibase http://dx.doi.org/10.1016/j.physrep.2014.07.001} {\bibfield
  {journal} {\bibinfo  {journal} {Physics Reports}\ }\textbf {\bibinfo {volume}
  {544}},\ \bibinfo {pages} {1}}\BibitemShut {NoStop}%
\bibitem [{\citenamefont {Boccaletti}\ \emph {et~al.}(2006)\citenamefont
  {Boccaletti}, \citenamefont {Latora}, \citenamefont {Moreno}, \citenamefont
  {Chavez},\ and\ \citenamefont {Hwang}}]{boccaletti2006cns}%
  \BibitemOpen
  \bibfield  {author} {\bibinfo {author} {\bibnamefont {Boccaletti},
  \bibfnamefont {S.}}, \bibinfo {author} {\bibfnamefont {V.}~\bibnamefont
  {Latora}}, \bibinfo {author} {\bibfnamefont {Y.}~\bibnamefont {Moreno}},
  \bibinfo {author} {\bibfnamefont {M.}~\bibnamefont {Chavez}}, \ and\ \bibinfo
  {author} {\bibfnamefont {D.}~\bibnamefont {Hwang}}} (\bibinfo {year}
  {2006}),\ \href@noop {} {\bibfield  {journal} {\bibinfo  {journal} {Phys.
  Rep.}\ }\textbf {\bibinfo {volume} {424}},\ \bibinfo {pages}
  {175}}\BibitemShut {NoStop}%
\bibitem [{\citenamefont {Bogu\~n\'a}\ \emph {et~al.}(2013)\citenamefont
  {Bogu\~n\'a}, \citenamefont {Castellano},\ and\ \citenamefont
  {Pastor-Satorras}}]{PhysRevLett.111.068701}%
  \BibitemOpen
  \bibfield  {author} {\bibinfo {author} {\bibnamefont {Bogu\~n\'a},
  \bibfnamefont {M.}}, \bibinfo {author} {\bibfnamefont {C.}~\bibnamefont
  {Castellano}}, \ and\ \bibinfo {author} {\bibfnamefont {R.}~\bibnamefont
  {Pastor-Satorras}}} (\bibinfo {year} {2013}),\ \href {\doibase
  10.1103/PhysRevLett.111.068701} {\bibfield  {journal} {\bibinfo  {journal}
  {Phys. Rev. Lett.}\ }\textbf {\bibinfo {volume} {111}},\ \bibinfo {pages}
  {068701}}\BibitemShut {NoStop}%
\bibitem [{\citenamefont {Bogu\~n\'a}\ \emph {et~al.}(2014)\citenamefont
  {Bogu\~n\'a}, \citenamefont {Lafuerza}, \citenamefont {Toral},\ and\
  \citenamefont {Serrano}}]{boguna_simulating_2013}%
  \BibitemOpen
  \bibfield  {author} {\bibinfo {author} {\bibnamefont {Bogu\~n\'a},
  \bibfnamefont {M.}}, \bibinfo {author} {\bibfnamefont {L.~F.}\ \bibnamefont
  {Lafuerza}}, \bibinfo {author} {\bibfnamefont {R.}~\bibnamefont {Toral}}, \
  and\ \bibinfo {author} {\bibfnamefont {M.~A.}\ \bibnamefont {Serrano}}}
  (\bibinfo {year} {2014}),\ \href@noop {} {\bibfield  {journal} {\bibinfo
  {journal} {Phys. Rev. E}\ }\textbf {\bibinfo {volume} {90}},\ \bibinfo
  {pages} {042108}}\BibitemShut {NoStop}%
\bibitem [{\citenamefont {{Bogu\~{n}\'{a}}}\ and\ \citenamefont
  {Pastor-Satorras}(2002)}]{marian1}%
  \BibitemOpen
  \bibfield  {author} {\bibinfo {author} {\bibnamefont {{Bogu\~{n}\'{a}}},
  \bibfnamefont {M.}}, \ and\ \bibinfo {author} {\bibfnamefont
  {R.}~\bibnamefont {Pastor-Satorras}}} (\bibinfo {year} {2002}),\ \href@noop
  {} {\bibfield  {journal} {\bibinfo  {journal} {Phys. Rev. E}\ }\textbf
  {\bibinfo {volume} {66}},\ \bibinfo {pages} {047104}}\BibitemShut {NoStop}%
\bibitem [{\citenamefont {Bogu\~n\'a}\ and\ \citenamefont
  {Serrano}(2005)}]{Boguna2005}%
  \BibitemOpen
  \bibfield  {author} {\bibinfo {author} {\bibnamefont {Bogu\~n\'a},
  \bibfnamefont {M.}}, \ and\ \bibinfo {author} {\bibfnamefont {M.~A.}\
  \bibnamefont {Serrano}}} (\bibinfo {year} {2005}),\ \href {\doibase
  10.1103/PhysRevE.72.016106} {\bibfield  {journal} {\bibinfo  {journal} {Phys.
  Rev. E}\ }\textbf {\bibinfo {volume} {72}},\ \bibinfo {pages}
  {016106}}\BibitemShut {NoStop}%
\bibitem [{\citenamefont {Bogu{\~n}{\'a}}\ \emph {et~al.}(2009)\citenamefont
  {Bogu{\~n}{\'a}}, \citenamefont {Castellano},\ and\ \citenamefont
  {Pastor-Satorras}}]{Boguna09}%
  \BibitemOpen
  \bibfield  {author} {\bibinfo {author} {\bibnamefont {Bogu{\~n}{\'a}},
  \bibfnamefont {M.}}, \bibinfo {author} {\bibfnamefont {C.}~\bibnamefont
  {Castellano}}, \ and\ \bibinfo {author} {\bibfnamefont {R.}~\bibnamefont
  {Pastor-Satorras}}} (\bibinfo {year} {2009}),\ \href@noop {} {\bibfield
  {journal} {\bibinfo  {journal} {Phys. Rev. E}\ }\textbf {\bibinfo {volume}
  {79}},\ \bibinfo {pages} {036110}}\BibitemShut {NoStop}%
\bibitem [{\citenamefont {Bogu{\~n}{\'a}}\ \emph
  {et~al.}(2003{\natexlab{a}})\citenamefont {Bogu{\~n}{\'a}}, \citenamefont
  {Pastor-Satorras},\ and\ \citenamefont {Vespignani}}]{marian3}%
  \BibitemOpen
  \bibfield  {author} {\bibinfo {author} {\bibnamefont {Bogu{\~n}{\'a}},
  \bibfnamefont {M.}}, \bibinfo {author} {\bibfnamefont {R.}~\bibnamefont
  {Pastor-Satorras}}, \ and\ \bibinfo {author} {\bibfnamefont {A.}~\bibnamefont
  {Vespignani}}} (\bibinfo {year} {2003}{\natexlab{a}}),\ \href@noop {}
  {\bibfield  {journal} {\bibinfo  {journal} {Phys. Rev. Lett.}\ }\textbf
  {\bibinfo {volume} {90}},\ \bibinfo {pages} {028701}}\BibitemShut {NoStop}%
\bibitem [{\citenamefont {Bogu{\~n}{\'a}}\ \emph
  {et~al.}(2003{\natexlab{b}})\citenamefont {Bogu{\~n}{\'a}}, \citenamefont
  {Pastor-Satorras},\ and\ \citenamefont {Vespignani}}]{marianproc}%
  \BibitemOpen
  \bibfield  {author} {\bibinfo {author} {\bibnamefont {Bogu{\~n}{\'a}},
  \bibfnamefont {M.}}, \bibinfo {author} {\bibfnamefont {R.}~\bibnamefont
  {Pastor-Satorras}}, \ and\ \bibinfo {author} {\bibfnamefont {A.}~\bibnamefont
  {Vespignani}}} (\bibinfo {year} {2003}{\natexlab{b}}),\ in\ \href@noop {}
  {\emph {\bibinfo {booktitle} {Statistical Mechanics of Complex Networks}}},\
  \bibinfo {series} {Lecture Notes in Physics}, Vol.\ \bibinfo {volume} {625},\
  \bibinfo {editor} {edited by\ \bibinfo {editor} {\bibfnamefont
  {R.}~\bibnamefont {Pastor-Satorras}}, \bibinfo {editor} {\bibfnamefont
  {J.~M.}\ \bibnamefont {Rub{\'\i}}}, \ and\ \bibinfo {editor} {\bibfnamefont
  {A.}~\bibnamefont {D{\'\i}az-Guilera}}}\ (\bibinfo  {publisher} {Springer
  Verlag},\ \bibinfo {address} {Berlin})\ pp.\ \bibinfo {pages}
  {127--147}\BibitemShut {NoStop}%
\bibitem [{\citenamefont {Bogu{\~n}{\'a}}\ \emph {et~al.}(2004)\citenamefont
  {Bogu{\~n}{\'a}}, \citenamefont {Pastor-Satorras},\ and\ \citenamefont
  {Vespignani}}]{mariancutofss}%
  \BibitemOpen
  \bibfield  {author} {\bibinfo {author} {\bibnamefont {Bogu{\~n}{\'a}},
  \bibfnamefont {M.}}, \bibinfo {author} {\bibfnamefont {R.}~\bibnamefont
  {Pastor-Satorras}}, \ and\ \bibinfo {author} {\bibfnamefont {A.}~\bibnamefont
  {Vespignani}}} (\bibinfo {year} {2004}),\ \href@noop {} {\bibfield  {journal}
  {\bibinfo  {journal} {Euro. Phys. J. B}\ }\textbf {\bibinfo {volume} {38}},\
  \bibinfo {pages} {205}}\BibitemShut {NoStop}%
\bibitem [{\citenamefont {Bonaccorsi}\ \emph {et~al.}(2014)\citenamefont
  {Bonaccorsi}, \citenamefont {Ottaviano}, \citenamefont {De~Pellegrini},
  \citenamefont {Socievole},\ and\ \citenamefont
  {Van~Mieghem}}]{PVM_SIS_communityNetworks2014}%
  \BibitemOpen
  \bibfield  {author} {\bibinfo {author} {\bibnamefont {Bonaccorsi},
  \bibfnamefont {S.}}, \bibinfo {author} {\bibfnamefont {S.}~\bibnamefont
  {Ottaviano}}, \bibinfo {author} {\bibfnamefont {F.}~\bibnamefont
  {De~Pellegrini}}, \bibinfo {author} {\bibfnamefont {A.}~\bibnamefont
  {Socievole}}, \ and\ \bibinfo {author} {\bibfnamefont {P.}~\bibnamefont
  {Van~Mieghem}}} (\bibinfo {year} {2014}),\ \href@noop {} {\bibfield
  {journal} {\bibinfo  {journal} {Physical Review E}\ }\textbf {\bibinfo
  {volume} {90}},\ \bibinfo {pages} {012810}}\BibitemShut {NoStop}%
\bibitem [{\citenamefont {Borge-Holthoefer}\ \emph
  {et~al.}(2012{\natexlab{a}})\citenamefont {Borge-Holthoefer}, \citenamefont
  {Meloni}, \citenamefont {Gon\c{c}alves},\ and\ \citenamefont
  {Moreno}}]{Borge-Holthoefer2012c}%
  \BibitemOpen
  \bibfield  {author} {\bibinfo {author} {\bibnamefont {Borge-Holthoefer},
  \bibfnamefont {J.}}, \bibinfo {author} {\bibfnamefont {S.}~\bibnamefont
  {Meloni}}, \bibinfo {author} {\bibfnamefont {B.}~\bibnamefont
  {Gon\c{c}alves}}, \ and\ \bibinfo {author} {\bibfnamefont {Y.}~\bibnamefont
  {Moreno}}} (\bibinfo {year} {2012}{\natexlab{a}}),\ \href {\doibase
  10.1007/s10955-012-0595-6} {\bibfield  {journal} {\bibinfo  {journal}
  {Journal of Statistical Physics}\ }\textbf {\bibinfo {volume} {151}},\
  \bibinfo {pages} {383}}\BibitemShut {NoStop}%
\bibitem [{\citenamefont {Borge-Holthoefer}\ and\ \citenamefont
  {Moreno}(2012)}]{Borge-Holthoefer2012a}%
  \BibitemOpen
  \bibfield  {author} {\bibinfo {author} {\bibnamefont {Borge-Holthoefer},
  \bibfnamefont {J.}}, \ and\ \bibinfo {author} {\bibfnamefont
  {Y.}~\bibnamefont {Moreno}}} (\bibinfo {year} {2012}),\ \href@noop {}
  {\bibfield  {journal} {\bibinfo  {journal} {Physical Review E}\ }\textbf
  {\bibinfo {volume} {85}}}\BibitemShut {NoStop}%
\bibitem [{\citenamefont {Borge-Holthoefer}\ \emph
  {et~al.}(2012{\natexlab{b}})\citenamefont {Borge-Holthoefer}, \citenamefont
  {Rivero},\ and\ \citenamefont {Moreno}}]{Borge-Holthoefer2012b}%
  \BibitemOpen
  \bibfield  {author} {\bibinfo {author} {\bibnamefont {Borge-Holthoefer},
  \bibfnamefont {J.}}, \bibinfo {author} {\bibfnamefont {A.}~\bibnamefont
  {Rivero}}, \ and\ \bibinfo {author} {\bibfnamefont {Y.}~\bibnamefont
  {Moreno}}} (\bibinfo {year} {2012}{\natexlab{b}}),\ \href@noop {} {\bibfield
  {journal} {\bibinfo  {journal} {Physical Review E}\ }\textbf {\bibinfo
  {volume} {85}},\ \bibinfo {pages} {066123}}\BibitemShut {NoStop}%
\bibitem [{\citenamefont {van~de Bovenkamp}\ \emph {et~al.}(2014)\citenamefont
  {van~de Bovenkamp}, \citenamefont {Kuipers},\ and\ \citenamefont
  {Van~Mieghem}}]{PVM_SIS_competing_virus_PRE2014}%
  \BibitemOpen
  \bibfield  {author} {\bibinfo {author} {\bibnamefont {van~de Bovenkamp},
  \bibfnamefont {R.}}, \bibinfo {author} {\bibfnamefont {F.}~\bibnamefont
  {Kuipers}}, \ and\ \bibinfo {author} {\bibfnamefont {P.}~\bibnamefont
  {Van~Mieghem}}} (\bibinfo {year} {2014}),\ \href@noop {} {\bibfield
  {journal} {\bibinfo  {journal} {Physical Review E}\ }\textbf {\bibinfo
  {volume} {89}},\ \bibinfo {pages} {042818}}\BibitemShut {NoStop}%
\bibitem [{\citenamefont {Brauer}\ and\ \citenamefont
  {{Castillo-Chavez}}(2010)}]{brauer2010}%
  \BibitemOpen
  \bibfield  {author} {\bibinfo {author} {\bibnamefont {Brauer}, \bibfnamefont
  {F.}}, \ and\ \bibinfo {author} {\bibfnamefont {C.}~\bibnamefont
  {{Castillo-Chavez}}}} (\bibinfo {year} {2010}),\ \href@noop {} {\emph
  {\bibinfo {title} {Mathematical Models in Population Biology and
  Epidemiology}}},\ \bibinfo {edition} {2nd}\ ed.,\ \bibinfo {series} {Textsin
  Applied Mathematics}, Vol.~\bibinfo {volume} {40}\ (\bibinfo  {publisher}
  {Springer US},\ \bibinfo {address} {New York})\BibitemShut {NoStop}%
\bibitem [{\citenamefont {Braunstein}\ \emph {et~al.}(2003)\citenamefont
  {Braunstein}, \citenamefont {Buldyrev}, \citenamefont {Cohen}, \citenamefont
  {Havlin},\ and\ \citenamefont {Stanley}}]{Braunstein03}%
  \BibitemOpen
  \bibfield  {author} {\bibinfo {author} {\bibnamefont {Braunstein},
  \bibfnamefont {L.~A.}}, \bibinfo {author} {\bibfnamefont {S.~V.}\
  \bibnamefont {Buldyrev}}, \bibinfo {author} {\bibfnamefont {R.}~\bibnamefont
  {Cohen}}, \bibinfo {author} {\bibfnamefont {S.}~\bibnamefont {Havlin}}, \
  and\ \bibinfo {author} {\bibfnamefont {H.~E.}\ \bibnamefont {Stanley}}}
  (\bibinfo {year} {2003}),\ \href {\doibase 10.1103/PhysRevLett.91.168701}
  {\bibfield  {journal} {\bibinfo  {journal} {Phys. Rev. Lett.}\ }\textbf
  {\bibinfo {volume} {91}},\ \bibinfo {pages} {168701}}\BibitemShut {NoStop}%
\bibitem [{\citenamefont {Britton}\ \emph {et~al.}(2011)\citenamefont
  {Britton}, \citenamefont {Deijfen},\ and\ \citenamefont
  {Liljeros}}]{brittonweight2011}%
  \BibitemOpen
  \bibfield  {author} {\bibinfo {author} {\bibnamefont {Britton}, \bibfnamefont
  {T.}}, \bibinfo {author} {\bibfnamefont {M.}~\bibnamefont {Deijfen}}, \ and\
  \bibinfo {author} {\bibfnamefont {F.}~\bibnamefont {Liljeros}}} (\bibinfo
  {year} {2011}),\ \href {\doibase 10.1007/s10955-011-0343-3} {\bibfield
  {journal} {\bibinfo  {journal} {Journal of Statistical Physics}\ }\textbf
  {\bibinfo {volume} {145}},\ \bibinfo {pages} {1368}}\BibitemShut {NoStop}%
\bibitem [{\citenamefont {Brockmann}\ and\ \citenamefont
  {Helbing}(2013)}]{Brockmann2013}%
  \BibitemOpen
  \bibfield  {author} {\bibinfo {author} {\bibnamefont {Brockmann},
  \bibfnamefont {D.}}, \ and\ \bibinfo {author} {\bibfnamefont
  {D.}~\bibnamefont {Helbing}}} (\bibinfo {year} {2013}),\ \href@noop {}
  {\bibfield  {journal} {\bibinfo  {journal} {Science}\ }\textbf {\bibinfo
  {volume} {342}},\ \bibinfo {pages} {1337}}\BibitemShut {NoStop}%
\bibitem [{\citenamefont {Broeck}\ \emph {et~al.}(2011)\citenamefont {Broeck},
  \citenamefont {Gioannini}, \citenamefont {Goncalves}, \citenamefont
  {Quaggiotto}, \citenamefont {Colizza},\ and\ \citenamefont
  {Vespignani}}]{Broeck2011}%
  \BibitemOpen
  \bibfield  {author} {\bibinfo {author} {\bibnamefont {Broeck}, \bibfnamefont
  {W.}}, \bibinfo {author} {\bibfnamefont {C.}~\bibnamefont {Gioannini}},
  \bibinfo {author} {\bibfnamefont {B.}~\bibnamefont {Goncalves}}, \bibinfo
  {author} {\bibfnamefont {M.}~\bibnamefont {Quaggiotto}}, \bibinfo {author}
  {\bibfnamefont {V.}~\bibnamefont {Colizza}}, \ and\ \bibinfo {author}
  {\bibfnamefont {A.}~\bibnamefont {Vespignani}}} (\bibinfo {year} {2011}),\
  \href {\doibase 10.1186/1471-2334-11-37} {\bibfield  {journal} {\bibinfo
  {journal} {BMC Infectious Diseases}\ }\textbf {\bibinfo {volume} {11}},\
  \bibinfo {pages} {37}}\BibitemShut {NoStop}%
\bibitem [{\citenamefont {Brummitt}\ \emph {et~al.}(2012)\citenamefont
  {Brummitt}, \citenamefont {Lee},\ and\ \citenamefont {Goh}}]{Brummitt2012}%
  \BibitemOpen
  \bibfield  {author} {\bibinfo {author} {\bibnamefont {Brummitt},
  \bibfnamefont {C.~D.}}, \bibinfo {author} {\bibfnamefont {K.-M.}\
  \bibnamefont {Lee}}, \ and\ \bibinfo {author} {\bibfnamefont {K.-I.}\
  \bibnamefont {Goh}}} (\bibinfo {year} {2012}),\ \href {\doibase
  10.1103/PhysRevE.85.045102} {\bibfield  {journal} {\bibinfo  {journal}
  {Physical Review E}\ }\textbf {\bibinfo {volume} {85}},\ \bibinfo {pages}
  {045102}}\BibitemShut {NoStop}%
\bibitem [{\citenamefont {Buono}\ \emph {et~al.}(2014)\citenamefont {Buono},
  \citenamefont {Alvarez-Zuzek}, \citenamefont {Macri},\ and\ \citenamefont
  {Braunstein}}]{Buono2014}%
  \BibitemOpen
  \bibfield  {author} {\bibinfo {author} {\bibnamefont {Buono}, \bibfnamefont
  {C.}}, \bibinfo {author} {\bibfnamefont {L.~G.}\ \bibnamefont
  {Alvarez-Zuzek}}, \bibinfo {author} {\bibfnamefont {P.~A.}\ \bibnamefont
  {Macri}}, \ and\ \bibinfo {author} {\bibfnamefont {L.~A.}\ \bibnamefont
  {Braunstein}}} (\bibinfo {year} {2014}),\ \href {\doibase
  10.1371/journal.pone.0092200} {\bibfield  {journal} {\bibinfo  {journal}
  {PLoS ONE}\ }\textbf {\bibinfo {volume} {9}},\ \bibinfo {pages}
  {e92200}}\BibitemShut {NoStop}%
\bibitem [{\citenamefont {Buono}\ \emph {et~al.}(2013)\citenamefont {Buono},
  \citenamefont {Vazquez}, \citenamefont {Macri},\ and\ \citenamefont
  {Braunstein}}]{Buono13}%
  \BibitemOpen
  \bibfield  {author} {\bibinfo {author} {\bibnamefont {Buono}, \bibfnamefont
  {C.}}, \bibinfo {author} {\bibfnamefont {F.}~\bibnamefont {Vazquez}},
  \bibinfo {author} {\bibfnamefont {P.~A.}\ \bibnamefont {Macri}}, \ and\
  \bibinfo {author} {\bibfnamefont {L.~A.}\ \bibnamefont {Braunstein}}}
  (\bibinfo {year} {2013}),\ \href {\doibase 10.1103/PhysRevE.88.022813}
  {\bibfield  {journal} {\bibinfo  {journal} {Phys. Rev. E}\ }\textbf {\bibinfo
  {volume} {88}},\ \bibinfo {pages} {022813}}\BibitemShut {NoStop}%
\bibitem [{\citenamefont {Butts}(2009)}]{butts:revisiting}%
  \BibitemOpen
  \bibfield  {author} {\bibinfo {author} {\bibnamefont {Butts}, \bibfnamefont
  {C.~T.}}} (\bibinfo {year} {2009}),\ \href@noop {} {\bibfield  {journal}
  {\bibinfo  {journal} {Science}\ }\textbf {\bibinfo {volume} {325}},\ \bibinfo
  {pages} {414}}\BibitemShut {NoStop}%
\bibitem [{\citenamefont {Caldarelli}(2007)}]{caldarelli2007sfn}%
  \BibitemOpen
  \bibfield  {author} {\bibinfo {author} {\bibnamefont {Caldarelli},
  \bibfnamefont {G.}}} (\bibinfo {year} {2007}),\ \href@noop {} {\emph
  {\bibinfo {title} {{Scale-Free Networks: Complex Webs in Nature and
  Technology}}}}\ (\bibinfo  {publisher} {Oxford University Press},\ \bibinfo
  {address} {Oxford})\BibitemShut {NoStop}%
\bibitem [{\citenamefont {Callaway}\ \emph {et~al.}(2000)\citenamefont
  {Callaway}, \citenamefont {Newman}, \citenamefont {Strogatz},\ and\
  \citenamefont {Watts}}]{Callaway2000}%
  \BibitemOpen
  \bibfield  {author} {\bibinfo {author} {\bibnamefont {Callaway},
  \bibfnamefont {D.~S.}}, \bibinfo {author} {\bibfnamefont {M.~E.}\
  \bibnamefont {Newman}}, \bibinfo {author} {\bibfnamefont {S.~H.}\
  \bibnamefont {Strogatz}}, \ and\ \bibinfo {author} {\bibfnamefont {D.~J.}\
  \bibnamefont {Watts}}} (\bibinfo {year} {2000}),\ \href
  {http://www.ncbi.nlm.nih.gov/pubmed/11136023} {\bibfield  {journal} {\bibinfo
   {journal} {Phys. Rev. Lett.}\ }\textbf {\bibinfo {volume} {85}},\ \bibinfo
  {pages} {5468}}\BibitemShut {NoStop}%
\bibitem [{\citenamefont {Castellano}\ and\ \citenamefont
  {Pastor-Satorras}(2006)}]{Castellano2006}%
  \BibitemOpen
  \bibfield  {author} {\bibinfo {author} {\bibnamefont {Castellano},
  \bibfnamefont {C.}}, \ and\ \bibinfo {author} {\bibfnamefont
  {R.}~\bibnamefont {Pastor-Satorras}}} (\bibinfo {year} {2006}),\ \href
  {\doibase 10.1103/PhysRevLett.96.038701} {\bibfield  {journal} {\bibinfo
  {journal} {Phys. Rev. Lett.}\ }\textbf {\bibinfo {volume} {96}},\ \bibinfo
  {pages} {038701}}\BibitemShut {NoStop}%
\bibitem [{\citenamefont {Castellano}\ and\ \citenamefont
  {Pastor-Satorras}(2010)}]{Castellano2010}%
  \BibitemOpen
  \bibfield  {author} {\bibinfo {author} {\bibnamefont {Castellano},
  \bibfnamefont {C.}}, \ and\ \bibinfo {author} {\bibfnamefont
  {R.}~\bibnamefont {Pastor-Satorras}}} (\bibinfo {year} {2010}),\ \href
  {\doibase 10.1103/PhysRevLett.105.218701} {\bibfield  {journal} {\bibinfo
  {journal} {Phys. Rev. Lett.}\ }\textbf {\bibinfo {volume} {105}},\ \bibinfo
  {pages} {218701}}\BibitemShut {NoStop}%
\bibitem [{\citenamefont {Castellano}\ and\ \citenamefont
  {Pastor-Satorras}(2012)}]{Castellano2012}%
  \BibitemOpen
  \bibfield  {author} {\bibinfo {author} {\bibnamefont {Castellano},
  \bibfnamefont {C.}}, \ and\ \bibinfo {author} {\bibfnamefont
  {R.}~\bibnamefont {Pastor-Satorras}}} (\bibinfo {year} {2012}),\ \href
  {\doibase 10.1038/srep00371} {\bibfield  {journal} {\bibinfo  {journal}
  {Scientific Reports}\ }\textbf {\bibinfo {volume} {2}},\
  10.1038/srep00371}\BibitemShut {NoStop}%
\bibitem [{\citenamefont {Catanzaro}\ \emph {et~al.}(2005)\citenamefont
  {Catanzaro}, \citenamefont {Bogu{\~n}{\'a}},\ and\ \citenamefont
  {Pastor-Satorras}}]{Catanzaro05}%
  \BibitemOpen
  \bibfield  {author} {\bibinfo {author} {\bibnamefont {Catanzaro},
  \bibfnamefont {M.}}, \bibinfo {author} {\bibfnamefont {M.}~\bibnamefont
  {Bogu{\~n}{\'a}}}, \ and\ \bibinfo {author} {\bibfnamefont {R.}~\bibnamefont
  {Pastor-Satorras}}} (\bibinfo {year} {2005}),\ \href@noop {} {\bibfield
  {journal} {\bibinfo  {journal} {Phys. Rev. E}\ }\textbf {\bibinfo {volume}
  {71}},\ \bibinfo {pages} {027103}}\BibitemShut {NoStop}%
\bibitem [{\citenamefont {Cator}\ \emph {et~al.}(2013)\citenamefont {Cator},
  \citenamefont {van~de Bovenkamp},\ and\ \citenamefont
  {Van~Mieghem}}]{PVM_nonMarkovianSIS_NIMFA_2013}%
  \BibitemOpen
  \bibfield  {author} {\bibinfo {author} {\bibnamefont {Cator}, \bibfnamefont
  {E.}}, \bibinfo {author} {\bibfnamefont {R.}~\bibnamefont {van~de
  Bovenkamp}}, \ and\ \bibinfo {author} {\bibfnamefont {P.}~\bibnamefont
  {Van~Mieghem}}} (\bibinfo {year} {2013}),\ \href@noop {} {\bibfield
  {journal} {\bibinfo  {journal} {Physical Review E}\ }\textbf {\bibinfo
  {volume} {87}},\ \bibinfo {pages} {062816}}\BibitemShut {NoStop}%
\bibitem [{\citenamefont {Cator}\ and\ \citenamefont
  {Van~Mieghem}(2012)}]{PVM_secondorder_SISmeanfield_PRE2012}%
  \BibitemOpen
  \bibfield  {author} {\bibinfo {author} {\bibnamefont {Cator}, \bibfnamefont
  {E.}}, \ and\ \bibinfo {author} {\bibfnamefont {P.}~\bibnamefont
  {Van~Mieghem}}} (\bibinfo {year} {2012}),\ \href@noop {} {\bibfield
  {journal} {\bibinfo  {journal} {Physical Review E}\ }\textbf {\bibinfo
  {volume} {85}},\ \bibinfo {pages} {056111}}\BibitemShut {NoStop}%
\bibitem [{\citenamefont {Cator}\ and\ \citenamefont
  {Van~Mieghem}(2013)}]{PVM_MSIS_star_PRE2012}%
  \BibitemOpen
  \bibfield  {author} {\bibinfo {author} {\bibnamefont {Cator}, \bibfnamefont
  {E.}}, \ and\ \bibinfo {author} {\bibfnamefont {P.}~\bibnamefont
  {Van~Mieghem}}} (\bibinfo {year} {2013}),\ \href@noop {} {\bibfield
  {journal} {\bibinfo  {journal} {Physical Review E}\ }\textbf {\bibinfo
  {volume} {87}},\ \bibinfo {pages} {012811}}\BibitemShut {NoStop}%
\bibitem [{\citenamefont {Cattuto}\ \emph {et~al.}(2010)\citenamefont
  {Cattuto}, \citenamefont {Van~den Broeck}, \citenamefont {Barrat},
  \citenamefont {Colizza}, \citenamefont {Pinton},\ and\ \citenamefont
  {Vespignani}}]{10.1371/journal.pone.0011596}%
  \BibitemOpen
  \bibfield  {author} {\bibinfo {author} {\bibnamefont {Cattuto}, \bibfnamefont
  {C.}}, \bibinfo {author} {\bibfnamefont {W.}~\bibnamefont {Van~den Broeck}},
  \bibinfo {author} {\bibfnamefont {A.}~\bibnamefont {Barrat}}, \bibinfo
  {author} {\bibfnamefont {V.}~\bibnamefont {Colizza}}, \bibinfo {author}
  {\bibfnamefont {J.-F.}\ \bibnamefont {Pinton}}, \ and\ \bibinfo {author}
  {\bibfnamefont {A.}~\bibnamefont {Vespignani}}} (\bibinfo {year} {2010}),\
  \href@noop {} {\bibfield  {journal} {\bibinfo  {journal} {PLoS ONE}\ }\textbf
  {\bibinfo {volume} {5}},\ \bibinfo {pages} {e11596}}\BibitemShut {NoStop}%
\bibitem [{\citenamefont {Centola}(2010)}]{Centola2010}%
  \BibitemOpen
  \bibfield  {author} {\bibinfo {author} {\bibnamefont {Centola}, \bibfnamefont
  {D.}}} (\bibinfo {year} {2010}),\ \href {\doibase 10.1126/science.1185231}
  {\bibfield  {journal} {\bibinfo  {journal} {Science}\ }\textbf {\bibinfo
  {volume} {329}},\ \bibinfo {pages} {1194}}\BibitemShut {NoStop}%
\bibitem [{\citenamefont {Centola}\ \emph {et~al.}(2007)\citenamefont
  {Centola}, \citenamefont {Egu\'{\i}luz},\ and\ \citenamefont
  {Macy}}]{Centola2007}%
  \BibitemOpen
  \bibfield  {author} {\bibinfo {author} {\bibnamefont {Centola}, \bibfnamefont
  {D.}}, \bibinfo {author} {\bibfnamefont {V.~M.}\ \bibnamefont
  {Egu\'{\i}luz}}, \ and\ \bibinfo {author} {\bibfnamefont {M.~W.}\
  \bibnamefont {Macy}}} (\bibinfo {year} {2007}),\ \href {\doibase
  10.1016/j.physa.2006.06.018} {\bibfield  {journal} {\bibinfo  {journal}
  {Physica A}\ }\textbf {\bibinfo {volume} {374}},\ \bibinfo {pages}
  {449}}\BibitemShut {NoStop}%
\bibitem [{\citenamefont {Centola}\ and\ \citenamefont
  {Macy}(2007)}]{Centola2007a}%
  \BibitemOpen
  \bibfield  {author} {\bibinfo {author} {\bibnamefont {Centola}, \bibfnamefont
  {D.}}, \ and\ \bibinfo {author} {\bibfnamefont {M.}~\bibnamefont {Macy}}}
  (\bibinfo {year} {2007}),\ \href
  {http://medcontent.metapress.com/index/A65RM03P4874243N.pdf
  http://www.jstor.org/stable/10.1086/521848} {\bibfield  {journal} {\bibinfo
  {journal} {American Journal of Sociology}\ }\textbf {\bibinfo {volume}
  {113}},\ \bibinfo {pages} {702}}\BibitemShut {NoStop}%
\bibitem [{\citenamefont {Chakrabarti}\ \emph {et~al.}(2008)\citenamefont
  {Chakrabarti}, \citenamefont {Wang}, \citenamefont {Wang}, \citenamefont
  {Leskovec},\ and\ \citenamefont {Faloutsos}}]{Chakrabarti_2008}%
  \BibitemOpen
  \bibfield  {author} {\bibinfo {author} {\bibnamefont {Chakrabarti},
  \bibfnamefont {D.}}, \bibinfo {author} {\bibfnamefont {Y.}~\bibnamefont
  {Wang}}, \bibinfo {author} {\bibfnamefont {C.}~\bibnamefont {Wang}}, \bibinfo
  {author} {\bibfnamefont {J.}~\bibnamefont {Leskovec}}, \ and\ \bibinfo
  {author} {\bibfnamefont {C.}~\bibnamefont {Faloutsos}}} (\bibinfo {year}
  {2008}),\ \href@noop {} {\bibfield  {journal} {\bibinfo  {journal} {ACM
  Transactions on Information and System Security (TISSEC)}\ }\textbf {\bibinfo
  {volume} {10}},\ \bibinfo {pages} {1}}\BibitemShut {NoStop}%
\bibitem [{\citenamefont {Chao}\ \emph {et~al.}(2010)\citenamefont {Chao},
  \citenamefont {Halloran}, \citenamefont {Obenchain},\ and\ \citenamefont
  {Longini}}]{Chao2010}%
  \BibitemOpen
  \bibfield  {author} {\bibinfo {author} {\bibnamefont {Chao}, \bibfnamefont
  {D.~L.}}, \bibinfo {author} {\bibfnamefont {M.~E.}\ \bibnamefont {Halloran}},
  \bibinfo {author} {\bibfnamefont {V.~J.}\ \bibnamefont {Obenchain}}, \ and\
  \bibinfo {author} {\bibfnamefont {I.~M.}\ \bibnamefont {Longini},
  \bibfnamefont {Jr}}} (\bibinfo {year} {2010}),\ \href {\doibase
  10.1371/journal.pcbi.1000656} {\bibfield  {journal} {\bibinfo  {journal}
  {PLoS Comput Biol}\ }\textbf {\bibinfo {volume} {6}},\ \bibinfo {pages}
  {e1000656}}\BibitemShut {NoStop}%
\bibitem [{\citenamefont {Chatterjee}\ and\ \citenamefont
  {Durrett}(2009)}]{Chatterjee_Durret2009}%
  \BibitemOpen
  \bibfield  {author} {\bibinfo {author} {\bibnamefont {Chatterjee},
  \bibfnamefont {S.}}, \ and\ \bibinfo {author} {\bibfnamefont
  {R.}~\bibnamefont {Durrett}}} (\bibinfo {year} {2009}),\ \href@noop {}
  {\bibfield  {journal} {\bibinfo  {journal} {Annals of Probability}\ }\textbf
  {\bibinfo {volume} {37}},\ \bibinfo {pages} {2332}}\BibitemShut {NoStop}%
\bibitem [{\citenamefont {Chen}\ \emph {et~al.}(2012)\citenamefont {Chen},
  \citenamefont {L\"{u}}, \citenamefont {Shang}, \citenamefont {Zhang},\ and\
  \citenamefont {Zhou}}]{Chen2012}%
  \BibitemOpen
  \bibfield  {author} {\bibinfo {author} {\bibnamefont {Chen}, \bibfnamefont
  {D.}}, \bibinfo {author} {\bibfnamefont {L.}~\bibnamefont {L\"{u}}}, \bibinfo
  {author} {\bibfnamefont {M.-S.}\ \bibnamefont {Shang}}, \bibinfo {author}
  {\bibfnamefont {Y.-C.}\ \bibnamefont {Zhang}}, \ and\ \bibinfo {author}
  {\bibfnamefont {T.}~\bibnamefont {Zhou}}} (\bibinfo {year} {2012}),\ \href
  {\doibase 10.1016/j.physa.2011.09.017} {\bibfield  {journal} {\bibinfo
  {journal} {Physica A}\ }\textbf {\bibinfo {volume} {391}},\ \bibinfo {pages}
  {1777}}\BibitemShut {NoStop}%
\bibitem [{\citenamefont {Chen}\ \emph {et~al.}(2013)\citenamefont {Chen},
  \citenamefont {Xiao}, \citenamefont {Zeng},\ and\ \citenamefont
  {Zhang}}]{Chen2013}%
  \BibitemOpen
  \bibfield  {author} {\bibinfo {author} {\bibnamefont {Chen}, \bibfnamefont
  {D.-B.}}, \bibinfo {author} {\bibfnamefont {R.}~\bibnamefont {Xiao}},
  \bibinfo {author} {\bibfnamefont {A.}~\bibnamefont {Zeng}}, \ and\ \bibinfo
  {author} {\bibfnamefont {Y.-C.}\ \bibnamefont {Zhang}}} (\bibinfo {year}
  {2013}),\ \href {http://stacks.iop.org/0295-5075/104/i=6/a=68006} {\bibfield
  {journal} {\bibinfo  {journal} {Europhysics Letters}\ }\textbf {\bibinfo
  {volume} {104}},\ \bibinfo {pages} {68006}}\BibitemShut {NoStop}%
\bibitem [{\citenamefont {Chen}\ \emph {et~al.}(2008)\citenamefont {Chen},
  \citenamefont {Paul}, \citenamefont {Havlin}, \citenamefont {Liljeros},\ and\
  \citenamefont {Stanley}}]{PhysRevLett.101.058701}%
  \BibitemOpen
  \bibfield  {author} {\bibinfo {author} {\bibnamefont {Chen}, \bibfnamefont
  {Y.}}, \bibinfo {author} {\bibfnamefont {G.}~\bibnamefont {Paul}}, \bibinfo
  {author} {\bibfnamefont {S.}~\bibnamefont {Havlin}}, \bibinfo {author}
  {\bibfnamefont {F.}~\bibnamefont {Liljeros}}, \ and\ \bibinfo {author}
  {\bibfnamefont {H.~E.}\ \bibnamefont {Stanley}}} (\bibinfo {year} {2008}),\
  \href {\doibase 10.1103/PhysRevLett.101.058701} {\bibfield  {journal}
  {\bibinfo  {journal} {Phys. Rev. Lett.}\ }\textbf {\bibinfo {volume} {101}},\
  \bibinfo {pages} {058701}}\BibitemShut {NoStop}%
\bibitem [{\citenamefont {Christakis}\ and\ \citenamefont
  {Fowler}(2007)}]{Christakis2007}%
  \BibitemOpen
  \bibfield  {author} {\bibinfo {author} {\bibnamefont {Christakis},
  \bibfnamefont {N.~A.}}, \ and\ \bibinfo {author} {\bibfnamefont {J.~H.}\
  \bibnamefont {Fowler}}} (\bibinfo {year} {2007}),\ \href@noop {} {\bibfield
  {journal} {\bibinfo  {journal} {New England Journal of Medicine}\ }\textbf
  {\bibinfo {volume} {357}},\ \bibinfo {pages} {370}}\BibitemShut {NoStop}%
\bibitem [{\citenamefont {Christakis}\ and\ \citenamefont
  {Fowler}(2008)}]{Christakis2008}%
  \BibitemOpen
  \bibfield  {author} {\bibinfo {author} {\bibnamefont {Christakis},
  \bibfnamefont {N.~A.}}, \ and\ \bibinfo {author} {\bibfnamefont {J.~H.}\
  \bibnamefont {Fowler}}} (\bibinfo {year} {2008}),\ \href@noop {} {\bibfield
  {journal} {\bibinfo  {journal} {New England Journal of Medicine}\ }\textbf
  {\bibinfo {volume} {358}},\ \bibinfo {pages} {2249}}\BibitemShut {NoStop}%
\bibitem [{\citenamefont {Christakis}\ and\ \citenamefont
  {Fowler}(2010)}]{Christakis2010}%
  \BibitemOpen
  \bibfield  {author} {\bibinfo {author} {\bibnamefont {Christakis},
  \bibfnamefont {N.~a.}}, \ and\ \bibinfo {author} {\bibfnamefont {J.~H.}\
  \bibnamefont {Fowler}}} (\bibinfo {year} {2010}),\ \href {\doibase
  10.1371/journal.pone.0012948} {\bibfield  {journal} {\bibinfo  {journal}
  {PLoS ONE}\ }\textbf {\bibinfo {volume} {5}},\ \bibinfo {pages}
  {e12948}}\BibitemShut {NoStop}%
\bibitem [{\citenamefont {Chu}\ \emph {et~al.}(2011)\citenamefont {Chu},
  \citenamefont {Zhang}, \citenamefont {Guan},\ and\ \citenamefont
  {Zhou}}]{Chu2011471}%
  \BibitemOpen
  \bibfield  {author} {\bibinfo {author} {\bibnamefont {Chu}, \bibfnamefont
  {X.}}, \bibinfo {author} {\bibfnamefont {Z.}~\bibnamefont {Zhang}}, \bibinfo
  {author} {\bibfnamefont {J.}~\bibnamefont {Guan}}, \ and\ \bibinfo {author}
  {\bibfnamefont {S.}~\bibnamefont {Zhou}}} (\bibinfo {year} {2011}),\ \href
  {\doibase http://dx.doi.org/10.1016/j.physa.2010.09.038} {\bibfield
  {journal} {\bibinfo  {journal} {Physica A}\ }\textbf {\bibinfo {volume}
  {390}},\ \bibinfo {pages} {471 }}\BibitemShut {NoStop}%
\bibitem [{\citenamefont {Chung}\ \emph {et~al.}(2003)\citenamefont {Chung},
  \citenamefont {Lu},\ and\ \citenamefont {Vu}}]{Chung03}%
  \BibitemOpen
  \bibfield  {author} {\bibinfo {author} {\bibnamefont {Chung}, \bibfnamefont
  {F.}}, \bibinfo {author} {\bibfnamefont {L.}~\bibnamefont {Lu}}, \ and\
  \bibinfo {author} {\bibfnamefont {V.}~\bibnamefont {Vu}}} (\bibinfo {year}
  {2003}),\ \href@noop {} {\bibfield  {journal} {\bibinfo  {journal} {Proc.
  Natl. Acad. Sci. USA}\ }\textbf {\bibinfo {volume} {100}},\ \bibinfo {pages}
  {6313}}\BibitemShut {NoStop}%
\bibitem [{\citenamefont {Chung}\ \emph {et~al.}(2012)\citenamefont {Chung},
  \citenamefont {Chew}, \citenamefont {Zhou},\ and\ \citenamefont
  {Lai}}]{Chung2012}%
  \BibitemOpen
  \bibfield  {author} {\bibinfo {author} {\bibnamefont {Chung}, \bibfnamefont
  {N.~N.}}, \bibinfo {author} {\bibfnamefont {L.~Y.}\ \bibnamefont {Chew}},
  \bibinfo {author} {\bibfnamefont {J.}~\bibnamefont {Zhou}}, \ and\ \bibinfo
  {author} {\bibfnamefont {C.~H.}\ \bibnamefont {Lai}}} (\bibinfo {year}
  {2012}),\ \href {\doibase 10.1209/0295-5075/98/58004} {\bibfield  {journal}
  {\bibinfo  {journal} {Europhysics Letters}\ }\textbf {\bibinfo {volume}
  {98}},\ \bibinfo {pages} {58004}}\BibitemShut {NoStop}%
\bibitem [{\citenamefont {Cohen}\ \emph {et~al.}(2002)\citenamefont {Cohen},
  \citenamefont {{ben-Avraham}},\ and\ \citenamefont {Havlin}}]{Cohen02}%
  \BibitemOpen
  \bibfield  {author} {\bibinfo {author} {\bibnamefont {Cohen}, \bibfnamefont
  {R.}}, \bibinfo {author} {\bibfnamefont {D.}~\bibnamefont {{ben-Avraham}}}, \
  and\ \bibinfo {author} {\bibfnamefont {S.}~\bibnamefont {Havlin}}} (\bibinfo
  {year} {2002}),\ \href {\doibase 10.1103/PhysRevE.66.036113} {\bibfield
  {journal} {\bibinfo  {journal} {Phys. Rev. E}\ }\textbf {\bibinfo {volume}
  {66}},\ \bibinfo {pages} {036113}}\BibitemShut {NoStop}%
\bibitem [{\citenamefont {Cohen}\ \emph {et~al.}(2000)\citenamefont {Cohen},
  \citenamefont {Erez}, \citenamefont {{ben-Avraham}},\ and\ \citenamefont
  {Havlin}}]{Cohen00}%
  \BibitemOpen
  \bibfield  {author} {\bibinfo {author} {\bibnamefont {Cohen}, \bibfnamefont
  {R.}}, \bibinfo {author} {\bibfnamefont {K.}~\bibnamefont {Erez}}, \bibinfo
  {author} {\bibfnamefont {D.}~\bibnamefont {{ben-Avraham}}}, \ and\ \bibinfo
  {author} {\bibfnamefont {S.}~\bibnamefont {Havlin}}} (\bibinfo {year}
  {2000}),\ \href {\doibase 10.1103/PhysRevLett.85.4626} {\bibfield  {journal}
  {\bibinfo  {journal} {Phys. Rev. Lett.}\ }\textbf {\bibinfo {volume} {85}},\
  \bibinfo {pages} {4626}}\BibitemShut {NoStop}%
\bibitem [{\citenamefont {Cohen}\ \emph {et~al.}(2001)\citenamefont {Cohen},
  \citenamefont {Erez}, \citenamefont {{ben-Avraham}},\ and\ \citenamefont
  {Havlin}}]{havlin01}%
  \BibitemOpen
  \bibfield  {author} {\bibinfo {author} {\bibnamefont {Cohen}, \bibfnamefont
  {R.}}, \bibinfo {author} {\bibfnamefont {K.}~\bibnamefont {Erez}}, \bibinfo
  {author} {\bibfnamefont {D.}~\bibnamefont {{ben-Avraham}}}, \ and\ \bibinfo
  {author} {\bibfnamefont {S.}~\bibnamefont {Havlin}}} (\bibinfo {year}
  {2001}),\ \href@noop {} {\bibfield  {journal} {\bibinfo  {journal} {Phys.
  Rev. Lett.}\ }\textbf {\bibinfo {volume} {86}},\ \bibinfo {pages}
  {3682}}\BibitemShut {NoStop}%
\bibitem [{\citenamefont {Cohen}\ and\ \citenamefont
  {Havlin}(2010)}]{havlinbook}%
  \BibitemOpen
  \bibfield  {author} {\bibinfo {author} {\bibnamefont {Cohen}, \bibfnamefont
  {R.}}, \ and\ \bibinfo {author} {\bibfnamefont {S.}~\bibnamefont {Havlin}}}
  (\bibinfo {year} {2010}),\ \href@noop {} {\emph {\bibinfo {title} {Complex
  Networks: Structure, Robustness and Function}}}\ (\bibinfo  {publisher}
  {Cambridge University Press},\ \bibinfo {address} {Cambridge})\BibitemShut
  {NoStop}%
\bibitem [{\citenamefont {Cohen}\ \emph {et~al.}(2003)\citenamefont {Cohen},
  \citenamefont {Havlin},\ and\ \citenamefont {{ben-Avraham}}}]{Cohen03}%
  \BibitemOpen
  \bibfield  {author} {\bibinfo {author} {\bibnamefont {Cohen}, \bibfnamefont
  {R.}}, \bibinfo {author} {\bibfnamefont {S.}~\bibnamefont {Havlin}}, \ and\
  \bibinfo {author} {\bibfnamefont {D.}~\bibnamefont {{ben-Avraham}}}}
  (\bibinfo {year} {2003}),\ \href@noop {} {\bibfield  {journal} {\bibinfo
  {journal} {Physical Review Letters}\ }\textbf {\bibinfo {volume} {91}},\
  \bibinfo {pages} {247901}}\BibitemShut {NoStop}%
\bibitem [{\citenamefont {Colizza}\ \emph {et~al.}(2006)\citenamefont
  {Colizza}, \citenamefont {A.Barrat}, \citenamefont {Barthelemy},\ and\
  \citenamefont {Vespignani}}]{colizza06:_predic}%
  \BibitemOpen
  \bibfield  {author} {\bibinfo {author} {\bibnamefont {Colizza}, \bibfnamefont
  {V.}}, \bibinfo {author} {\bibnamefont {A.Barrat}}, \bibinfo {author}
  {\bibfnamefont {M.}~\bibnamefont {Barthelemy}}, \ and\ \bibinfo {author}
  {\bibfnamefont {A.}~\bibnamefont {Vespignani}}} (\bibinfo {year} {2006}),\
  \href@noop {} {\bibfield  {journal} {\bibinfo  {journal} {Proc. Natl. Acad.
  Sci. USA}\ }\textbf {\bibinfo {volume} {103}},\ \bibinfo {pages}
  {2015}}\BibitemShut {NoStop}%
\bibitem [{\citenamefont {Colizza}\ \emph
  {et~al.}(2007{\natexlab{a}})\citenamefont {Colizza}, \citenamefont {Barrat},
  \citenamefont {Barthelemy}, \citenamefont {Valleron},\ and\ \citenamefont
  {Vespignani}}]{Colizza:2007}%
  \BibitemOpen
  \bibfield  {author} {\bibinfo {author} {\bibnamefont {Colizza}, \bibfnamefont
  {V.}}, \bibinfo {author} {\bibfnamefont {A.}~\bibnamefont {Barrat}}, \bibinfo
  {author} {\bibfnamefont {M.}~\bibnamefont {Barthelemy}}, \bibinfo {author}
  {\bibfnamefont {A.-J.}\ \bibnamefont {Valleron}}, \ and\ \bibinfo {author}
  {\bibfnamefont {A.}~\bibnamefont {Vespignani}}} (\bibinfo {year}
  {2007}{\natexlab{a}}),\ \href {\doibase 10.1371/journal.pmed.0040013}
  {\bibfield  {journal} {\bibinfo  {journal} {PLoS Med}\ }\textbf {\bibinfo
  {volume} {4}},\ \bibinfo {pages} {e13}}\BibitemShut {NoStop}%
\bibitem [{\citenamefont {Colizza}\ \emph
  {et~al.}(2007{\natexlab{b}})\citenamefont {Colizza}, \citenamefont
  {Pastor-Satorras},\ and\ \citenamefont {Vespignani}}]{v.07:_react}%
  \BibitemOpen
  \bibfield  {author} {\bibinfo {author} {\bibnamefont {Colizza}, \bibfnamefont
  {V.}}, \bibinfo {author} {\bibfnamefont {R.}~\bibnamefont {Pastor-Satorras}},
  \ and\ \bibinfo {author} {\bibfnamefont {A.}~\bibnamefont {Vespignani}}}
  (\bibinfo {year} {2007}{\natexlab{b}}),\ \href@noop {} {\bibfield  {journal}
  {\bibinfo  {journal} {Nature Physics}\ }\textbf {\bibinfo {volume} {3}},\
  \bibinfo {pages} {276}}\BibitemShut {NoStop}%
\bibitem [{\citenamefont {Colizza}\ and\ \citenamefont
  {Vespignani}(2007)}]{colizza07:_invas_thres}%
  \BibitemOpen
  \bibfield  {author} {\bibinfo {author} {\bibnamefont {Colizza}, \bibfnamefont
  {V.}}, \ and\ \bibinfo {author} {\bibfnamefont {A.}~\bibnamefont
  {Vespignani}}} (\bibinfo {year} {2007}),\ \href@noop {} {\bibfield  {journal}
  {\bibinfo  {journal} {Phys. Rev. Lett.}\ }\textbf {\bibinfo {volume} {99}},\
  \bibinfo {pages} {148701}}\BibitemShut {NoStop}%
\bibitem [{\citenamefont {Colizza}\ and\ \citenamefont
  {Vespignani}(2008)}]{colizza07:_epidem_model_published}%
  \BibitemOpen
  \bibfield  {author} {\bibinfo {author} {\bibnamefont {Colizza}, \bibfnamefont
  {V.}}, \ and\ \bibinfo {author} {\bibfnamefont {A.}~\bibnamefont
  {Vespignani}}} (\bibinfo {year} {2008}),\ \href@noop {} {\bibfield  {journal}
  {\bibinfo  {journal} {J. Theor. Biol.}\ }\textbf {\bibinfo {volume} {251}},\
  \bibinfo {pages} {450}}\BibitemShut {NoStop}%
\bibitem [{\citenamefont {Comin}\ and\ \citenamefont {{da Fontoura
  Costa}}(2011)}]{Comin2011}%
  \BibitemOpen
  \bibfield  {author} {\bibinfo {author} {\bibnamefont {Comin}, \bibfnamefont
  {C.~H.}}, \ and\ \bibinfo {author} {\bibfnamefont {L.}~\bibnamefont {{da
  Fontoura Costa}}}} (\bibinfo {year} {2011}),\ \href {\doibase
  10.1103/PhysRevE.84.056105} {\bibfield  {journal} {\bibinfo  {journal}
  {Physical Review E}\ }\textbf {\bibinfo {volume} {84}},\ \bibinfo {pages}
  {056105}}\BibitemShut {NoStop}%
\bibitem [{\citenamefont {Cooper}\ \emph {et~al.}(2006)\citenamefont {Cooper},
  \citenamefont {Pitman}, \citenamefont {Edmunds},\ and\ \citenamefont
  {Gay}}]{Cooper:2006}%
  \BibitemOpen
  \bibfield  {author} {\bibinfo {author} {\bibnamefont {Cooper}, \bibfnamefont
  {B.}}, \bibinfo {author} {\bibfnamefont {R.}~\bibnamefont {Pitman}}, \bibinfo
  {author} {\bibfnamefont {W.}~\bibnamefont {Edmunds}}, \ and\ \bibinfo
  {author} {\bibfnamefont {N.}~\bibnamefont {Gay}}} (\bibinfo {year} {2006}),\
  \href@noop {} {\bibfield  {journal} {\bibinfo  {journal} {PLoS Med}\ }\textbf
  {\bibinfo {volume} {3}},\ \bibinfo {pages} {845}}\BibitemShut {NoStop}%
\bibitem [{\citenamefont {Costa}\ \emph {et~al.}(2007)\citenamefont {Costa},
  \citenamefont {Rodrigues}, \citenamefont {Travieso},\ and\ \citenamefont
  {Villas~Boas}}]{fontourareview}%
  \BibitemOpen
  \bibfield  {author} {\bibinfo {author} {\bibnamefont {Costa}, \bibfnamefont
  {L.~d.~F.}}, \bibinfo {author} {\bibfnamefont {F.~A.}\ \bibnamefont
  {Rodrigues}}, \bibinfo {author} {\bibfnamefont {G.}~\bibnamefont {Travieso}},
  \ and\ \bibinfo {author} {\bibfnamefont {P.~R.}\ \bibnamefont {Villas~Boas}}}
  (\bibinfo {year} {2007}),\ \href@noop {} {\bibfield  {journal} {\bibinfo
  {journal} {Advances in Physics}\ }\textbf {\bibinfo {volume} {56}},\ \bibinfo
  {pages} {167}}\BibitemShut {NoStop}%
\bibitem [{\citenamefont {Cox}(1967)}]{renewal}%
  \BibitemOpen
  \bibfield  {author} {\bibinfo {author} {\bibnamefont {Cox}, \bibfnamefont
  {D.~R.}}} (\bibinfo {year} {1967}),\ \href@noop {} {\emph {\bibinfo {title}
  {Renewal Theory}}}\ (\bibinfo  {publisher} {Methuen},\ \bibinfo {address}
  {London})\BibitemShut {NoStop}%
\bibitem [{\citenamefont {Cozzo}\ \emph {et~al.}(2013)\citenamefont {Cozzo},
  \citenamefont {Banos}, \citenamefont {Meloni},\ and\ \citenamefont
  {Moreno}}]{Cozzo2013}%
  \BibitemOpen
  \bibfield  {author} {\bibinfo {author} {\bibnamefont {Cozzo}, \bibfnamefont
  {E.}}, \bibinfo {author} {\bibfnamefont {R.~A.}\ \bibnamefont {Banos}},
  \bibinfo {author} {\bibfnamefont {S.}~\bibnamefont {Meloni}}, \ and\ \bibinfo
  {author} {\bibfnamefont {Y.}~\bibnamefont {Moreno}}} (\bibinfo {year}
  {2013}),\ \href@noop {} {\bibfield  {journal} {\bibinfo  {journal} {Physical
  Review E}\ }\textbf {\bibinfo {volume} {88}},\ \bibinfo {pages}
  {050801}}\BibitemShut {NoStop}%
\bibitem [{\citenamefont {Cross}\ \emph {et~al.}(2007)\citenamefont {Cross},
  \citenamefont {Johnson}, \citenamefont {Lloyd-Smith},\ and\ \citenamefont
  {Wayne}}]{Cross:2007}%
  \BibitemOpen
  \bibfield  {author} {\bibinfo {author} {\bibnamefont {Cross}, \bibfnamefont
  {P.}}, \bibinfo {author} {\bibfnamefont {P.}~\bibnamefont {Johnson}},
  \bibinfo {author} {\bibfnamefont {J.}~\bibnamefont {Lloyd-Smith}}, \ and\
  \bibinfo {author} {\bibfnamefont {M.}~\bibnamefont {Wayne}}} (\bibinfo {year}
  {2007}),\ \href {\doibase 10.1098/rsif.2006.0185} {\bibfield  {journal}
  {\bibinfo  {journal} {J R Soc Interface}\ }\textbf {\bibinfo {volume} {4}},\
  \bibinfo {pages} {315}}\BibitemShut {NoStop}%
\bibitem [{\citenamefont {Cross}\ \emph {et~al.}(2005)\citenamefont {Cross},
  \citenamefont {Lloyd-Smith}, \citenamefont {Johnson},\ and\ \citenamefont
  {Wayne}}]{Cross:2005}%
  \BibitemOpen
  \bibfield  {author} {\bibinfo {author} {\bibnamefont {Cross}, \bibfnamefont
  {P.}}, \bibinfo {author} {\bibfnamefont {J.}~\bibnamefont {Lloyd-Smith}},
  \bibinfo {author} {\bibfnamefont {P.}~\bibnamefont {Johnson}}, \ and\
  \bibinfo {author} {\bibfnamefont {M.}~\bibnamefont {Wayne}}} (\bibinfo {year}
  {2005}),\ \href {\doibase 10.1111/j.1461-0248.2005.00760.x} {\bibfield
  {journal} {\bibinfo  {journal} {Ecol Lett}\ }\textbf {\bibinfo {volume}
  {8}},\ \bibinfo {pages} {587}}\BibitemShut {NoStop}%
\bibitem [{\citenamefont {Daley}\ and\ \citenamefont
  {Kendall}(1964)}]{Daley1964}%
  \BibitemOpen
  \bibfield  {author} {\bibinfo {author} {\bibnamefont {Daley}, \bibfnamefont
  {D.~J.}}, \ and\ \bibinfo {author} {\bibfnamefont {D.~G.}\ \bibnamefont
  {Kendall}}} (\bibinfo {year} {1964}),\ \href@noop {} {\bibfield  {journal}
  {\bibinfo  {journal} {Nature}\ }\textbf {\bibinfo {volume} {2004}},\ \bibinfo
  {pages} {1118}}\BibitemShut {NoStop}%
\bibitem [{\citenamefont {Daley}\ and\ \citenamefont
  {Kendall}(1965)}]{Daley1965}%
  \BibitemOpen
  \bibfield  {author} {\bibinfo {author} {\bibnamefont {Daley}, \bibfnamefont
  {D.~J.}}, \ and\ \bibinfo {author} {\bibfnamefont {D.~G.}\ \bibnamefont
  {Kendall}}} (\bibinfo {year} {1965}),\ \href {\doibase 10.1093/imamat/1.1.42}
  {\bibfield  {journal} {\bibinfo  {journal} {IMA Journal of Applied
  Mathematics}\ }\textbf {\bibinfo {volume} {1}}~(\bibinfo {number} {1}),\
  \bibinfo {pages} {42}}\BibitemShut {NoStop}%
\bibitem [{\citenamefont {Danon}\ \emph {et~al.}(2011)\citenamefont {Danon},
  \citenamefont {Ford}, \citenamefont {House}, \citenamefont {Jewell},
  \citenamefont {Keeling}, \citenamefont {Roberts}, \citenamefont {Ross},\ and\
  \citenamefont {Vernon}}]{danonreview}%
  \BibitemOpen
  \bibfield  {author} {\bibinfo {author} {\bibnamefont {Danon}, \bibfnamefont
  {L.}}, \bibinfo {author} {\bibfnamefont {A.~P.}\ \bibnamefont {Ford}},
  \bibinfo {author} {\bibfnamefont {T.}~\bibnamefont {House}}, \bibinfo
  {author} {\bibfnamefont {C.~P.}\ \bibnamefont {Jewell}}, \bibinfo {author}
  {\bibfnamefont {M.~J.}\ \bibnamefont {Keeling}}, \bibinfo {author}
  {\bibfnamefont {G.~O.}\ \bibnamefont {Roberts}}, \bibinfo {author}
  {\bibfnamefont {J.~V.}\ \bibnamefont {Ross}}, \ and\ \bibinfo {author}
  {\bibfnamefont {M.~C.}\ \bibnamefont {Vernon}}} (\bibinfo {year} {2011}),\
  \href@noop {} {\bibfield  {journal} {\bibinfo  {journal} {Interdisciplinary
  Perspectives on Infectious Diseases}\ }\textbf {\bibinfo {volume} {2011}},\
  \bibinfo {pages} {284909}}\BibitemShut {NoStop}%
\bibitem [{\citenamefont {Darabi~Sahneh}\ \emph {et~al.}(2013)\citenamefont
  {Darabi~Sahneh}, \citenamefont {Scoglio},\ and\ \citenamefont
  {Van~Mieghem}}]{PVM_GEMF}%
  \BibitemOpen
  \bibfield  {author} {\bibinfo {author} {\bibnamefont {Darabi~Sahneh},
  \bibfnamefont {F.}}, \bibinfo {author} {\bibfnamefont {C.}~\bibnamefont
  {Scoglio}}, \ and\ \bibinfo {author} {\bibfnamefont {P.}~\bibnamefont
  {Van~Mieghem}}} (\bibinfo {year} {2013}),\ \href@noop {} {\bibfield
  {journal} {\bibinfo  {journal} {IEEE/ACM Transaction on Networking}\ }\textbf
  {\bibinfo {volume} {21}}~(\bibinfo {number} {5}),\ \bibinfo {pages}
  {1609}}\BibitemShut {NoStop}%
\bibitem [{\citenamefont {Decreusefond}\ \emph {et~al.}(2012)\citenamefont
  {Decreusefond}, \citenamefont {Dhersin}, \citenamefont {Moyal},\ and\
  \citenamefont {Tran}}]{Decreusefond2012}%
  \BibitemOpen
  \bibfield  {author} {\bibinfo {author} {\bibnamefont {Decreusefond},
  \bibfnamefont {L.}}, \bibinfo {author} {\bibfnamefont {J.-S.}\ \bibnamefont
  {Dhersin}}, \bibinfo {author} {\bibfnamefont {P.}~\bibnamefont {Moyal}}, \
  and\ \bibinfo {author} {\bibfnamefont {V.~C.}\ \bibnamefont {Tran}}}
  (\bibinfo {year} {2012}),\ \href {\doibase 10.1214/11-AAP773} {\bibfield
  {journal} {\bibinfo  {journal} {The Annals of Applied Probability}\ }\textbf
  {\bibinfo {volume} {22}}~(\bibinfo {number} {2}),\ \bibinfo {pages}
  {541}}\BibitemShut {NoStop}%
\bibitem [{\citenamefont {Deijfen}(2011)}]{Deijfen201157}%
  \BibitemOpen
  \bibfield  {author} {\bibinfo {author} {\bibnamefont {Deijfen}, \bibfnamefont
  {M.}}} (\bibinfo {year} {2011}),\ \href {\doibase
  http://dx.doi.org/10.1016/j.mbs.2011.04.003} {\bibfield  {journal} {\bibinfo
  {journal} {Mathematical Biosciences}\ }\textbf {\bibinfo {volume}
  {232}}~(\bibinfo {number} {1}),\ \bibinfo {pages} {57 }}\BibitemShut
  {NoStop}%
\bibitem [{\citenamefont {Dezs{\"o}}\ and\ \citenamefont
  {Barab{\'a}si}(2002)}]{aidsbar}%
  \BibitemOpen
  \bibfield  {author} {\bibinfo {author} {\bibnamefont {Dezs{\"o}},
  \bibfnamefont {Z.}}, \ and\ \bibinfo {author} {\bibfnamefont {A.-L.}\
  \bibnamefont {Barab{\'a}si}}} (\bibinfo {year} {2002}),\ \href@noop {}
  {\bibfield  {journal} {\bibinfo  {journal} {Phys. Rev. E}\ }\textbf {\bibinfo
  {volume} {65}},\ \bibinfo {pages} {055103}}\BibitemShut {NoStop}%
\bibitem [{\citenamefont {Dickison}\ \emph {et~al.}(2012)\citenamefont
  {Dickison}, \citenamefont {Havlin},\ and\ \citenamefont
  {Stanley}}]{Dickison2012}%
  \BibitemOpen
  \bibfield  {author} {\bibinfo {author} {\bibnamefont {Dickison},
  \bibfnamefont {M.}}, \bibinfo {author} {\bibfnamefont {S.}~\bibnamefont
  {Havlin}}, \ and\ \bibinfo {author} {\bibfnamefont {H.~E.}\ \bibnamefont
  {Stanley}}} (\bibinfo {year} {2012}),\ \href@noop {} {\bibfield  {journal}
  {\bibinfo  {journal} {Physical Review E}\ }\textbf {\bibinfo {volume}
  {85}}~(\bibinfo {number} {6}),\ \bibinfo {pages} {066109}}\BibitemShut
  {NoStop}%
\bibitem [{\citenamefont {Diekmann}\ \emph {et~al.}(2012)\citenamefont
  {Diekmann}, \citenamefont {Heesterbeek},\ and\ \citenamefont
  {Britton}}]{Diekmann_Heesterbeek_Britton_boek2012}%
  \BibitemOpen
  \bibfield  {author} {\bibinfo {author} {\bibnamefont {Diekmann},
  \bibfnamefont {O.}}, \bibinfo {author} {\bibfnamefont {H.}~\bibnamefont
  {Heesterbeek}}, \ and\ \bibinfo {author} {\bibfnamefont {T.}~\bibnamefont
  {Britton}}} (\bibinfo {year} {2012}),\ \href@noop {} {\emph {\bibinfo {title}
  {Mathematical Tools for Understanding Infectious Disease Dynamics}}}\
  (\bibinfo  {publisher} {Princeton University Press},\ \bibinfo {address}
  {Princeton, USA})\BibitemShut {NoStop}%
\bibitem [{\citenamefont {Diekmann}\ and\ \citenamefont
  {Heesterbeek}(2000)}]{epidemics}%
  \BibitemOpen
  \bibfield  {author} {\bibinfo {author} {\bibnamefont {Diekmann},
  \bibfnamefont {O.}}, \ and\ \bibinfo {author} {\bibfnamefont
  {J.}~\bibnamefont {Heesterbeek}}} (\bibinfo {year} {2000}),\ \href@noop {}
  {\emph {\bibinfo {title} {Mathematical epidemiology of infectious diseases:
  model building, analysis and interpretation}}}\ (\bibinfo  {publisher} {John
  Wiley \& Sons},\ \bibinfo {address} {New York})\BibitemShut {NoStop}%
\bibitem [{\citenamefont {Dodds}\ and\ \citenamefont
  {Payne}(2009)}]{Dodds2009}%
  \BibitemOpen
  \bibfield  {author} {\bibinfo {author} {\bibnamefont {Dodds}, \bibfnamefont
  {P.}}, \ and\ \bibinfo {author} {\bibfnamefont {J.}~\bibnamefont {Payne}}}
  (\bibinfo {year} {2009}),\ \href {\doibase 10.1103/PhysRevE.79.066115}
  {\bibfield  {journal} {\bibinfo  {journal} {Physical Review E}\ }\textbf
  {\bibinfo {volume} {79}}~(\bibinfo {number} {6}),\ \bibinfo {pages}
  {066115}}\BibitemShut {NoStop}%
\bibitem [{\citenamefont {Dodds}\ and\ \citenamefont
  {Watts}(2004)}]{Dodds2004}%
  \BibitemOpen
  \bibfield  {author} {\bibinfo {author} {\bibnamefont {Dodds}, \bibfnamefont
  {P.}}, \ and\ \bibinfo {author} {\bibfnamefont {D.~J.}\ \bibnamefont
  {Watts}}} (\bibinfo {year} {2004}),\ \href@noop {} {\bibfield  {journal}
  {\bibinfo  {journal} {Physical Review Letters}\ }\textbf {\bibinfo {volume}
  {92}}~(\bibinfo {number} {21}),\ \bibinfo {pages} {218701}}\BibitemShut
  {NoStop}%
\bibitem [{\citenamefont {Dodds}\ \emph {et~al.}(2011)\citenamefont {Dodds},
  \citenamefont {Harris},\ and\ \citenamefont {Payne}}]{Dodds2011}%
  \BibitemOpen
  \bibfield  {author} {\bibinfo {author} {\bibnamefont {Dodds}, \bibfnamefont
  {P.~S.}}, \bibinfo {author} {\bibfnamefont {K.~D.}\ \bibnamefont {Harris}}, \
  and\ \bibinfo {author} {\bibfnamefont {J.~L.}\ \bibnamefont {Payne}}}
  (\bibinfo {year} {2011}),\ \href {\doibase 10.1103/PhysRevE.83.056122}
  {\bibfield  {journal} {\bibinfo  {journal} {Phys. Rev. E}\ }\textbf {\bibinfo
  {volume} {83}},\ \bibinfo {pages} {056122}}\BibitemShut {NoStop}%
\bibitem [{\citenamefont {Doerr}\ \emph {et~al.}(2012)\citenamefont {Doerr},
  \citenamefont {Blenn}, \citenamefont {Tang},\ and\ \citenamefont {{Van
  Mieghem}}}]{Doerr2012}%
  \BibitemOpen
  \bibfield  {author} {\bibinfo {author} {\bibnamefont {Doerr}, \bibfnamefont
  {C.}}, \bibinfo {author} {\bibfnamefont {N.}~\bibnamefont {Blenn}}, \bibinfo
  {author} {\bibfnamefont {S.}~\bibnamefont {Tang}}, \ and\ \bibinfo {author}
  {\bibfnamefont {P.}~\bibnamefont {{Van Mieghem}}}} (\bibinfo {year} {2012}),\
  \href {\doibase 10.1016/j.comcom.2012.02.001} {\bibfield  {journal} {\bibinfo
   {journal} {Computer Communications}\ }\textbf {\bibinfo {volume}
  {35}}~(\bibinfo {number} {7}),\ \bibinfo {pages} {796}}\BibitemShut {NoStop}%
\bibitem [{\citenamefont {Doerr}\ \emph {et~al.}(2013)\citenamefont {Doerr},
  \citenamefont {Blenn},\ and\ \citenamefont
  {Van~Mieghem}}]{PVM_Twitter_lognormal}%
  \BibitemOpen
  \bibfield  {author} {\bibinfo {author} {\bibnamefont {Doerr}, \bibfnamefont
  {C.}}, \bibinfo {author} {\bibfnamefont {N.}~\bibnamefont {Blenn}}, \ and\
  \bibinfo {author} {\bibfnamefont {P.}~\bibnamefont {Van~Mieghem}}} (\bibinfo
  {year} {2013}),\ \href@noop {} {\bibfield  {journal} {\bibinfo  {journal}
  {PLoS ONE}\ }\textbf {\bibinfo {volume} {8}}~(\bibinfo {number} {5}),\
  \bibinfo {pages} {e64349}}\BibitemShut {NoStop}%
\bibitem [{\citenamefont {Domingos}\ and\ \citenamefont
  {Richardson}(2001)}]{Domingos2001}%
  \BibitemOpen
  \bibfield  {author} {\bibinfo {author} {\bibnamefont {Domingos},
  \bibfnamefont {P.}}, \ and\ \bibinfo {author} {\bibfnamefont
  {M.}~\bibnamefont {Richardson}}} (\bibinfo {year} {2001}),\ in\ \href
  {\doibase 10.1145/502512.502525} {\emph {\bibinfo {booktitle} {Proceedings of
  the seventh ACM SIGKDD international conference on Knowledge discovery and
  data mining}}},\ \bibinfo {series and number} {KDD '01}\ (\bibinfo
  {publisher} {ACM},\ \bibinfo {address} {New York, NY, USA})\ pp.\ \bibinfo
  {pages} {57--66}\BibitemShut {NoStop}%
\bibitem [{\citenamefont {Dorogovtsev}(2010)}]{Dorogobook2010}%
  \BibitemOpen
  \bibfield  {author} {\bibinfo {author} {\bibnamefont {Dorogovtsev},
  \bibfnamefont {S.~N.}}} (\bibinfo {year} {2010}),\ \href@noop {} {\emph
  {\bibinfo {title} {Lectures on complex networks}}},\ Oxford Master Series in
  Physics\ (\bibinfo  {publisher} {Oxford University Press},\ \bibinfo
  {address} {Oxford})\BibitemShut {NoStop}%
\bibitem [{\citenamefont {Dorogovtsev}\ \emph {et~al.}(2008)\citenamefont
  {Dorogovtsev}, \citenamefont {Goltsev},\ and\ \citenamefont
  {Mendes}}]{dorogovtsev07:_critic_phenom}%
  \BibitemOpen
  \bibfield  {author} {\bibinfo {author} {\bibnamefont {Dorogovtsev},
  \bibfnamefont {S.~N.}}, \bibinfo {author} {\bibfnamefont {A.~V.}\
  \bibnamefont {Goltsev}}, \ and\ \bibinfo {author} {\bibfnamefont {J.~F.~F.}\
  \bibnamefont {Mendes}}} (\bibinfo {year} {2008}),\ \href@noop {} {\bibfield
  {journal} {\bibinfo  {journal} {Rev. Mod. Phys.}\ }\textbf {\bibinfo {volume}
  {80}},\ \bibinfo {pages} {1275}}\BibitemShut {NoStop}%
\bibitem [{\citenamefont {Dorogovtsev}\ and\ \citenamefont
  {Mendes}(2002)}]{Dorogovtsev:2002}%
  \BibitemOpen
  \bibfield  {author} {\bibinfo {author} {\bibnamefont {Dorogovtsev},
  \bibfnamefont {S.~N.}}, \ and\ \bibinfo {author} {\bibfnamefont {J.~F.~F.}\
  \bibnamefont {Mendes}}} (\bibinfo {year} {2002}),\ \href@noop {} {\bibfield
  {journal} {\bibinfo  {journal} {Advances in Physics}\ }\textbf {\bibinfo
  {volume} {51}},\ \bibinfo {pages} {1079}}\BibitemShut {NoStop}%
\bibitem [{\citenamefont {Dorogovtsev}\ and\ \citenamefont
  {Mendes}(2003)}]{mendesbook}%
  \BibitemOpen
  \bibfield  {author} {\bibinfo {author} {\bibnamefont {Dorogovtsev},
  \bibfnamefont {S.~N.}}, \ and\ \bibinfo {author} {\bibfnamefont {J.~F.~F.}\
  \bibnamefont {Mendes}}} (\bibinfo {year} {2003}),\ \href@noop {} {\emph
  {\bibinfo {title} {Evolution of networks: From biological nets to the
  {I}nternet and {WWW}}}}\ (\bibinfo  {publisher} {Oxford University Press},\
  \bibinfo {address} {Oxford})\BibitemShut {NoStop}%
\bibitem [{\citenamefont {Dorogovtsev}\ \emph {et~al.}(2000)\citenamefont
  {Dorogovtsev}, \citenamefont {Mendes},\ and\ \citenamefont
  {Samukhin}}]{mendes99}%
  \BibitemOpen
  \bibfield  {author} {\bibinfo {author} {\bibnamefont {Dorogovtsev},
  \bibfnamefont {S.~N.}}, \bibinfo {author} {\bibfnamefont {J.~F.~F.}\
  \bibnamefont {Mendes}}, \ and\ \bibinfo {author} {\bibfnamefont {A.~N.}\
  \bibnamefont {Samukhin}}} (\bibinfo {year} {2000}),\ \href@noop {} {\bibfield
   {journal} {\bibinfo  {journal} {Phys. Rev. Lett.}\ }\textbf {\bibinfo
  {volume} {85}},\ \bibinfo {pages} {4633}}\BibitemShut {NoStop}%
\bibitem [{\citenamefont {Dorogovtsev}\ \emph {et~al.}(2001)\citenamefont
  {Dorogovtsev}, \citenamefont {Mendes},\ and\ \citenamefont
  {Samukhin}}]{dorogodirect01}%
  \BibitemOpen
  \bibfield  {author} {\bibinfo {author} {\bibnamefont {Dorogovtsev},
  \bibfnamefont {S.~N.}}, \bibinfo {author} {\bibfnamefont {J.~F.~F.}\
  \bibnamefont {Mendes}}, \ and\ \bibinfo {author} {\bibfnamefont {A.~N.}\
  \bibnamefont {Samukhin}}} (\bibinfo {year} {2001}),\ \href@noop {} {\bibfield
   {journal} {\bibinfo  {journal} {Phys. Rev. E}\ }\textbf {\bibinfo {volume}
  {64}},\ \bibinfo {pages} {025101}}\BibitemShut {NoStop}%
\bibitem [{\citenamefont {Dunbar}(1998)}]{dunbar1998social}%
  \BibitemOpen
  \bibfield  {author} {\bibinfo {author} {\bibnamefont {Dunbar}, \bibfnamefont
  {R.~I.}}} (\bibinfo {year} {1998}),\ \href@noop {} {\bibfield  {journal}
  {\bibinfo  {journal} {Evolutionary Anthropology}\ }\textbf {\bibinfo {volume}
  {9}},\ \bibinfo {pages} {178}}\BibitemShut {NoStop}%
\bibitem [{\citenamefont {Durrett}(2010)}]{Durrett_PNAS2010}%
  \BibitemOpen
  \bibfield  {author} {\bibinfo {author} {\bibnamefont {Durrett}, \bibfnamefont
  {R.}}} (\bibinfo {year} {2010}),\ \href@noop {} {\bibfield  {journal}
  {\bibinfo  {journal} {Proc. Natl. Acad. Sci. USA}\ }\textbf {\bibinfo
  {volume} {107}},\ \bibinfo {pages} {4491}}\BibitemShut {NoStop}%
\bibitem [{\citenamefont {Eames}\ \emph {et~al.}(2009)\citenamefont {Eames},
  \citenamefont {Read},\ and\ \citenamefont {Edmunds}}]{Eames200970}%
  \BibitemOpen
  \bibfield  {author} {\bibinfo {author} {\bibnamefont {Eames}, \bibfnamefont
  {K.~T.}}, \bibinfo {author} {\bibfnamefont {J.~M.}\ \bibnamefont {Read}}, \
  and\ \bibinfo {author} {\bibfnamefont {W.~J.}\ \bibnamefont {Edmunds}}}
  (\bibinfo {year} {2009}),\ \href@noop {} {\bibfield  {journal} {\bibinfo
  {journal} {Epidemics}\ }\textbf {\bibinfo {volume} {1}}~(\bibinfo {number}
  {1}),\ \bibinfo {pages} {70 }}\BibitemShut {NoStop}%
\bibitem [{\citenamefont {Eames}\ and\ \citenamefont
  {Keeling}(2002)}]{Eames2002}%
  \BibitemOpen
  \bibfield  {author} {\bibinfo {author} {\bibnamefont {Eames}, \bibfnamefont
  {K.~T.~D.}}, \ and\ \bibinfo {author} {\bibfnamefont {M.~J.}\ \bibnamefont
  {Keeling}}} (\bibinfo {year} {2002}),\ \href {\doibase
  10.1073/pnas.202244299} {\bibfield  {journal} {\bibinfo  {journal} {Proc.
  Natl. Acad. Sci. USA}\ }\textbf {\bibinfo {volume} {99}},\ \bibinfo {pages}
  {13330}}\BibitemShut {NoStop}%
\bibitem [{\citenamefont {Easley}\ and\ \citenamefont
  {Kleinberg}(2010)}]{Easley2010}%
  \BibitemOpen
  \bibfield  {author} {\bibinfo {author} {\bibnamefont {Easley}, \bibfnamefont
  {D.}}, \ and\ \bibinfo {author} {\bibfnamefont {J.}~\bibnamefont
  {Kleinberg}}} (\bibinfo {year} {2010}),\ \href@noop {} {\emph {\bibinfo
  {title} {Networks, crowds, and markets}}}\ (\bibinfo  {publisher} {Cambridge
  University Press},\ \bibinfo {address} {Cambridge, U.K.})\BibitemShut
  {NoStop}%
\bibitem [{\citenamefont {Egu{\'\i}luz}\ and\ \citenamefont
  {Klemm}(2002)}]{structured}%
  \BibitemOpen
  \bibfield  {author} {\bibinfo {author} {\bibnamefont {Egu{\'\i}luz},
  \bibfnamefont {V.~M.}}, \ and\ \bibinfo {author} {\bibfnamefont
  {K.}~\bibnamefont {Klemm}}} (\bibinfo {year} {2002}),\ \href@noop {}
  {\bibfield  {journal} {\bibinfo  {journal} {Phys. Rev. Lett.}\ }\textbf
  {\bibinfo {volume} {89}},\ \bibinfo {pages} {108701}}\BibitemShut {NoStop}%
\bibitem [{\citenamefont {Erd\H{o}s}\ and\ \citenamefont
  {R{\'e}nyi}(1959)}]{erdos59}%
  \BibitemOpen
  \bibfield  {author} {\bibinfo {author} {\bibnamefont {Erd\H{o}s},
  \bibfnamefont {P.}}, \ and\ \bibinfo {author} {\bibfnamefont
  {P.}~\bibnamefont {R{\'e}nyi}}} (\bibinfo {year} {1959}),\ \href@noop {}
  {\bibfield  {journal} {\bibinfo  {journal} {Publicationes Mathematicae}\
  }\textbf {\bibinfo {volume} {6}},\ \bibinfo {pages} {290}}\BibitemShut
  {NoStop}%
\bibitem [{\citenamefont {Eubank}\ \emph {et~al.}(2004)\citenamefont {Eubank},
  \citenamefont {Guclu}, \citenamefont {Kumar}, \citenamefont {Marathe},
  \citenamefont {Srinivasan}, \citenamefont {Toroczkai},\ and\ \citenamefont
  {Wang}}]{Eubank2004}%
  \BibitemOpen
  \bibfield  {author} {\bibinfo {author} {\bibnamefont {Eubank}, \bibfnamefont
  {S.}}, \bibinfo {author} {\bibfnamefont {H.}~\bibnamefont {Guclu}}, \bibinfo
  {author} {\bibfnamefont {V.}~\bibnamefont {Kumar}}, \bibinfo {author}
  {\bibfnamefont {M.}~\bibnamefont {Marathe}}, \bibinfo {author} {\bibfnamefont
  {A.}~\bibnamefont {Srinivasan}}, \bibinfo {author} {\bibfnamefont
  {Z.}~\bibnamefont {Toroczkai}}, \ and\ \bibinfo {author} {\bibfnamefont
  {N.}~\bibnamefont {Wang}}} (\bibinfo {year} {2004}),\ \href {\doibase
  10.1038/nature02541} {\bibfield  {journal} {\bibinfo  {journal} {Nature}\
  }\textbf {\bibinfo {volume} {429}},\ \bibinfo {pages} {180}}\BibitemShut
  {NoStop}%
\bibitem [{\citenamefont {Ferguson}\ \emph {et~al.}(2005)\citenamefont
  {Ferguson}, \citenamefont {Cummings}, \citenamefont {Cauchemez},
  \citenamefont {Fraser}, \citenamefont {Riley}, \citenamefont {Meeyai},
  \citenamefont {Iamsirithaworn},\ and\ \citenamefont {Burke}}]{Ferguson2005}%
  \BibitemOpen
  \bibfield  {author} {\bibinfo {author} {\bibnamefont {Ferguson},
  \bibfnamefont {N.~M.}}, \bibinfo {author} {\bibfnamefont {D.~A.}\
  \bibnamefont {Cummings}}, \bibinfo {author} {\bibfnamefont {S.}~\bibnamefont
  {Cauchemez}}, \bibinfo {author} {\bibfnamefont {C.}~\bibnamefont {Fraser}},
  \bibinfo {author} {\bibfnamefont {S.}~\bibnamefont {Riley}}, \bibinfo
  {author} {\bibfnamefont {A.}~\bibnamefont {Meeyai}}, \bibinfo {author}
  {\bibfnamefont {S.}~\bibnamefont {Iamsirithaworn}}, \ and\ \bibinfo {author}
  {\bibfnamefont {D.~S.}\ \bibnamefont {Burke}}} (\bibinfo {year} {2005}),\
  \href@noop {} {\bibfield  {journal} {\bibinfo  {journal} {Nature}\ }\textbf
  {\bibinfo {volume} {437}},\ \bibinfo {pages} {209}}\BibitemShut {NoStop}%
\bibitem [{\citenamefont {Ferreira}\ \emph {et~al.}(2012)\citenamefont
  {Ferreira}, \citenamefont {Castellano},\ and\ \citenamefont
  {Pastor-Satorras}}]{Ferreira12}%
  \BibitemOpen
  \bibfield  {author} {\bibinfo {author} {\bibnamefont {Ferreira},
  \bibfnamefont {S.~C.}}, \bibinfo {author} {\bibfnamefont {C.}~\bibnamefont
  {Castellano}}, \ and\ \bibinfo {author} {\bibfnamefont {R.}~\bibnamefont
  {Pastor-Satorras}}} (\bibinfo {year} {2012}),\ \href@noop {} {\bibfield
  {journal} {\bibinfo  {journal} {Phys. Rev. E}\ }\textbf {\bibinfo {volume}
  {86}},\ \bibinfo {pages} {041125}}\BibitemShut {NoStop}%
\bibitem [{\citenamefont {Ferreira}\ \emph {et~al.}(2011)\citenamefont
  {Ferreira}, \citenamefont {Ferreira}, \citenamefont {Castellano},\ and\
  \citenamefont {Pastor-Satorras}}]{FFCR11}%
  \BibitemOpen
  \bibfield  {author} {\bibinfo {author} {\bibnamefont {Ferreira},
  \bibfnamefont {S.~C.}}, \bibinfo {author} {\bibfnamefont {R.~S.}\
  \bibnamefont {Ferreira}}, \bibinfo {author} {\bibfnamefont {C.}~\bibnamefont
  {Castellano}}, \ and\ \bibinfo {author} {\bibfnamefont {R.}~\bibnamefont
  {Pastor-Satorras}}} (\bibinfo {year} {2011}),\ \href@noop {} {\bibfield
  {journal} {\bibinfo  {journal} {Phys. Rev. E}\ }\textbf {\bibinfo {volume}
  {84}},\ \bibinfo {pages} {066102}}\BibitemShut {NoStop}%
\bibitem [{\citenamefont {Flahault}\ and\ \citenamefont
  {Valleron}(1991)}]{Flahault:1991}%
  \BibitemOpen
  \bibfield  {author} {\bibinfo {author} {\bibnamefont {Flahault},
  \bibfnamefont {A.}}, \ and\ \bibinfo {author} {\bibfnamefont {A.-J.}\
  \bibnamefont {Valleron}}} (\bibinfo {year} {1991}),\ \href {\doibase
  10.1080/08898489109525319} {\bibfield  {journal} {\bibinfo  {journal} {Math
  Popul Stud}\ }\textbf {\bibinfo {volume} {3}},\ \bibinfo {pages}
  {1}}\BibitemShut {NoStop}%
\bibitem [{\citenamefont {Fortunato}(2010)}]{Fortunato201075}%
  \BibitemOpen
  \bibfield  {author} {\bibinfo {author} {\bibnamefont {Fortunato},
  \bibfnamefont {S.}}} (\bibinfo {year} {2010}),\ \href {\doibase
  http://dx.doi.org/10.1016/j.physrep.2009.11.002} {\bibfield  {journal}
  {\bibinfo  {journal} {Physics Reports}\ }\textbf {\bibinfo {volume}
  {486}}~(\bibinfo {number} {3--5}),\ \bibinfo {pages} {75 }}\BibitemShut
  {NoStop}%
\bibitem [{\citenamefont {Fowler}\ and\ \citenamefont
  {Christakis}(2008)}]{Fowler2008}%
  \BibitemOpen
  \bibfield  {author} {\bibinfo {author} {\bibnamefont {Fowler}, \bibfnamefont
  {J.~H.}}, \ and\ \bibinfo {author} {\bibfnamefont {N.~A.}\ \bibnamefont
  {Christakis}}} (\bibinfo {year} {2008}),\ \href@noop {} {\bibfield  {journal}
  {\bibinfo  {journal} {BMJ: British Medical Journal}\ }\textbf {\bibinfo
  {volume} {337}},\ \bibinfo {pages} {a2338}}\BibitemShut {NoStop}%
\bibitem [{\citenamefont {Freeman}(1977)}]{freeman77}%
  \BibitemOpen
  \bibfield  {author} {\bibinfo {author} {\bibnamefont {Freeman}, \bibfnamefont
  {L.~C.}}} (\bibinfo {year} {1977}),\ \href@noop {} {\bibfield  {journal}
  {\bibinfo  {journal} {Sociometry}\ }\textbf {\bibinfo {volume} {40}},\
  \bibinfo {pages} {35}}\BibitemShut {NoStop}%
\bibitem [{\citenamefont {Fujiwara}\ \emph {et~al.}(2011)\citenamefont
  {Fujiwara}, \citenamefont {Kurths},\ and\ \citenamefont
  {D{\'\i}az-Guilera}}]{albert2011sync}%
  \BibitemOpen
  \bibfield  {author} {\bibinfo {author} {\bibnamefont {Fujiwara},
  \bibfnamefont {N.}}, \bibinfo {author} {\bibfnamefont {J.}~\bibnamefont
  {Kurths}}, \ and\ \bibinfo {author} {\bibfnamefont {A.}~\bibnamefont
  {D{\'\i}az-Guilera}}} (\bibinfo {year} {2011}),\ \href@noop {} {\bibfield
  {journal} {\bibinfo  {journal} {Physical Review E}\ }\textbf {\bibinfo
  {volume} {83}}~(\bibinfo {number} {2}),\ \bibinfo {pages}
  {025101}}\BibitemShut {NoStop}%
\bibitem [{\citenamefont {Fumanelli}\ \emph {et~al.}(2012)\citenamefont
  {Fumanelli}, \citenamefont {Ajelli}, \citenamefont {Manfredi}, \citenamefont
  {Vespignani},\ and\ \citenamefont {Merler}}]{Fumanelli2012}%
  \BibitemOpen
  \bibfield  {author} {\bibinfo {author} {\bibnamefont {Fumanelli},
  \bibfnamefont {L.}}, \bibinfo {author} {\bibfnamefont {M.}~\bibnamefont
  {Ajelli}}, \bibinfo {author} {\bibfnamefont {P.}~\bibnamefont {Manfredi}},
  \bibinfo {author} {\bibfnamefont {A.}~\bibnamefont {Vespignani}}, \ and\
  \bibinfo {author} {\bibfnamefont {S.}~\bibnamefont {Merler}}} (\bibinfo
  {year} {2012}),\ \href {\doibase 10.1371/journal.pcbi.1002673} {\bibfield
  {journal} {\bibinfo  {journal} {PLoS Comput Biol}\ }\textbf {\bibinfo
  {volume} {8}}~(\bibinfo {number} {9}),\ \bibinfo {pages}
  {e1002673}}\BibitemShut {NoStop}%
\bibitem [{\citenamefont {Funk}\ \emph {et~al.}(2009)\citenamefont {Funk},
  \citenamefont {Gilad}, \citenamefont {Watkins},\ and\ \citenamefont
  {Jansen}}]{Funk2009}%
  \BibitemOpen
  \bibfield  {author} {\bibinfo {author} {\bibnamefont {Funk}, \bibfnamefont
  {S.}}, \bibinfo {author} {\bibfnamefont {E.}~\bibnamefont {Gilad}}, \bibinfo
  {author} {\bibfnamefont {C.}~\bibnamefont {Watkins}}, \ and\ \bibinfo
  {author} {\bibfnamefont {V.~A.~A.}\ \bibnamefont {Jansen}}} (\bibinfo {year}
  {2009}),\ \href@noop {} {\bibfield  {journal} {\bibinfo  {journal}
  {Proceedings of the National Academy of Sciences}\ }\textbf {\bibinfo
  {volume} {106}}~(\bibinfo {number} {16}),\ \bibinfo {pages}
  {6872}}\BibitemShut {NoStop}%
\bibitem [{\citenamefont {Funk}\ and\ \citenamefont
  {Jansen}(2010)}]{Funk2010b}%
  \BibitemOpen
  \bibfield  {author} {\bibinfo {author} {\bibnamefont {Funk}, \bibfnamefont
  {S.}}, \ and\ \bibinfo {author} {\bibfnamefont {V.~A.~A.}\ \bibnamefont
  {Jansen}}} (\bibinfo {year} {2010}),\ \href {\doibase
  10.1103/PhysRevE.81.036118} {\bibfield  {journal} {\bibinfo  {journal} {Phys.
  Rev. E}\ }\textbf {\bibinfo {volume} {81}},\ \bibinfo {pages}
  {036118}}\BibitemShut {NoStop}%
\bibitem [{\citenamefont {Funk}\ \emph {et~al.}(2010)\citenamefont {Funk},
  \citenamefont {Salath\'{e}},\ and\ \citenamefont {Jansen}}]{Funk2010}%
  \BibitemOpen
  \bibfield  {author} {\bibinfo {author} {\bibnamefont {Funk}, \bibfnamefont
  {S.}}, \bibinfo {author} {\bibfnamefont {M.}~\bibnamefont {Salath\'{e}}}, \
  and\ \bibinfo {author} {\bibfnamefont {V.~a.~a.}\ \bibnamefont {Jansen}}}
  (\bibinfo {year} {2010}),\ \href {\doibase 10.1098/rsif.2010.0142} {\bibfield
   {journal} {\bibinfo  {journal} {Journal of the Royal Society, Interface}\
  }\textbf {\bibinfo {volume} {7}},\ \bibinfo {pages} {1247}}\BibitemShut
  {NoStop}%
\bibitem [{\citenamefont {Gallos}\ and\ \citenamefont
  {Argyrakis}(2004)}]{gallos2004absence}%
  \BibitemOpen
  \bibfield  {author} {\bibinfo {author} {\bibnamefont {Gallos}, \bibfnamefont
  {L.~K.}}, \ and\ \bibinfo {author} {\bibfnamefont {P.}~\bibnamefont
  {Argyrakis}}} (\bibinfo {year} {2004}),\ \href@noop {} {\bibfield  {journal}
  {\bibinfo  {journal} {Physical Review Letters}\ }\textbf {\bibinfo {volume}
  {92}}~(\bibinfo {number} {13}),\ \bibinfo {pages} {138301}}\BibitemShut
  {NoStop}%
\bibitem [{\citenamefont {Galstyan}\ and\ \citenamefont
  {Cohen}(2007)}]{Galstyan2007}%
  \BibitemOpen
  \bibfield  {author} {\bibinfo {author} {\bibnamefont {Galstyan},
  \bibfnamefont {A.}}, \ and\ \bibinfo {author} {\bibfnamefont
  {P.}~\bibnamefont {Cohen}}} (\bibinfo {year} {2007}),\ \href {\doibase
  10.1103/PhysRevE.75.036109} {\bibfield  {journal} {\bibinfo  {journal} {Phys.
  Rev. E}\ }\textbf {\bibinfo {volume} {75}},\ \bibinfo {pages}
  {036109}}\BibitemShut {NoStop}%
\bibitem [{\citenamefont {Ganesh}\ \emph {et~al.}(2005)\citenamefont {Ganesh},
  \citenamefont {Massoulie},\ and\ \citenamefont {Towsley}}]{Ganesh05}%
  \BibitemOpen
  \bibfield  {author} {\bibinfo {author} {\bibnamefont {Ganesh}, \bibfnamefont
  {A.}}, \bibinfo {author} {\bibfnamefont {L.}~\bibnamefont {Massoulie}}, \
  and\ \bibinfo {author} {\bibfnamefont {D.}~\bibnamefont {Towsley}}} (\bibinfo
  {year} {2005}),\ in\ \href {\doibase 10.1109/INFCOM.2005.1498374} {\emph
  {\bibinfo {booktitle} {INFOCOM 2005. 24th Annual Joint Conference of the IEEE
  Computer and Communications Societies. Proceedings IEEE}}},\ Vol.~\bibinfo
  {volume} {2},\ pp.\ \bibinfo {pages} {1455--1466}\BibitemShut {NoStop}%
\bibitem [{\citenamefont {Gang}\ \emph {et~al.}(2005)\citenamefont {Gang},
  \citenamefont {Tao}, \citenamefont {Jie}, \citenamefont {Zhong-Qian},\ and\
  \citenamefont {Bing-Hong}}]{0256-307X-22-2-068}%
  \BibitemOpen
  \bibfield  {author} {\bibinfo {author} {\bibnamefont {Gang}, \bibfnamefont
  {Y.}}, \bibinfo {author} {\bibfnamefont {Z.}~\bibnamefont {Tao}}, \bibinfo
  {author} {\bibfnamefont {W.}~\bibnamefont {Jie}}, \bibinfo {author}
  {\bibfnamefont {F.}~\bibnamefont {Zhong-Qian}}, \ and\ \bibinfo {author}
  {\bibfnamefont {W.}~\bibnamefont {Bing-Hong}}} (\bibinfo {year} {2005}),\
  \href {http://stacks.iop.org/0256-307X/22/i=2/a=068} {\bibfield  {journal}
  {\bibinfo  {journal} {Chinese Physics Letters}\ }\textbf {\bibinfo {volume}
  {22}}~(\bibinfo {number} {2}),\ \bibinfo {pages} {510}}\BibitemShut {NoStop}%
\bibitem [{\citenamefont {Gantmacher}(1974)}]{Gantmacher}%
  \BibitemOpen
  \bibfield  {author} {\bibinfo {author} {\bibnamefont {Gantmacher},
  \bibfnamefont {F.~R.}}} (\bibinfo {year} {1974}),\ \href@noop {} {\emph
  {\bibinfo {title} {The theory of matrices}}},\ Vol.~\bibinfo {volume} {II}\
  (\bibinfo  {publisher} {Chelsea Publishing Company},\ \bibinfo {address} {New
  York})\BibitemShut {NoStop}%
\bibitem [{\citenamefont {Garas}\ \emph {et~al.}(2012)\citenamefont {Garas},
  \citenamefont {Schweitzer},\ and\ \citenamefont {Havlin}}]{Garas2012}%
  \BibitemOpen
  \bibfield  {author} {\bibinfo {author} {\bibnamefont {Garas}, \bibfnamefont
  {A.}}, \bibinfo {author} {\bibfnamefont {F.}~\bibnamefont {Schweitzer}}, \
  and\ \bibinfo {author} {\bibfnamefont {S.}~\bibnamefont {Havlin}}} (\bibinfo
  {year} {2012}),\ \href {\doibase 10.1088/1367-2630/14/8/083030} {\bibfield
  {journal} {\bibinfo  {journal} {New Journal of Physics}\ }\textbf {\bibinfo
  {volume} {14}}~(\bibinfo {number} {8}),\ \bibinfo {pages}
  {083030}}\BibitemShut {NoStop}%
\bibitem [{\citenamefont {Garcia-Herranz}\ \emph {et~al.}(2014)\citenamefont
  {Garcia-Herranz}, \citenamefont {Moro}, \citenamefont {Cebrian},
  \citenamefont {Christakis},\ and\ \citenamefont {Fowler}}]{Garcia-Herranz14}%
  \BibitemOpen
  \bibfield  {author} {\bibinfo {author} {\bibnamefont {Garcia-Herranz},
  \bibfnamefont {M.}}, \bibinfo {author} {\bibfnamefont {E.}~\bibnamefont
  {Moro}}, \bibinfo {author} {\bibfnamefont {M.}~\bibnamefont {Cebrian}},
  \bibinfo {author} {\bibfnamefont {N.~A.}\ \bibnamefont {Christakis}}, \ and\
  \bibinfo {author} {\bibfnamefont {J.~H.}\ \bibnamefont {Fowler}}} (\bibinfo
  {year} {2014}),\ \href {\doibase 10.1371/journal.pone.0092413} {\bibfield
  {journal} {\bibinfo  {journal} {PLoS ONE}\ }\textbf {\bibinfo {volume}
  {9}}~(\bibinfo {number} {4}),\ \bibinfo {pages} {e92413}}\BibitemShut
  {NoStop}%
\bibitem [{\citenamefont {Gautreau}\ \emph {et~al.}(2008)\citenamefont
  {Gautreau}, \citenamefont {Barrat},\ and\ \citenamefont
  {Barth{\'e}lemy}}]{Bart:2008}%
  \BibitemOpen
  \bibfield  {author} {\bibinfo {author} {\bibnamefont {Gautreau},
  \bibfnamefont {A.}}, \bibinfo {author} {\bibfnamefont {A.}~\bibnamefont
  {Barrat}}, \ and\ \bibinfo {author} {\bibfnamefont {M.}~\bibnamefont
  {Barth{\'e}lemy}}} (\bibinfo {year} {2008}),\ \href@noop {} {\bibfield
  {journal} {\bibinfo  {journal} {Journal of Theoretical Biology}\ }\textbf
  {\bibinfo {volume} {251}},\ \bibinfo {pages} {509}}\BibitemShut {NoStop}%
\bibitem [{\citenamefont {Gil}\ and\ \citenamefont
  {Zanette}(2005)}]{gil05:_optim_disor}%
  \BibitemOpen
  \bibfield  {author} {\bibinfo {author} {\bibnamefont {Gil}, \bibfnamefont
  {S.}}, \ and\ \bibinfo {author} {\bibfnamefont {D.}~\bibnamefont {Zanette}}}
  (\bibinfo {year} {2005}),\ \href@noop {} {\bibfield  {journal} {\bibinfo
  {journal} {Eur. Phys. J. B}\ }\textbf {\bibinfo {volume} {47}},\ \bibinfo
  {pages} {265}}\BibitemShut {NoStop}%
\bibitem [{\citenamefont {Gilbert}(1959)}]{gilbert59}%
  \BibitemOpen
  \bibfield  {author} {\bibinfo {author} {\bibnamefont {Gilbert}, \bibfnamefont
  {E.~N.}}} (\bibinfo {year} {1959}),\ \href@noop {} {\bibfield  {journal}
  {\bibinfo  {journal} {Annals of Mathematical Statistics}\ }\textbf {\bibinfo
  {volume} {30}},\ \bibinfo {pages} {1141}}\BibitemShut {NoStop}%
\bibitem [{\citenamefont {Gillespie}(1977)}]{gillespie_exact_1977}%
  \BibitemOpen
  \bibfield  {author} {\bibinfo {author} {\bibnamefont {Gillespie},
  \bibfnamefont {D.~T.}}} (\bibinfo {year} {1977}),\ \href
  {http://pubs.acs.org/doi/abs/10.1021/j100540a008} {\bibfield  {journal}
  {\bibinfo  {journal} {The Journal of Physical Chemistry}\ }\textbf {\bibinfo
  {volume} {81}},\ \bibinfo {pages} {2340}}\BibitemShut {NoStop}%
\bibitem [{\citenamefont {Givan}\ \emph {et~al.}(2011)\citenamefont {Givan},
  \citenamefont {Schwartz}, \citenamefont {Cygelberg},\ and\ \citenamefont
  {Stone}}]{Givan2011}%
  \BibitemOpen
  \bibfield  {author} {\bibinfo {author} {\bibnamefont {Givan}, \bibfnamefont
  {O.}}, \bibinfo {author} {\bibfnamefont {N.}~\bibnamefont {Schwartz}},
  \bibinfo {author} {\bibfnamefont {A.}~\bibnamefont {Cygelberg}}, \ and\
  \bibinfo {author} {\bibfnamefont {L.}~\bibnamefont {Stone}}} (\bibinfo {year}
  {2011}),\ \href {\doibase 10.1016/j.jtbi.2011.07.015} {\bibfield  {journal}
  {\bibinfo  {journal} {Journal of Theoretical Biology}\ }\textbf {\bibinfo
  {volume} {288}},\ \bibinfo {pages} {21}}\BibitemShut {NoStop}%
\bibitem [{\citenamefont {Gleeson}\ and\ \citenamefont
  {Cahalane}(2007)}]{Gleeson2007}%
  \BibitemOpen
  \bibfield  {author} {\bibinfo {author} {\bibnamefont {Gleeson}, \bibfnamefont
  {J.}}, \ and\ \bibinfo {author} {\bibfnamefont {D.}~\bibnamefont {Cahalane}}}
  (\bibinfo {year} {2007}),\ \href {\doibase 10.1103/PhysRevE.75.056103}
  {\bibfield  {journal} {\bibinfo  {journal} {Physical Review E}\ }\textbf
  {\bibinfo {volume} {75}}~(\bibinfo {number} {5}),\ \bibinfo {pages}
  {056103}}\BibitemShut {NoStop}%
\bibitem [{\citenamefont {Gleeson}(2008)}]{Gleeson2008}%
  \BibitemOpen
  \bibfield  {author} {\bibinfo {author} {\bibnamefont {Gleeson}, \bibfnamefont
  {J.~P.}}} (\bibinfo {year} {2008}),\ \href {\doibase
  10.1103/PhysRevE.77.046117} {\bibfield  {journal} {\bibinfo  {journal}
  {Physical Review E}\ }\textbf {\bibinfo {volume} {77}}~(\bibinfo {number}
  {4}),\ \bibinfo {pages} {046117}}\BibitemShut {NoStop}%
\bibitem [{\citenamefont {Gleeson}(2011)}]{Gleeson11}%
  \BibitemOpen
  \bibfield  {author} {\bibinfo {author} {\bibnamefont {Gleeson}, \bibfnamefont
  {J.~P.}}} (\bibinfo {year} {2011}),\ \href {\doibase
  10.1103/PhysRevLett.107.068701} {\bibfield  {journal} {\bibinfo  {journal}
  {Phys. Rev. Lett.}\ }\textbf {\bibinfo {volume} {107}},\ \bibinfo {pages}
  {068701}}\BibitemShut {NoStop}%
\bibitem [{\citenamefont {Gleeson}(2013)}]{PhysRevX.3.021004}%
  \BibitemOpen
  \bibfield  {author} {\bibinfo {author} {\bibnamefont {Gleeson}, \bibfnamefont
  {J.~P.}}} (\bibinfo {year} {2013}),\ \href {\doibase
  10.1103/PhysRevX.3.021004} {\bibfield  {journal} {\bibinfo  {journal} {Phys.
  Rev. X}\ }\textbf {\bibinfo {volume} {3}},\ \bibinfo {pages}
  {021004}}\BibitemShut {NoStop}%
\bibitem [{\citenamefont {Goffman}(1966)}]{Goffman1966}%
  \BibitemOpen
  \bibfield  {author} {\bibinfo {author} {\bibnamefont {Goffman}, \bibfnamefont
  {W.}}} (\bibinfo {year} {1966}),\ \href@noop {} {\bibfield  {journal}
  {\bibinfo  {journal} {Nature}\ }\textbf {\bibinfo {volume} {212}}~(\bibinfo
  {number} {5061}),\ \bibinfo {pages} {449}}\BibitemShut {NoStop}%
\bibitem [{\citenamefont {Goffman}\ and\ \citenamefont
  {Newill}(1964)}]{Goffman1964}%
  \BibitemOpen
  \bibfield  {author} {\bibinfo {author} {\bibnamefont {Goffman}, \bibfnamefont
  {W.}}, \ and\ \bibinfo {author} {\bibfnamefont {V.~A.}\ \bibnamefont
  {Newill}}} (\bibinfo {year} {1964}),\ \href {\doibase 10.1038/204225a0}
  {\bibfield  {journal} {\bibinfo  {journal} {Nature}\ }\textbf {\bibinfo
  {volume} {204}},\ \bibinfo {pages} {225}}\BibitemShut {NoStop}%
\bibitem [{\citenamefont {Goltsev}\ \emph {et~al.}(2008)\citenamefont
  {Goltsev}, \citenamefont {Dorogovtsev},\ and\ \citenamefont
  {Mendes}}]{PhysRevE.78.051105}%
  \BibitemOpen
  \bibfield  {author} {\bibinfo {author} {\bibnamefont {Goltsev}, \bibfnamefont
  {A.~V.}}, \bibinfo {author} {\bibfnamefont {S.~N.}\ \bibnamefont
  {Dorogovtsev}}, \ and\ \bibinfo {author} {\bibfnamefont {J.~F.~F.}\
  \bibnamefont {Mendes}}} (\bibinfo {year} {2008}),\ \href {\doibase
  10.1103/PhysRevE.78.051105} {\bibfield  {journal} {\bibinfo  {journal} {Phys.
  Rev. E}\ }\textbf {\bibinfo {volume} {78}},\ \bibinfo {pages}
  {051105}}\BibitemShut {NoStop}%
\bibitem [{\citenamefont {{Goltsev}}\ \emph {et~al.}(2012)\citenamefont
  {{Goltsev}}, \citenamefont {{Dorogovtsev}}, \citenamefont {{Oliveira}},\ and\
  \citenamefont {{Mendes}}}]{Goltsev12}%
  \BibitemOpen
  \bibfield  {author} {\bibinfo {author} {\bibnamefont {{Goltsev}},
  \bibfnamefont {A.~V.}}, \bibinfo {author} {\bibfnamefont {S.~N.}\
  \bibnamefont {{Dorogovtsev}}}, \bibinfo {author} {\bibfnamefont {J.~G.}\
  \bibnamefont {{Oliveira}}}, \ and\ \bibinfo {author} {\bibfnamefont
  {J.~F.~F.}\ \bibnamefont {{Mendes}}}} (\bibinfo {year} {2012}),\ \href@noop
  {} {\bibfield  {journal} {\bibinfo  {journal} {Phys. Rev. Lett}\ }\textbf
  {\bibinfo {volume} {109}},\ \bibinfo {pages} {128702}}\BibitemShut {NoStop}%
\bibitem [{\citenamefont {G{\'o}mez}\ \emph {et~al.}(2010)\citenamefont
  {G{\'o}mez}, \citenamefont {Arenas}, \citenamefont {Borge-Holthoefer},
  \citenamefont {Meloni},\ and\ \citenamefont {Moreno}}]{Gomez10}%
  \BibitemOpen
  \bibfield  {author} {\bibinfo {author} {\bibnamefont {G{\'o}mez},
  \bibfnamefont {S.}}, \bibinfo {author} {\bibfnamefont {A.}~\bibnamefont
  {Arenas}}, \bibinfo {author} {\bibfnamefont {J.}~\bibnamefont
  {Borge-Holthoefer}}, \bibinfo {author} {\bibfnamefont {S.}~\bibnamefont
  {Meloni}}, \ and\ \bibinfo {author} {\bibfnamefont {Y.}~\bibnamefont
  {Moreno}}} (\bibinfo {year} {2010}),\ \href@noop {} {\bibfield  {journal}
  {\bibinfo  {journal} {Europhys. Lett.}\ }\textbf {\bibinfo {volume}
  {89}}~(\bibinfo {number} {3}),\ \bibinfo {pages} {38009}}\BibitemShut
  {NoStop}%
\bibitem [{\citenamefont {Gomez-Gardenes}\ \emph {et~al.}(2006)\citenamefont
  {Gomez-Gardenes}, \citenamefont {Echenique},\ and\ \citenamefont
  {Moreno}}]{gomez2006}%
  \BibitemOpen
  \bibfield  {author} {\bibinfo {author} {\bibnamefont {Gomez-Gardenes},
  \bibfnamefont {J.}}, \bibinfo {author} {\bibfnamefont {P.}~\bibnamefont
  {Echenique}}, \ and\ \bibinfo {author} {\bibfnamefont {Y.}~\bibnamefont
  {Moreno}}} (\bibinfo {year} {2006}),\ \href
  {http://dx.doi.org/10.1140/epjb/e2006-00041-1} {\bibfield  {journal}
  {\bibinfo  {journal} {The European Physical Journal B - Condensed Matter and
  Complex Systems}\ }\textbf {\bibinfo {volume} {49}}~(\bibinfo {number} {2}),\
  \bibinfo {pages} {259}}\BibitemShut {NoStop}%
\bibitem [{\citenamefont {Gomez-Gardenes}\ \emph {et~al.}(2008)\citenamefont
  {Gomez-Gardenes}, \citenamefont {Latora}, \citenamefont {Moreno},\ and\
  \citenamefont {Profumo}}]{Gomez-Gardenes05022008}%
  \BibitemOpen
  \bibfield  {author} {\bibinfo {author} {\bibnamefont {Gomez-Gardenes},
  \bibfnamefont {J.}}, \bibinfo {author} {\bibfnamefont {V.}~\bibnamefont
  {Latora}}, \bibinfo {author} {\bibfnamefont {Y.}~\bibnamefont {Moreno}}, \
  and\ \bibinfo {author} {\bibfnamefont {E.}~\bibnamefont {Profumo}}} (\bibinfo
  {year} {2008}),\ \href@noop {} {\bibfield  {journal} {\bibinfo  {journal}
  {Proc. Natl. Acad. Sci. USA}\ }\textbf {\bibinfo {volume} {105}},\ \bibinfo
  {pages} {1399}}\BibitemShut {NoStop}%
\bibitem [{\citenamefont {Gon{\c{c}}alves}\ \emph {et~al.}(2011)\citenamefont
  {Gon{\c{c}}alves}, \citenamefont {Perra},\ and\ \citenamefont
  {Vespignani}}]{Goncalves2011Dunbar}%
  \BibitemOpen
  \bibfield  {author} {\bibinfo {author} {\bibnamefont {Gon{\c{c}}alves},
  \bibfnamefont {B.}}, \bibinfo {author} {\bibfnamefont {N.}~\bibnamefont
  {Perra}}, \ and\ \bibinfo {author} {\bibfnamefont {A.}~\bibnamefont
  {Vespignani}}} (\bibinfo {year} {2011}),\ \href@noop {} {\bibfield  {journal}
  {\bibinfo  {journal} {PloS one}\ }\textbf {\bibinfo {volume} {6}}~(\bibinfo
  {number} {8}),\ \bibinfo {pages} {e22656}}\BibitemShut {NoStop}%
\bibitem [{\citenamefont {Gourdin}\ \emph {et~al.}(2011)\citenamefont
  {Gourdin}, \citenamefont {Omic},\ and\ \citenamefont
  {Mieghem}}]{PVM_Gourdin_Networkprotection_DRCN2011}%
  \BibitemOpen
  \bibfield  {author} {\bibinfo {author} {\bibnamefont {Gourdin}, \bibfnamefont
  {E.}}, \bibinfo {author} {\bibfnamefont {J.}~\bibnamefont {Omic}}, \ and\
  \bibinfo {author} {\bibfnamefont {P.~V.}\ \bibnamefont {Mieghem}}} (\bibinfo
  {year} {2011}),\ \href@noop {} {\bibinfo  {journal} {8th International
  Workshop on Design of Reliable Communication Networks (DRCN 2011), Krakow,
  Poland}\ }\BibitemShut {NoStop}%
\bibitem [{\citenamefont {Grais}\ \emph {et~al.}(2004)\citenamefont {Grais},
  \citenamefont {Eliis}, \citenamefont {Kress},\ and\ \citenamefont
  {Glass}}]{Grais:2004}%
  \BibitemOpen
\bibfield  {journal} {  }\bibfield  {author} {\bibinfo {author} {\bibnamefont
  {Grais}, \bibfnamefont {R.}}, \bibinfo {author} {\bibfnamefont
  {J.}~\bibnamefont {Eliis}}, \bibinfo {author} {\bibfnamefont
  {A.}~\bibnamefont {Kress}}, \ and\ \bibinfo {author} {\bibfnamefont
  {G.}~\bibnamefont {Glass}}} (\bibinfo {year} {2004}),\ \href@noop {}
  {\bibfield  {journal} {\bibinfo  {journal} {Health Care Manag Sci}\ }\textbf
  {\bibinfo {volume} {7}},\ \bibinfo {pages} {127}}\BibitemShut {NoStop}%
\bibitem [{\citenamefont {Granell}\ \emph {et~al.}(2013)\citenamefont
  {Granell}, \citenamefont {G\'omez},\ and\ \citenamefont
  {Arenas}}]{Granell2013}%
  \BibitemOpen
  \bibfield  {author} {\bibinfo {author} {\bibnamefont {Granell}, \bibfnamefont
  {C.}}, \bibinfo {author} {\bibfnamefont {S.}~\bibnamefont {G\'omez}}, \ and\
  \bibinfo {author} {\bibfnamefont {A.}~\bibnamefont {Arenas}}} (\bibinfo
  {year} {2013}),\ \href {\doibase 10.1103/PhysRevLett.111.128701} {\bibfield
  {journal} {\bibinfo  {journal} {Phys. Rev. Lett.}\ }\textbf {\bibinfo
  {volume} {111}},\ \bibinfo {pages} {128701}}\BibitemShut {NoStop}%
\bibitem [{\citenamefont {Granovetter}(1978)}]{Granovetter1978}%
  \BibitemOpen
  \bibfield  {author} {\bibinfo {author} {\bibnamefont {Granovetter},
  \bibfnamefont {M.}}} (\bibinfo {year} {1978}),\ \href@noop {} {\bibfield
  {journal} {\bibinfo  {journal} {American Journal of Sociology}\ }\textbf
  {\bibinfo {volume} {83}},\ \bibinfo {pages} {1420}}\BibitemShut {NoStop}%
\bibitem [{\citenamefont {Granovetter}(1973)}]{Granovetter1973}%
  \BibitemOpen
  \bibfield  {author} {\bibinfo {author} {\bibnamefont {Granovetter},
  \bibfnamefont {M.~S.}}} (\bibinfo {year} {1973}),\ \href@noop {} {\bibfield
  {journal} {\bibinfo  {journal} {American Journal of Sociology}\ }\textbf
  {\bibinfo {volume} {78}},\ \bibinfo {pages} {1360}}\BibitemShut {NoStop}%
\bibitem [{\citenamefont {Grassberger}(1983)}]{Grassberger1983}%
  \BibitemOpen
  \bibfield  {author} {\bibinfo {author} {\bibnamefont {Grassberger},
  \bibfnamefont {P.}}} (\bibinfo {year} {1983}),\ \href {\doibase
  10.1016/0025-5564(82)90036-0} {\bibfield  {journal} {\bibinfo  {journal}
  {Mathematical Biosciences}\ }\textbf {\bibinfo {volume} {63}}~(\bibinfo
  {number} {2}),\ \bibinfo {pages} {157}}\BibitemShut {NoStop}%
\bibitem [{\citenamefont {Grenfell}\ and\ \citenamefont
  {Harwood}(1997)}]{grenfell1997meta}%
  \BibitemOpen
  \bibfield  {author} {\bibinfo {author} {\bibnamefont {Grenfell},
  \bibfnamefont {B.}}, \ and\ \bibinfo {author} {\bibfnamefont
  {J.}~\bibnamefont {Harwood}}} (\bibinfo {year} {1997}),\ \href@noop {}
  {\bibfield  {journal} {\bibinfo  {journal} {Trends in ecology \& evolution}\
  }\textbf {\bibinfo {volume} {12}}~(\bibinfo {number} {10}),\ \bibinfo {pages}
  {395}}\BibitemShut {NoStop}%
\bibitem [{\citenamefont {Gross}\ and\ \citenamefont
  {Blasius}(2008)}]{Gross2008}%
  \BibitemOpen
  \bibfield  {author} {\bibinfo {author} {\bibnamefont {Gross}, \bibfnamefont
  {T.}}, \ and\ \bibinfo {author} {\bibfnamefont {B.}~\bibnamefont {Blasius}}}
  (\bibinfo {year} {2008}),\ \href@noop {} {\bibfield  {journal} {\bibinfo
  {journal} {Journal of the Royal Society Interface}\ }\textbf {\bibinfo
  {volume} {5}}~(\bibinfo {number} {20}),\ \bibinfo {pages} {259}}\BibitemShut
  {NoStop}%
\bibitem [{\citenamefont {Gross}\ \emph {et~al.}(2006)\citenamefont {Gross},
  \citenamefont {D'Lima},\ and\ \citenamefont {Blasius}}]{Gross2006}%
  \BibitemOpen
  \bibfield  {author} {\bibinfo {author} {\bibnamefont {Gross}, \bibfnamefont
  {T.}}, \bibinfo {author} {\bibfnamefont {C.}~\bibnamefont {D'Lima}}, \ and\
  \bibinfo {author} {\bibfnamefont {B.}~\bibnamefont {Blasius}}} (\bibinfo
  {year} {2006}),\ \href {\doibase 10.1103/PhysRevLett.96.208701} {\bibfield
  {journal} {\bibinfo  {journal} {Physical Review Letters}\ }\textbf {\bibinfo
  {volume} {96}}~(\bibinfo {number} {20}),\ \bibinfo {pages}
  {208701}}\BibitemShut {NoStop}%
\bibitem [{\citenamefont {Gruhl}\ \emph {et~al.}(2004)\citenamefont {Gruhl},
  \citenamefont {Guha}, \citenamefont {Liben-Nowell},\ and\ \citenamefont
  {Tomkins}}]{Gruhl2004}%
  \BibitemOpen
  \bibfield  {author} {\bibinfo {author} {\bibnamefont {Gruhl}, \bibfnamefont
  {D.}}, \bibinfo {author} {\bibfnamefont {R.}~\bibnamefont {Guha}}, \bibinfo
  {author} {\bibfnamefont {D.}~\bibnamefont {Liben-Nowell}}, \ and\ \bibinfo
  {author} {\bibfnamefont {A.}~\bibnamefont {Tomkins}}} (\bibinfo {year}
  {2004}),\ in\ \href {\doibase 10.1145/988672.988739} {\emph {\bibinfo
  {booktitle} {Proceedings of the 13th conference on World Wide Web - WWW
  '04}}}\ (\bibinfo  {publisher} {ACM Press},\ \bibinfo {address} {New York,
  New York, USA})\ p.\ \bibinfo {pages} {491}\BibitemShut {NoStop}%
\bibitem [{\citenamefont {Guo}\ \emph {et~al.}(2013)\citenamefont {Guo},
  \citenamefont {Trajanovski}, \citenamefont {van~de Bovenkamp}, \citenamefont
  {Wang},\ and\ \citenamefont {{Van Mieghem}}}]{Guo2013}%
  \BibitemOpen
  \bibfield  {author} {\bibinfo {author} {\bibnamefont {Guo}, \bibfnamefont
  {D.}}, \bibinfo {author} {\bibfnamefont {S.}~\bibnamefont {Trajanovski}},
  \bibinfo {author} {\bibfnamefont {R.}~\bibnamefont {van~de Bovenkamp}},
  \bibinfo {author} {\bibfnamefont {H.}~\bibnamefont {Wang}}, \ and\ \bibinfo
  {author} {\bibfnamefont {P.}~\bibnamefont {{Van Mieghem}}}} (\bibinfo {year}
  {2013}),\ \href {\doibase 10.1103/PhysRevE.88.042802} {\bibfield  {journal}
  {\bibinfo  {journal} {Physical Review E}\ }\textbf {\bibinfo {volume}
  {88}}~(\bibinfo {number} {4}),\ \bibinfo {pages} {042802}}\BibitemShut
  {NoStop}%
\bibitem [{\citenamefont {Hackett}\ \emph {et~al.}(2011)\citenamefont
  {Hackett}, \citenamefont {Melnik},\ and\ \citenamefont
  {Gleeson}}]{Hackett2011}%
  \BibitemOpen
  \bibfield  {author} {\bibinfo {author} {\bibnamefont {Hackett}, \bibfnamefont
  {A.}}, \bibinfo {author} {\bibfnamefont {S.}~\bibnamefont {Melnik}}, \ and\
  \bibinfo {author} {\bibfnamefont {J.}~\bibnamefont {Gleeson}}} (\bibinfo
  {year} {2011}),\ \href {\doibase 10.1103/PhysRevE.83.056107} {\bibfield
  {journal} {\bibinfo  {journal} {Physical Review E}\ }\textbf {\bibinfo
  {volume} {83}}~(\bibinfo {number} {5}),\ \bibinfo {pages} {35}}\BibitemShut
  {NoStop}%
\bibitem [{\citenamefont {Halloran}\ \emph {et~al.}(2008)\citenamefont
  {Halloran}, \citenamefont {Ferguson}, \citenamefont {Eubank}, \citenamefont
  {Longini}, \citenamefont {Cummings}, \citenamefont {Lewis}, \citenamefont
  {Xu}, \citenamefont {Fraser}, \citenamefont {Vullikanti}, \citenamefont
  {Germann}, \citenamefont {Wagener}, \citenamefont {Beckman}, \citenamefont
  {Kadau}, \citenamefont {Barrett}, \citenamefont {Macken}, \citenamefont
  {Burke},\ and\ \citenamefont {Cooley}}]{Halloran2008}%
  \BibitemOpen
  \bibfield  {author} {\bibinfo {author} {\bibnamefont {Halloran},
  \bibfnamefont {M.~E.}}, \bibinfo {author} {\bibfnamefont {N.~M.}\
  \bibnamefont {Ferguson}}, \bibinfo {author} {\bibfnamefont {S.}~\bibnamefont
  {Eubank}}, \bibinfo {author} {\bibfnamefont {I.~M.}\ \bibnamefont {Longini}},
  \bibinfo {author} {\bibfnamefont {D.~A.~T.}\ \bibnamefont {Cummings}},
  \bibinfo {author} {\bibfnamefont {B.}~\bibnamefont {Lewis}}, \bibinfo
  {author} {\bibfnamefont {S.}~\bibnamefont {Xu}}, \bibinfo {author}
  {\bibfnamefont {C.}~\bibnamefont {Fraser}}, \bibinfo {author} {\bibfnamefont
  {A.}~\bibnamefont {Vullikanti}}, \bibinfo {author} {\bibfnamefont {T.~C.}\
  \bibnamefont {Germann}}, \bibinfo {author} {\bibfnamefont {D.}~\bibnamefont
  {Wagener}}, \bibinfo {author} {\bibfnamefont {R.}~\bibnamefont {Beckman}},
  \bibinfo {author} {\bibfnamefont {K.}~\bibnamefont {Kadau}}, \bibinfo
  {author} {\bibfnamefont {C.}~\bibnamefont {Barrett}}, \bibinfo {author}
  {\bibfnamefont {C.~A.}\ \bibnamefont {Macken}}, \bibinfo {author}
  {\bibfnamefont {D.~S.}\ \bibnamefont {Burke}}, \ and\ \bibinfo {author}
  {\bibfnamefont {P.}~\bibnamefont {Cooley}}} (\bibinfo {year} {2008}),\
  \href@noop {} {\bibfield  {journal} {\bibinfo  {journal} {Proc. Natl. Acad.
  Sci. USA}\ }\textbf {\bibinfo {volume} {105}},\ \bibinfo {pages}
  {4639}}\BibitemShut {NoStop}%
\bibitem [{\citenamefont {Hamilton}\ and\ \citenamefont
  {Pryadko}(2014)}]{Hamilton2014}%
  \BibitemOpen
  \bibfield  {author} {\bibinfo {author} {\bibnamefont {Hamilton},
  \bibfnamefont {K.~E.}}, \ and\ \bibinfo {author} {\bibfnamefont {L.~P.}\
  \bibnamefont {Pryadko}}} (\bibinfo {year} {2014}),\ \href {\doibase
  10.1103/PhysRevLett.113.208701} {\bibfield  {journal} {\bibinfo  {journal}
  {Phys. Rev. Lett.}\ }\textbf {\bibinfo {volume} {113}},\ \bibinfo {pages}
  {208701}}\BibitemShut {NoStop}%
\bibitem [{\citenamefont {Hammersley}\ and\ \citenamefont
  {Welsh}(1965)}]{hammersley1965first}%
  \BibitemOpen
  \bibfield  {author} {\bibinfo {author} {\bibnamefont {Hammersley},
  \bibfnamefont {J.~M.}}, \ and\ \bibinfo {author} {\bibfnamefont
  {D.}~\bibnamefont {Welsh}}} (\bibinfo {year} {1965}),\ in\ \href@noop {}
  {\emph {\bibinfo {booktitle} {Bernoulli 1713, Bayes 1763, Laplace 1813}}}\
  (\bibinfo  {publisher} {Springer})\ pp.\ \bibinfo {pages}
  {61--110}\BibitemShut {NoStop}%
\bibitem [{\citenamefont {Hanski}\ and\ \citenamefont
  {Gaggiotti}(2004)}]{Hanski:2004}%
  \BibitemOpen
  \bibfield  {author} {\bibinfo {author} {\bibnamefont {Hanski}, \bibfnamefont
  {I.}}, \ and\ \bibinfo {author} {\bibfnamefont {O.}~\bibnamefont
  {Gaggiotti}}} (\bibinfo {year} {2004}),\ \href@noop {} {\emph {\bibinfo
  {title} {Ecology, Genetics and Evolution of Metapopulations}}}\ (\bibinfo
  {publisher} {Elsevier Science},\ \bibinfo {address} {Princeton})\BibitemShut
  {NoStop}%
\bibitem [{\citenamefont {Harris}(1974)}]{harris74}%
  \BibitemOpen
  \bibfield  {author} {\bibinfo {author} {\bibnamefont {Harris}, \bibfnamefont
  {T.~E.}}} (\bibinfo {year} {1974}),\ \href@noop {} {\bibfield  {journal}
  {\bibinfo  {journal} {Ann. Prob.}\ }\textbf {\bibinfo {volume} {2}},\
  \bibinfo {pages} {969}}\BibitemShut {NoStop}%
\bibitem [{\citenamefont {H\'{e}bert-Dufresne}\ \emph
  {et~al.}(2013)\citenamefont {H\'{e}bert-Dufresne}, \citenamefont {Allard},
  \citenamefont {Young},\ and\ \citenamefont {Dub\'{e}}}]{HebertDufresne2012}%
  \BibitemOpen
  \bibfield  {author} {\bibinfo {author} {\bibnamefont {H\'{e}bert-Dufresne},
  \bibfnamefont {L.}}, \bibinfo {author} {\bibfnamefont {A.}~\bibnamefont
  {Allard}}, \bibinfo {author} {\bibfnamefont {J.-g.}\ \bibnamefont {Young}}, \
  and\ \bibinfo {author} {\bibfnamefont {L.~J.}\ \bibnamefont {Dub\'{e}}}}
  (\bibinfo {year} {2013}),\ \href@noop {} {\bibfield  {journal} {\bibinfo
  {journal} {Scientific Reports}\ }\textbf {\bibinfo {volume} {3}},\ \bibinfo
  {pages} {2171}}\BibitemShut {NoStop}%
\bibitem [{\citenamefont {Henkel}\ \emph {et~al.}(2008)\citenamefont {Henkel},
  \citenamefont {Hinrichsen},\ and\ \citenamefont {L\"ubeck}}]{Henkel}%
  \BibitemOpen
  \bibfield  {author} {\bibinfo {author} {\bibnamefont {Henkel}, \bibfnamefont
  {M.}}, \bibinfo {author} {\bibfnamefont {H.}~\bibnamefont {Hinrichsen}}, \
  and\ \bibinfo {author} {\bibfnamefont {S.}~\bibnamefont {L\"ubeck}}}
  (\bibinfo {year} {2008}),\ \href@noop {} {\emph {\bibinfo {title}
  {Non-equilibrium phase transition: Absorbing Phase Transitions}}}\ (\bibinfo
  {publisher} {Springer Verlag},\ \bibinfo {address} {Netherlands})\BibitemShut
  {NoStop}%
\bibitem [{\citenamefont {Hern\'andez}\ and\ \citenamefont
  {Risau-Gusman}(2013)}]{Hernandez2013}%
  \BibitemOpen
  \bibfield  {author} {\bibinfo {author} {\bibnamefont {Hern\'andez},
  \bibfnamefont {D.~G.}}, \ and\ \bibinfo {author} {\bibfnamefont
  {S.}~\bibnamefont {Risau-Gusman}}} (\bibinfo {year} {2013}),\ \href {\doibase
  10.1103/PhysRevE.88.052801} {\bibfield  {journal} {\bibinfo  {journal} {Phys.
  Rev. E}\ }\textbf {\bibinfo {volume} {88}},\ \bibinfo {pages}
  {052801}}\BibitemShut {NoStop}%
\bibitem [{\citenamefont {Hethcote}\ and\ \citenamefont
  {Yorke}(1984)}]{hethcote1984gonorrhea}%
  \BibitemOpen
  \bibfield  {author} {\bibinfo {author} {\bibnamefont {Hethcote},
  \bibfnamefont {H.}}, \ and\ \bibinfo {author} {\bibfnamefont
  {J.}~\bibnamefont {Yorke}}} (\bibinfo {year} {1984}),\ \href
  {http://books.google.es/books?id=YfHVMQEACAAJ} {\emph {\bibinfo {title}
  {Gonorrhea Transmission Dynamics and Control}}},\ Lecture Notes in
  Biomathematics\ (\bibinfo  {publisher} {Springer-Verlag})\BibitemShut
  {NoStop}%
\bibitem [{\citenamefont {Hethcote}(2000)}]{hethcote2000}%
  \BibitemOpen
  \bibfield  {author} {\bibinfo {author} {\bibnamefont {Hethcote},
  \bibfnamefont {H.~W.}}} (\bibinfo {year} {2000}),\ \href@noop {} {\bibfield
  {journal} {\bibinfo  {journal} {SIAM Review}\ }\textbf {\bibinfo {volume}
  {42}},\ \bibinfo {pages} {599}}\BibitemShut {NoStop}%
\bibitem [{\citenamefont {Hoffmann}\ \emph {et~al.}(2012)\citenamefont
  {Hoffmann}, \citenamefont {Porter},\ and\ \citenamefont
  {Lambiotte}}]{hoffmann_generalized_2012}%
  \BibitemOpen
  \bibfield  {author} {\bibinfo {author} {\bibnamefont {Hoffmann},
  \bibfnamefont {T.}}, \bibinfo {author} {\bibfnamefont {M.~A.}\ \bibnamefont
  {Porter}}, \ and\ \bibinfo {author} {\bibfnamefont {R.}~\bibnamefont
  {Lambiotte}}} (\bibinfo {year} {2012}),\ \href {\doibase
  10.1103/PhysRevE.86.046102} {\bibfield  {journal} {\bibinfo  {journal}
  {Physical Review E}\ }\textbf {\bibinfo {volume} {86}}~(\bibinfo {number}
  {4}),\ \bibinfo {pages} {046102}}\BibitemShut {NoStop}%
\bibitem [{\citenamefont {Hollingsworth}\ \emph {et~al.}(2006)\citenamefont
  {Hollingsworth}, \citenamefont {Ferguson},\ and\ \citenamefont
  {Anderson}}]{Hollingsworth:2006}%
  \BibitemOpen
  \bibfield  {author} {\bibinfo {author} {\bibnamefont {Hollingsworth},
  \bibfnamefont {T.}}, \bibinfo {author} {\bibfnamefont {N.}~\bibnamefont
  {Ferguson}}, \ and\ \bibinfo {author} {\bibfnamefont {R.}~\bibnamefont
  {Anderson}}} (\bibinfo {year} {2006}),\ \href {\doibase 10.1038/nm0506-497}
  {\bibfield  {journal} {\bibinfo  {journal} {Nat Med}\ }\textbf {\bibinfo
  {volume} {12}},\ \bibinfo {pages} {497}}\BibitemShut {NoStop}%
\bibitem [{\citenamefont {Holme}(2004)}]{0295-5075-68-6-908}%
  \BibitemOpen
  \bibfield  {author} {\bibinfo {author} {\bibnamefont {Holme}, \bibfnamefont
  {P.}}} (\bibinfo {year} {2004}),\ \href
  {http://stacks.iop.org/0295-5075/68/i=6/a=908} {\bibfield  {journal}
  {\bibinfo  {journal} {Europhysics Letters}\ }\textbf {\bibinfo {volume}
  {68}}~(\bibinfo {number} {6}),\ \bibinfo {pages} {908}}\BibitemShut {NoStop}%
\bibitem [{\citenamefont {Holme}(2005)}]{PhysRevE.71.046119}%
  \BibitemOpen
  \bibfield  {author} {\bibinfo {author} {\bibnamefont {Holme}, \bibfnamefont
  {P.}}} (\bibinfo {year} {2005}),\ \href@noop {} {\bibfield  {journal}
  {\bibinfo  {journal} {Phys. Rev. E}\ }\textbf {\bibinfo {volume} {71}},\
  \bibinfo {pages} {046119}}\BibitemShut {NoStop}%
\bibitem [{\citenamefont {Holme}\ \emph {et~al.}(2002)\citenamefont {Holme},
  \citenamefont {Kim}, \citenamefont {Yoon},\ and\ \citenamefont
  {Han}}]{PhysRevE.65.056109}%
  \BibitemOpen
  \bibfield  {author} {\bibinfo {author} {\bibnamefont {Holme}, \bibfnamefont
  {P.}}, \bibinfo {author} {\bibfnamefont {B.~J.}\ \bibnamefont {Kim}},
  \bibinfo {author} {\bibfnamefont {C.~N.}\ \bibnamefont {Yoon}}, \ and\
  \bibinfo {author} {\bibfnamefont {S.~K.}\ \bibnamefont {Han}}} (\bibinfo
  {year} {2002}),\ \href@noop {} {\bibfield  {journal} {\bibinfo  {journal}
  {Phys. Rev. E}\ }\textbf {\bibinfo {volume} {65}},\ \bibinfo {pages}
  {056109}}\BibitemShut {NoStop}%
\bibitem [{\citenamefont {Holme}\ and\ \citenamefont
  {Saram{\"a}ki}(2012)}]{Holme:2011fk}%
  \BibitemOpen
  \bibfield  {author} {\bibinfo {author} {\bibnamefont {Holme}, \bibfnamefont
  {P.}}, \ and\ \bibinfo {author} {\bibfnamefont {J.}~\bibnamefont
  {Saram{\"a}ki}}} (\bibinfo {year} {2012}),\ \href@noop {} {\bibfield
  {journal} {\bibinfo  {journal} {Physics Reports}\ }\textbf {\bibinfo {volume}
  {519}},\ \bibinfo {pages} {97}}\BibitemShut {NoStop}%
\bibitem [{\citenamefont {Holme}\ and\ \citenamefont
  {Saram{\"a}ki}(2013)}]{temporalnetworksbook}%
  \BibitemOpen
  \bibinfo {editor} {\bibnamefont {Holme}, \bibfnamefont {P.}}, \ and\ \bibinfo
  {editor} {\bibfnamefont {J.}~\bibnamefont {Saram{\"a}ki}},\ Eds. (\bibinfo
  {year} {2013}),\ \href@noop {} {\emph {\bibinfo {title} {Temporal
  networks}}}\ (\bibinfo  {publisher} {Springer},\ \bibinfo {address}
  {Berlin})\BibitemShut {NoStop}%
\bibitem [{\citenamefont {Hou}\ \emph {et~al.}(2012)\citenamefont {Hou},
  \citenamefont {Yao},\ and\ \citenamefont {Liao}}]{Hou2012}%
  \BibitemOpen
  \bibfield  {author} {\bibinfo {author} {\bibnamefont {Hou}, \bibfnamefont
  {B.}}, \bibinfo {author} {\bibfnamefont {Y.}~\bibnamefont {Yao}}, \ and\
  \bibinfo {author} {\bibfnamefont {D.}~\bibnamefont {Liao}}} (\bibinfo {year}
  {2012}),\ \href {\doibase 10.1016/j.physa.2012.02.033} {\bibfield  {journal}
  {\bibinfo  {journal} {Physica A}\ }\textbf {\bibinfo {volume}
  {391}}~(\bibinfo {number} {15}),\ \bibinfo {pages} {4012}}\BibitemShut
  {NoStop}%
\bibitem [{\citenamefont {House}\ and\ \citenamefont
  {Keeling}(2011)}]{House2011}%
  \BibitemOpen
  \bibfield  {author} {\bibinfo {author} {\bibnamefont {House}, \bibfnamefont
  {T.}}, \ and\ \bibinfo {author} {\bibfnamefont {M.~J.}\ \bibnamefont
  {Keeling}}} (\bibinfo {year} {2011}),\ \href@noop {} {\bibfield  {journal}
  {\bibinfo  {journal} {Journal of The Royal Society Interface}\ }\textbf
  {\bibinfo {volume} {8}}~(\bibinfo {number} {54}),\ \bibinfo {pages}
  {67}}\BibitemShut {NoStop}%
\bibitem [{\citenamefont {Huang}\ and\ \citenamefont {Li}(2007)}]{Huang2007}%
  \BibitemOpen
  \bibfield  {author} {\bibinfo {author} {\bibnamefont {Huang}, \bibfnamefont
  {W.}}, \ and\ \bibinfo {author} {\bibfnamefont {C.}~\bibnamefont {Li}}}
  (\bibinfo {year} {2007}),\ \href
  {http://stacks.iop.org/1742-5468/2007/i=01/a=P01014} {\bibfield  {journal}
  {\bibinfo  {journal} {Journal of Statistical Mechanics: Theory and
  Experiment}\ }\textbf {\bibinfo {volume} {2007}}~(\bibinfo {number} {01}),\
  \bibinfo {pages} {P01014}}\BibitemShut {NoStop}%
\bibitem [{\citenamefont {Hufnagel}\ \emph {et~al.}(2004)\citenamefont
  {Hufnagel}, \citenamefont {Brockmann},\ and\ \citenamefont
  {Geisel}}]{Hufnagel:2004}%
  \BibitemOpen
  \bibfield  {author} {\bibinfo {author} {\bibnamefont {Hufnagel},
  \bibfnamefont {L.}}, \bibinfo {author} {\bibfnamefont {D.}~\bibnamefont
  {Brockmann}}, \ and\ \bibinfo {author} {\bibfnamefont {T.}~\bibnamefont
  {Geisel}}} (\bibinfo {year} {2004}),\ \href {\doibase
  10.1073/pnas.0308344101} {\bibfield  {journal} {\bibinfo  {journal} {Proc
  Natl Acad Sci USA}\ }\textbf {\bibinfo {volume} {101}},\ \bibinfo {pages}
  {15124}}\BibitemShut {NoStop}%
\bibitem [{\citenamefont {Hui}\ \emph {et~al.}(2005)\citenamefont {Hui},
  \citenamefont {Chaintreau}, \citenamefont {Scott}, \citenamefont {Gass},
  \citenamefont {Crowcroft},\ and\ \citenamefont {Diot}}]{Hui:2005}%
  \BibitemOpen
  \bibfield  {author} {\bibinfo {author} {\bibnamefont {Hui}, \bibfnamefont
  {P.}}, \bibinfo {author} {\bibfnamefont {A.}~\bibnamefont {Chaintreau}},
  \bibinfo {author} {\bibfnamefont {J.}~\bibnamefont {Scott}}, \bibinfo
  {author} {\bibfnamefont {R.}~\bibnamefont {Gass}}, \bibinfo {author}
  {\bibfnamefont {J.}~\bibnamefont {Crowcroft}}, \ and\ \bibinfo {author}
  {\bibfnamefont {C.}~\bibnamefont {Diot}}} (\bibinfo {year} {2005}),\ in\
  \href {\doibase http://doi.acm.org/10.1145/1080139.1080142} {\emph {\bibinfo
  {booktitle} {WDTN '05: Proceedings of the 2005 ACM SIGCOMM workshop on
  Delay-tolerant networking}}}\ (\bibinfo  {publisher} {ACM},\ \bibinfo
  {address} {New York, NY, USA})\ pp.\ \bibinfo {pages} {244--251}\BibitemShut
  {NoStop}%
\bibitem [{\citenamefont {Iribarren}\ and\ \citenamefont
  {Moro}(2009)}]{Iribarren2009}%
  \BibitemOpen
  \bibfield  {author} {\bibinfo {author} {\bibnamefont {Iribarren},
  \bibfnamefont {J.}}, \ and\ \bibinfo {author} {\bibfnamefont
  {E.}~\bibnamefont {Moro}}} (\bibinfo {year} {2009}),\ \href {\doibase
  10.1103/PhysRevLett.103.038702} {\bibfield  {journal} {\bibinfo  {journal}
  {Physical Review Letters}\ }\textbf {\bibinfo {volume} {103}}~(\bibinfo
  {number} {3}),\ \bibinfo {pages} {038702}}\BibitemShut {NoStop}%
\bibitem [{\citenamefont {Jackson}(2010)}]{Jackson2010}%
  \BibitemOpen
  \bibfield  {author} {\bibinfo {author} {\bibnamefont {Jackson}, \bibfnamefont
  {M.}}} (\bibinfo {year} {2010}),\ \href@noop {} {\emph {\bibinfo {title}
  {Social and Economic Networks}}}\ (\bibinfo  {publisher} {Princeton
  University Press},\ \bibinfo {address} {Princeton})\BibitemShut {NoStop}%
\bibitem [{\citenamefont {Janson}\ \emph {et~al.}(2014)\citenamefont {Janson},
  \citenamefont {Luczak},\ and\ \citenamefont {Windridge}}]{Janson2013}%
  \BibitemOpen
  \bibfield  {author} {\bibinfo {author} {\bibnamefont {Janson}, \bibfnamefont
  {S.}}, \bibinfo {author} {\bibfnamefont {M.}~\bibnamefont {Luczak}}, \ and\
  \bibinfo {author} {\bibfnamefont {P.}~\bibnamefont {Windridge}}} (\bibinfo
  {year} {2014}),\ \href@noop {} {\bibfield  {journal} {\bibinfo  {journal}
  {Random Struct. Algorithms}\ }\textbf {\bibinfo {volume} {45}},\ \bibinfo
  {pages} {724}}\BibitemShut {NoStop}%
\bibitem [{\citenamefont {{Jo}}\ \emph {et~al.}(2014)\citenamefont {{Jo}},
  \citenamefont {{Perotti}}, \citenamefont {{Kaski}},\ and\ \citenamefont
  {{Kert{\'e}sz}}}]{2013arXiv1309.0701J}%
  \BibitemOpen
  \bibfield  {author} {\bibinfo {author} {\bibnamefont {{Jo}}, \bibfnamefont
  {H.-H.}}, \bibinfo {author} {\bibfnamefont {J.~I.}\ \bibnamefont
  {{Perotti}}}, \bibinfo {author} {\bibfnamefont {K.}~\bibnamefont {{Kaski}}},
  \ and\ \bibinfo {author} {\bibfnamefont {J.}~\bibnamefont {{Kert{\'e}sz}}}}
  (\bibinfo {year} {2014}),\ \href@noop {} {\bibfield  {journal} {\bibinfo
  {journal} {Phys. Rev. X}\ }\textbf {\bibinfo {volume} {4}},\ \bibinfo {pages}
  {011041}}\BibitemShut {NoStop}%
\bibitem [{\citenamefont {Joh}\ \emph {et~al.}(2009)\citenamefont {Joh},
  \citenamefont {Wang}, \citenamefont {Weiss},\ and\ \citenamefont
  {Weitz}}]{Joh2009}%
  \BibitemOpen
  \bibfield  {author} {\bibinfo {author} {\bibnamefont {Joh}, \bibfnamefont
  {R.~I.}}, \bibinfo {author} {\bibfnamefont {H.}~\bibnamefont {Wang}},
  \bibinfo {author} {\bibfnamefont {H.}~\bibnamefont {Weiss}}, \ and\ \bibinfo
  {author} {\bibfnamefont {J.~S.}\ \bibnamefont {Weitz}}} (\bibinfo {year}
  {2009}),\ \href {\doibase 10.1007/s11538-008-9384-4} {\bibfield  {journal}
  {\bibinfo  {journal} {Bulletin of Mathematical Biology}\ }\textbf {\bibinfo
  {volume} {71}},\ \bibinfo {pages} {845}}\BibitemShut {NoStop}%
\bibitem [{\citenamefont {Joo}\ and\ \citenamefont {Lebowitz}(2004)}]{Joo2004}%
  \BibitemOpen
  \bibfield  {author} {\bibinfo {author} {\bibnamefont {Joo}, \bibfnamefont
  {J.}}, \ and\ \bibinfo {author} {\bibfnamefont {J.~L.}\ \bibnamefont
  {Lebowitz}}} (\bibinfo {year} {2004}),\ \href {\doibase
  10.1103/PhysRevE.69.066105} {\bibfield  {journal} {\bibinfo  {journal} {Phys.
  Rev. E}\ }\textbf {\bibinfo {volume} {69}},\ \bibinfo {pages}
  {066105}}\BibitemShut {NoStop}%
\bibitem [{\citenamefont {Juh\'asz}\ \emph {et~al.}(2012)\citenamefont
  {Juh\'asz}, \citenamefont {\'Odor}, \citenamefont {Castellano},\ and\
  \citenamefont {Mu\~noz}}]{Juhasz2012}%
  \BibitemOpen
  \bibfield  {author} {\bibinfo {author} {\bibnamefont {Juh\'asz},
  \bibfnamefont {R.}}, \bibinfo {author} {\bibfnamefont {G.}~\bibnamefont
  {\'Odor}}, \bibinfo {author} {\bibfnamefont {C.}~\bibnamefont {Castellano}},
  \ and\ \bibinfo {author} {\bibfnamefont {M.~A.}\ \bibnamefont {Mu\~noz}}}
  (\bibinfo {year} {2012}),\ \href {\doibase 10.1103/PhysRevE.85.066125}
  {\bibfield  {journal} {\bibinfo  {journal} {Phys. Rev. E}\ }\textbf {\bibinfo
  {volume} {85}},\ \bibinfo {pages} {066125}}\BibitemShut {NoStop}%
\bibitem [{\citenamefont {Karimi}\ and\ \citenamefont
  {Holme}(2013)}]{Karimi2013}%
  \BibitemOpen
  \bibfield  {author} {\bibinfo {author} {\bibnamefont {Karimi}, \bibfnamefont
  {F.}}, \ and\ \bibinfo {author} {\bibfnamefont {P.}~\bibnamefont {Holme}}}
  (\bibinfo {year} {2013}),\ \href {\doibase 10.1016/j.physa.2013.03.050}
  {\bibfield  {journal} {\bibinfo  {journal} {Physica A}\ }\textbf {\bibinfo
  {volume} {392}}~(\bibinfo {number} {16}),\ \bibinfo {pages}
  {3476}}\BibitemShut {NoStop}%
\bibitem [{\citenamefont {Karrer}\ and\ \citenamefont
  {Newman}(2010)}]{Karrer2010}%
  \BibitemOpen
  \bibfield  {author} {\bibinfo {author} {\bibnamefont {Karrer}, \bibfnamefont
  {B.}}, \ and\ \bibinfo {author} {\bibfnamefont {M.~E.~J.}\ \bibnamefont
  {Newman}}} (\bibinfo {year} {2010}),\ \href {\doibase
  10.1103/PhysRevE.82.016101} {\bibfield  {journal} {\bibinfo  {journal} {Phys.
  Rev. E}\ }\textbf {\bibinfo {volume} {82}}~(\bibinfo {number} {1}),\ \bibinfo
  {pages} {016101}}\BibitemShut {NoStop}%
\bibitem [{\citenamefont {Karrer}\ and\ \citenamefont
  {Newman}(2011)}]{Newmancompeting2011}%
  \BibitemOpen
  \bibfield  {author} {\bibinfo {author} {\bibnamefont {Karrer}, \bibfnamefont
  {B.}}, \ and\ \bibinfo {author} {\bibfnamefont {M.~E.~J.}\ \bibnamefont
  {Newman}}} (\bibinfo {year} {2011}),\ \href {\doibase
  10.1103/PhysRevE.84.036106} {\bibfield  {journal} {\bibinfo  {journal} {Phys.
  Rev. E}\ }\textbf {\bibinfo {volume} {84}},\ \bibinfo {pages}
  {036106}}\BibitemShut {NoStop}%
\bibitem [{\citenamefont {Karrer}\ \emph {et~al.}(2014)\citenamefont {Karrer},
  \citenamefont {Newman},\ and\ \citenamefont {Zdeborov\'a}}]{Karrer2014}%
  \BibitemOpen
  \bibfield  {author} {\bibinfo {author} {\bibnamefont {Karrer}, \bibfnamefont
  {B.}}, \bibinfo {author} {\bibfnamefont {M.~E.~J.}\ \bibnamefont {Newman}}, \
  and\ \bibinfo {author} {\bibfnamefont {L.}~\bibnamefont {Zdeborov\'a}}}
  (\bibinfo {year} {2014}),\ \href {\doibase 10.1103/PhysRevLett.113.208702}
  {\bibfield  {journal} {\bibinfo  {journal} {Phys. Rev. Lett.}\ }\textbf
  {\bibinfo {volume} {113}},\ \bibinfo {pages} {208702}}\BibitemShut {NoStop}%
\bibitem [{\citenamefont {Karsai}\ \emph {et~al.}(2006)\citenamefont {Karsai},
  \citenamefont {Juh\'asz},\ and\ \citenamefont {Igl\'oi}}]{karsai:036116}%
  \BibitemOpen
  \bibfield  {author} {\bibinfo {author} {\bibnamefont {Karsai}, \bibfnamefont
  {M.}}, \bibinfo {author} {\bibfnamefont {R.}~\bibnamefont {Juh\'asz}}, \ and\
  \bibinfo {author} {\bibfnamefont {F.}~\bibnamefont {Igl\'oi}}} (\bibinfo
  {year} {2006}),\ \href {\doibase 10.1103/PhysRevE.73.036116} {\bibfield
  {journal} {\bibinfo  {journal} {Phys. Rev. E}\ }\textbf {\bibinfo {volume}
  {73}},\ \bibinfo {pages} {036116}}\BibitemShut {NoStop}%
\bibitem [{\citenamefont {Karsai}\ \emph {et~al.}(2011)\citenamefont {Karsai},
  \citenamefont {Kivel\"a}, \citenamefont {Pan}, \citenamefont {Kaski},
  \citenamefont {Kert\'esz}, \citenamefont {Barab\'asi},\ and\ \citenamefont
  {Saram\"aki}}]{PhysRevE.83.025102}%
  \BibitemOpen
  \bibfield  {author} {\bibinfo {author} {\bibnamefont {Karsai}, \bibfnamefont
  {M.}}, \bibinfo {author} {\bibfnamefont {M.}~\bibnamefont {Kivel\"a}},
  \bibinfo {author} {\bibfnamefont {R.~K.}\ \bibnamefont {Pan}}, \bibinfo
  {author} {\bibfnamefont {K.}~\bibnamefont {Kaski}}, \bibinfo {author}
  {\bibfnamefont {J.}~\bibnamefont {Kert\'esz}}, \bibinfo {author}
  {\bibfnamefont {A.-L.}\ \bibnamefont {Barab\'asi}}, \ and\ \bibinfo {author}
  {\bibfnamefont {J.}~\bibnamefont {Saram\"aki}}} (\bibinfo {year} {2011}),\
  \href {\doibase 10.1103/PhysRevE.83.025102} {\bibfield  {journal} {\bibinfo
  {journal} {Phys. Rev. E}\ }\textbf {\bibinfo {volume} {83}},\ \bibinfo
  {pages} {025102}}\BibitemShut {NoStop}%
\bibitem [{\citenamefont {Ke}\ and\ \citenamefont
  {Yi}(2006)}]{1009-1963-15-12-003}%
  \BibitemOpen
  \bibfield  {author} {\bibinfo {author} {\bibnamefont {Ke}, \bibfnamefont
  {H.}}, \ and\ \bibinfo {author} {\bibfnamefont {T.}~\bibnamefont {Yi}}}
  (\bibinfo {year} {2006}),\ \href {\doibase 10.1088/1009-1963/15/12/003}
  {\bibfield  {journal} {\bibinfo  {journal} {Chinese Phys.}\ }\textbf
  {\bibinfo {volume} {15}}~(\bibinfo {number} {12}),\ \bibinfo {pages}
  {2782}}\BibitemShut {NoStop}%
\bibitem [{\citenamefont {Keeling}\ and\ \citenamefont
  {Rohani}(2007)}]{Keeling07book}%
  \BibitemOpen
  \bibfield  {author} {\bibinfo {author} {\bibnamefont {Keeling}, \bibfnamefont
  {M.}}, \ and\ \bibinfo {author} {\bibfnamefont {P.}~\bibnamefont {Rohani}}}
  (\bibinfo {year} {2007}),\ \href
  {http://books.google.it/books?id=LxzILSuKDhUC} {\emph {\bibinfo {title}
  {Modeling Infectious Diseases in Humans and Animals}}}\ (\bibinfo
  {publisher} {Princeton University Press},\ \bibinfo {address}
  {Princeton})\BibitemShut {NoStop}%
\bibitem [{\citenamefont {Keeling}(1999)}]{keeling99}%
  \BibitemOpen
  \bibfield  {author} {\bibinfo {author} {\bibnamefont {Keeling}, \bibfnamefont
  {M.~J.}}} (\bibinfo {year} {1999}),\ \href@noop {} {\bibfield  {journal}
  {\bibinfo  {journal} {Proc. R. Soc. Lond. B}\ }\textbf {\bibinfo {volume}
  {266}},\ \bibinfo {pages} {859}}\BibitemShut {NoStop}%
\bibitem [{\citenamefont {Keeling}\ and\ \citenamefont
  {Eames}(2005)}]{keeling05:_networ}%
  \BibitemOpen
  \bibfield  {author} {\bibinfo {author} {\bibnamefont {Keeling}, \bibfnamefont
  {M.~J.}}, \ and\ \bibinfo {author} {\bibfnamefont {K.~T.~D.}\ \bibnamefont
  {Eames}}} (\bibinfo {year} {2005}),\ \href@noop {} {\bibfield  {journal}
  {\bibinfo  {journal} {J. R. Soc. Interface}\ }\textbf {\bibinfo {volume}
  {2}},\ \bibinfo {pages} {295}}\BibitemShut {NoStop}%
\bibitem [{\citenamefont {Keeling}\ and\ \citenamefont
  {Grenfell}(1997)}]{Keeling03011997}%
  \BibitemOpen
  \bibfield  {author} {\bibinfo {author} {\bibnamefont {Keeling}, \bibfnamefont
  {M.~J.}}, \ and\ \bibinfo {author} {\bibfnamefont {B.~T.}\ \bibnamefont
  {Grenfell}}} (\bibinfo {year} {1997}),\ \href {\doibase
  10.1126/science.275.5296.65} {\bibfield  {journal} {\bibinfo  {journal}
  {Science}\ }\textbf {\bibinfo {volume} {275}}~(\bibinfo {number} {5296}),\
  \bibinfo {pages} {65}}\BibitemShut {NoStop}%
\bibitem [{\citenamefont {Keeling}\ and\ \citenamefont
  {Rohani}(2002)}]{Keeling:2002}%
  \BibitemOpen
  \bibfield  {author} {\bibinfo {author} {\bibnamefont {Keeling}, \bibfnamefont
  {M.~J.}}, \ and\ \bibinfo {author} {\bibfnamefont {P.}~\bibnamefont
  {Rohani}}} (\bibinfo {year} {2002}),\ \href {\doibase
  10.1046/j.1461-0248.2002.00268.x} {\bibfield  {journal} {\bibinfo  {journal}
  {Ecology Letters}\ }\textbf {\bibinfo {volume} {5}}~(\bibinfo {number} {1}),\
  \bibinfo {pages} {20}}\BibitemShut {NoStop}%
\bibitem [{\citenamefont {Kempe}\ \emph {et~al.}(2003)\citenamefont {Kempe},
  \citenamefont {Kleinberg},\ and\ \citenamefont {Tardos}}]{Kempe2003}%
  \BibitemOpen
  \bibfield  {author} {\bibinfo {author} {\bibnamefont {Kempe}, \bibfnamefont
  {D.}}, \bibinfo {author} {\bibfnamefont {J.}~\bibnamefont {Kleinberg}}, \
  and\ \bibinfo {author} {\bibfnamefont {E.}~\bibnamefont {Tardos}}} (\bibinfo
  {year} {2003}),\ in\ \href@noop {} {\emph {\bibinfo {booktitle} {Proceedings
  of the ninth ACM SIGKDD international conference on Knowledge discovery and
  data mining}}},\ \bibinfo {editor} {edited by\ \bibinfo {editor}
  {\bibnamefont {ACM}}},\ pp.\ \bibinfo {pages} {137--146}\BibitemShut
  {NoStop}%
\bibitem [{\citenamefont {Kenah}\ and\ \citenamefont
  {Miller}(2011)}]{Kenah2011}%
  \BibitemOpen
  \bibfield  {author} {\bibinfo {author} {\bibnamefont {Kenah}, \bibfnamefont
  {E.}}, \ and\ \bibinfo {author} {\bibfnamefont {J.~C.}\ \bibnamefont
  {Miller}}} (\bibinfo {year} {2011}),\ \href {\doibase 10.1155/2011/543520}
  {\bibfield  {journal} {\bibinfo  {journal} {Interdisciplinary Perspectives on
  Infectious Diseases}\ }\textbf {\bibinfo {volume} {2011}},\ \bibinfo {pages}
  {543520}}\BibitemShut {NoStop}%
\bibitem [{\citenamefont {Kenah}\ and\ \citenamefont
  {Robins}(2007)}]{Kenah2007}%
  \BibitemOpen
  \bibfield  {author} {\bibinfo {author} {\bibnamefont {Kenah}, \bibfnamefont
  {E.}}, \ and\ \bibinfo {author} {\bibfnamefont {J.~M.}\ \bibnamefont
  {Robins}}} (\bibinfo {year} {2007}),\ \href {\doibase
  10.1103/PhysRevE.76.036113} {\bibfield  {journal} {\bibinfo  {journal} {Phys.
  Rev. E}\ }\textbf {\bibinfo {volume} {76}},\ \bibinfo {pages}
  {036113}}\BibitemShut {NoStop}%
\bibitem [{\citenamefont {Kermack}\ and\ \citenamefont
  {McKendrick}(1927)}]{siroriginal}%
  \BibitemOpen
  \bibfield  {author} {\bibinfo {author} {\bibnamefont {Kermack}, \bibfnamefont
  {W.~O.}}, \ and\ \bibinfo {author} {\bibfnamefont {A.~G.}\ \bibnamefont
  {McKendrick}}} (\bibinfo {year} {1927}),\ \href@noop {} {\bibfield  {journal}
  {\bibinfo  {journal} {Proc. R. Soc. Lond. A}\ }\textbf {\bibinfo {volume}
  {115}},\ \bibinfo {pages} {700}}\BibitemShut {NoStop}%
\bibitem [{\citenamefont {Kesten}(2003)}]{kesten2003first}%
  \BibitemOpen
  \bibfield  {author} {\bibinfo {author} {\bibnamefont {Kesten}, \bibfnamefont
  {H.}}} (\bibinfo {year} {2003}),\ in\ \href@noop {} {\emph {\bibinfo
  {booktitle} {From classical to modern probability}}}\ (\bibinfo  {publisher}
  {Springer})\ pp.\ \bibinfo {pages} {93--143}\BibitemShut {NoStop}%
\bibitem [{\citenamefont {Kimura}\ \emph {et~al.}(2009)\citenamefont {Kimura},
  \citenamefont {Saito}, \citenamefont {Nakano},\ and\ \citenamefont
  {Motoda}}]{Kimura2009}%
  \BibitemOpen
  \bibfield  {author} {\bibinfo {author} {\bibnamefont {Kimura}, \bibfnamefont
  {M.}}, \bibinfo {author} {\bibfnamefont {K.}~\bibnamefont {Saito}}, \bibinfo
  {author} {\bibfnamefont {R.}~\bibnamefont {Nakano}}, \ and\ \bibinfo {author}
  {\bibfnamefont {H.}~\bibnamefont {Motoda}}} (\bibinfo {year} {2009}),\ \href
  {\doibase 10.1007/s10618-009-0150-5} {\bibfield  {journal} {\bibinfo
  {journal} {Data Mining and Knowledge Discovery}\ }\textbf {\bibinfo {volume}
  {20}}~(\bibinfo {number} {1}),\ \bibinfo {pages} {70}}\BibitemShut {NoStop}%
\bibitem [{\citenamefont {{Kiss}}\ \emph {et~al.}(2015)\citenamefont {{Kiss}},
  \citenamefont {{Morris}}, \citenamefont {{S{\'e}lley}}, \citenamefont
  {{Simon}},\ and\ \citenamefont {{Wilkinson}}}]{2013arXiv1307.7737K}%
  \BibitemOpen
  \bibfield  {author} {\bibinfo {author} {\bibnamefont {{Kiss}}, \bibfnamefont
  {I.~Z.}}, \bibinfo {author} {\bibfnamefont {C.~G.}\ \bibnamefont {{Morris}}},
  \bibinfo {author} {\bibfnamefont {F.}~\bibnamefont {{S{\'e}lley}}}, \bibinfo
  {author} {\bibfnamefont {P.~L.}\ \bibnamefont {{Simon}}}, \ and\ \bibinfo
  {author} {\bibfnamefont {R.~R.}\ \bibnamefont {{Wilkinson}}}} (\bibinfo
  {year} {2015}),\ \href@noop {} {\bibfield  {journal} {\bibinfo  {journal}
  {Journal of Mathematical Biology}\ }\textbf {\bibinfo {volume} {70}},\
  \bibinfo {pages} {437}}\BibitemShut {NoStop}%
\bibitem [{\citenamefont {Kitsak}\ \emph {et~al.}(2010)\citenamefont {Kitsak},
  \citenamefont {Gallos}, \citenamefont {Havlin}, \citenamefont {Liljeros},
  \citenamefont {Muchnik}, \citenamefont {Stanley},\ and\ \citenamefont
  {Makse}}]{kitsak2010}%
  \BibitemOpen
  \bibfield  {author} {\bibinfo {author} {\bibnamefont {Kitsak}, \bibfnamefont
  {M.}}, \bibinfo {author} {\bibfnamefont {L.~K.}\ \bibnamefont {Gallos}},
  \bibinfo {author} {\bibfnamefont {S.}~\bibnamefont {Havlin}}, \bibinfo
  {author} {\bibfnamefont {F.}~\bibnamefont {Liljeros}}, \bibinfo {author}
  {\bibfnamefont {L.}~\bibnamefont {Muchnik}}, \bibinfo {author} {\bibfnamefont
  {H.~E.}\ \bibnamefont {Stanley}}, \ and\ \bibinfo {author} {\bibfnamefont
  {H.~A.}\ \bibnamefont {Makse}}} (\bibinfo {year} {2010}),\ \href@noop {}
  {\bibfield  {journal} {\bibinfo  {journal} {Nature Physics}\ }\textbf
  {\bibinfo {volume} {6}},\ \bibinfo {pages} {888}}\BibitemShut {NoStop}%
\bibitem [{\citenamefont {Kivel\"{a}}\ \emph {et~al.}(2014)\citenamefont
  {Kivel\"{a}}, \citenamefont {Arenas}, \citenamefont {Barthelemy},
  \citenamefont {Gleeson}, \citenamefont {Moreno},\ and\ \citenamefont
  {Porter}}]{Kivela2013}%
  \BibitemOpen
  \bibfield  {author} {\bibinfo {author} {\bibnamefont {Kivel\"{a}},
  \bibfnamefont {M.}}, \bibinfo {author} {\bibfnamefont {A.}~\bibnamefont
  {Arenas}}, \bibinfo {author} {\bibfnamefont {M.}~\bibnamefont {Barthelemy}},
  \bibinfo {author} {\bibfnamefont {J.~P.}\ \bibnamefont {Gleeson}}, \bibinfo
  {author} {\bibfnamefont {Y.}~\bibnamefont {Moreno}}, \ and\ \bibinfo {author}
  {\bibfnamefont {M.~A.}\ \bibnamefont {Porter}}} (\bibinfo {year} {2014}),\
  \href@noop {} {\bibfield  {journal} {\bibinfo  {journal} {J. Complex
  Networks}\ }\textbf {\bibinfo {volume} {2}},\ \bibinfo {pages}
  {203}}\BibitemShut {NoStop}%
\bibitem [{\citenamefont {Kivel{\"a}}\ \emph {et~al.}(2012)\citenamefont
  {Kivel{\"a}}, \citenamefont {Pan}, \citenamefont {Kaski}, \citenamefont
  {Kert{\'e}sz}, \citenamefont {ari Saram{\"a}ki},\ and\ \citenamefont
  {Karsai}}]{dynnetkaski2011}%
  \BibitemOpen
  \bibfield  {author} {\bibinfo {author} {\bibnamefont {Kivel{\"a}},
  \bibfnamefont {M.}}, \bibinfo {author} {\bibfnamefont {R.~K.}\ \bibnamefont
  {Pan}}, \bibinfo {author} {\bibfnamefont {K.}~\bibnamefont {Kaski}}, \bibinfo
  {author} {\bibfnamefont {J.}~\bibnamefont {Kert{\'e}sz}}, \bibinfo {author}
  {\bibfnamefont {J.}~\bibnamefont {ari Saram{\"a}ki}}, \ and\ \bibinfo
  {author} {\bibfnamefont {M.}~\bibnamefont {Karsai}}} (\bibinfo {year}
  {2012}),\ \href {http://stacks.iop.org/1742-5468/2012/i=03/a=P03005}
  {\bibfield  {journal} {\bibinfo  {journal} {Journal of Statistical Mechanics:
  Theory and Experiment}\ }\textbf {\bibinfo {volume} {2012}}~(\bibinfo
  {number} {03}),\ \bibinfo {pages} {P03005}}\BibitemShut {NoStop}%
\bibitem [{\citenamefont {Klemm}\ \emph {et~al.}(2012)\citenamefont {Klemm},
  \citenamefont {Serrano}, \citenamefont {Egu{\'\i}luz},\ and\ \citenamefont
  {San~Miguel}}]{Klemm2012}%
  \BibitemOpen
  \bibfield  {author} {\bibinfo {author} {\bibnamefont {Klemm}, \bibfnamefont
  {K.}}, \bibinfo {author} {\bibfnamefont {M.~{\'A}.}\ \bibnamefont {Serrano}},
  \bibinfo {author} {\bibfnamefont {V.~M.}\ \bibnamefont {Egu{\'\i}luz}}, \
  and\ \bibinfo {author} {\bibfnamefont {M.}~\bibnamefont {San~Miguel}}}
  (\bibinfo {year} {2012}),\ \href@noop {} {\bibfield  {journal} {\bibinfo
  {journal} {Scientific reports}\ }\textbf {\bibinfo {volume} {2}}}\BibitemShut
  {NoStop}%
\bibitem [{\citenamefont {Kooij}\ \emph {et~al.}(2009)\citenamefont {Kooij},
  \citenamefont {Schumm}, \citenamefont {Scoglio},\ and\ \citenamefont
  {Youssef}}]{Rob_VC_networking2009}%
  \BibitemOpen
  \bibfield  {author} {\bibinfo {author} {\bibnamefont {Kooij}, \bibfnamefont
  {R.}}, \bibinfo {author} {\bibfnamefont {P.}~\bibnamefont {Schumm}}, \bibinfo
  {author} {\bibfnamefont {C.}~\bibnamefont {Scoglio}}, \ and\ \bibinfo
  {author} {\bibfnamefont {M.}~\bibnamefont {Youssef}}} (\bibinfo {year}
  {2009}),\ \href@noop {} {\bibinfo  {journal} {Networking 2009, LNCS 5550}\ ,\
  \bibinfo {pages} {562 }}\BibitemShut {NoStop}%
\bibitem [{\citenamefont {Krone}(1999)}]{1999}%
  \BibitemOpen
\bibfield  {journal} {  }\bibfield  {author} {\bibinfo {author} {\bibnamefont
  {Krone}, \bibfnamefont {S.~M.}}} (\bibinfo {year} {1999}),\ \href
  {http://www.jstor.org/stable/2667335} {\bibfield  {journal} {\bibinfo
  {journal} {The Annals of Applied Probability}\ }\textbf {\bibinfo {volume}
  {9}}~(\bibinfo {number} {2}),\ \bibinfo {pages} {pp. 331}}\BibitemShut
  {NoStop}%
\bibitem [{\citenamefont {Lagorio}\ \emph {et~al.}(2011)\citenamefont
  {Lagorio}, \citenamefont {Dickison}, \citenamefont {Vazquez}, \citenamefont
  {Braunstein}, \citenamefont {Macri}, \citenamefont {Migueles}, \citenamefont
  {Havlin},\ and\ \citenamefont {Stanley}}]{Lagorio2011}%
  \BibitemOpen
  \bibfield  {author} {\bibinfo {author} {\bibnamefont {Lagorio}, \bibfnamefont
  {C.}}, \bibinfo {author} {\bibfnamefont {M.}~\bibnamefont {Dickison}},
  \bibinfo {author} {\bibfnamefont {F.}~\bibnamefont {Vazquez}}, \bibinfo
  {author} {\bibfnamefont {L.~A.}\ \bibnamefont {Braunstein}}, \bibinfo
  {author} {\bibfnamefont {P.~A.}\ \bibnamefont {Macri}}, \bibinfo {author}
  {\bibfnamefont {M.~V.}\ \bibnamefont {Migueles}}, \bibinfo {author}
  {\bibfnamefont {S.}~\bibnamefont {Havlin}}, \ and\ \bibinfo {author}
  {\bibfnamefont {H.~E.}\ \bibnamefont {Stanley}}} (\bibinfo {year} {2011}),\
  \href {\doibase 10.1103/PhysRevE.83.026102} {\bibfield  {journal} {\bibinfo
  {journal} {Phys. Rev. E}\ }\textbf {\bibinfo {volume} {83}},\ \bibinfo
  {pages} {026102}}\BibitemShut {NoStop}%
\bibitem [{\citenamefont {Lagorio}\ \emph {et~al.}(2009)\citenamefont
  {Lagorio}, \citenamefont {Migueles}, \citenamefont {Braunstein},
  \citenamefont {L{\'o}pez},\ and\ \citenamefont {Macri}}]{Lagorio2009755}%
  \BibitemOpen
  \bibfield  {author} {\bibinfo {author} {\bibnamefont {Lagorio}, \bibfnamefont
  {C.}}, \bibinfo {author} {\bibfnamefont {M.}~\bibnamefont {Migueles}},
  \bibinfo {author} {\bibfnamefont {L.}~\bibnamefont {Braunstein}}, \bibinfo
  {author} {\bibfnamefont {E.}~\bibnamefont {L{\'o}pez}}, \ and\ \bibinfo
  {author} {\bibfnamefont {P.}~\bibnamefont {Macri}}} (\bibinfo {year}
  {2009}),\ \href {\doibase http://dx.doi.org/10.1016/j.physa.2008.10.045}
  {\bibfield  {journal} {\bibinfo  {journal} {Physica A: Statistical Mechanics
  and its Applications}\ }\textbf {\bibinfo {volume} {388}}~(\bibinfo {number}
  {5}),\ \bibinfo {pages} {755 }}\BibitemShut {NoStop}%
\bibitem [{\citenamefont {Lambiotte}\ \emph {et~al.}(2013)\citenamefont
  {Lambiotte}, \citenamefont {Tabourier},\ and\ \citenamefont
  {Delvenne}}]{burstylambiotte2013}%
  \BibitemOpen
  \bibfield  {author} {\bibinfo {author} {\bibnamefont {Lambiotte},
  \bibfnamefont {R.}}, \bibinfo {author} {\bibfnamefont {L.}~\bibnamefont
  {Tabourier}}, \ and\ \bibinfo {author} {\bibfnamefont {J.-C.}\ \bibnamefont
  {Delvenne}}} (\bibinfo {year} {2013}),\ \href@noop {} {\bibfield  {journal}
  {\bibinfo  {journal} {Eur. Phys. J. B}\ }\textbf {\bibinfo {volume} {86}},\
  \bibinfo {pages} {320}}\BibitemShut {NoStop}%
\bibitem [{\citenamefont {Lee}\ \emph {et~al.}(2013)\citenamefont {Lee},
  \citenamefont {Shim},\ and\ \citenamefont {Noh}}]{Lee2013}%
  \BibitemOpen
  \bibfield  {author} {\bibinfo {author} {\bibnamefont {Lee}, \bibfnamefont
  {H.~K.}}, \bibinfo {author} {\bibfnamefont {P.-S.}\ \bibnamefont {Shim}}, \
  and\ \bibinfo {author} {\bibfnamefont {J.~D.}\ \bibnamefont {Noh}}} (\bibinfo
  {year} {2013}),\ \href {\doibase 10.1103/PhysRevE.87.062812} {\bibfield
  {journal} {\bibinfo  {journal} {Phys. Rev. E}\ }\textbf {\bibinfo {volume}
  {87}},\ \bibinfo {pages} {062812}}\BibitemShut {NoStop}%
\bibitem [{\citenamefont {Lee}\ \emph {et~al.}(2012)\citenamefont {Lee},
  \citenamefont {Rocha}, \citenamefont {Liljeros},\ and\ \citenamefont
  {Holme}}]{Lee:2010fk}%
  \BibitemOpen
  \bibfield  {author} {\bibinfo {author} {\bibnamefont {Lee}, \bibfnamefont
  {S.}}, \bibinfo {author} {\bibfnamefont {L.~E.~C.}\ \bibnamefont {Rocha}},
  \bibinfo {author} {\bibfnamefont {F.}~\bibnamefont {Liljeros}}, \ and\
  \bibinfo {author} {\bibfnamefont {P.}~\bibnamefont {Holme}}} (\bibinfo {year}
  {2012}),\ \href@noop {} {\bibfield  {journal} {\bibinfo  {journal} {PLoS
  ONE}\ }\textbf {\bibinfo {volume} {7}}~(\bibinfo {number} {5}),\ \bibinfo
  {pages} {e36439}}\BibitemShut {NoStop}%
\bibitem [{\citenamefont {Lerman}\ and\ \citenamefont
  {Ghosh}(2010)}]{Lerman10icwsm}%
  \BibitemOpen
  \bibfield  {author} {\bibinfo {author} {\bibnamefont {Lerman}, \bibfnamefont
  {K.}}, \ and\ \bibinfo {author} {\bibfnamefont {R.}~\bibnamefont {Ghosh}}}
  (\bibinfo {year} {2010}),\ in\ \href@noop {} {\emph {\bibinfo {booktitle}
  {Proceedings of 4th International Conference on Weblogs and Social Media
  (ICWSM)}}}\ (\bibinfo  {publisher} {The AAAI Press},\ \bibinfo {address}
  {Menlo Park, California})\ p.~\bibinfo {pages} {90}\BibitemShut {NoStop}%
\bibitem [{\citenamefont {Leskovec}\ \emph
  {et~al.}(2007{\natexlab{a}})\citenamefont {Leskovec}, \citenamefont
  {Adamic},\ and\ \citenamefont {Huberman}}]{Leskovec2007a}%
  \BibitemOpen
  \bibfield  {author} {\bibinfo {author} {\bibnamefont {Leskovec},
  \bibfnamefont {J.}}, \bibinfo {author} {\bibfnamefont {L.}~\bibnamefont
  {Adamic}}, \ and\ \bibinfo {author} {\bibfnamefont {B.}~\bibnamefont
  {Huberman}}} (\bibinfo {year} {2007}{\natexlab{a}}),\ \href
  {http://dl.acm.org/citation.cfm?id=1232727} {\bibfield  {journal} {\bibinfo
  {journal} {ACM Transactions on the Web}\ }\textbf {\bibinfo {volume} {1}},\
  \bibinfo {pages} {5}}\BibitemShut {NoStop}%
\bibitem [{\citenamefont {Leskovec}\ \emph
  {et~al.}(2007{\natexlab{b}})\citenamefont {Leskovec}, \citenamefont
  {Mcglohon}, \citenamefont {Faloutsos}, \citenamefont {Glance},\ and\
  \citenamefont {Hurst}}]{Leskovec2007}%
  \BibitemOpen
  \bibfield  {author} {\bibinfo {author} {\bibnamefont {Leskovec},
  \bibfnamefont {J.}}, \bibinfo {author} {\bibfnamefont {M.}~\bibnamefont
  {Mcglohon}}, \bibinfo {author} {\bibfnamefont {C.}~\bibnamefont {Faloutsos}},
  \bibinfo {author} {\bibfnamefont {N.}~\bibnamefont {Glance}}, \ and\ \bibinfo
  {author} {\bibfnamefont {M.}~\bibnamefont {Hurst}}} (\bibinfo {year}
  {2007}{\natexlab{b}}),\ \enquote {\bibinfo {title} {Patterns of cascading
  behavior in large blog graphs},}\ in\ \href@noop {} {\emph {\bibinfo
  {booktitle} {Proceedings of the 2007 SIAM International Conference on Data
  Mining}}},\ Chap.~\bibinfo {chapter} {60}\ (\bibinfo  {publisher} {SIAM})\
  pp.\ \bibinfo {pages} {551--556}\BibitemShut {NoStop}%
\bibitem [{\citenamefont {Levins}(1970)}]{Levins:1970}%
  \BibitemOpen
  \bibfield  {author} {\bibinfo {author} {\bibnamefont {Levins}, \bibfnamefont
  {R.}}} (\bibinfo {year} {1970}),\ \href@noop {} {\bibfield  {journal}
  {\bibinfo  {journal} {Lecture Notes in Mathematics}\ }\textbf {\bibinfo
  {volume} {2}},\ \bibinfo {pages} {75}}\BibitemShut {NoStop}%
\bibitem [{\citenamefont {Li}\ \emph {et~al.}(2012{\natexlab{a}})\citenamefont
  {Li}, \citenamefont {van~de Bovenkamp},\ and\ \citenamefont
  {Van~Mieghem}}]{PVM_comparisonSIS_meanfield_PRE2012}%
  \BibitemOpen
  \bibfield  {author} {\bibinfo {author} {\bibnamefont {Li}, \bibfnamefont
  {C.}}, \bibinfo {author} {\bibfnamefont {R.}~\bibnamefont {van~de
  Bovenkamp}}, \ and\ \bibinfo {author} {\bibfnamefont {P.}~\bibnamefont
  {Van~Mieghem}}} (\bibinfo {year} {2012}{\natexlab{a}}),\ \href@noop {}
  {\bibfield  {journal} {\bibinfo  {journal} {Physical Review E}\ }\textbf
  {\bibinfo {volume} {86}},\ \bibinfo {pages} {026116}}\BibitemShut {NoStop}%
\bibitem [{\citenamefont {Li}\ \emph {et~al.}(2013)\citenamefont {Li},
  \citenamefont {Wang},\ and\ \citenamefont
  {Van~Mieghem}}]{PhysRevE.88.062802}%
  \BibitemOpen
  \bibfield  {author} {\bibinfo {author} {\bibnamefont {Li}, \bibfnamefont
  {C.}}, \bibinfo {author} {\bibfnamefont {H.}~\bibnamefont {Wang}}, \ and\
  \bibinfo {author} {\bibfnamefont {P.}~\bibnamefont {Van~Mieghem}}} (\bibinfo
  {year} {2013}),\ \href {\doibase 10.1103/PhysRevE.88.062802} {\bibfield
  {journal} {\bibinfo  {journal} {Phys. Rev. E}\ }\textbf {\bibinfo {volume}
  {88}},\ \bibinfo {pages} {062802}}\BibitemShut {NoStop}%
\bibitem [{\citenamefont {Li}\ \emph {et~al.}(2012{\natexlab{b}})\citenamefont
  {Li}, \citenamefont {Zhang}, \citenamefont {Xu},\ and\ \citenamefont
  {Small}}]{Li2012}%
  \BibitemOpen
  \bibfield  {author} {\bibinfo {author} {\bibnamefont {Li}, \bibfnamefont
  {P.}}, \bibinfo {author} {\bibfnamefont {J.}~\bibnamefont {Zhang}}, \bibinfo
  {author} {\bibfnamefont {X.-K.}\ \bibnamefont {Xu}}, \ and\ \bibinfo {author}
  {\bibfnamefont {M.}~\bibnamefont {Small}}} (\bibinfo {year}
  {2012}{\natexlab{b}}),\ \href {\doibase 10.1088/0256-307X/29/4/048903}
  {\bibfield  {journal} {\bibinfo  {journal} {Chinese Physics Letters}\
  }\textbf {\bibinfo {volume} {29}}~(\bibinfo {number} {4}),\ \bibinfo {pages}
  {048903}}\BibitemShut {NoStop}%
\bibitem [{\citenamefont {Liben-Nowell}\ and\ \citenamefont
  {Kleinberg}(2008)}]{Liben-Nowell2008}%
  \BibitemOpen
  \bibfield  {author} {\bibinfo {author} {\bibnamefont {Liben-Nowell},
  \bibfnamefont {D.}}, \ and\ \bibinfo {author} {\bibfnamefont
  {J.}~\bibnamefont {Kleinberg}}} (\bibinfo {year} {2008}),\ \href@noop {}
  {\bibfield  {journal} {\bibinfo  {journal} {Proc. Natl. Acad. Sci. USA}\
  }\textbf {\bibinfo {volume} {105}},\ \bibinfo {pages} {4633}}\BibitemShut
  {NoStop}%
\bibitem [{\citenamefont {Liljeros}\ \emph {et~al.}(2001)\citenamefont
  {Liljeros}, \citenamefont {Edling}, \citenamefont {Amaral}, \citenamefont
  {Stanley},\ and\ \citenamefont {{\AA}berg}}]{liljeros_web_2001}%
  \BibitemOpen
  \bibfield  {author} {\bibinfo {author} {\bibnamefont {Liljeros},
  \bibfnamefont {F.}}, \bibinfo {author} {\bibfnamefont {C.~R.}\ \bibnamefont
  {Edling}}, \bibinfo {author} {\bibfnamefont {L.~A.~N.}\ \bibnamefont
  {Amaral}}, \bibinfo {author} {\bibfnamefont {H.~E.}\ \bibnamefont {Stanley}},
  \ and\ \bibinfo {author} {\bibfnamefont {Y.}~\bibnamefont {{\AA}berg}}}
  (\bibinfo {year} {2001}),\ \href {\doibase 10.1038/35082140} {\bibfield
  {journal} {\bibinfo  {journal} {Nature}\ }\textbf {\bibinfo {volume}
  {411}}~(\bibinfo {number} {6840}),\ \bibinfo {pages} {907}}\BibitemShut
  {NoStop}%
\bibitem [{\citenamefont {Lindquist}\ \emph {et~al.}(2011)\citenamefont
  {Lindquist}, \citenamefont {Ma}, \citenamefont {Driessche},\ and\
  \citenamefont {Willeboordse}}]{Lindquist2011}%
  \BibitemOpen
  \bibfield  {author} {\bibinfo {author} {\bibnamefont {Lindquist},
  \bibfnamefont {J.}}, \bibinfo {author} {\bibfnamefont {J.}~\bibnamefont
  {Ma}}, \bibinfo {author} {\bibfnamefont {P.}~\bibnamefont {Driessche}}, \
  and\ \bibinfo {author} {\bibfnamefont {F.}~\bibnamefont {Willeboordse}}}
  (\bibinfo {year} {2011}),\ \href {\doibase 10.1007/s00285-010-0331-2}
  {\bibfield  {journal} {\bibinfo  {journal} {Journal of Mathematical Biology}\
  }\textbf {\bibinfo {volume} {62}}~(\bibinfo {number} {2}),\ \bibinfo {pages}
  {143}}\BibitemShut {NoStop}%
\bibitem [{\citenamefont {Liu}\ \emph {et~al.}(2004)\citenamefont {Liu},
  \citenamefont {Tang},\ and\ \citenamefont {Yang}}]{liu04:_spread}%
  \BibitemOpen
  \bibfield  {author} {\bibinfo {author} {\bibnamefont {Liu}, \bibfnamefont
  {J.}}, \bibinfo {author} {\bibfnamefont {Y.}~\bibnamefont {Tang}}, \ and\
  \bibinfo {author} {\bibfnamefont {Z.}~\bibnamefont {Yang}}} (\bibinfo {year}
  {2004}),\ \href@noop {} {\bibinfo  {journal} {J. Stat. Mech.}\ ,\ \bibinfo
  {pages} {P08008}}\BibitemShut {NoStop}%
\bibitem [{\citenamefont {Liu}\ \emph {et~al.}(2013)\citenamefont {Liu},
  \citenamefont {Ren},\ and\ \citenamefont {Guo}}]{Liu2013}%
  \BibitemOpen
\bibfield  {journal} {  }\bibfield  {author} {\bibinfo {author} {\bibnamefont
  {Liu}, \bibfnamefont {J.-G.}}, \bibinfo {author} {\bibfnamefont {Z.-M.}\
  \bibnamefont {Ren}}, \ and\ \bibinfo {author} {\bibfnamefont
  {Q.}~\bibnamefont {Guo}}} (\bibinfo {year} {2013}),\ \href {\doibase
  http://dx.doi.org/10.1016/j.physa.2013.04.037} {\bibfield  {journal}
  {\bibinfo  {journal} {Physica A: Statistical Mechanics and its Applications}\
  }\textbf {\bibinfo {volume} {392}}~(\bibinfo {number} {18}),\ \bibinfo
  {pages} {4154 }}\BibitemShut {NoStop}%
\bibitem [{\citenamefont {{Liu}}\ \emph {et~al.}(2014)\citenamefont {{Liu}},
  \citenamefont {{Perra}}, \citenamefont {{Karsai}},\ and\ \citenamefont
  {{Vespignani}}}]{2013arXiv1309.7031L}%
  \BibitemOpen
  \bibfield  {author} {\bibinfo {author} {\bibnamefont {{Liu}}, \bibfnamefont
  {S.}}, \bibinfo {author} {\bibfnamefont {N.}~\bibnamefont {{Perra}}},
  \bibinfo {author} {\bibfnamefont {M.}~\bibnamefont {{Karsai}}}, \ and\
  \bibinfo {author} {\bibfnamefont {A.}~\bibnamefont {{Vespignani}}}} (\bibinfo
  {year} {2014}),\ \href@noop {} {\bibfield  {journal} {\bibinfo  {journal}
  {Phys. Rev. Lett.}\ }\textbf {\bibinfo {volume} {112}},\ \bibinfo {pages}
  {118702}}\BibitemShut {NoStop}%
\bibitem [{\citenamefont {Liu}\ and\ \citenamefont {Hu}(2005)}]{Liu2005}%
  \BibitemOpen
  \bibfield  {author} {\bibinfo {author} {\bibnamefont {Liu}, \bibfnamefont
  {Z.}}, \ and\ \bibinfo {author} {\bibfnamefont {B.}~\bibnamefont {Hu}}}
  (\bibinfo {year} {2005}),\ \href
  {http://stacks.iop.org/0295-5075/72/i=2/a=315} {\bibfield  {journal}
  {\bibinfo  {journal} {Europhys. Lett.}\ }\textbf {\bibinfo {volume} {72}},\
  \bibinfo {pages} {315}}\BibitemShut {NoStop}%
\bibitem [{\citenamefont {Liu}\ \emph {et~al.}(2003)\citenamefont {Liu},
  \citenamefont {Lai},\ and\ \citenamefont {Ye}}]{Liu2003}%
  \BibitemOpen
  \bibfield  {author} {\bibinfo {author} {\bibnamefont {Liu}, \bibfnamefont
  {Z.}}, \bibinfo {author} {\bibfnamefont {Y.-C.}\ \bibnamefont {Lai}}, \ and\
  \bibinfo {author} {\bibfnamefont {N.}~\bibnamefont {Ye}}} (\bibinfo {year}
  {2003}),\ \href {\doibase 10.1103/PhysRevE.67.031911} {\bibfield  {journal}
  {\bibinfo  {journal} {Physical Review E}\ }\textbf {\bibinfo {volume}
  {67}}~(\bibinfo {number} {3}),\ \bibinfo {pages} {031911}}\BibitemShut
  {NoStop}%
\bibitem [{\citenamefont {Lloyd}(2001{\natexlab{a}})}]{Lloyd07052001}%
  \BibitemOpen
  \bibfield  {author} {\bibinfo {author} {\bibnamefont {Lloyd}, \bibfnamefont
  {A.~L.}}} (\bibinfo {year} {2001}{\natexlab{a}}),\ \href {\doibase
  10.1098/rspb.2001.1599} {\bibfield  {journal} {\bibinfo  {journal}
  {Proceedings of the Royal Society of London. Series B: Biological Sciences}\
  }\textbf {\bibinfo {volume} {268}}~(\bibinfo {number} {1470}),\ \bibinfo
  {pages} {985}}\BibitemShut {NoStop}%
\bibitem [{\citenamefont {Lloyd}(2001{\natexlab{b}})}]{Lloyd200159}%
  \BibitemOpen
  \bibfield  {author} {\bibinfo {author} {\bibnamefont {Lloyd}, \bibfnamefont
  {A.~L.}}} (\bibinfo {year} {2001}{\natexlab{b}}),\ \href {\doibase
  http://dx.doi.org/10.1006/tpbi.2001.1525} {\bibfield  {journal} {\bibinfo
  {journal} {Theoretical Population Biology}\ }\textbf {\bibinfo {volume}
  {60}}~(\bibinfo {number} {1}),\ \bibinfo {pages} {59 }}\BibitemShut {NoStop}%
\bibitem [{\citenamefont {Lloyd}\ and\ \citenamefont {May}(2001)}]{lloyd01}%
  \BibitemOpen
  \bibfield  {author} {\bibinfo {author} {\bibnamefont {Lloyd}, \bibfnamefont
  {A.~L.}}, \ and\ \bibinfo {author} {\bibfnamefont {R.~M.}\ \bibnamefont
  {May}}} (\bibinfo {year} {2001}),\ \href@noop {} {\bibfield  {journal}
  {\bibinfo  {journal} {Science}\ }\textbf {\bibinfo {volume} {292}},\ \bibinfo
  {pages} {1316}}\BibitemShut {NoStop}%
\bibitem [{\citenamefont {Lofgren}\ \emph {et~al.}(2014)\citenamefont
  {Lofgren}, \citenamefont {Halloran}, \citenamefont {Rivers}, \citenamefont
  {Drake}, \citenamefont {Porco}, \citenamefont {Lewis}, \citenamefont {Yang},
  \citenamefont {Vespignani}, \citenamefont {Shaman}, \citenamefont
  {Eisenberg}, \citenamefont {Eisenberg}, \citenamefont {Marathe},
  \citenamefont {Scarpino}, \citenamefont {Alexander}, \citenamefont {Meza},
  \citenamefont {Ferrari}, \citenamefont {Hyman}, \citenamefont {Meyers},\ and\
  \citenamefont {Eubank}}]{Lofgren2014}%
  \BibitemOpen
  \bibfield  {author} {\bibinfo {author} {\bibnamefont {Lofgren}, \bibfnamefont
  {E.~T.}}, \bibinfo {author} {\bibfnamefont {M.~E.}\ \bibnamefont {Halloran}},
  \bibinfo {author} {\bibfnamefont {C.~M.}\ \bibnamefont {Rivers}}, \bibinfo
  {author} {\bibfnamefont {J.~M.}\ \bibnamefont {Drake}}, \bibinfo {author}
  {\bibfnamefont {T.~C.}\ \bibnamefont {Porco}}, \bibinfo {author}
  {\bibfnamefont {B.}~\bibnamefont {Lewis}}, \bibinfo {author} {\bibfnamefont
  {W.}~\bibnamefont {Yang}}, \bibinfo {author} {\bibfnamefont {A.}~\bibnamefont
  {Vespignani}}, \bibinfo {author} {\bibfnamefont {J.}~\bibnamefont {Shaman}},
  \bibinfo {author} {\bibfnamefont {J.~N.~S.}\ \bibnamefont {Eisenberg}},
  \bibinfo {author} {\bibfnamefont {M.~C.}\ \bibnamefont {Eisenberg}}, \bibinfo
  {author} {\bibfnamefont {M.}~\bibnamefont {Marathe}}, \bibinfo {author}
  {\bibfnamefont {S.~V.}\ \bibnamefont {Scarpino}}, \bibinfo {author}
  {\bibfnamefont {K.~A.}\ \bibnamefont {Alexander}}, \bibinfo {author}
  {\bibfnamefont {R.}~\bibnamefont {Meza}}, \bibinfo {author} {\bibfnamefont
  {M.~J.}\ \bibnamefont {Ferrari}}, \bibinfo {author} {\bibfnamefont {J.~M.}\
  \bibnamefont {Hyman}}, \bibinfo {author} {\bibfnamefont {L.~A.}\ \bibnamefont
  {Meyers}}, \ and\ \bibinfo {author} {\bibfnamefont {S.}~\bibnamefont
  {Eubank}}} (\bibinfo {year} {2014}),\ \href {\doibase
  10.1073/pnas.1421551111} {\bibfield  {journal} {\bibinfo  {journal} {Proc.
  Natl. Acad. Sci. USA}\ }\textbf {\bibinfo {volume} {111}},\ \bibinfo {pages}
  {18095}}\BibitemShut {NoStop}%
\bibitem [{\citenamefont {Lokhov}\ \emph {et~al.}(2014)\citenamefont {Lokhov},
  \citenamefont {M\'{e}zard}, \citenamefont {Ohta},\ and\ \citenamefont
  {Zdeborov\'{a}}}]{Lokhov2013}%
  \BibitemOpen
  \bibfield  {author} {\bibinfo {author} {\bibnamefont {Lokhov}, \bibfnamefont
  {A.~Y.}}, \bibinfo {author} {\bibfnamefont {M.}~\bibnamefont {M\'{e}zard}},
  \bibinfo {author} {\bibfnamefont {H.}~\bibnamefont {Ohta}}, \ and\ \bibinfo
  {author} {\bibfnamefont {L.}~\bibnamefont {Zdeborov\'{a}}}} (\bibinfo {year}
  {2014}),\ \href {http://arxiv.org/abs/1303.5315} {\bibfield  {journal}
  {\bibinfo  {journal} {Phys. Rev. E}\ }\textbf {\bibinfo {volume} {90}},\
  \bibinfo {pages} {012801}}\BibitemShut {NoStop}%
\bibitem [{\citenamefont {Longini}\ \emph {et~al.}(2005)\citenamefont
  {Longini}, \citenamefont {Nizam}, \citenamefont {Xu}, \citenamefont
  {Ungchusak}, \citenamefont {Hanshaoworakul}, \citenamefont {Cummings},\ and\
  \citenamefont {Halloran}}]{Longini2005}%
  \BibitemOpen
  \bibfield  {author} {\bibinfo {author} {\bibnamefont {Longini}, \bibfnamefont
  {I.~M.}}, \bibinfo {author} {\bibfnamefont {A.}~\bibnamefont {Nizam}},
  \bibinfo {author} {\bibfnamefont {S.}~\bibnamefont {Xu}}, \bibinfo {author}
  {\bibfnamefont {K.}~\bibnamefont {Ungchusak}}, \bibinfo {author}
  {\bibfnamefont {W.}~\bibnamefont {Hanshaoworakul}}, \bibinfo {author}
  {\bibfnamefont {D.~A.~T.}\ \bibnamefont {Cummings}}, \ and\ \bibinfo {author}
  {\bibfnamefont {M.~E.}\ \bibnamefont {Halloran}}} (\bibinfo {year} {2005}),\
  \href@noop {} {\bibfield  {journal} {\bibinfo  {journal} {Science}\ }\textbf
  {\bibinfo {volume} {309}},\ \bibinfo {pages} {1083}}\BibitemShut {NoStop}%
\bibitem [{\citenamefont {Lorenz}\ \emph {et~al.}(2009)\citenamefont {Lorenz},
  \citenamefont {Battiston},\ and\ \citenamefont {Schweitzer}}]{Lorenz2009}%
  \BibitemOpen
  \bibfield  {author} {\bibinfo {author} {\bibnamefont {Lorenz}, \bibfnamefont
  {J.}}, \bibinfo {author} {\bibfnamefont {S.}~\bibnamefont {Battiston}}, \
  and\ \bibinfo {author} {\bibfnamefont {F.}~\bibnamefont {Schweitzer}}}
  (\bibinfo {year} {2009}),\ \href {\doibase 10.1140/epjb/e2009-00347-4}
  {\bibfield  {journal} {\bibinfo  {journal} {The European Physical Journal B}\
  }\textbf {\bibinfo {volume} {71}}~(\bibinfo {number} {4}),\ \bibinfo {pages}
  {441}}\BibitemShut {NoStop}%
\bibitem [{\citenamefont {Ludwig}(1975)}]{Ludwig1975}%
  \BibitemOpen
  \bibfield  {author} {\bibinfo {author} {\bibnamefont {Ludwig}, \bibfnamefont
  {D.}}} (\bibinfo {year} {1975}),\ \href
  {http://www.sciencedirect.com/science/article/pii/0025556475901194}
  {\bibfield  {journal} {\bibinfo  {journal} {Mathematical Biosciences}\
  }\textbf {\bibinfo {volume} {23}},\ \bibinfo {pages} {33}}\BibitemShut
  {NoStop}%
\bibitem [{\citenamefont {Maki}\ and\ \citenamefont
  {Thompson}(1973)}]{Maki1973}%
  \BibitemOpen
  \bibfield  {author} {\bibinfo {author} {\bibnamefont {Maki}, \bibfnamefont
  {D.~P.}}, \ and\ \bibinfo {author} {\bibfnamefont {M.}~\bibnamefont
  {Thompson}}} (\bibinfo {year} {1973}),\ \href@noop {} {\emph {\bibinfo
  {title} {Mathematical models and applications: with emphasis on the social,
  life, and management sciences}}}\ (\bibinfo  {publisher} {Prentice-Hall
  Englewood Cliffs})\BibitemShut {NoStop}%
\bibitem [{\citenamefont {Marathe}\ and\ \citenamefont
  {Vullikanti}(2013)}]{Marathe2013}%
  \BibitemOpen
  \bibfield  {author} {\bibinfo {author} {\bibnamefont {Marathe}, \bibfnamefont
  {M.}}, \ and\ \bibinfo {author} {\bibfnamefont {A.~K.~S.}\ \bibnamefont
  {Vullikanti}}} (\bibinfo {year} {2013}),\ \href {\doibase
  10.1145/2483852.2483871} {\bibfield  {journal} {\bibinfo  {journal} {Commun.
  ACM}\ }\textbf {\bibinfo {volume} {56}}~(\bibinfo {number} {7}),\ \bibinfo
  {pages} {88}}\BibitemShut {NoStop}%
\bibitem [{\citenamefont {Marceau}\ \emph {et~al.}(2010)\citenamefont
  {Marceau}, \citenamefont {No\"el}, \citenamefont {H\'ebert-Dufresne},
  \citenamefont {Allard},\ and\ \citenamefont {Dub\'e}}]{Marceau2010}%
  \BibitemOpen
  \bibfield  {author} {\bibinfo {author} {\bibnamefont {Marceau}, \bibfnamefont
  {V.}}, \bibinfo {author} {\bibfnamefont {P.-A.}\ \bibnamefont {No\"el}},
  \bibinfo {author} {\bibfnamefont {L.}~\bibnamefont {H\'ebert-Dufresne}},
  \bibinfo {author} {\bibfnamefont {A.}~\bibnamefont {Allard}}, \ and\ \bibinfo
  {author} {\bibfnamefont {L.~J.}\ \bibnamefont {Dub\'e}}} (\bibinfo {year}
  {2010}),\ \href {\doibase 10.1103/PhysRevE.82.036116} {\bibfield  {journal}
  {\bibinfo  {journal} {Phys. Rev. E}\ }\textbf {\bibinfo {volume} {82}},\
  \bibinfo {pages} {036116}}\BibitemShut {NoStop}%
\bibitem [{\citenamefont {Marceau}\ \emph {et~al.}(2011)\citenamefont
  {Marceau}, \citenamefont {No\"el}, \citenamefont {H\'ebert-Dufresne},
  \citenamefont {Allard},\ and\ \citenamefont {Dub\'e}}]{Marceau11}%
  \BibitemOpen
  \bibfield  {author} {\bibinfo {author} {\bibnamefont {Marceau}, \bibfnamefont
  {V.}}, \bibinfo {author} {\bibfnamefont {P.-A.}\ \bibnamefont {No\"el}},
  \bibinfo {author} {\bibfnamefont {L.}~\bibnamefont {H\'ebert-Dufresne}},
  \bibinfo {author} {\bibfnamefont {A.}~\bibnamefont {Allard}}, \ and\ \bibinfo
  {author} {\bibfnamefont {L.~J.}\ \bibnamefont {Dub\'e}}} (\bibinfo {year}
  {2011}),\ \href {\doibase 10.1103/PhysRevE.84.026105} {\bibfield  {journal}
  {\bibinfo  {journal} {Phys. Rev. E}\ }\textbf {\bibinfo {volume} {84}},\
  \bibinfo {pages} {026105}}\BibitemShut {NoStop}%
\bibitem [{\citenamefont {Marder}(2007)}]{marder_dynamics_2007}%
  \BibitemOpen
  \bibfield  {author} {\bibinfo {author} {\bibnamefont {Marder}, \bibfnamefont
  {M.}}} (\bibinfo {year} {2007}),\ \href@noop {} {\bibfield  {journal}
  {\bibinfo  {journal} {Physical Review E}\ }\textbf {\bibinfo {volume}
  {75}}~(\bibinfo {number} {6}),\ \bibinfo {pages} {066103}}\BibitemShut
  {NoStop}%
\bibitem [{\citenamefont {Marro}\ and\ \citenamefont
  {Dickman}(1999)}]{Marrobook}%
  \BibitemOpen
  \bibfield  {author} {\bibinfo {author} {\bibnamefont {Marro}, \bibfnamefont
  {J.}}, \ and\ \bibinfo {author} {\bibfnamefont {R.}~\bibnamefont {Dickman}}}
  (\bibinfo {year} {1999}),\ \href@noop {} {\emph {\bibinfo {title}
  {Nonequilibrium phase transitions in lattice models}}}\ (\bibinfo
  {publisher} {Cambridge University Press},\ \bibinfo {address}
  {Cambridge})\BibitemShut {NoStop}%
\bibitem [{\citenamefont {{Martin}}\ \emph {et~al.}(2014)\citenamefont
  {{Martin}}, \citenamefont {{Zhang}},\ and\ \citenamefont
  {{Newman}}}]{2014arXiv1401.5093M}%
  \BibitemOpen
  \bibfield  {author} {\bibinfo {author} {\bibnamefont {{Martin}},
  \bibfnamefont {T.}}, \bibinfo {author} {\bibfnamefont {X.}~\bibnamefont
  {{Zhang}}}, \ and\ \bibinfo {author} {\bibfnamefont {M.~E.~J.}\ \bibnamefont
  {{Newman}}}} (\bibinfo {year} {2014}),\ \href@noop {} {\bibfield  {journal}
  {\bibinfo  {journal} {Phys. Rev. E}\ }\textbf {\bibinfo {volume} {90}},\
  \bibinfo {pages} {052808}}\BibitemShut {NoStop}%
\bibitem [{\citenamefont {Maslov}\ and\ \citenamefont
  {Sneppen}(2002)}]{maslov02}%
  \BibitemOpen
  \bibfield  {author} {\bibinfo {author} {\bibnamefont {Maslov}, \bibfnamefont
  {S.}}, \ and\ \bibinfo {author} {\bibfnamefont {K.}~\bibnamefont {Sneppen}}}
  (\bibinfo {year} {2002}),\ \href@noop {} {\bibfield  {journal} {\bibinfo
  {journal} {Science}\ }\textbf {\bibinfo {volume} {296}},\ \bibinfo {pages}
  {910}}\BibitemShut {NoStop}%
\bibitem [{\citenamefont {Maslov}\ \emph {et~al.}(2004)\citenamefont {Maslov},
  \citenamefont {Sneppen},\ and\ \citenamefont {Zaliznyak}}]{maslovcorr}%
  \BibitemOpen
  \bibfield  {author} {\bibinfo {author} {\bibnamefont {Maslov}, \bibfnamefont
  {S.}}, \bibinfo {author} {\bibfnamefont {K.}~\bibnamefont {Sneppen}}, \ and\
  \bibinfo {author} {\bibfnamefont {A.}~\bibnamefont {Zaliznyak}}} (\bibinfo
  {year} {2004}),\ \href@noop {} {\bibfield  {journal} {\bibinfo  {journal}
  {Physica A}\ }\textbf {\bibinfo {volume} {333}},\ \bibinfo {pages}
  {529}}\BibitemShut {NoStop}%
\bibitem [{\citenamefont {Masuda}\ and\ \citenamefont
  {Konno}(2006)}]{Masuda200664}%
  \BibitemOpen
  \bibfield  {author} {\bibinfo {author} {\bibnamefont {Masuda}, \bibfnamefont
  {N.}}, \ and\ \bibinfo {author} {\bibfnamefont {N.}~\bibnamefont {Konno}}}
  (\bibinfo {year} {2006}),\ \href {\doibase
  http://dx.doi.org/10.1016/j.jtbi.2006.06.010} {\bibfield  {journal} {\bibinfo
   {journal} {Journal of Theoretical Biology}\ }\textbf {\bibinfo {volume}
  {243}}~(\bibinfo {number} {1}),\ \bibinfo {pages} {64 }}\BibitemShut
  {NoStop}%
\bibitem [{\citenamefont {Mata}\ and\ \citenamefont
  {Ferreira}(2013)}]{0295-5075-103-4-48003}%
  \BibitemOpen
  \bibfield  {author} {\bibinfo {author} {\bibnamefont {Mata}, \bibfnamefont
  {A.~S.}}, \ and\ \bibinfo {author} {\bibfnamefont {S.~C.}\ \bibnamefont
  {Ferreira}}} (\bibinfo {year} {2013}),\ \href
  {http://stacks.iop.org/0295-5075/103/i=4/a=48003} {\bibfield  {journal}
  {\bibinfo  {journal} {Europhysics Letters}\ }\textbf {\bibinfo {volume}
  {103}}~(\bibinfo {number} {4}),\ \bibinfo {pages} {48003}}\BibitemShut
  {NoStop}%
\bibitem [{\citenamefont {May}\ and\ \citenamefont
  {Lloyd}(2001)}]{may2001infection}%
  \BibitemOpen
  \bibfield  {author} {\bibinfo {author} {\bibnamefont {May}, \bibfnamefont
  {R.~M.}}, \ and\ \bibinfo {author} {\bibfnamefont {A.~L.}\ \bibnamefont
  {Lloyd}}} (\bibinfo {year} {2001}),\ \href@noop {} {\bibfield  {journal}
  {\bibinfo  {journal} {Physical Review E}\ }\textbf {\bibinfo {volume}
  {64}}~(\bibinfo {number} {6}),\ \bibinfo {pages} {066112}}\BibitemShut
  {NoStop}%
\bibitem [{\citenamefont {Meloni}\ \emph {et~al.}(2011)\citenamefont {Meloni},
  \citenamefont {Perra}, \citenamefont {Arenas}, \citenamefont {Gomes},
  \citenamefont {Moreno},\ and\ \citenamefont {Vespignani}}]{Meloni:2011}%
  \BibitemOpen
  \bibfield  {author} {\bibinfo {author} {\bibnamefont {Meloni}, \bibfnamefont
  {S.}}, \bibinfo {author} {\bibfnamefont {N.}~\bibnamefont {Perra}}, \bibinfo
  {author} {\bibfnamefont {A.}~\bibnamefont {Arenas}}, \bibinfo {author}
  {\bibfnamefont {S.}~\bibnamefont {Gomes}}, \bibinfo {author} {\bibfnamefont
  {Y.}~\bibnamefont {Moreno}}, \ and\ \bibinfo {author} {\bibfnamefont
  {A.}~\bibnamefont {Vespignani}}} (\bibinfo {year} {2011}),\ \href@noop {}
  {\bibfield  {journal} {\bibinfo  {journal} {Sci Rep}\ }\textbf {\bibinfo
  {volume} {1}},\ \bibinfo {pages} {62}}\BibitemShut {NoStop}%
\bibitem [{\citenamefont {Merler}\ \emph {et~al.}(2011)\citenamefont {Merler},
  \citenamefont {Ajelli}, \citenamefont {Pugliese},\ and\ \citenamefont
  {Ferguson}}]{Merler2011}%
  \BibitemOpen
  \bibfield  {author} {\bibinfo {author} {\bibnamefont {Merler}, \bibfnamefont
  {S.}}, \bibinfo {author} {\bibfnamefont {M.}~\bibnamefont {Ajelli}}, \bibinfo
  {author} {\bibfnamefont {A.}~\bibnamefont {Pugliese}}, \ and\ \bibinfo
  {author} {\bibfnamefont {N.~M.}\ \bibnamefont {Ferguson}}} (\bibinfo {year}
  {2011}),\ \href {\doibase 10.1371/journal.pcbi.1002205} {\bibfield  {journal}
  {\bibinfo  {journal} {PLoS Comput Biol}\ }\textbf {\bibinfo {volume}
  {7}}~(\bibinfo {number} {9}),\ \bibinfo {pages} {e1002205}}\BibitemShut
  {NoStop}%
\bibitem [{\citenamefont {Meyers}\ \emph {et~al.}(2006)\citenamefont {Meyers},
  \citenamefont {Newman},\ and\ \citenamefont {Pourbohloul}}]{Meyers2006}%
  \BibitemOpen
  \bibfield  {author} {\bibinfo {author} {\bibnamefont {Meyers}, \bibfnamefont
  {L.~A.}}, \bibinfo {author} {\bibfnamefont {M.}~\bibnamefont {Newman}}, \
  and\ \bibinfo {author} {\bibfnamefont {B.}~\bibnamefont {Pourbohloul}}}
  (\bibinfo {year} {2006}),\ \href {\doibase
  http://dx.doi.org/10.1016/j.jtbi.2005.10.004} {\bibfield  {journal} {\bibinfo
   {journal} {Journal of Theoretical Biology}\ }\textbf {\bibinfo {volume}
  {240}}~(\bibinfo {number} {3}),\ \bibinfo {pages} {400 }}\BibitemShut
  {NoStop}%
\bibitem [{\citenamefont {Miller}(2007)}]{Miller2007}%
  \BibitemOpen
  \bibfield  {author} {\bibinfo {author} {\bibnamefont {Miller}, \bibfnamefont
  {J.}}} (\bibinfo {year} {2007}),\ \href {\doibase 10.1103/PhysRevE.76.010101}
  {\bibfield  {journal} {\bibinfo  {journal} {Physical Review E}\ }\textbf
  {\bibinfo {volume} {76}}~(\bibinfo {number} {1}),\ \bibinfo {pages}
  {010101}}\BibitemShut {NoStop}%
\bibitem [{\citenamefont {Miller}(2009{\natexlab{a}})}]{Miller2009}%
  \BibitemOpen
  \bibfield  {author} {\bibinfo {author} {\bibnamefont {Miller}, \bibfnamefont
  {J.}}} (\bibinfo {year} {2009}{\natexlab{a}}),\ \href {\doibase
  10.1103/PhysRevE.80.020901} {\bibfield  {journal} {\bibinfo  {journal}
  {Physical Review E}\ }\textbf {\bibinfo {volume} {80}}~(\bibinfo {number}
  {2}),\ \bibinfo {pages} {020901}}\BibitemShut {NoStop}%
\bibitem [{\citenamefont {Miller}(2011)}]{Miller2011}%
  \BibitemOpen
  \bibfield  {author} {\bibinfo {author} {\bibnamefont {Miller}, \bibfnamefont
  {J.}}} (\bibinfo {year} {2011}),\ \href {\doibase 10.1007/s00285-010-0337-9}
  {\bibfield  {journal} {\bibinfo  {journal} {Journal of Mathematical Biology}\
  }\textbf {\bibinfo {volume} {62}}~(\bibinfo {number} {3}),\ \bibinfo {pages}
  {349}}\BibitemShut {NoStop}%
\bibitem [{\citenamefont {Miller}(2009{\natexlab{b}})}]{Miller06122009}%
  \BibitemOpen
  \bibfield  {author} {\bibinfo {author} {\bibnamefont {Miller}, \bibfnamefont
  {J.~C.}}} (\bibinfo {year} {2009}{\natexlab{b}}),\ \href {\doibase
  10.1098/rsif.2008.0524} {\bibfield  {journal} {\bibinfo  {journal} {Journal
  of The Royal Society Interface}\ }\textbf {\bibinfo {volume} {6}}~(\bibinfo
  {number} {41}),\ \bibinfo {pages} {1121}}\BibitemShut {NoStop}%
\bibitem [{\citenamefont {Miller}(2013)}]{Miller13}%
  \BibitemOpen
  \bibfield  {author} {\bibinfo {author} {\bibnamefont {Miller}, \bibfnamefont
  {J.~C.}}} (\bibinfo {year} {2013}),\ \href {\doibase
  10.1103/PhysRevE.87.060801} {\bibfield  {journal} {\bibinfo  {journal} {Phys.
  Rev. E}\ }\textbf {\bibinfo {volume} {87}},\ \bibinfo {pages}
  {060801}}\BibitemShut {NoStop}%
\bibitem [{\citenamefont {Miller}\ \emph {et~al.}(2012)\citenamefont {Miller},
  \citenamefont {Slim},\ and\ \citenamefont {Volz}}]{Miller2012}%
  \BibitemOpen
  \bibfield  {author} {\bibinfo {author} {\bibnamefont {Miller}, \bibfnamefont
  {J.~C.}}, \bibinfo {author} {\bibfnamefont {A.~C.}\ \bibnamefont {Slim}}, \
  and\ \bibinfo {author} {\bibfnamefont {E.~M.}\ \bibnamefont {Volz}}}
  (\bibinfo {year} {2012}),\ \href@noop {} {\bibfield  {journal} {\bibinfo
  {journal} {Journal of The Royal Society Interface}\ }\textbf {\bibinfo
  {volume} {9}},\ \bibinfo {pages} {890}}\BibitemShut {NoStop}%
\bibitem [{\citenamefont {Miller}\ and\ \citenamefont
  {Volz}(2013)}]{10.1371/journal.pone.0069162}%
  \BibitemOpen
  \bibfield  {author} {\bibinfo {author} {\bibnamefont {Miller}, \bibfnamefont
  {J.~C.}}, \ and\ \bibinfo {author} {\bibfnamefont {E.~M.}\ \bibnamefont
  {Volz}}} (\bibinfo {year} {2013}),\ \href {\doibase
  10.1371/journal.pone.0069162} {\bibfield  {journal} {\bibinfo  {journal}
  {PLoS ONE}\ }\textbf {\bibinfo {volume} {8}}~(\bibinfo {number} {8}),\
  \bibinfo {pages} {e69162}}\BibitemShut {NoStop}%
\bibitem [{\citenamefont {Min}\ \emph {et~al.}(2013)\citenamefont {Min},
  \citenamefont {Goh},\ and\ \citenamefont {Kim}}]{min_suppression_2013}%
  \BibitemOpen
  \bibfield  {author} {\bibinfo {author} {\bibnamefont {Min}, \bibfnamefont
  {B.}}, \bibinfo {author} {\bibfnamefont {K.-I.}\ \bibnamefont {Goh}}, \ and\
  \bibinfo {author} {\bibfnamefont {I.-M.}\ \bibnamefont {Kim}}} (\bibinfo
  {year} {2013}),\ \href {\doibase 10.1209/0295-5075/103/50002} {\bibfield
  {journal} {\bibinfo  {journal} {Europhys. Lett.}\ }\textbf {\bibinfo {volume}
  {103}},\ \bibinfo {pages} {50002}}\BibitemShut {NoStop}%
\bibitem [{\citenamefont {Min}\ \emph {et~al.}(2011)\citenamefont {Min},
  \citenamefont {Goh},\ and\ \citenamefont {Vazquez}}]{min_spreading_2011}%
  \BibitemOpen
  \bibfield  {author} {\bibinfo {author} {\bibnamefont {Min}, \bibfnamefont
  {B.}}, \bibinfo {author} {\bibfnamefont {K.-I.}\ \bibnamefont {Goh}}, \ and\
  \bibinfo {author} {\bibfnamefont {A.}~\bibnamefont {Vazquez}}} (\bibinfo
  {year} {2011}),\ \href {\doibase 10.1103/PhysRevE.83.036102} {\bibfield
  {journal} {\bibinfo  {journal} {Physical Review E}\ }\textbf {\bibinfo
  {volume} {83}}~(\bibinfo {number} {3}),\ \bibinfo {pages}
  {036102}}\BibitemShut {NoStop}%
\bibitem [{\citenamefont {Miritello}\ \emph {et~al.}(2013)\citenamefont
  {Miritello}, \citenamefont {Moro}, \citenamefont {Lara}, \citenamefont
  {Mart{\'\i}nez-L{\'o}pez}, \citenamefont {Belchamber}, \citenamefont
  {Roberts},\ and\ \citenamefont {Dunbar}}]{miritello2013time}%
  \BibitemOpen
  \bibfield  {author} {\bibinfo {author} {\bibnamefont {Miritello},
  \bibfnamefont {G.}}, \bibinfo {author} {\bibfnamefont {E.}~\bibnamefont
  {Moro}}, \bibinfo {author} {\bibfnamefont {R.}~\bibnamefont {Lara}}, \bibinfo
  {author} {\bibfnamefont {R.}~\bibnamefont {Mart{\'\i}nez-L{\'o}pez}},
  \bibinfo {author} {\bibfnamefont {J.}~\bibnamefont {Belchamber}}, \bibinfo
  {author} {\bibfnamefont {S.~G.}\ \bibnamefont {Roberts}}, \ and\ \bibinfo
  {author} {\bibfnamefont {R.~I.}\ \bibnamefont {Dunbar}}} (\bibinfo {year}
  {2013}),\ \href@noop {} {\bibfield  {journal} {\bibinfo  {journal} {Social
  Networks}\ }\textbf {\bibinfo {volume} {35}}~(\bibinfo {number} {1}),\
  \bibinfo {pages} {89}}\BibitemShut {NoStop}%
\bibitem [{\citenamefont {Molloy}\ and\ \citenamefont {Reed}(1995)}]{molloy95}%
  \BibitemOpen
  \bibfield  {author} {\bibinfo {author} {\bibnamefont {Molloy}, \bibfnamefont
  {M.}}, \ and\ \bibinfo {author} {\bibfnamefont {B.}~\bibnamefont {Reed}}}
  (\bibinfo {year} {1995}),\ \href@noop {} {\bibfield  {journal} {\bibinfo
  {journal} {Random Struct. Algorithms}\ }\textbf {\bibinfo {volume} {6}},\
  \bibinfo {pages} {161}}\BibitemShut {NoStop}%
\bibitem [{\citenamefont {{Monasson}}(1999)}]{Monasson-1999}%
  \BibitemOpen
  \bibfield  {author} {\bibinfo {author} {\bibnamefont {{Monasson}},
  \bibfnamefont {R.}}} (\bibinfo {year} {1999}),\ \href {\doibase
  10.1007/s100510051038} {\bibfield  {journal} {\bibinfo  {journal} {European
  Physical Journal B}\ }\textbf {\bibinfo {volume} {12}},\ \bibinfo {pages}
  {555}}\BibitemShut {NoStop}%
\bibitem [{\citenamefont {Moore}\ and\ \citenamefont {Newman}(2000)}]{moore00}%
  \BibitemOpen
  \bibfield  {author} {\bibinfo {author} {\bibnamefont {Moore}, \bibfnamefont
  {C.}}, \ and\ \bibinfo {author} {\bibfnamefont {M.~E.~J.}\ \bibnamefont
  {Newman}}} (\bibinfo {year} {2000}),\ \href@noop {} {\bibfield  {journal}
  {\bibinfo  {journal} {Phys. Rev. E}\ }\textbf {\bibinfo {volume} {61}},\
  \bibinfo {pages} {5678}}\BibitemShut {NoStop}%
\bibitem [{\citenamefont {Moreno}\ \emph {et~al.}(2003)\citenamefont {Moreno},
  \citenamefont {G\'omez},\ and\ \citenamefont {Pacheco}}]{PhysRevE.68.035103}%
  \BibitemOpen
  \bibfield  {author} {\bibinfo {author} {\bibnamefont {Moreno}, \bibfnamefont
  {Y.}}, \bibinfo {author} {\bibfnamefont {J.~B.}\ \bibnamefont {G\'omez}}, \
  and\ \bibinfo {author} {\bibfnamefont {A.~F.}\ \bibnamefont {Pacheco}}}
  (\bibinfo {year} {2003}),\ \href {\doibase 10.1103/PhysRevE.68.035103}
  {\bibfield  {journal} {\bibinfo  {journal} {Phys. Rev. E}\ }\textbf {\bibinfo
  {volume} {68}},\ \bibinfo {pages} {035103}}\BibitemShut {NoStop}%
\bibitem [{\citenamefont {Moreno}\ \emph
  {et~al.}(2004{\natexlab{a}})\citenamefont {Moreno}, \citenamefont {Nekovee},\
  and\ \citenamefont {Pacheco}}]{Moreno2004b}%
  \BibitemOpen
  \bibfield  {author} {\bibinfo {author} {\bibnamefont {Moreno}, \bibfnamefont
  {Y.}}, \bibinfo {author} {\bibfnamefont {M.}~\bibnamefont {Nekovee}}, \ and\
  \bibinfo {author} {\bibfnamefont {A.~F.}\ \bibnamefont {Pacheco}}} (\bibinfo
  {year} {2004}{\natexlab{a}}),\ \href {\doibase 10.1103/PhysRevE.69.066130}
  {\bibfield  {journal} {\bibinfo  {journal} {Physical Review E}\ }\textbf
  {\bibinfo {volume} {69}}~(\bibinfo {number} {6}),\ \bibinfo {pages}
  {066130}}\BibitemShut {NoStop}%
\bibitem [{\citenamefont {Moreno}\ \emph
  {et~al.}(2004{\natexlab{b}})\citenamefont {Moreno}, \citenamefont {Nekovee},\
  and\ \citenamefont {Vespignani}}]{Moreno2004a}%
  \BibitemOpen
  \bibfield  {author} {\bibinfo {author} {\bibnamefont {Moreno}, \bibfnamefont
  {Y.}}, \bibinfo {author} {\bibfnamefont {M.}~\bibnamefont {Nekovee}}, \ and\
  \bibinfo {author} {\bibfnamefont {A.}~\bibnamefont {Vespignani}}} (\bibinfo
  {year} {2004}{\natexlab{b}}),\ \href {\doibase 10.1103/PhysRevE.69.055101}
  {\bibfield  {journal} {\bibinfo  {journal} {Physical Review E}\ }\textbf
  {\bibinfo {volume} {69}}~(\bibinfo {number} {5}),\ \bibinfo {pages}
  {055101}}\BibitemShut {NoStop}%
\bibitem [{\citenamefont {Moreno}\ \emph {et~al.}(2002)\citenamefont {Moreno},
  \citenamefont {Pastor-Satorras},\ and\ \citenamefont {Vespignani}}]{refId0}%
  \BibitemOpen
  \bibfield  {author} {\bibinfo {author} {\bibnamefont {Moreno}, \bibfnamefont
  {Y.}}, \bibinfo {author} {\bibfnamefont {R.}~\bibnamefont {Pastor-Satorras}},
  \ and\ \bibinfo {author} {\bibfnamefont {A.}~\bibnamefont {Vespignani}}}
  (\bibinfo {year} {2002}),\ \href@noop {} {\bibfield  {journal} {\bibinfo
  {journal} {Eur. Phys. J. B}\ }\textbf {\bibinfo {volume} {26}}~(\bibinfo
  {number} {4}),\ \bibinfo {pages} {521}}\BibitemShut {NoStop}%
\bibitem [{\citenamefont {Morris}(2000)}]{Morris2000}%
  \BibitemOpen
  \bibfield  {author} {\bibinfo {author} {\bibnamefont {Morris}, \bibfnamefont
  {S.}}} (\bibinfo {year} {2000}),\ \href {\doibase 10.1111/1467-937X.00121}
  {\bibfield  {journal} {\bibinfo  {journal} {Review of Economic Studies}\
  }\textbf {\bibinfo {volume} {67}}~(\bibinfo {number} {1}),\ \bibinfo {pages}
  {57}}\BibitemShut {NoStop}%
\bibitem [{\citenamefont {Motter}\ and\ \citenamefont
  {Lai}(2002)}]{Motter2002}%
  \BibitemOpen
  \bibfield  {author} {\bibinfo {author} {\bibnamefont {Motter}, \bibfnamefont
  {A.}}, \ and\ \bibinfo {author} {\bibfnamefont {Y.-C.}\ \bibnamefont {Lai}}}
  (\bibinfo {year} {2002}),\ \href {\doibase 10.1103/PhysRevE.66.065102}
  {\bibfield  {journal} {\bibinfo  {journal} {Physical Review E}\ }\textbf
  {\bibinfo {volume} {66}}~(\bibinfo {number} {6}),\ \bibinfo {pages}
  {065102}}\BibitemShut {NoStop}%
\bibitem [{\citenamefont {Mountford}\ \emph {et~al.}(2013)\citenamefont
  {Mountford}, \citenamefont {Mourrat}, \citenamefont {Valesin},\ and\
  \citenamefont {Yao}}]{Mountford2013}%
  \BibitemOpen
  \bibfield  {author} {\bibinfo {author} {\bibnamefont {Mountford},
  \bibfnamefont {T.}}, \bibinfo {author} {\bibfnamefont {J.-C.}\ \bibnamefont
  {Mourrat}}, \bibinfo {author} {\bibfnamefont {D.}~\bibnamefont {Valesin}}, \
  and\ \bibinfo {author} {\bibfnamefont {Q.}~\bibnamefont {Yao}}} (\bibinfo
  {year} {2013}),\ \href@noop {} {\bibinfo  {journal} {arXiv:1203.2972v1}\
  }\BibitemShut {NoStop}%
\bibitem [{\citenamefont {Mu\~noz}\ \emph {et~al.}(2010)\citenamefont
  {Mu\~noz}, \citenamefont {Juh\'asz}, \citenamefont {Castellano},\ and\
  \citenamefont {\'Odor}}]{Munoz2010}%
  \BibitemOpen
\bibfield  {journal} {  }\bibfield  {author} {\bibinfo {author} {\bibnamefont
  {Mu\~noz}, \bibfnamefont {M.~A.}}, \bibinfo {author} {\bibfnamefont
  {R.}~\bibnamefont {Juh\'asz}}, \bibinfo {author} {\bibfnamefont
  {C.}~\bibnamefont {Castellano}}, \ and\ \bibinfo {author} {\bibfnamefont
  {G.}~\bibnamefont {\'Odor}}} (\bibinfo {year} {2010}),\ \href {\doibase
  10.1103/PhysRevLett.105.128701} {\bibfield  {journal} {\bibinfo  {journal}
  {Phys. Rev. Lett.}\ }\textbf {\bibinfo {volume} {105}},\ \bibinfo {pages}
  {128701}}\BibitemShut {NoStop}%
\bibitem [{\citenamefont {Nardini}\ \emph {et~al.}(2008)\citenamefont
  {Nardini}, \citenamefont {Kozma},\ and\ \citenamefont
  {Barrat}}]{Nardini2008}%
  \BibitemOpen
  \bibfield  {author} {\bibinfo {author} {\bibnamefont {Nardini}, \bibfnamefont
  {C.}}, \bibinfo {author} {\bibfnamefont {B.}~\bibnamefont {Kozma}}, \ and\
  \bibinfo {author} {\bibfnamefont {A.}~\bibnamefont {Barrat}}} (\bibinfo
  {year} {2008}),\ \href {\doibase 10.1103/PhysRevLett.100.158701} {\bibfield
  {journal} {\bibinfo  {journal} {Phys. Rev. Lett.}\ }\textbf {\bibinfo
  {volume} {100}},\ \bibinfo {pages} {158701}}\BibitemShut {NoStop}%
\bibitem [{\citenamefont {Nekovee}\ \emph {et~al.}(2007)\citenamefont
  {Nekovee}, \citenamefont {Moreno}, \citenamefont {Bianconi},\ and\
  \citenamefont {Marsili}}]{Nekovee2007}%
  \BibitemOpen
  \bibfield  {author} {\bibinfo {author} {\bibnamefont {Nekovee}, \bibfnamefont
  {M.}}, \bibinfo {author} {\bibfnamefont {Y.}~\bibnamefont {Moreno}}, \bibinfo
  {author} {\bibfnamefont {G.}~\bibnamefont {Bianconi}}, \ and\ \bibinfo
  {author} {\bibfnamefont {M.}~\bibnamefont {Marsili}}} (\bibinfo {year}
  {2007}),\ \href {\doibase 10.1016/j.physa.2006.07.017} {\bibfield  {journal}
  {\bibinfo  {journal} {Physica A}\ }\textbf {\bibinfo {volume}
  {374}}~(\bibinfo {number} {1}),\ \bibinfo {pages} {457}}\BibitemShut
  {NoStop}%
\bibitem [{\citenamefont {Newman}(2010)}]{Newman10}%
  \BibitemOpen
  \bibfield  {author} {\bibinfo {author} {\bibnamefont {Newman}, \bibfnamefont
  {M.}}} (\bibinfo {year} {2010}),\ \href@noop {} {\emph {\bibinfo {title}
  {Networks: An Introduction}}}\ (\bibinfo  {publisher} {Oxford University
  Press},\ \bibinfo {address} {New York, NY})\BibitemShut {NoStop}%
\bibitem [{\citenamefont {Newman}(2002{\natexlab{a}})}]{assortative}%
  \BibitemOpen
  \bibfield  {author} {\bibinfo {author} {\bibnamefont {Newman}, \bibfnamefont
  {M.~E.~J.}}} (\bibinfo {year} {2002}{\natexlab{a}}),\ \href@noop {}
  {\bibfield  {journal} {\bibinfo  {journal} {Phys. Rev. Lett.}\ }\textbf
  {\bibinfo {volume} {89}},\ \bibinfo {pages} {208701}}\BibitemShut {NoStop}%
\bibitem [{\citenamefont {Newman}(2002{\natexlab{b}})}]{newman02}%
  \BibitemOpen
  \bibfield  {author} {\bibinfo {author} {\bibnamefont {Newman}, \bibfnamefont
  {M.~E.~J.}}} (\bibinfo {year} {2002}{\natexlab{b}}),\ \href {\doibase
  10.1103/PhysRevE.66.016128} {\bibfield  {journal} {\bibinfo  {journal} {Phys.
  Rev. E}\ }\textbf {\bibinfo {volume} {66}},\ \bibinfo {pages}
  {016128}}\BibitemShut {NoStop}%
\bibitem [{\citenamefont {Newman}(2003{\natexlab{a}})}]{PhysRevE.68.026121}%
  \BibitemOpen
  \bibfield  {author} {\bibinfo {author} {\bibnamefont {Newman}, \bibfnamefont
  {M.~E.~J.}}} (\bibinfo {year} {2003}{\natexlab{a}}),\ \href {\doibase
  10.1103/PhysRevE.68.026121} {\bibfield  {journal} {\bibinfo  {journal} {Phys.
  Rev. E}\ }\textbf {\bibinfo {volume} {68}},\ \bibinfo {pages}
  {026121}}\BibitemShut {NoStop}%
\bibitem [{\citenamefont {Newman}(2003{\natexlab{b}})}]{newman-review}%
  \BibitemOpen
  \bibfield  {author} {\bibinfo {author} {\bibnamefont {Newman}, \bibfnamefont
  {M.~E.~J.}}} (\bibinfo {year} {2003}{\natexlab{b}}),\ \href@noop {}
  {\bibfield  {journal} {\bibinfo  {journal} {SIAM Review}\ }\textbf {\bibinfo
  {volume} {45}},\ \bibinfo {pages} {167}}\BibitemShut {NoStop}%
\bibitem [{\citenamefont {Newman}(2005)}]{Newman2005}%
  \BibitemOpen
  \bibfield  {author} {\bibinfo {author} {\bibnamefont {Newman}, \bibfnamefont
  {M.~E.~J.}}} (\bibinfo {year} {2005}),\ \href {\doibase
  10.1103/PhysRevLett.95.108701} {\bibfield  {journal} {\bibinfo  {journal}
  {Phys. Rev. Lett.}\ }\textbf {\bibinfo {volume} {95}},\ \bibinfo {pages}
  {108701}}\BibitemShut {NoStop}%
\bibitem [{\citenamefont {Newman}\ and\ \citenamefont
  {Ferrario}(2013)}]{Newmaninteracting2013}%
  \BibitemOpen
  \bibfield  {author} {\bibinfo {author} {\bibnamefont {Newman}, \bibfnamefont
  {M.~E.~J.}}, \ and\ \bibinfo {author} {\bibfnamefont {C.~R.}\ \bibnamefont
  {Ferrario}}} (\bibinfo {year} {2013}),\ \href {\doibase
  10.1371/journal.pone.0071321} {\bibfield  {journal} {\bibinfo  {journal}
  {PLoS ONE}\ }\textbf {\bibinfo {volume} {8}}~(\bibinfo {number} {8}),\
  \bibinfo {pages} {e71321}}\BibitemShut {NoStop}%
\bibitem [{\citenamefont {Newman}\ \emph {et~al.}(2001)\citenamefont {Newman},
  \citenamefont {Strogatz},\ and\ \citenamefont {Watts}}]{Newman2001}%
  \BibitemOpen
  \bibfield  {author} {\bibinfo {author} {\bibnamefont {Newman}, \bibfnamefont
  {M.~E.~J.}}, \bibinfo {author} {\bibfnamefont {S.~H.}\ \bibnamefont
  {Strogatz}}, \ and\ \bibinfo {author} {\bibfnamefont {D.~J.}\ \bibnamefont
  {Watts}}} (\bibinfo {year} {2001}),\ \href {\doibase
  10.1103/PhysRevE.64.026118} {\bibfield  {journal} {\bibinfo  {journal}
  {Physical Review E}\ }\textbf {\bibinfo {volume} {64}}~(\bibinfo {number}
  {2}),\ \bibinfo {pages} {026118}}\BibitemShut {NoStop}%
\bibitem [{\citenamefont {Ni}\ and\ \citenamefont {Weng}(2009)}]{Ni:2009}%
  \BibitemOpen
  \bibfield  {author} {\bibinfo {author} {\bibnamefont {Ni}, \bibfnamefont
  {S.}}, \ and\ \bibinfo {author} {\bibfnamefont {W.}~\bibnamefont {Weng}}}
  (\bibinfo {year} {2009}),\ \href {\doibase 10.1103/PhysRevE.79.016111}
  {\bibfield  {journal} {\bibinfo  {journal} {Phys. Rev. E}\ }\textbf {\bibinfo
  {volume} {79}},\ \bibinfo {pages} {016111}}\BibitemShut {NoStop}%
\bibitem [{\citenamefont {Nian}\ and\ \citenamefont {Wang}(2010)}]{nian2010}%
  \BibitemOpen
  \bibfield  {author} {\bibinfo {author} {\bibnamefont {Nian}, \bibfnamefont
  {F.}}, \ and\ \bibinfo {author} {\bibfnamefont {X.}~\bibnamefont {Wang}}}
  (\bibinfo {year} {2010}),\ \href@noop {} {\bibfield  {journal} {\bibinfo
  {journal} {J. Theor. Biol.}\ }\textbf {\bibinfo {volume} {264}},\ \bibinfo
  {pages} {77}}\BibitemShut {NoStop}%
\bibitem [{\citenamefont {Nicolaides}\ \emph {et~al.}(2013)\citenamefont
  {Nicolaides}, \citenamefont {Cueto-Felgueroso},\ and\ \citenamefont
  {Juanes}}]{Nicolaides:2013}%
  \BibitemOpen
  \bibfield  {author} {\bibinfo {author} {\bibnamefont {Nicolaides},
  \bibfnamefont {C.}}, \bibinfo {author} {\bibfnamefont {L.}~\bibnamefont
  {Cueto-Felgueroso}}, \ and\ \bibinfo {author} {\bibfnamefont
  {R.}~\bibnamefont {Juanes}}} (\bibinfo {year} {2013}),\ \href@noop {}
  {\bibfield  {journal} {\bibinfo  {journal} {Journal of The Royal Society
  Interface}\ }\textbf {\bibinfo {volume} {10}}~(\bibinfo {number}
  {87})}\BibitemShut {NoStop}%
\bibitem [{\citenamefont {Nishiura}(2011)}]{Nishiura2011}%
  \BibitemOpen
  \bibfield  {author} {\bibinfo {author} {\bibnamefont {Nishiura},
  \bibfnamefont {N.}}} (\bibinfo {year} {2011}),\ \href {\doibase
  10.1007/s10654-011-9597-y} {\bibfield  {journal} {\bibinfo  {journal} {Eur J
  Epidemiol}\ }\textbf {\bibinfo {volume} {26}},\ \bibinfo {pages}
  {583}}\BibitemShut {NoStop}%
\bibitem [{\citenamefont {No\"el}\ \emph {et~al.}(2012)\citenamefont {No\"el},
  \citenamefont {Allard}, \citenamefont {H\'ebert-Dufresne}, \citenamefont
  {Marceau},\ and\ \citenamefont {Dub\'e}}]{Noel2012}%
  \BibitemOpen
  \bibfield  {author} {\bibinfo {author} {\bibnamefont {No\"el}, \bibfnamefont
  {P.-A.}}, \bibinfo {author} {\bibfnamefont {A.}~\bibnamefont {Allard}},
  \bibinfo {author} {\bibfnamefont {L.~u.}\ \bibnamefont {H\'ebert-Dufresne}},
  \bibinfo {author} {\bibfnamefont {V.}~\bibnamefont {Marceau}}, \ and\
  \bibinfo {author} {\bibfnamefont {L.~J.}\ \bibnamefont {Dub\'e}}} (\bibinfo
  {year} {2012}),\ \href {\doibase 10.1103/PhysRevE.85.031118} {\bibfield
  {journal} {\bibinfo  {journal} {Phys. Rev. E}\ }\textbf {\bibinfo {volume}
  {85}},\ \bibinfo {pages} {031118}}\BibitemShut {NoStop}%
\bibitem [{\citenamefont {No{\"e}l}\ \emph {et~al.}(2009)\citenamefont
  {No{\"e}l}, \citenamefont {Davoudi}, \citenamefont {Brunham}, \citenamefont
  {Dub{\'e}},\ and\ \citenamefont {Pourbohloul}}]{noel_time_2009}%
  \BibitemOpen
  \bibfield  {author} {\bibinfo {author} {\bibnamefont {No{\"e}l},
  \bibfnamefont {P.-A.}}, \bibinfo {author} {\bibfnamefont {B.}~\bibnamefont
  {Davoudi}}, \bibinfo {author} {\bibfnamefont {R.~C.}\ \bibnamefont
  {Brunham}}, \bibinfo {author} {\bibfnamefont {L.~J.}\ \bibnamefont
  {Dub{\'e}}}, \ and\ \bibinfo {author} {\bibfnamefont {B.}~\bibnamefont
  {Pourbohloul}}} (\bibinfo {year} {2009}),\ \href {\doibase
  10.1103/PhysRevE.79.026101} {\bibfield  {journal} {\bibinfo  {journal}
  {Physical Review E}\ }\textbf {\bibinfo {volume} {79}}~(\bibinfo {number}
  {2}),\ \bibinfo {pages} {026101}}\BibitemShut {NoStop}%
\bibitem [{\citenamefont {Noh}\ and\ \citenamefont
  {Rieger}(2004)}]{PhysRevLett.92.118701}%
  \BibitemOpen
  \bibfield  {author} {\bibinfo {author} {\bibnamefont {Noh}, \bibfnamefont
  {J.~D.}}, \ and\ \bibinfo {author} {\bibfnamefont {H.}~\bibnamefont
  {Rieger}}} (\bibinfo {year} {2004}),\ \href {\doibase
  10.1103/PhysRevLett.92.118701} {\bibfield  {journal} {\bibinfo  {journal}
  {Phys. Rev. Lett.}\ }\textbf {\bibinfo {volume} {92}}~(\bibinfo {number}
  {11}),\ \bibinfo {pages} {118701}}\BibitemShut {NoStop}%
\bibitem [{\citenamefont {Nsoesie}\ \emph {et~al.}(2013)\citenamefont
  {Nsoesie}, \citenamefont {Brownstein}, \citenamefont {Ramakrishnan},\ and\
  \citenamefont {Marathe}}]{Nsoesie2013}%
  \BibitemOpen
  \bibfield  {author} {\bibinfo {author} {\bibnamefont {Nsoesie}, \bibfnamefont
  {E.~O.}}, \bibinfo {author} {\bibfnamefont {J.~S.}\ \bibnamefont
  {Brownstein}}, \bibinfo {author} {\bibfnamefont {N.}~\bibnamefont
  {Ramakrishnan}}, \ and\ \bibinfo {author} {\bibfnamefont {M.~V.}\
  \bibnamefont {Marathe}}} (\bibinfo {year} {2013}),\ \href@noop {} {\bibfield
  {journal} {\bibinfo  {journal} {Influenza and Other Respiratory Viruses}\
  }\textbf {\bibinfo {volume} {8}},\ \bibinfo {pages} {309}}\BibitemShut
  {NoStop}%
\bibitem [{\citenamefont {\'Odor}(2013{\natexlab{a}})}]{PhysRevE.87.042132}%
  \BibitemOpen
  \bibfield  {author} {\bibinfo {author} {\bibnamefont {\'Odor}, \bibfnamefont
  {G.}}} (\bibinfo {year} {2013}{\natexlab{a}}),\ \href {\doibase
  10.1103/PhysRevE.87.042132} {\bibfield  {journal} {\bibinfo  {journal} {Phys.
  Rev. E}\ }\textbf {\bibinfo {volume} {87}},\ \bibinfo {pages}
  {042132}}\BibitemShut {NoStop}%
\bibitem [{\citenamefont {\'Odor}(2013{\natexlab{b}})}]{PhysRevE.88.032109}%
  \BibitemOpen
  \bibfield  {author} {\bibinfo {author} {\bibnamefont {\'Odor}, \bibfnamefont
  {G.}}} (\bibinfo {year} {2013}{\natexlab{b}}),\ \href {\doibase
  10.1103/PhysRevE.88.032109} {\bibfield  {journal} {\bibinfo  {journal} {Phys.
  Rev. E}\ }\textbf {\bibinfo {volume} {88}},\ \bibinfo {pages}
  {032109}}\BibitemShut {NoStop}%
\bibitem [{\citenamefont {\'Odor}\ and\ \citenamefont
  {Pastor-Satorras}(2012)}]{PhysRevE.86.026117}%
  \BibitemOpen
  \bibfield  {author} {\bibinfo {author} {\bibnamefont {\'Odor}, \bibfnamefont
  {G.}}, \ and\ \bibinfo {author} {\bibfnamefont {R.}~\bibnamefont
  {Pastor-Satorras}}} (\bibinfo {year} {2012}),\ \href {\doibase
  10.1103/PhysRevE.86.026117} {\bibfield  {journal} {\bibinfo  {journal} {Phys.
  Rev. E}\ }\textbf {\bibinfo {volume} {86}},\ \bibinfo {pages}
  {026117}}\BibitemShut {NoStop}%
\bibitem [{\citenamefont {Olinky}\ and\ \citenamefont
  {Stone}(2004)}]{Olinky2004}%
  \BibitemOpen
  \bibfield  {author} {\bibinfo {author} {\bibnamefont {Olinky}, \bibfnamefont
  {R.}}, \ and\ \bibinfo {author} {\bibfnamefont {L.}~\bibnamefont {Stone}}}
  (\bibinfo {year} {2004}),\ \href {\doibase 10.1103/PhysRevE.70.030902}
  {\bibfield  {journal} {\bibinfo  {journal} {Phys. Rev. E}\ }\textbf {\bibinfo
  {volume} {70}},\ \bibinfo {pages} {030902}}\BibitemShut {NoStop}%
\bibitem [{\citenamefont {Oliveira}\ and\ \citenamefont
  {Barabasi}(2005)}]{Oliveira:2005fk}%
  \BibitemOpen
  \bibfield  {author} {\bibinfo {author} {\bibnamefont {Oliveira},
  \bibfnamefont {J.~G.}}, \ and\ \bibinfo {author} {\bibfnamefont {A.-L.}\
  \bibnamefont {Barabasi}}} (\bibinfo {year} {2005}),\ \href@noop {} {\bibfield
   {journal} {\bibinfo  {journal} {Nature}\ }\textbf {\bibinfo {volume}
  {437}}~(\bibinfo {number} {7063}),\ \bibinfo {pages} {1251}}\BibitemShut
  {NoStop}%
\bibitem [{\citenamefont {de~Oliveira}\ and\ \citenamefont
  {Dickman}(2005)}]{DeOliveira05}%
  \BibitemOpen
  \bibfield  {author} {\bibinfo {author} {\bibnamefont {de~Oliveira},
  \bibfnamefont {M.~M.}}, \ and\ \bibinfo {author} {\bibfnamefont
  {R.}~\bibnamefont {Dickman}}} (\bibinfo {year} {2005}),\ \href {\doibase
  10.1103/PhysRevE.71.016129} {\bibfield  {journal} {\bibinfo  {journal} {Phys.
  Rev. E}\ }\textbf {\bibinfo {volume} {71}},\ \bibinfo {pages}
  {016129}}\BibitemShut {NoStop}%
\bibitem [{\citenamefont {Omic}\ \emph {et~al.}(2009)\citenamefont {Omic},
  \citenamefont {Orda},\ and\ \citenamefont
  {Van~Mieghem}}]{PVM_Jasmina_Game_protection_INFOCOM2009}%
  \BibitemOpen
  \bibfield  {author} {\bibinfo {author} {\bibnamefont {Omic}, \bibfnamefont
  {J.}}, \bibinfo {author} {\bibfnamefont {A.}~\bibnamefont {Orda}}, \ and\
  \bibinfo {author} {\bibfnamefont {P.}~\bibnamefont {Van~Mieghem}}} (\bibinfo
  {year} {2009}),\ in\ \href {\doibase 10.1109/INFCOM.2009.5062065} {\emph
  {\bibinfo {booktitle} {INFOCOM 2009, IEEE}}},\ pp.\ \bibinfo {pages}
  {1485--1493}\BibitemShut {NoStop}%
\bibitem [{\citenamefont {Onnela}\ \emph {et~al.}(2005)\citenamefont {Onnela},
  \citenamefont {Saram{\"a}ki}, \citenamefont {Kert\'esz},\ and\ \citenamefont
  {Kaski}}]{onnela05:_inten}%
  \BibitemOpen
  \bibfield  {author} {\bibinfo {author} {\bibnamefont {Onnela}, \bibfnamefont
  {J.}}, \bibinfo {author} {\bibfnamefont {J.}~\bibnamefont {Saram{\"a}ki}},
  \bibinfo {author} {\bibfnamefont {J.}~\bibnamefont {Kert\'esz}}, \ and\
  \bibinfo {author} {\bibfnamefont {K.}~\bibnamefont {Kaski}}} (\bibinfo {year}
  {2005}),\ \href@noop {} {\bibfield  {journal} {\bibinfo  {journal} {Phys.
  Rev. E}\ }\textbf {\bibinfo {volume} {71}},\ \bibinfo {pages}
  {065103}}\BibitemShut {NoStop}%
\bibitem [{\citenamefont {Onnela}\ \emph {et~al.}(2007)\citenamefont {Onnela},
  \citenamefont {Saram{\"a}ki}, \citenamefont {Hyv\"onen}, \citenamefont
  {Szab\'o}, \citenamefont {Lazer}, \citenamefont {Kaski}, \citenamefont
  {Kert\'esz},\ and\ \citenamefont {Barab\'asi}}]{Onnela:2007}%
  \BibitemOpen
  \bibfield  {author} {\bibinfo {author} {\bibnamefont {Onnela}, \bibfnamefont
  {J.-P.}}, \bibinfo {author} {\bibfnamefont {J.}~\bibnamefont {Saram{\"a}ki},
  \bibfnamefont {Jari}}, \bibinfo {author} {\bibfnamefont {J.}~\bibnamefont
  {Hyv\"onen}}, \bibinfo {author} {\bibfnamefont {G.}~\bibnamefont {Szab\'o}},
  \bibinfo {author} {\bibfnamefont {D.}~\bibnamefont {Lazer}}, \bibinfo
  {author} {\bibfnamefont {K.}~\bibnamefont {Kaski}}, \bibinfo {author}
  {\bibfnamefont {J.}~\bibnamefont {Kert\'esz}}, \ and\ \bibinfo {author}
  {\bibfnamefont {A.-L.}\ \bibnamefont {Barab\'asi}}} (\bibinfo {year}
  {2007}),\ \href {\doibase 10.1073/pnas.0610245104} {\bibfield  {journal}
  {\bibinfo  {journal} {Proceedings of the National Academy of Sciences}\
  }\textbf {\bibinfo {volume} {104}}~(\bibinfo {number} {18}),\ \bibinfo
  {pages} {7332}}\BibitemShut {NoStop}%
\bibitem [{\citenamefont {Park}\ and\ \citenamefont
  {Newman}(2003)}]{PhysRevE.68.026112}%
  \BibitemOpen
  \bibfield  {author} {\bibinfo {author} {\bibnamefont {Park}, \bibfnamefont
  {J.}}, \ and\ \bibinfo {author} {\bibfnamefont {M.~E.~J.}\ \bibnamefont
  {Newman}}} (\bibinfo {year} {2003}),\ \href {\doibase
  10.1103/PhysRevE.68.026112} {\bibfield  {journal} {\bibinfo  {journal} {Phys.
  Rev. E}\ }\textbf {\bibinfo {volume} {68}}~(\bibinfo {number} {2}),\ \bibinfo
  {pages} {026112}}\BibitemShut {NoStop}%
\bibitem [{\citenamefont {Parshani}\ \emph {et~al.}(2010)\citenamefont
  {Parshani}, \citenamefont {Carmi},\ and\ \citenamefont
  {Havlin}}]{Parshani2010}%
  \BibitemOpen
  \bibfield  {author} {\bibinfo {author} {\bibnamefont {Parshani},
  \bibfnamefont {R.}}, \bibinfo {author} {\bibfnamefont {S.}~\bibnamefont
  {Carmi}}, \ and\ \bibinfo {author} {\bibfnamefont {S.}~\bibnamefont
  {Havlin}}} (\bibinfo {year} {2010}),\ \href {\doibase
  10.1103/PhysRevLett.104.258701} {\bibfield  {journal} {\bibinfo  {journal}
  {Phys. Rev. Lett.}\ }\textbf {\bibinfo {volume} {104}},\ \bibinfo {pages}
  {258701}}\BibitemShut {NoStop}%
\bibitem [{\citenamefont {Pastor-Satorras}\ \emph {et~al.}(2001)\citenamefont
  {Pastor-Satorras}, \citenamefont {V{\'a}zquez},\ and\ \citenamefont
  {Vespignani}}]{alexei}%
  \BibitemOpen
  \bibfield  {author} {\bibinfo {author} {\bibnamefont {Pastor-Satorras},
  \bibfnamefont {R.}}, \bibinfo {author} {\bibfnamefont {A.}~\bibnamefont
  {V{\'a}zquez}}, \ and\ \bibinfo {author} {\bibfnamefont {A.}~\bibnamefont
  {Vespignani}}} (\bibinfo {year} {2001}),\ \href@noop {} {\bibfield  {journal}
  {\bibinfo  {journal} {Phys. Rev. Lett.}\ }\textbf {\bibinfo {volume} {87}},\
  \bibinfo {pages} {258701}}\BibitemShut {NoStop}%
\bibitem [{\citenamefont {Pastor-Satorras}\ and\ \citenamefont
  {Vespignani}(2001{\natexlab{a}})}]{Pastor01b}%
  \BibitemOpen
  \bibfield  {author} {\bibinfo {author} {\bibnamefont {Pastor-Satorras},
  \bibfnamefont {R.}}, \ and\ \bibinfo {author} {\bibfnamefont
  {A.}~\bibnamefont {Vespignani}}} (\bibinfo {year} {2001}{\natexlab{a}}),\
  \href {\doibase 10.1103/PhysRevE.63.066117} {\bibfield  {journal} {\bibinfo
  {journal} {Phys. Rev. E}\ }\textbf {\bibinfo {volume} {63}},\ \bibinfo
  {pages} {066117}}\BibitemShut {NoStop}%
\bibitem [{\citenamefont {Pastor-Satorras}\ and\ \citenamefont
  {Vespignani}(2001{\natexlab{b}})}]{pv01a}%
  \BibitemOpen
  \bibfield  {author} {\bibinfo {author} {\bibnamefont {Pastor-Satorras},
  \bibfnamefont {R.}}, \ and\ \bibinfo {author} {\bibfnamefont
  {A.}~\bibnamefont {Vespignani}}} (\bibinfo {year} {2001}{\natexlab{b}}),\
  \href@noop {} {\bibfield  {journal} {\bibinfo  {journal} {Phys. Rev. Lett.}\
  }\textbf {\bibinfo {volume} {86}},\ \bibinfo {pages} {3200}}\BibitemShut
  {NoStop}%
\bibitem [{\citenamefont {Pastor-Satorras}\ and\ \citenamefont
  {Vespignani}(2002{\natexlab{a}})}]{pastor2002fs}%
  \BibitemOpen
  \bibfield  {author} {\bibinfo {author} {\bibnamefont {Pastor-Satorras},
  \bibfnamefont {R.}}, \ and\ \bibinfo {author} {\bibfnamefont
  {A.}~\bibnamefont {Vespignani}}} (\bibinfo {year} {2002}{\natexlab{a}}),\
  \href@noop {} {\bibfield  {journal} {\bibinfo  {journal} {Physical Review E}\
  }\textbf {\bibinfo {volume} {65}}~(\bibinfo {number} {3}),\ \bibinfo {pages}
  {035108}}\BibitemShut {NoStop}%
\bibitem [{\citenamefont {Pastor-Satorras}\ and\ \citenamefont
  {Vespignani}(2002{\natexlab{b}})}]{PhysRevE.65.036104}%
  \BibitemOpen
  \bibfield  {author} {\bibinfo {author} {\bibnamefont {Pastor-Satorras},
  \bibfnamefont {R.}}, \ and\ \bibinfo {author} {\bibfnamefont
  {A.}~\bibnamefont {Vespignani}}} (\bibinfo {year} {2002}{\natexlab{b}}),\
  \href {\doibase 10.1103/PhysRevE.65.036104} {\bibfield  {journal} {\bibinfo
  {journal} {Phys. Rev. E}\ }\textbf {\bibinfo {volume} {65}}~(\bibinfo
  {number} {3}),\ \bibinfo {pages} {036104}}\BibitemShut {NoStop}%
\bibitem [{\citenamefont {Pastor-Satorras}\ and\ \citenamefont
  {Vespignani}(2004)}]{romuvespibook}%
  \BibitemOpen
  \bibfield  {author} {\bibinfo {author} {\bibnamefont {Pastor-Satorras},
  \bibfnamefont {R.}}, \ and\ \bibinfo {author} {\bibfnamefont
  {A.}~\bibnamefont {Vespignani}}} (\bibinfo {year} {2004}),\ \href@noop {}
  {\emph {\bibinfo {title} {Evolution and structure of the Internet: A
  statistical physics approach}}}\ (\bibinfo  {publisher} {Cambridge University
  Press},\ \bibinfo {address} {Cambridge})\BibitemShut {NoStop}%
\bibitem [{\citenamefont {Payne}\ \emph {et~al.}(2011)\citenamefont {Payne},
  \citenamefont {Harris},\ and\ \citenamefont {Dodds}}]{Payne2011}%
  \BibitemOpen
  \bibfield  {author} {\bibinfo {author} {\bibnamefont {Payne}, \bibfnamefont
  {J.~L.}}, \bibinfo {author} {\bibfnamefont {K.~D.}\ \bibnamefont {Harris}}, \
  and\ \bibinfo {author} {\bibfnamefont {P.~S.}\ \bibnamefont {Dodds}}}
  (\bibinfo {year} {2011}),\ \href {\doibase 10.1103/PhysRevE.84.016110}
  {\bibfield  {journal} {\bibinfo  {journal} {Phys. Rev. E}\ }\textbf {\bibinfo
  {volume} {84}},\ \bibinfo {pages} {016110}}\BibitemShut {NoStop}%
\bibitem [{\citenamefont {Peng}\ \emph {et~al.}(2010)\citenamefont {Peng},
  \citenamefont {Jin},\ and\ \citenamefont {Shi}}]{Peng2010549}%
  \BibitemOpen
  \bibfield  {author} {\bibinfo {author} {\bibnamefont {Peng}, \bibfnamefont
  {C.}}, \bibinfo {author} {\bibfnamefont {X.}~\bibnamefont {Jin}}, \ and\
  \bibinfo {author} {\bibfnamefont {M.}~\bibnamefont {Shi}}} (\bibinfo {year}
  {2010}),\ \href {\doibase http://dx.doi.org/10.1016/j.physa.2009.09.047}
  {\bibfield  {journal} {\bibinfo  {journal} {Physica A}\ }\textbf {\bibinfo
  {volume} {389}}~(\bibinfo {number} {3}),\ \bibinfo {pages} {549}}\BibitemShut
  {NoStop}%
\bibitem [{\citenamefont {Peng}\ and\ \citenamefont {Li}(2009)}]{5365379}%
  \BibitemOpen
  \bibfield  {author} {\bibinfo {author} {\bibnamefont {Peng}, \bibfnamefont
  {S.}}, \ and\ \bibinfo {author} {\bibfnamefont {C.}~\bibnamefont {Li}}}
  (\bibinfo {year} {2009}),\ in\ \href {\doibase 10.1109/CISE.2009.5365379}
  {\emph {\bibinfo {booktitle} {Computational Intelligence and Software
  Engineering, 2009. CiSE 2009. International Conference on}}},\ pp.\ \bibinfo
  {pages} {1--4}\BibitemShut {NoStop}%
\bibitem [{\citenamefont {Perra}\ \emph {et~al.}(2011)\citenamefont {Perra},
  \citenamefont {Balcan}, \citenamefont {Gon{\c c}alves},\ and\ \citenamefont
  {Vespignani}}]{Perra2011}%
  \BibitemOpen
  \bibfield  {author} {\bibinfo {author} {\bibnamefont {Perra}, \bibfnamefont
  {N.}}, \bibinfo {author} {\bibfnamefont {D.}~\bibnamefont {Balcan}}, \bibinfo
  {author} {\bibfnamefont {B.}~\bibnamefont {Gon{\c c}alves}}, \ and\ \bibinfo
  {author} {\bibfnamefont {A.}~\bibnamefont {Vespignani}}} (\bibinfo {year}
  {2011}),\ \href {\doibase 10.1371/journal.pone.0023084} {\bibfield  {journal}
  {\bibinfo  {journal} {PLoS ONE}\ }\textbf {\bibinfo {volume} {6}}~(\bibinfo
  {number} {8}),\ \bibinfo {pages} {e23084}}\BibitemShut {NoStop}%
\bibitem [{\citenamefont {Perra}\ \emph
  {et~al.}(2012{\natexlab{a}})\citenamefont {Perra}, \citenamefont
  {Baronchelli}, \citenamefont {Mocanu}, \citenamefont {Gon{\c c}alves},
  \citenamefont {Pastor-Satorras},\ and\ \citenamefont
  {Vespignani}}]{perra_random_2012}%
  \BibitemOpen
  \bibfield  {author} {\bibinfo {author} {\bibnamefont {Perra}, \bibfnamefont
  {N.}}, \bibinfo {author} {\bibfnamefont {A.}~\bibnamefont {Baronchelli}},
  \bibinfo {author} {\bibfnamefont {D.}~\bibnamefont {Mocanu}}, \bibinfo
  {author} {\bibfnamefont {B.}~\bibnamefont {Gon{\c c}alves}}, \bibinfo
  {author} {\bibfnamefont {R.}~\bibnamefont {Pastor-Satorras}}, \ and\ \bibinfo
  {author} {\bibfnamefont {A.}~\bibnamefont {Vespignani}}} (\bibinfo {year}
  {2012}{\natexlab{a}}),\ \href {\doibase 10.1103/PhysRevLett.109.238701}
  {\bibfield  {journal} {\bibinfo  {journal} {Physical Review Letters}\
  }\textbf {\bibinfo {volume} {109}}~(\bibinfo {number} {23}),\ \bibinfo
  {pages} {238701}}\BibitemShut {NoStop}%
\bibitem [{\citenamefont {Perra}\ \emph
  {et~al.}(2012{\natexlab{b}})\citenamefont {Perra}, \citenamefont
  {Gon{\c{c}}alves}, \citenamefont {Pastor-Satorras},\ and\ \citenamefont
  {Vespignani}}]{2012arXiv1203.5351P}%
  \BibitemOpen
  \bibfield  {author} {\bibinfo {author} {\bibnamefont {Perra}, \bibfnamefont
  {N.}}, \bibinfo {author} {\bibfnamefont {B.}~\bibnamefont {Gon{\c{c}}alves}},
  \bibinfo {author} {\bibfnamefont {R.}~\bibnamefont {Pastor-Satorras}}, \ and\
  \bibinfo {author} {\bibfnamefont {A.}~\bibnamefont {Vespignani}}} (\bibinfo
  {year} {2012}{\natexlab{b}}),\ \href@noop {} {\bibfield  {journal} {\bibinfo
  {journal} {Scientific reports}\ }\textbf {\bibinfo {volume} {2}}}\BibitemShut
  {NoStop}%
\bibitem [{\citenamefont {Pinto}\ \emph {et~al.}(2012)\citenamefont {Pinto},
  \citenamefont {Thiran},\ and\ \citenamefont {Vetterli}}]{Pinto2012}%
  \BibitemOpen
  \bibfield  {author} {\bibinfo {author} {\bibnamefont {Pinto}, \bibfnamefont
  {P.~C.}}, \bibinfo {author} {\bibfnamefont {P.}~\bibnamefont {Thiran}}, \
  and\ \bibinfo {author} {\bibfnamefont {M.}~\bibnamefont {Vetterli}}}
  (\bibinfo {year} {2012}),\ \href {\doibase 10.1103/PhysRevLett.109.068702}
  {\bibfield  {journal} {\bibinfo  {journal} {Phys. Rev. Lett.}\ }\textbf
  {\bibinfo {volume} {109}},\ \bibinfo {pages} {068702}}\BibitemShut {NoStop}%
\bibitem [{\citenamefont {Poletto}\ \emph {et~al.}(2013)\citenamefont
  {Poletto}, \citenamefont {Meloni}, \citenamefont {Colizza}, \citenamefont
  {Moreno},\ and\ \citenamefont {Vespignani}}]{Poletto2013}%
  \BibitemOpen
  \bibfield  {author} {\bibinfo {author} {\bibnamefont {Poletto}, \bibfnamefont
  {C.}}, \bibinfo {author} {\bibfnamefont {S.}~\bibnamefont {Meloni}}, \bibinfo
  {author} {\bibfnamefont {V.}~\bibnamefont {Colizza}}, \bibinfo {author}
  {\bibfnamefont {Y.}~\bibnamefont {Moreno}}, \ and\ \bibinfo {author}
  {\bibfnamefont {A.}~\bibnamefont {Vespignani}}} (\bibinfo {year} {2013}),\
  \href {\doibase 10.1371/journal.pcbi.1003169} {\bibfield  {journal} {\bibinfo
   {journal} {PLoS Comput Biol}\ }\textbf {\bibinfo {volume} {9}}~(\bibinfo
  {number} {8}),\ \bibinfo {pages} {e1003169}}\BibitemShut {NoStop}%
\bibitem [{\citenamefont {Poletto}\ \emph {et~al.}(2012)\citenamefont
  {Poletto}, \citenamefont {Tizzoni},\ and\ \citenamefont
  {Colizza}}]{Poletto:2012}%
  \BibitemOpen
  \bibfield  {author} {\bibinfo {author} {\bibnamefont {Poletto}, \bibfnamefont
  {C.}}, \bibinfo {author} {\bibfnamefont {M.}~\bibnamefont {Tizzoni}}, \ and\
  \bibinfo {author} {\bibfnamefont {V.}~\bibnamefont {Colizza}}} (\bibinfo
  {year} {2012}),\ \href@noop {} {\bibfield  {journal} {\bibinfo  {journal}
  {Sci Rep}\ }\textbf {\bibinfo {volume} {2}},\ \bibinfo {pages}
  {476}}\BibitemShut {NoStop}%
\bibitem [{\citenamefont {Prakash}\ \emph {et~al.}(2012)\citenamefont
  {Prakash}, \citenamefont {Chakrabarti}, \citenamefont {Valler}, \citenamefont
  {Faloutsos},\ and\ \citenamefont {Faloutsos}}]{Prakash2012}%
  \BibitemOpen
  \bibfield  {author} {\bibinfo {author} {\bibnamefont {Prakash}, \bibfnamefont
  {B.}}, \bibinfo {author} {\bibfnamefont {D.}~\bibnamefont {Chakrabarti}},
  \bibinfo {author} {\bibfnamefont {N.}~\bibnamefont {Valler}}, \bibinfo
  {author} {\bibfnamefont {M.}~\bibnamefont {Faloutsos}}, \ and\ \bibinfo
  {author} {\bibfnamefont {C.}~\bibnamefont {Faloutsos}}} (\bibinfo {year}
  {2012}),\ \href {\doibase 10.1007/s10115-012-0520-y} {\bibfield  {journal}
  {\bibinfo  {journal} {Knowledge and Information Systems}\ }\textbf {\bibinfo
  {volume} {33}}~(\bibinfo {number} {3}),\ \bibinfo {pages} {549}}\BibitemShut
  {NoStop}%
\bibitem [{\citenamefont {{Rattana}}\ \emph {et~al.}(2013)\citenamefont
  {{Rattana}}, \citenamefont {{Blyuss}}, \citenamefont {{Eames}},\ and\
  \citenamefont {{Kiss}}}]{2012arXiv1208.6036R}%
  \BibitemOpen
  \bibfield  {author} {\bibinfo {author} {\bibnamefont {{Rattana}},
  \bibfnamefont {P.}}, \bibinfo {author} {\bibfnamefont {K.~B.}\ \bibnamefont
  {{Blyuss}}}, \bibinfo {author} {\bibfnamefont {K.~T.~D.}\ \bibnamefont
  {{Eames}}}, \ and\ \bibinfo {author} {\bibfnamefont {I.~Z.}\ \bibnamefont
  {{Kiss}}}} (\bibinfo {year} {2013}),\ \href@noop {} {\bibfield  {journal}
  {\bibinfo  {journal} {Bulletin of Mathematical Biology}\ }\textbf {\bibinfo
  {volume} {75}},\ \bibinfo {pages} {466}}\BibitemShut {NoStop}%
\bibitem [{\citenamefont {Ravasz}\ and\ \citenamefont
  {Barab{\'a}si}(2003)}]{ravasz_hierarchical_2003}%
  \BibitemOpen
  \bibfield  {author} {\bibinfo {author} {\bibnamefont {Ravasz}, \bibfnamefont
  {E.}}, \ and\ \bibinfo {author} {\bibfnamefont {A.-L.}\ \bibnamefont
  {Barab{\'a}si}}} (\bibinfo {year} {2003}),\ \href {\doibase
  10.1103/PhysRevE.67.026112} {\bibfield  {journal} {\bibinfo  {journal}
  {Physical Review E}\ }\textbf {\bibinfo {volume} {67}}~(\bibinfo {number}
  {2}),\ \bibinfo {pages} {026112}}\BibitemShut {NoStop}%
\bibitem [{\citenamefont {Riley}(2007)}]{Riley2007}%
  \BibitemOpen
  \bibfield  {author} {\bibinfo {author} {\bibnamefont {Riley}, \bibfnamefont
  {S.}}} (\bibinfo {year} {2007}),\ \href {\doibase 10.1126/science.1134695}
  {\bibfield  {journal} {\bibinfo  {journal} {Science}\ }\textbf {\bibinfo
  {volume} {316}},\ \bibinfo {pages} {1298}}\BibitemShut {NoStop}%
\bibitem [{\citenamefont {Risau-Gusman}\ and\ \citenamefont
  {Zanette}(2009)}]{Risau-Gusman2009}%
  \BibitemOpen
  \bibfield  {author} {\bibinfo {author} {\bibnamefont {Risau-Gusman},
  \bibfnamefont {S.}}, \ and\ \bibinfo {author} {\bibfnamefont {D.~H.}\
  \bibnamefont {Zanette}}} (\bibinfo {year} {2009}),\ \href {\doibase
  10.1016/j.jtbi.2008.10.027} {\bibfield  {journal} {\bibinfo  {journal}
  {Journal of Theoretical Biology}\ }\textbf {\bibinfo {volume} {257}},\
  \bibinfo {pages} {52}}\BibitemShut {NoStop}%
\bibitem [{\citenamefont {Rocha}\ and\ \citenamefont
  {Blondel}(2013)}]{Rocha:2013}%
  \BibitemOpen
  \bibfield  {author} {\bibinfo {author} {\bibnamefont {Rocha}, \bibfnamefont
  {L.~E.~C.}}, \ and\ \bibinfo {author} {\bibfnamefont {V.~D.}\ \bibnamefont
  {Blondel}}} (\bibinfo {year} {2013}),\ \href {\doibase
  10.1371/journal.pcbi.1002974} {\bibfield  {journal} {\bibinfo  {journal}
  {PLoS Comput Biol}\ }\textbf {\bibinfo {volume} {9}}~(\bibinfo {number}
  {3}),\ \bibinfo {pages} {e1002974}}\BibitemShut {NoStop}%
\bibitem [{\citenamefont {Rocha}\ \emph {et~al.}(2011)\citenamefont {Rocha},
  \citenamefont {Liljeros},\ and\ \citenamefont {Holme}}]{Rocha:2010}%
  \BibitemOpen
  \bibfield  {author} {\bibinfo {author} {\bibnamefont {Rocha}, \bibfnamefont
  {L.~E.~C.}}, \bibinfo {author} {\bibfnamefont {F.}~\bibnamefont {Liljeros}},
  \ and\ \bibinfo {author} {\bibfnamefont {P.}~\bibnamefont {Holme}}} (\bibinfo
  {year} {2011}),\ \href {\doibase 10.1371/journal.pcbi.1001109} {\bibfield
  {journal} {\bibinfo  {journal} {PLoS Comput Biol}\ }\textbf {\bibinfo
  {volume} {7}}~(\bibinfo {number} {3}),\ \bibinfo {pages}
  {e1001109}}\BibitemShut {NoStop}%
\bibitem [{\citenamefont {Rogers}(2010)}]{Rogers2010}%
  \BibitemOpen
  \bibfield  {author} {\bibinfo {author} {\bibnamefont {Rogers}, \bibfnamefont
  {E.~M.}}} (\bibinfo {year} {2010}),\ \href@noop {} {\emph {\bibinfo {title}
  {Diffusion of innovations}}}\ (\bibinfo  {publisher} {Simon and Schuster},\
  \bibinfo {address} {New York})\BibitemShut {NoStop}%
\bibitem [{\citenamefont {Ross}(1996)}]{Ross1996}%
  \BibitemOpen
  \bibfield  {author} {\bibinfo {author} {\bibnamefont {Ross}, \bibfnamefont
  {S.~M.}}} (\bibinfo {year} {1996}),\ \href@noop {} {\emph {\bibinfo {title}
  {Stochastic Processes}}}\ (\bibinfo  {publisher} {John Wiley \& Sons},\
  \bibinfo {address} {New York})\BibitemShut {NoStop}%
\bibitem [{\citenamefont {Rvachev}\ and\ \citenamefont
  {Longini}(1985)}]{Rvachev:1985}%
  \BibitemOpen
  \bibfield  {author} {\bibinfo {author} {\bibnamefont {Rvachev}, \bibfnamefont
  {L.}}, \ and\ \bibinfo {author} {\bibfnamefont {I.}~\bibnamefont {Longini}}}
  (\bibinfo {year} {1985}),\ \href {\doibase 10.1016/0025-5564(85)90064-1}
  {\bibfield  {journal} {\bibinfo  {journal} {Math Biosci}\ }\textbf {\bibinfo
  {volume} {75}},\ \bibinfo {pages} {3}}\BibitemShut {NoStop}%
\bibitem [{\citenamefont {Sahneh}\ \emph {et~al.}(2013)\citenamefont {Sahneh},
  \citenamefont {Scoglio},\ and\ \citenamefont
  {Chowdhury}}]{DarabiSahneh_Scogio_interdependentnets2013}%
  \BibitemOpen
  \bibfield  {author} {\bibinfo {author} {\bibnamefont {Sahneh}, \bibfnamefont
  {F.~D.}}, \bibinfo {author} {\bibfnamefont {C.}~\bibnamefont {Scoglio}}, \
  and\ \bibinfo {author} {\bibfnamefont {F.~N.}\ \bibnamefont {Chowdhury}}}
  (\bibinfo {year} {2013}),\ in\ \href@noop {} {\emph {\bibinfo {booktitle}
  {American Control Conference (ACC), 2013}}},\ pp.\ \bibinfo {pages}
  {2307--2312}\BibitemShut {NoStop}%
\bibitem [{\citenamefont {Salath{\'e}}\ and\ \citenamefont
  {Bonhoeffer}(2008)}]{Salathe2008}%
  \BibitemOpen
  \bibfield  {author} {\bibinfo {author} {\bibnamefont {Salath{\'e}},
  \bibfnamefont {M.}}, \ and\ \bibinfo {author} {\bibfnamefont
  {S.}~\bibnamefont {Bonhoeffer}}} (\bibinfo {year} {2008}),\ \href {\doibase
  10.1098/rsif.2008.0271} {\bibfield  {journal} {\bibinfo  {journal} {Journal
  of The Royal Society Interface}\ }~(\bibinfo {number} {29}),\ \bibinfo
  {pages} {1505}}\BibitemShut {NoStop}%
\bibitem [{\citenamefont {Salath{\'e}}\ and\ \citenamefont
  {Jones}(2010)}]{Salathe2010}%
  \BibitemOpen
  \bibfield  {author} {\bibinfo {author} {\bibnamefont {Salath{\'e}},
  \bibfnamefont {M.}}, \ and\ \bibinfo {author} {\bibfnamefont {J.~H.}\
  \bibnamefont {Jones}}} (\bibinfo {year} {2010}),\ \href@noop {} {\bibfield
  {journal} {\bibinfo  {journal} {PLoS Computational Biology}\ }\textbf
  {\bibinfo {volume} {6}}~(\bibinfo {number} {4}),\ \bibinfo {pages}
  {e1000736}}\BibitemShut {NoStop}%
\bibitem [{\citenamefont {Salda\~na}(2008)}]{Saldana:2008}%
  \BibitemOpen
  \bibfield  {author} {\bibinfo {author} {\bibnamefont {Salda\~na},
  \bibfnamefont {J.}}} (\bibinfo {year} {2008}),\ \href {\doibase
  10.1103/PhysRevE.78.012902} {\bibfield  {journal} {\bibinfo  {journal} {Phys.
  Rev. E}\ }\textbf {\bibinfo {volume} {78}},\ \bibinfo {pages}
  {012902}}\BibitemShut {NoStop}%
\bibitem [{\citenamefont {{Santos}}\ \emph {et~al.}(2013)\citenamefont
  {{Santos}}, \citenamefont {{Moura}},\ and\ \citenamefont
  {{Xavier}}}]{2013arXiv1306.6812S}%
  \BibitemOpen
  \bibfield  {author} {\bibinfo {author} {\bibnamefont {{Santos}},
  \bibfnamefont {A.}}, \bibinfo {author} {\bibfnamefont {J.~M.~F.}\
  \bibnamefont {{Moura}}}, \ and\ \bibinfo {author} {\bibfnamefont
  {J.}~\bibnamefont {{Xavier}}}} (\bibinfo {year} {2013}),\ \href@noop {}
  {\bibfield  {journal} {\bibinfo  {journal} {ArXiv e-prints}\ }}\Eprint
  {http://arxiv.org/abs/1306.6812} {arXiv:1306.6812} \BibitemShut {NoStop}%
\bibitem [{\citenamefont {Sattenspiel}\ and\ \citenamefont
  {Dietz}(1995)}]{Sattenspiel:1995}%
  \BibitemOpen
  \bibfield  {author} {\bibinfo {author} {\bibnamefont {Sattenspiel},
  \bibfnamefont {L.}}, \ and\ \bibinfo {author} {\bibfnamefont
  {K.}~\bibnamefont {Dietz}}} (\bibinfo {year} {1995}),\ \href {\doibase
  10.1016/0025-5564(94)00068-B} {\bibfield  {journal} {\bibinfo  {journal}
  {Math Biosci}\ }\textbf {\bibinfo {volume} {128}},\ \bibinfo {pages}
  {71}}\BibitemShut {NoStop}%
\bibitem [{\citenamefont {Saumell-Mendiola}\ \emph {et~al.}(2012)\citenamefont
  {Saumell-Mendiola}, \citenamefont {Serrano},\ and\ \citenamefont
  {Bogu{\~n}{\'a}}}]{Saumell-Mendiola2012}%
  \BibitemOpen
  \bibfield  {author} {\bibinfo {author} {\bibnamefont {Saumell-Mendiola},
  \bibfnamefont {A.}}, \bibinfo {author} {\bibfnamefont {M.~{\'A}.}\
  \bibnamefont {Serrano}}, \ and\ \bibinfo {author} {\bibfnamefont
  {M.}~\bibnamefont {Bogu{\~n}{\'a}}}} (\bibinfo {year} {2012}),\ \href@noop {}
  {\bibfield  {journal} {\bibinfo  {journal} {Physical Review E}\ }\textbf
  {\bibinfo {volume} {86}}~(\bibinfo {number} {2}),\ \bibinfo {pages}
  {026106}}\BibitemShut {NoStop}%
\bibitem [{\citenamefont {Schneider}\ \emph {et~al.}(2011)\citenamefont
  {Schneider}, \citenamefont {Mihaljev}, \citenamefont {Havlin},\ and\
  \citenamefont {Herrmann}}]{PhysRevE.84.061911}%
  \BibitemOpen
  \bibfield  {author} {\bibinfo {author} {\bibnamefont {Schneider},
  \bibfnamefont {C.~M.}}, \bibinfo {author} {\bibfnamefont {T.}~\bibnamefont
  {Mihaljev}}, \bibinfo {author} {\bibfnamefont {S.}~\bibnamefont {Havlin}}, \
  and\ \bibinfo {author} {\bibfnamefont {H.~J.}\ \bibnamefont {Herrmann}}}
  (\bibinfo {year} {2011}),\ \href {\doibase 10.1103/PhysRevE.84.061911}
  {\bibfield  {journal} {\bibinfo  {journal} {Phys. Rev. E}\ }\textbf {\bibinfo
  {volume} {84}},\ \bibinfo {pages} {061911}}\BibitemShut {NoStop}%
\bibitem [{\citenamefont {Schumm}\ \emph {et~al.}(2007)\citenamefont {Schumm},
  \citenamefont {Scoglio}, \citenamefont {Gruenbacher},\ and\ \citenamefont
  {Easton}}]{4610111}%
  \BibitemOpen
  \bibfield  {author} {\bibinfo {author} {\bibnamefont {Schumm}, \bibfnamefont
  {P.}}, \bibinfo {author} {\bibfnamefont {C.}~\bibnamefont {Scoglio}},
  \bibinfo {author} {\bibfnamefont {D.}~\bibnamefont {Gruenbacher}}, \ and\
  \bibinfo {author} {\bibfnamefont {T.}~\bibnamefont {Easton}}} (\bibinfo
  {year} {2007}),\ in\ \href {\doibase 10.1109/BIMNICS.2007.4610111} {\emph
  {\bibinfo {booktitle} {Bio-Inspired Models of Network, Information and
  Computing Systems, 2007. Bionetics 2007. 2nd}}},\ pp.\ \bibinfo {pages}
  {201--208}\BibitemShut {NoStop}%
\bibitem [{\citenamefont {Schwartz}\ \emph {et~al.}(2002)\citenamefont
  {Schwartz}, \citenamefont {Cohen}, \citenamefont {{ben-Avraham}},
  \citenamefont {Barab\'asi},\ and\ \citenamefont
  {Havlin}}]{PhysRevE.66.015104}%
  \BibitemOpen
  \bibfield  {author} {\bibinfo {author} {\bibnamefont {Schwartz},
  \bibfnamefont {N.}}, \bibinfo {author} {\bibfnamefont {R.}~\bibnamefont
  {Cohen}}, \bibinfo {author} {\bibfnamefont {D.}~\bibnamefont
  {{ben-Avraham}}}, \bibinfo {author} {\bibfnamefont {A.-L.}\ \bibnamefont
  {Barab\'asi}}, \ and\ \bibinfo {author} {\bibfnamefont {S.}~\bibnamefont
  {Havlin}}} (\bibinfo {year} {2002}),\ \href {\doibase
  10.1103/PhysRevE.66.015104} {\bibfield  {journal} {\bibinfo  {journal} {Phys.
  Rev. E}\ }\textbf {\bibinfo {volume} {66}},\ \bibinfo {pages}
  {015104}}\BibitemShut {NoStop}%
\bibitem [{\citenamefont {Schwartz}\ and\ \citenamefont
  {Stone}(2013)}]{PhysRevE.87.042815}%
  \BibitemOpen
  \bibfield  {author} {\bibinfo {author} {\bibnamefont {Schwartz},
  \bibfnamefont {N.}}, \ and\ \bibinfo {author} {\bibfnamefont
  {L.}~\bibnamefont {Stone}}} (\bibinfo {year} {2013}),\ \href {\doibase
  10.1103/PhysRevE.87.042815} {\bibfield  {journal} {\bibinfo  {journal} {Phys.
  Rev. E}\ }\textbf {\bibinfo {volume} {87}},\ \bibinfo {pages}
  {042815}}\BibitemShut {NoStop}%
\bibitem [{\citenamefont {Seidman}(1983)}]{Seidman1983269}%
  \BibitemOpen
  \bibfield  {author} {\bibinfo {author} {\bibnamefont {Seidman}, \bibfnamefont
  {S.~B.}}} (\bibinfo {year} {1983}),\ \href {\doibase DOI:
  10.1016/0378-8733(83)90028-X} {\bibfield  {journal} {\bibinfo  {journal}
  {Social Networks}\ }\textbf {\bibinfo {volume} {5}},\ \bibinfo {pages} {269
  }}\BibitemShut {NoStop}%
\bibitem [{\citenamefont {Serrano}\ and\ \citenamefont
  {Bogu\~n\'a}(2006)}]{PhysRevLett.97.088701}%
  \BibitemOpen
  \bibfield  {author} {\bibinfo {author} {\bibnamefont {Serrano}, \bibfnamefont
  {M.~A.}}, \ and\ \bibinfo {author} {\bibfnamefont {M.}~\bibnamefont
  {Bogu\~n\'a}}} (\bibinfo {year} {2006}),\ \href {\doibase
  10.1103/PhysRevLett.97.088701} {\bibfield  {journal} {\bibinfo  {journal}
  {Phys. Rev. Lett.}\ }\textbf {\bibinfo {volume} {97}},\ \bibinfo {pages}
  {088701}}\BibitemShut {NoStop}%
\bibitem [{\citenamefont {Serrano}\ \emph {et~al.}(2006)\citenamefont
  {Serrano}, \citenamefont {Bogu\~n\'a},\ and\ \citenamefont
  {Pastor-Satorras}}]{PhysRevE.74.055101}%
  \BibitemOpen
  \bibfield  {author} {\bibinfo {author} {\bibnamefont {Serrano}, \bibfnamefont
  {M.~A.}}, \bibinfo {author} {\bibfnamefont {M.}~\bibnamefont {Bogu\~n\'a}}, \
  and\ \bibinfo {author} {\bibfnamefont {R.}~\bibnamefont {Pastor-Satorras}}}
  (\bibinfo {year} {2006}),\ \href {\doibase 10.1103/PhysRevE.74.055101}
  {\bibfield  {journal} {\bibinfo  {journal} {Phys. Rev. E}\ }\textbf {\bibinfo
  {volume} {74}},\ \bibinfo {pages} {055101}}\BibitemShut {NoStop}%
\bibitem [{\citenamefont {Serrano}\ and\ \citenamefont
  {Bogu{\~n}{\'a}}(2005)}]{PhysRevE.72.036133}%
  \BibitemOpen
  \bibfield  {author} {\bibinfo {author} {\bibnamefont {Serrano}, \bibfnamefont
  {M.~{\'A}.}}, \ and\ \bibinfo {author} {\bibfnamefont {M.}~\bibnamefont
  {Bogu{\~n}{\'a}}}} (\bibinfo {year} {2005}),\ \href {\doibase
  10.1103/PhysRevE.72.036133} {\bibfield  {journal} {\bibinfo  {journal} {Phys.
  Rev. E}\ }\textbf {\bibinfo {volume} {72}},\ \bibinfo {pages}
  {036133}}\BibitemShut {NoStop}%
\bibitem [{\citenamefont {Shalizi}\ and\ \citenamefont
  {Thomas}(2011)}]{Shalizi2011}%
  \BibitemOpen
  \bibfield  {author} {\bibinfo {author} {\bibnamefont {Shalizi}, \bibfnamefont
  {C.~R.}}, \ and\ \bibinfo {author} {\bibfnamefont {A.~C.}\ \bibnamefont
  {Thomas}}} (\bibinfo {year} {2011}),\ \href {\doibase
  10.1177/0049124111404820} {\bibfield  {journal} {\bibinfo  {journal}
  {Sociological Methods \& Research}\ }\textbf {\bibinfo {volume} {40}},\
  \bibinfo {pages} {211}}\BibitemShut {NoStop}%
\bibitem [{\citenamefont {Sharkey}(2008)}]{Sharkey2008}%
  \BibitemOpen
  \bibfield  {author} {\bibinfo {author} {\bibnamefont {Sharkey}, \bibfnamefont
  {K.}}} (\bibinfo {year} {2008}),\ \href {\doibase 10.1007/s00285-008-0161-7}
  {\bibfield  {journal} {\bibinfo  {journal} {Journal of Mathematical Biology}\
  }\textbf {\bibinfo {volume} {57}}~(\bibinfo {number} {3}),\ \bibinfo {pages}
  {311}}\BibitemShut {NoStop}%
\bibitem [{\citenamefont {Sharkey}\ \emph {et~al.}(2013)\citenamefont
  {Sharkey}, \citenamefont {Kiss}, \citenamefont {Wilkinson},\ and\
  \citenamefont {Simon}}]{Sharkey2013}%
  \BibitemOpen
  \bibfield  {author} {\bibinfo {author} {\bibnamefont {Sharkey}, \bibfnamefont
  {K.}}, \bibinfo {author} {\bibfnamefont {I.}~\bibnamefont {Kiss}}, \bibinfo
  {author} {\bibfnamefont {R.}~\bibnamefont {Wilkinson}}, \ and\ \bibinfo
  {author} {\bibfnamefont {P.}~\bibnamefont {Simon}}} (\bibinfo {year}
  {2013}),\ \href {\doibase 10.1007/s11538-013-9923-5} {\bibinfo  {journal}
  {Bulletin of Mathematical Biology}\ ,\ \bibinfo {pages} {1}}\BibitemShut
  {NoStop}%
\bibitem [{\citenamefont {Sharkey}(2011)}]{Sharkey2011}%
  \BibitemOpen
\bibfield  {journal} {  }\bibfield  {author} {\bibinfo {author} {\bibnamefont
  {Sharkey}, \bibfnamefont {K.~J.}}} (\bibinfo {year} {2011}),\ \href {\doibase
  http://dx.doi.org/10.1016/j.tpb.2011.01.004} {\bibfield  {journal} {\bibinfo
  {journal} {Theoretical Population Biology}\ }\textbf {\bibinfo {volume}
  {79}},\ \bibinfo {pages} {115}}\BibitemShut {NoStop}%
\bibitem [{\citenamefont {Shaw}(2008)}]{Shaw2008}%
  \BibitemOpen
  \bibfield  {author} {\bibinfo {author} {\bibnamefont {Shaw}, \bibfnamefont
  {L.~B.}}} (\bibinfo {year} {2008}),\ \href {\doibase
  10.1103/PhysRevE.77.066101} {\bibfield  {journal} {\bibinfo  {journal}
  {Physical Review E}\ }\textbf {\bibinfo {volume} {77}}~(\bibinfo {number}
  {6}),\ \bibinfo {pages} {066101}}\BibitemShut {NoStop}%
\bibitem [{\citenamefont {Shaw}\ and\ \citenamefont
  {Schwartz}(2010)}]{Shaw2010}%
  \BibitemOpen
  \bibfield  {author} {\bibinfo {author} {\bibnamefont {Shaw}, \bibfnamefont
  {L.~B.}}, \ and\ \bibinfo {author} {\bibfnamefont {I.~B.}\ \bibnamefont
  {Schwartz}}} (\bibinfo {year} {2010}),\ \href {\doibase
  10.1103/PhysRevE.81.046120} {\bibfield  {journal} {\bibinfo  {journal}
  {Physical Review E}\ }\textbf {\bibinfo {volume} {81}}~(\bibinfo {number}
  {4}),\ \bibinfo {pages} {046120}}\BibitemShut {NoStop}%
\bibitem [{\citenamefont {da~Silva}\ \emph {et~al.}(2012)\citenamefont
  {da~Silva}, \citenamefont {Viana},\ and\ \citenamefont
  {da~Fontoura~Costa}}]{daSilva2012}%
  \BibitemOpen
  \bibfield  {author} {\bibinfo {author} {\bibnamefont {da~Silva},
  \bibfnamefont {R.~A.~P.}}, \bibinfo {author} {\bibfnamefont {M.~P.}\
  \bibnamefont {Viana}}, \ and\ \bibinfo {author} {\bibfnamefont
  {L.}~\bibnamefont {da~Fontoura~Costa}}} (\bibinfo {year} {2012}),\ \href
  {http://stacks.iop.org/1742-5468/2012/i=07/a=P07005} {\bibfield  {journal}
  {\bibinfo  {journal} {Journal of Statistical Mechanics: Theory and
  Experiment}\ }\textbf {\bibinfo {volume} {2012}}~(\bibinfo {number} {07}),\
  \bibinfo {pages} {P07005}}\BibitemShut {NoStop}%
\bibitem [{\citenamefont {Simon}\ \emph {et~al.}(2011)\citenamefont {Simon},
  \citenamefont {Taylor},\ and\ \citenamefont
  {Kiss}}]{Simon_Taylor_Kiss_MathBiol2011}%
  \BibitemOpen
  \bibfield  {author} {\bibinfo {author} {\bibnamefont {Simon}, \bibfnamefont
  {P.}}, \bibinfo {author} {\bibfnamefont {M.}~\bibnamefont {Taylor}}, \ and\
  \bibinfo {author} {\bibfnamefont {I.}~\bibnamefont {Kiss}}} (\bibinfo {year}
  {2011}),\ \href@noop {} {\bibfield  {journal} {\bibinfo  {journal} {Journal
  of Mathematical Biology}\ }\textbf {\bibinfo {volume} {62}},\ \bibinfo
  {pages} {479}}\BibitemShut {NoStop}%
\bibitem [{\citenamefont {Singh}\ \emph {et~al.}(2013)\citenamefont {Singh},
  \citenamefont {Sreenivasan}, \citenamefont {Szymanski},\ and\ \citenamefont
  {Korniss}}]{Singh2013}%
  \BibitemOpen
  \bibfield  {author} {\bibinfo {author} {\bibnamefont {Singh}, \bibfnamefont
  {P.}}, \bibinfo {author} {\bibfnamefont {S.}~\bibnamefont {Sreenivasan}},
  \bibinfo {author} {\bibfnamefont {B.~K.}\ \bibnamefont {Szymanski}}, \ and\
  \bibinfo {author} {\bibfnamefont {G.}~\bibnamefont {Korniss}}} (\bibinfo
  {year} {2013}),\ \href@noop {} {\bibfield  {journal} {\bibinfo  {journal}
  {Scientific Reports}\ }\textbf {\bibinfo {volume} {3}},\ \bibinfo {pages}
  {2330}}\BibitemShut {NoStop}%
\bibitem [{\citenamefont {Small}\ and\ \citenamefont
  {Tse}(2005)}]{doi:10.1142/S0218127405012776}%
  \BibitemOpen
  \bibfield  {author} {\bibinfo {author} {\bibnamefont {Small}, \bibfnamefont
  {M.}}, \ and\ \bibinfo {author} {\bibfnamefont {C.~K.}\ \bibnamefont {Tse}}}
  (\bibinfo {year} {2005}),\ \href@noop {} {\bibfield  {journal} {\bibinfo
  {journal} {International Journal of Bifurcation and Chaos}\ }\textbf
  {\bibinfo {volume} {15}}~(\bibinfo {number} {05}),\ \bibinfo {pages}
  {1745}}\BibitemShut {NoStop}%
\bibitem [{\citenamefont {Solomonoff}\ and\ \citenamefont
  {Rapoport}(1951)}]{solomonoff51}%
  \BibitemOpen
  \bibfield  {author} {\bibinfo {author} {\bibnamefont {Solomonoff},
  \bibfnamefont {R.}}, \ and\ \bibinfo {author} {\bibfnamefont
  {A.}~\bibnamefont {Rapoport}}} (\bibinfo {year} {1951}),\ \href@noop {}
  {\bibfield  {journal} {\bibinfo  {journal} {Bulletin of Mathematical
  Biophysics}\ }\textbf {\bibinfo {volume} {14}},\ \bibinfo {pages}
  {153}}\BibitemShut {NoStop}%
\bibitem [{\citenamefont {Son}\ \emph {et~al.}(2012)\citenamefont {Son},
  \citenamefont {Bizhani}, \citenamefont {Christensen}, \citenamefont
  {Grassberger},\ and\ \citenamefont {Paczuski}}]{interdependent12}%
  \BibitemOpen
  \bibfield  {author} {\bibinfo {author} {\bibnamefont {Son}, \bibfnamefont
  {S.-W.}}, \bibinfo {author} {\bibfnamefont {G.}~\bibnamefont {Bizhani}},
  \bibinfo {author} {\bibfnamefont {C.}~\bibnamefont {Christensen}}, \bibinfo
  {author} {\bibfnamefont {P.}~\bibnamefont {Grassberger}}, \ and\ \bibinfo
  {author} {\bibfnamefont {M.}~\bibnamefont {Paczuski}}} (\bibinfo {year}
  {2012}),\ \href {http://stacks.iop.org/0295-5075/97/i=1/a=16006} {\bibfield
  {journal} {\bibinfo  {journal} {Europhysics Letters}\ }\textbf {\bibinfo
  {volume} {97}}~(\bibinfo {number} {1}),\ \bibinfo {pages}
  {16006}}\BibitemShut {NoStop}%
\bibitem [{\citenamefont {Stanley}(1971)}]{stanley}%
  \BibitemOpen
  \bibfield  {author} {\bibinfo {author} {\bibnamefont {Stanley}, \bibfnamefont
  {H.~E.}}} (\bibinfo {year} {1971}),\ \href@noop {} {\emph {\bibinfo {title}
  {Introduction to phase transitions and critical phenomena}}}\ (\bibinfo
  {publisher} {Oxford University Press},\ \bibinfo {address}
  {Oxford})\BibitemShut {NoStop}%
\bibitem [{\citenamefont {Starnini}\ \emph {et~al.}(2012)\citenamefont
  {Starnini}, \citenamefont {Baronchelli}, \citenamefont {Barrat},\ and\
  \citenamefont {Pastor-Satorras}}]{PhysRevE.85.056115}%
  \BibitemOpen
  \bibfield  {author} {\bibinfo {author} {\bibnamefont {Starnini},
  \bibfnamefont {M.}}, \bibinfo {author} {\bibfnamefont {A.}~\bibnamefont
  {Baronchelli}}, \bibinfo {author} {\bibfnamefont {A.}~\bibnamefont {Barrat}},
  \ and\ \bibinfo {author} {\bibfnamefont {R.}~\bibnamefont {Pastor-Satorras}}}
  (\bibinfo {year} {2012}),\ \href {\doibase 10.1103/PhysRevE.85.056115}
  {\bibfield  {journal} {\bibinfo  {journal} {Phys. Rev. E}\ }\textbf {\bibinfo
  {volume} {85}},\ \bibinfo {pages} {056115}}\BibitemShut {NoStop}%
\bibitem [{\citenamefont {Starnini}\ \emph {et~al.}(2013)\citenamefont
  {Starnini}, \citenamefont {Machens}, \citenamefont {Cattuto}, \citenamefont
  {Barrat},\ and\ \citenamefont {Pastor-Satorras}}]{2013arXiv1305.2357S}%
  \BibitemOpen
  \bibfield  {author} {\bibinfo {author} {\bibnamefont {Starnini},
  \bibfnamefont {M.}}, \bibinfo {author} {\bibfnamefont {A.}~\bibnamefont
  {Machens}}, \bibinfo {author} {\bibfnamefont {C.}~\bibnamefont {Cattuto}},
  \bibinfo {author} {\bibfnamefont {A.}~\bibnamefont {Barrat}}, \ and\ \bibinfo
  {author} {\bibfnamefont {R.}~\bibnamefont {Pastor-Satorras}}} (\bibinfo
  {year} {2013}),\ \href {\doibase
  http://dx.doi.org/10.1016/j.jtbi.2013.07.004} {\bibfield  {journal} {\bibinfo
   {journal} {Journal of Theoretical Biology}\ }\textbf {\bibinfo {volume}
  {337}},\ \bibinfo {pages} {89 }}\BibitemShut {NoStop}%
\bibitem [{\citenamefont {Starnini}\ and\ \citenamefont
  {Pastor-Satorras}(2013)}]{PhysRevE.87.062807}%
  \BibitemOpen
  \bibfield  {author} {\bibinfo {author} {\bibnamefont {Starnini},
  \bibfnamefont {M.}}, \ and\ \bibinfo {author} {\bibfnamefont
  {R.}~\bibnamefont {Pastor-Satorras}}} (\bibinfo {year} {2013}),\ \href
  {\doibase 10.1103/PhysRevE.87.062807} {\bibfield  {journal} {\bibinfo
  {journal} {Phys. Rev. E}\ }\textbf {\bibinfo {volume} {87}},\ \bibinfo
  {pages} {062807}}\BibitemShut {NoStop}%
\bibitem [{\citenamefont {{Starnini}}\ and\ \citenamefont {{Pastor
  Satorras}}(2014)}]{2013arXiv1312.5259S}%
  \BibitemOpen
  \bibfield  {author} {\bibinfo {author} {\bibnamefont {{Starnini}},
  \bibfnamefont {M.}}, \ and\ \bibinfo {author} {\bibfnamefont
  {R.}~\bibnamefont {{Pastor Satorras}}}} (\bibinfo {year} {2014}),\ \href@noop
  {} {\bibfield  {journal} {\bibinfo  {journal} {Phys. Rev. E}\ }\textbf
  {\bibinfo {volume} {89}},\ \bibinfo {pages} {032807}}\BibitemShut {NoStop}%
\bibitem [{\citenamefont {Stauffer}\ and\ \citenamefont
  {Barbosa}(2006)}]{PhysRevE.74.056105}%
  \BibitemOpen
  \bibfield  {author} {\bibinfo {author} {\bibnamefont {Stauffer},
  \bibfnamefont {A.~O.}}, \ and\ \bibinfo {author} {\bibfnamefont {V.~C.}\
  \bibnamefont {Barbosa}}} (\bibinfo {year} {2006}),\ \href {\doibase
  10.1103/PhysRevE.74.056105} {\bibfield  {journal} {\bibinfo  {journal} {Phys.
  Rev. E}\ }\textbf {\bibinfo {volume} {74}},\ \bibinfo {pages}
  {056105}}\BibitemShut {NoStop}%
\bibitem [{\citenamefont {Stauffer}\ and\ \citenamefont
  {Aharony}(1994)}]{stauffer94}%
  \BibitemOpen
  \bibfield  {author} {\bibinfo {author} {\bibnamefont {Stauffer},
  \bibfnamefont {D.}}, \ and\ \bibinfo {author} {\bibfnamefont
  {A.}~\bibnamefont {Aharony}}} (\bibinfo {year} {1994}),\ \href@noop {} {\emph
  {\bibinfo {title} {Introduction to Percolation Theory}}},\ \bibinfo {edition}
  {2nd}\ ed.\ (\bibinfo  {publisher} {Taylor \& Francis},\ \bibinfo {address}
  {London})\BibitemShut {NoStop}%
\bibitem [{\citenamefont {Stauffer}\ and\ \citenamefont
  {Sahimi}(2005)}]{stauffer_annealed2005}%
  \BibitemOpen
  \bibfield  {author} {\bibinfo {author} {\bibnamefont {Stauffer},
  \bibfnamefont {D.}}, \ and\ \bibinfo {author} {\bibfnamefont
  {M.}~\bibnamefont {Sahimi}}} (\bibinfo {year} {2005}),\ \href@noop {}
  {\bibfield  {journal} {\bibinfo  {journal} {Phys. Rev. E}\ }\textbf {\bibinfo
  {volume} {72}},\ \bibinfo {pages} {46128}}\BibitemShut {NoStop}%
\bibitem [{\citenamefont {Stehle}\ \emph {et~al.}(2011)\citenamefont {Stehle},
  \citenamefont {Voirin}, \citenamefont {Barrat}, \citenamefont {Cattuto},
  \citenamefont {Colizza}, \citenamefont {Isella}, \citenamefont {Regis},
  \citenamefont {Pinton}, \citenamefont {Khanafer}, \citenamefont {Van~den
  Broeck},\ and\ \citenamefont {Vanhems}}]{Stehle:2011nx}%
  \BibitemOpen
  \bibfield  {author} {\bibinfo {author} {\bibnamefont {Stehle}, \bibfnamefont
  {J.}}, \bibinfo {author} {\bibfnamefont {N.}~\bibnamefont {Voirin}}, \bibinfo
  {author} {\bibfnamefont {A.}~\bibnamefont {Barrat}}, \bibinfo {author}
  {\bibfnamefont {C.}~\bibnamefont {Cattuto}}, \bibinfo {author} {\bibfnamefont
  {V.}~\bibnamefont {Colizza}}, \bibinfo {author} {\bibfnamefont
  {L.}~\bibnamefont {Isella}}, \bibinfo {author} {\bibfnamefont
  {C.}~\bibnamefont {Regis}}, \bibinfo {author} {\bibfnamefont {J.-F.}\
  \bibnamefont {Pinton}}, \bibinfo {author} {\bibfnamefont {N.}~\bibnamefont
  {Khanafer}}, \bibinfo {author} {\bibfnamefont {W.}~\bibnamefont {Van~den
  Broeck}}, \ and\ \bibinfo {author} {\bibfnamefont {P.}~\bibnamefont
  {Vanhems}}} (\bibinfo {year} {2011}),\ \href
  {http://www.biomedcentral.com/1741-7015/9/87} {\bibfield  {journal} {\bibinfo
   {journal} {BMC Medicine}\ }\textbf {\bibinfo {volume} {9}}~(\bibinfo
  {number} {87})}\BibitemShut {NoStop}%
\bibitem [{\citenamefont {Sudbury}(1985)}]{Sudbury1985}%
  \BibitemOpen
  \bibfield  {author} {\bibinfo {author} {\bibnamefont {Sudbury}, \bibfnamefont
  {A.}}} (\bibinfo {year} {1985}),\ \href@noop {} {\bibinfo  {journal} {Journal
  of applied probability}\ ,\ \bibinfo {pages} {443}}\BibitemShut {NoStop}%
\bibitem [{\citenamefont {Takaguchi}\ \emph {et~al.}(2013)\citenamefont
  {Takaguchi}, \citenamefont {Masuda},\ and\ \citenamefont
  {Holme}}]{Takaguchi2012}%
  \BibitemOpen
\bibfield  {journal} {  }\bibfield  {author} {\bibinfo {author} {\bibnamefont
  {Takaguchi}, \bibfnamefont {T.}}, \bibinfo {author} {\bibfnamefont
  {N.}~\bibnamefont {Masuda}}, \ and\ \bibinfo {author} {\bibfnamefont
  {P.}~\bibnamefont {Holme}}} (\bibinfo {year} {2013}),\ \href@noop {}
  {\bibfield  {journal} {\bibinfo  {journal} {PLoS One}\ }\textbf {\bibinfo
  {volume} {8}},\ \bibinfo {pages} {e68629}}\BibitemShut {NoStop}%
\bibitem [{\citenamefont {Tang}\ \emph {et~al.}(2010)\citenamefont {Tang},
  \citenamefont {Scellato}, \citenamefont {Musolesi}, \citenamefont {Mascolo},\
  and\ \citenamefont {Latora}}]{Tang:2010}%
  \BibitemOpen
  \bibfield  {author} {\bibinfo {author} {\bibnamefont {Tang}, \bibfnamefont
  {J.}}, \bibinfo {author} {\bibfnamefont {S.}~\bibnamefont {Scellato}},
  \bibinfo {author} {\bibfnamefont {M.}~\bibnamefont {Musolesi}}, \bibinfo
  {author} {\bibfnamefont {C.}~\bibnamefont {Mascolo}}, \ and\ \bibinfo
  {author} {\bibfnamefont {V.}~\bibnamefont {Latora}}} (\bibinfo {year}
  {2010}),\ \href {\doibase 10.1103/PhysRevE.81.055101} {\bibfield  {journal}
  {\bibinfo  {journal} {Phys. Rev. E}\ }\textbf {\bibinfo {volume} {81}},\
  \bibinfo {pages} {055101}}\BibitemShut {NoStop}%
\bibitem [{\citenamefont {Tanimoto}(2011)}]{tanimoto_epidemic_2011}%
  \BibitemOpen
  \bibfield  {author} {\bibinfo {author} {\bibnamefont {Tanimoto},
  \bibfnamefont {S.}}} (\bibinfo {year} {2011}),\ \href@noop {} {\bibinfo
  {journal} {{arXiv} preprint {arXiv:1103.1680}}\ }\BibitemShut {NoStop}%
\bibitem [{\citenamefont {Tijms}(2003)}]{tijms2003first}%
  \BibitemOpen
\bibfield  {journal} {  }\bibfield  {author} {\bibinfo {author} {\bibnamefont
  {Tijms}, \bibfnamefont {H.}}} (\bibinfo {year} {2003}),\ \href
  {http://books.google.es/books?id=WibF8iVHaiMC} {\emph {\bibinfo {title} {A
  First Course in Stochastic Models}}}\ (\bibinfo  {publisher} {Wiley},\
  \bibinfo {address} {Chichester})\BibitemShut {NoStop}%
\bibitem [{\citenamefont {Tizzoni}\ \emph {et~al.}(2012)\citenamefont
  {Tizzoni}, \citenamefont {Bajardi}, \citenamefont {Poletto}, \citenamefont
  {Ramasco}, \citenamefont {Balcan}, \citenamefont {Goncalves}, \citenamefont
  {Perra}, \citenamefont {Colizza},\ and\ \citenamefont
  {Vespignani}}]{Tizzoni2013}%
  \BibitemOpen
  \bibfield  {author} {\bibinfo {author} {\bibnamefont {Tizzoni}, \bibfnamefont
  {M.}}, \bibinfo {author} {\bibfnamefont {P.}~\bibnamefont {Bajardi}},
  \bibinfo {author} {\bibfnamefont {C.}~\bibnamefont {Poletto}}, \bibinfo
  {author} {\bibfnamefont {J.}~\bibnamefont {Ramasco}}, \bibinfo {author}
  {\bibfnamefont {D.}~\bibnamefont {Balcan}}, \bibinfo {author} {\bibfnamefont
  {B.}~\bibnamefont {Goncalves}}, \bibinfo {author} {\bibfnamefont
  {N.}~\bibnamefont {Perra}}, \bibinfo {author} {\bibfnamefont
  {V.}~\bibnamefont {Colizza}}, \ and\ \bibinfo {author} {\bibfnamefont
  {A.}~\bibnamefont {Vespignani}}} (\bibinfo {year} {2012}),\ \href {\doibase
  10.1186/1741-7015-10-165} {\bibfield  {journal} {\bibinfo  {journal} {BMC
  Medicine}\ }\textbf {\bibinfo {volume} {10}},\ \bibinfo {pages}
  {165}}\BibitemShut {NoStop}%
\bibitem [{\citenamefont {Trapman}(2007)}]{Trapman2007}%
  \BibitemOpen
  \bibfield  {author} {\bibinfo {author} {\bibnamefont {Trapman}, \bibfnamefont
  {P.}}} (\bibinfo {year} {2007}),\ \href {\doibase
  http://dx.doi.org/10.1016/j.tpb.2006.11.002} {\bibfield  {journal} {\bibinfo
  {journal} {Theoretical Population Biology}\ }\textbf {\bibinfo {volume}
  {71}}~(\bibinfo {number} {2}),\ \bibinfo {pages} {160 }}\BibitemShut
  {NoStop}%
\bibitem [{\citenamefont {Trpevski}\ \emph {et~al.}(2010)\citenamefont
  {Trpevski}, \citenamefont {Tang},\ and\ \citenamefont
  {Kocarev}}]{Trpevski2010}%
  \BibitemOpen
  \bibfield  {author} {\bibinfo {author} {\bibnamefont {Trpevski},
  \bibfnamefont {D.}}, \bibinfo {author} {\bibfnamefont {W.~K.~S.}\
  \bibnamefont {Tang}}, \ and\ \bibinfo {author} {\bibfnamefont
  {L.}~\bibnamefont {Kocarev}}} (\bibinfo {year} {2010}),\ \href {\doibase
  10.1103/PhysRevE.81.056102} {\bibfield  {journal} {\bibinfo  {journal} {Phys.
  Rev. E}\ }\textbf {\bibinfo {volume} {81}},\ \bibinfo {pages}
  {056102}}\BibitemShut {NoStop}%
\bibitem [{\citenamefont {Tunc}\ \emph {et~al.}(2013)\citenamefont {Tunc},
  \citenamefont {Shkarayev},\ and\ \citenamefont {Shaw}}]{Tunc2013}%
  \BibitemOpen
  \bibfield  {author} {\bibinfo {author} {\bibnamefont {Tunc}, \bibfnamefont
  {I.}}, \bibinfo {author} {\bibfnamefont {M.~S.}\ \bibnamefont {Shkarayev}}, \
  and\ \bibinfo {author} {\bibfnamefont {L.~B.}\ \bibnamefont {Shaw}}}
  (\bibinfo {year} {2013}),\ \href {\doibase 10.1007/s10955-012-0667-7}
  {\bibfield  {journal} {\bibinfo  {journal} {Journal of Statistical Physics}\
  }\textbf {\bibinfo {volume} {151}}~(\bibinfo {number} {1-2}),\ \bibinfo
  {pages} {355}}\BibitemShut {NoStop}%
\bibitem [{\citenamefont {Valdez}\ \emph
  {et~al.}(2012{\natexlab{a}})\citenamefont {Valdez}, \citenamefont {Macri},\
  and\ \citenamefont {Braunstein}}]{Valdez2012}%
  \BibitemOpen
  \bibfield  {author} {\bibinfo {author} {\bibnamefont {Valdez}, \bibfnamefont
  {L.}}, \bibinfo {author} {\bibfnamefont {P.~A.}\ \bibnamefont {Macri}}, \
  and\ \bibinfo {author} {\bibfnamefont {L.~A.}\ \bibnamefont {Braunstein}}}
  (\bibinfo {year} {2012}{\natexlab{a}}),\ \href@noop {} {\bibfield  {journal}
  {\bibinfo  {journal} {Physical Review E}\ }\textbf {\bibinfo {volume}
  {85}}~(\bibinfo {number} {3}),\ \bibinfo {pages} {036108}}\BibitemShut
  {NoStop}%
\bibitem [{\citenamefont {Valdez}\ \emph
  {et~al.}(2012{\natexlab{b}})\citenamefont {Valdez}, \citenamefont {Macri},\
  and\ \citenamefont {Braunstein}}]{Valdez2012b}%
  \BibitemOpen
  \bibfield  {author} {\bibinfo {author} {\bibnamefont {Valdez}, \bibfnamefont
  {L.~D.}}, \bibinfo {author} {\bibfnamefont {P.~A.}\ \bibnamefont {Macri}}, \
  and\ \bibinfo {author} {\bibfnamefont {L.~A.}\ \bibnamefont {Braunstein}}}
  (\bibinfo {year} {2012}{\natexlab{b}}),\ \href {\doibase
  10.1371/journal.pone.0044188} {\bibfield  {journal} {\bibinfo  {journal}
  {PLoS ONE}\ }\textbf {\bibinfo {volume} {7}}~(\bibinfo {number} {9}),\
  \bibinfo {pages} {e44188}}\BibitemShut {NoStop}%
\bibitem [{\citenamefont {{van Kampen}}(1981)}]{vankampen}%
  \BibitemOpen
  \bibfield  {author} {\bibinfo {author} {\bibnamefont {{van Kampen}},
  \bibfnamefont {N.~G.}}} (\bibinfo {year} {1981}),\ \href@noop {} {\emph
  {\bibinfo {title} {Stochastic processes in chemistry and physics}}}\
  (\bibinfo  {publisher} {North Holland},\ \bibinfo {address}
  {Amsterdam})\BibitemShut {NoStop}%
\bibitem [{\citenamefont {Van~Mieghem}(2011)}]{PVM_graphspectra}%
  \BibitemOpen
  \bibfield  {author} {\bibinfo {author} {\bibnamefont {Van~Mieghem},
  \bibfnamefont {P.}}} (\bibinfo {year} {2011}),\ \href@noop {} {\emph
  {\bibinfo {title} {Graph Spectra for Complex Networks}}}\ (\bibinfo
  {publisher} {Cambridge University Press},\ \bibinfo {address}
  {Cambridge})\BibitemShut {NoStop}%
\bibitem [{\citenamefont
  {Van~Mieghem}(2012{\natexlab{a}})}]{PVM_epidemic_phase_transition2011}%
  \BibitemOpen
  \bibfield  {author} {\bibinfo {author} {\bibnamefont {Van~Mieghem},
  \bibfnamefont {P.}}} (\bibinfo {year} {2012}{\natexlab{a}}),\ \href@noop {}
  {\bibfield  {journal} {\bibinfo  {journal} {Europhysics Letters}\ }\textbf
  {\bibinfo {volume} {97}},\ \bibinfo {pages} {48004}}\BibitemShut {NoStop}%
\bibitem [{\citenamefont
  {Van~Mieghem}(2012{\natexlab{b}})}]{PVM_viral_conductance}%
  \BibitemOpen
  \bibfield  {author} {\bibinfo {author} {\bibnamefont {Van~Mieghem},
  \bibfnamefont {P.}}} (\bibinfo {year} {2012}{\natexlab{b}}),\ \href@noop {}
  {\bibfield  {journal} {\bibinfo  {journal} {Computer Communications}\
  }\textbf {\bibinfo {volume} {35}}~(\bibinfo {number} {12}),\ \bibinfo {pages}
  {1494}}\BibitemShut {NoStop}%
\bibitem [{\citenamefont {Van~Mieghem}(2013)}]{PVM_decay_SIS2014}%
  \BibitemOpen
  \bibfield  {author} {\bibinfo {author} {\bibnamefont {Van~Mieghem},
  \bibfnamefont {P.}}} (\bibinfo {year} {2013}),\ \href@noop {} {\bibinfo
  {journal} {arXiv:1310.3980}\ }\BibitemShut {NoStop}%
\bibitem [{\citenamefont
  {Van~Mieghem}(2014{\natexlab{a}})}]{PVM_upperbound_SIS_epidemic_threshold}%
  \BibitemOpen
\bibfield  {journal} {  }\bibfield  {author} {\bibinfo {author} {\bibnamefont
  {Van~Mieghem}, \bibfnamefont {P.}}} (\bibinfo {year} {2014}{\natexlab{a}}),\
  \href@noop {} {\bibinfo  {journal} {arXiv:1402.1731}\ }\BibitemShut {NoStop}%
\bibitem [{\citenamefont
  {Van~Mieghem}(2014{\natexlab{b}})}]{PVM_PAComplexNetsCUP}%
  \BibitemOpen
\bibfield  {journal} {  }\bibfield  {author} {\bibinfo {author} {\bibnamefont
  {Van~Mieghem}, \bibfnamefont {P.}}} (\bibinfo {year} {2014}{\natexlab{b}}),\
  \href@noop {} {\emph {\bibinfo {title} {Performance Analysis of Complex
  Networks and Systems}}}\ (\bibinfo  {publisher} {Cambridge University
  Press},\ \bibinfo {address} {Cambridge})\BibitemShut {NoStop}%
\bibitem [{\citenamefont {Van~Mieghem}\ \emph {et~al.}(2011)\citenamefont
  {Van~Mieghem}, \citenamefont {Blenn},\ and\ \citenamefont
  {Doerr}}]{PVM_lognormal_Digg_EJPb2011}%
  \BibitemOpen
  \bibfield  {author} {\bibinfo {author} {\bibnamefont {Van~Mieghem},
  \bibfnamefont {P.}}, \bibinfo {author} {\bibfnamefont {N.}~\bibnamefont
  {Blenn}}, \ and\ \bibinfo {author} {\bibfnamefont {C.}~\bibnamefont {Doerr}}}
  (\bibinfo {year} {2011}),\ \href@noop {} {\bibfield  {journal} {\bibinfo
  {journal} {European Physical Journal B}\ }\textbf {\bibinfo {volume}
  {83}}~(\bibinfo {number} {2}),\ \bibinfo {pages} {252}}\BibitemShut {NoStop}%
\bibitem [{\citenamefont {Van~Mieghem}\ and\ \citenamefont {van~de
  Bovenkamp}(2013)}]{van_mieghem_non-markovian_2013}%
  \BibitemOpen
  \bibfield  {author} {\bibinfo {author} {\bibnamefont {Van~Mieghem},
  \bibfnamefont {P.}}, \ and\ \bibinfo {author} {\bibfnamefont
  {R.}~\bibnamefont {van~de Bovenkamp}}} (\bibinfo {year} {2013}),\ \href
  {\doibase 10.1103/PhysRevLett.110.108701} {\bibfield  {journal} {\bibinfo
  {journal} {Physical Review Letters}\ }\textbf {\bibinfo {volume}
  {110}}~(\bibinfo {number} {10}),\ 10.1103/PhysRevLett.110.108701}\BibitemShut
  {NoStop}%
\bibitem [{\citenamefont {Van~Mieghem}\ and\ \citenamefont
  {Cator}(2012)}]{PVM_EpsilonSIS_PRE2012}%
  \BibitemOpen
  \bibfield  {author} {\bibinfo {author} {\bibnamefont {Van~Mieghem},
  \bibfnamefont {P.}}, \ and\ \bibinfo {author} {\bibfnamefont
  {E.}~\bibnamefont {Cator}}} (\bibinfo {year} {2012}),\ \href@noop {}
  {\bibfield  {journal} {\bibinfo  {journal} {Physical Review E}\ }\textbf
  {\bibinfo {volume} {86}},\ \bibinfo {pages} {016116}}\BibitemShut {NoStop}%
\bibitem [{\citenamefont {Van~Mieghem}\ and\ \citenamefont
  {Omic}(2008)}]{PVM_heterogeneous_virusspread}%
  \BibitemOpen
  \bibfield  {author} {\bibinfo {author} {\bibnamefont {Van~Mieghem},
  \bibfnamefont {P.}}, \ and\ \bibinfo {author} {\bibfnamefont
  {J.}~\bibnamefont {Omic}}} (\bibinfo {year} {2008}),\ \href@noop {} {\bibinfo
   {journal} {arXiv:1306.2588}\ }\BibitemShut {NoStop}%
\bibitem [{\citenamefont {Van~Mieghem}\ \emph {et~al.}(2009)\citenamefont
  {Van~Mieghem}, \citenamefont {Omic},\ and\ \citenamefont
  {Kooij}}]{PVM_ToN_VirusSpread}%
  \BibitemOpen
\bibfield  {journal} {  }\bibfield  {author} {\bibinfo {author} {\bibnamefont
  {Van~Mieghem}, \bibfnamefont {P.}}, \bibinfo {author} {\bibfnamefont
  {J.}~\bibnamefont {Omic}}, \ and\ \bibinfo {author} {\bibfnamefont {R.~E.}\
  \bibnamefont {Kooij}}} (\bibinfo {year} {2009}),\ \href@noop {} {\bibfield
  {journal} {\bibinfo  {journal} {IEEE/ACM Transactions on Networking}\
  }\textbf {\bibinfo {volume} {17}}~(\bibinfo {number} {1}),\ \bibinfo {pages}
  {1}}\BibitemShut {NoStop}%
\bibitem [{\citenamefont {{Van Mieghem}}\ \emph {et~al.}(2014)\citenamefont
  {{Van Mieghem}}, \citenamefont {Sahneh},\ and\ \citenamefont
  {Scoglio}}]{XXXXXX}%
  \BibitemOpen
  \bibfield  {author} {\bibinfo {author} {\bibnamefont {{Van Mieghem}},
  \bibfnamefont {P.}}, \bibinfo {author} {\bibfnamefont {F.~D.}\ \bibnamefont
  {Sahneh}}, \ and\ \bibinfo {author} {\bibfnamefont {C.}~\bibnamefont
  {Scoglio}}} (\bibinfo {year} {2014}),\ in\ \href@noop {} {\emph {\bibinfo
  {booktitle} {Proceedings of the 53rd IEEE Conference on Decision and
  Control}}},\ \bibinfo {series and number} {CDC'14}\ (\bibinfo {address} {Los
  Angeles, CA})\BibitemShut {NoStop}%
\bibitem [{\citenamefont {Van~Mieghem}\ \emph {et~al.}(2010)\citenamefont
  {Van~Mieghem}, \citenamefont {Wang}, \citenamefont {Ge}, \citenamefont
  {Tang},\ and\ \citenamefont {Kuipers}}]{PVM_assortativity_EJB2010}%
  \BibitemOpen
  \bibfield  {author} {\bibinfo {author} {\bibnamefont {Van~Mieghem},
  \bibfnamefont {P.}}, \bibinfo {author} {\bibfnamefont {H.}~\bibnamefont
  {Wang}}, \bibinfo {author} {\bibfnamefont {X.}~\bibnamefont {Ge}}, \bibinfo
  {author} {\bibfnamefont {S.}~\bibnamefont {Tang}}, \ and\ \bibinfo {author}
  {\bibfnamefont {F.~A.}\ \bibnamefont {Kuipers}}} (\bibinfo {year} {2010}),\
  \href@noop {} {\bibfield  {journal} {\bibinfo  {journal} {The European
  Physical Journal B}\ }\textbf {\bibinfo {volume} {76}}~(\bibinfo {number}
  {4}),\ \bibinfo {pages} {643}}\BibitemShut {NoStop}%
\bibitem [{\citenamefont {Van~Segbroeck}\ \emph {et~al.}(2010)\citenamefont
  {Van~Segbroeck}, \citenamefont {Santos},\ and\ \citenamefont
  {Pacheco}}]{VanSegbroek2010}%
  \BibitemOpen
  \bibfield  {author} {\bibinfo {author} {\bibnamefont {Van~Segbroeck},
  \bibfnamefont {S.}}, \bibinfo {author} {\bibfnamefont {F.~C.}\ \bibnamefont
  {Santos}}, \ and\ \bibinfo {author} {\bibfnamefont {J.~M.}\ \bibnamefont
  {Pacheco}}} (\bibinfo {year} {2010}),\ \href@noop {} {\bibfield  {journal}
  {\bibinfo  {journal} {PLoS computational biology}\ }\textbf {\bibinfo
  {volume} {6}}~(\bibinfo {number} {8}),\ \bibinfo {pages}
  {e1000895}}\BibitemShut {NoStop}%
\bibitem [{\citenamefont {Vazquez}(2007)}]{Vazquez:2007}%
  \BibitemOpen
  \bibfield  {author} {\bibinfo {author} {\bibnamefont {Vazquez}, \bibfnamefont
  {A.}}} (\bibinfo {year} {2007}),\ \href {\doibase
  http://dx.doi.org/10.1016/j.jtbi.2006.09.018} {\bibfield  {journal} {\bibinfo
   {journal} {Journal of Theoretical Biology}\ }\textbf {\bibinfo {volume}
  {245}}~(\bibinfo {number} {1}),\ \bibinfo {pages} {125 }}\BibitemShut
  {NoStop}%
\bibitem [{\citenamefont {V{\'a}zquez}\ \emph {et~al.}(2003)\citenamefont
  {V{\'a}zquez}, \citenamefont {Bogu{\~n}{\'a}}, \citenamefont {Moreno},
  \citenamefont {Pastor-Satorras},\ and\ \citenamefont
  {Vespignani}}]{structurednets}%
  \BibitemOpen
  \bibfield  {author} {\bibinfo {author} {\bibnamefont {V{\'a}zquez},
  \bibfnamefont {A.}}, \bibinfo {author} {\bibfnamefont {M.}~\bibnamefont
  {Bogu{\~n}{\'a}}}, \bibinfo {author} {\bibfnamefont {Y.}~\bibnamefont
  {Moreno}}, \bibinfo {author} {\bibfnamefont {R.}~\bibnamefont
  {Pastor-Satorras}}, \ and\ \bibinfo {author} {\bibfnamefont {A.}~\bibnamefont
  {Vespignani}}} (\bibinfo {year} {2003}),\ \href@noop {} {\bibfield  {journal}
  {\bibinfo  {journal} {Phys. Rev. E}\ }\textbf {\bibinfo {volume} {67}},\
  \bibinfo {pages} {046111}}\BibitemShut {NoStop}%
\bibitem [{\citenamefont {V{\'a}zquez}\ and\ \citenamefont
  {Moreno}(2003)}]{morenopercolation}%
  \BibitemOpen
  \bibfield  {author} {\bibinfo {author} {\bibnamefont {V{\'a}zquez},
  \bibfnamefont {A.}}, \ and\ \bibinfo {author} {\bibfnamefont
  {Y.}~\bibnamefont {Moreno}}} (\bibinfo {year} {2003}),\ \href@noop {}
  {\bibfield  {journal} {\bibinfo  {journal} {Phys. Rev. E}\ }\textbf {\bibinfo
  {volume} {67}},\ \bibinfo {pages} {015101}}\BibitemShut {NoStop}%
\bibitem [{\citenamefont {V{\'a}zquez}\ \emph {et~al.}(2002)\citenamefont
  {V{\'a}zquez}, \citenamefont {Pastor-Satorras},\ and\ \citenamefont
  {Vespignani}}]{alexei02}%
  \BibitemOpen
  \bibfield  {author} {\bibinfo {author} {\bibnamefont {V{\'a}zquez},
  \bibfnamefont {A.}}, \bibinfo {author} {\bibfnamefont {R.}~\bibnamefont
  {Pastor-Satorras}}, \ and\ \bibinfo {author} {\bibfnamefont {A.}~\bibnamefont
  {Vespignani}}} (\bibinfo {year} {2002}),\ \href@noop {} {\bibfield  {journal}
  {\bibinfo  {journal} {Phys. Rev. E}\ }\textbf {\bibinfo {volume} {65}},\
  \bibinfo {pages} {066130}}\BibitemShut {NoStop}%
\bibitem [{\citenamefont {Vazquez}\ \emph {et~al.}(2007)\citenamefont
  {Vazquez}, \citenamefont {R\'acz}, \citenamefont {Luk\'acs},\ and\
  \citenamefont {Barab\'asi}}]{PhysRevLett.98.158702}%
  \BibitemOpen
  \bibfield  {author} {\bibinfo {author} {\bibnamefont {Vazquez}, \bibfnamefont
  {A.}}, \bibinfo {author} {\bibfnamefont {B.}~\bibnamefont {R\'acz}}, \bibinfo
  {author} {\bibfnamefont {A.}~\bibnamefont {Luk\'acs}}, \ and\ \bibinfo
  {author} {\bibfnamefont {A.-L.}\ \bibnamefont {Barab\'asi}}} (\bibinfo {year}
  {2007}),\ \href {\doibase 10.1103/PhysRevLett.98.158702} {\bibfield
  {journal} {\bibinfo  {journal} {Phys. Rev. Lett.}\ }\textbf {\bibinfo
  {volume} {98}},\ \bibinfo {pages} {158702}}\BibitemShut {NoStop}%
\bibitem [{\citenamefont {Ver~Steeg}\ \emph {et~al.}(2011)\citenamefont
  {Ver~Steeg}, \citenamefont {Ghosh},\ and\ \citenamefont
  {Lerman}}]{VerSteeg2011}%
  \BibitemOpen
  \bibfield  {author} {\bibinfo {author} {\bibnamefont {Ver~Steeg},
  \bibfnamefont {G.}}, \bibinfo {author} {\bibfnamefont {R.}~\bibnamefont
  {Ghosh}}, \ and\ \bibinfo {author} {\bibfnamefont {K.}~\bibnamefont
  {Lerman}}} (\bibinfo {year} {2011}),\ in\ \href
  {http://arxiv.org/abs/1102.1985} {\emph {\bibinfo {booktitle} {Proceedings of
  the Fifth International AAAI Conference on Weblogs and Social Media}}},\ pp.\
  \bibinfo {pages} {377--384}\BibitemShut {NoStop}%
\bibitem [{\citenamefont {Vespignani}(2009)}]{Vespgnani_2009}%
  \BibitemOpen
  \bibfield  {author} {\bibinfo {author} {\bibnamefont {Vespignani},
  \bibfnamefont {A.}}} (\bibinfo {year} {2009}),\ \href {\doibase
  10.1126/science.1171990} {\bibfield  {journal} {\bibinfo  {journal}
  {Science}\ }\textbf {\bibinfo {volume} {325}},\ \bibinfo {pages}
  {425}}\BibitemShut {NoStop}%
\bibitem [{\citenamefont {Vespignani}(2012)}]{Vespignani:2012fk}%
  \BibitemOpen
  \bibfield  {author} {\bibinfo {author} {\bibnamefont {Vespignani},
  \bibfnamefont {A.}}} (\bibinfo {year} {2012}),\ \href@noop {} {\bibfield
  {journal} {\bibinfo  {journal} {Nature Physics}\ }\textbf {\bibinfo {volume}
  {8}},\ \bibinfo {pages} {32}}\BibitemShut {NoStop}%
\bibitem [{\citenamefont {Viboud}\ \emph {et~al.}(2006)\citenamefont {Viboud},
  \citenamefont {Bjornstad}, \citenamefont {Smith}, \citenamefont {Simonsen},
  \citenamefont {Miller},\ and\ \citenamefont {Grenfell}}]{Viboud:2006}%
  \BibitemOpen
  \bibfield  {author} {\bibinfo {author} {\bibnamefont {Viboud}, \bibfnamefont
  {C.}}, \bibinfo {author} {\bibfnamefont {O.}~\bibnamefont {Bjornstad}},
  \bibinfo {author} {\bibfnamefont {D.}~\bibnamefont {Smith}}, \bibinfo
  {author} {\bibfnamefont {L.}~\bibnamefont {Simonsen}}, \bibinfo {author}
  {\bibfnamefont {M.}~\bibnamefont {Miller}}, \ and\ \bibinfo {author}
  {\bibfnamefont {B.}~\bibnamefont {Grenfell}}} (\bibinfo {year} {2006}),\
  \href {\doibase 10.1126/science.1125237} {\bibfield  {journal} {\bibinfo
  {journal} {Science}\ }\textbf {\bibinfo {volume} {312}},\ \bibinfo {pages}
  {447}}\BibitemShut {NoStop}%
\bibitem [{\citenamefont {Vojta}(2006)}]{Vojta2006}%
  \BibitemOpen
  \bibfield  {author} {\bibinfo {author} {\bibnamefont {Vojta}, \bibfnamefont
  {T.}}} (\bibinfo {year} {2006}),\ \href
  {http://stacks.iop.org/0305-4470/39/i=22/a=R01} {\bibfield  {journal}
  {\bibinfo  {journal} {Journal of Physics A: Mathematical and General}\
  }\textbf {\bibinfo {volume} {39}}~(\bibinfo {number} {22}),\ \bibinfo {pages}
  {R143}}\BibitemShut {NoStop}%
\bibitem [{\citenamefont {Volz}(2008)}]{Volz2008}%
  \BibitemOpen
  \bibfield  {author} {\bibinfo {author} {\bibnamefont {Volz}, \bibfnamefont
  {E.}}} (\bibinfo {year} {2008}),\ \href {\doibase 10.1007/s00285-007-0116-4}
  {\bibfield  {journal} {\bibinfo  {journal} {Journal of Mathematical Biology}\
  }\textbf {\bibinfo {volume} {56}}~(\bibinfo {number} {3}),\ \bibinfo {pages}
  {293}}\BibitemShut {NoStop}%
\bibitem [{\citenamefont {Volz}\ and\ \citenamefont
  {Meyers}(2009)}]{volz_epidemic_2009}%
  \BibitemOpen
  \bibfield  {author} {\bibinfo {author} {\bibnamefont {Volz}, \bibfnamefont
  {E.}}, \ and\ \bibinfo {author} {\bibfnamefont {L.~A.}\ \bibnamefont
  {Meyers}}} (\bibinfo {year} {2009}),\ \href {\doibase 10.1098/rsif.2008.0218}
  {\bibfield  {journal} {\bibinfo  {journal} {Journal of The Royal Society
  Interface}\ }\textbf {\bibinfo {volume} {6}}~(\bibinfo {number} {32}),\
  \bibinfo {pages} {233}}\BibitemShut {NoStop}%
\bibitem [{\citenamefont {Wang}\ \emph {et~al.}(2003)\citenamefont {Wang},
  \citenamefont {Chakrabarti}, \citenamefont {Wang},\ and\ \citenamefont
  {Faloutsos}}]{Wang03}%
  \BibitemOpen
  \bibfield  {author} {\bibinfo {author} {\bibnamefont {Wang}, \bibfnamefont
  {Y.}}, \bibinfo {author} {\bibfnamefont {D.}~\bibnamefont {Chakrabarti}},
  \bibinfo {author} {\bibfnamefont {C.}~\bibnamefont {Wang}}, \ and\ \bibinfo
  {author} {\bibfnamefont {C.}~\bibnamefont {Faloutsos}}} (\bibinfo {year}
  {2003}),\ in\ \href@noop {} {\emph {\bibinfo {booktitle} {22nd International
  Symposium on Reliable Distributed Systems (SRDS'03)}}}\ (\bibinfo
  {publisher} {IEEE Computer Society},\ \bibinfo {address} {Los Alamitos, CA})\
  pp.\ \bibinfo {pages} {25--34}\BibitemShut {NoStop}%
\bibitem [{\citenamefont {Warren}\ \emph {et~al.}(2002)\citenamefont {Warren},
  \citenamefont {Sander},\ and\ \citenamefont {Sokolov}}]{sander}%
  \BibitemOpen
  \bibfield  {author} {\bibinfo {author} {\bibnamefont {Warren}, \bibfnamefont
  {C.~P.}}, \bibinfo {author} {\bibfnamefont {L.~M.}\ \bibnamefont {Sander}}, \
  and\ \bibinfo {author} {\bibfnamefont {I.~M.}\ \bibnamefont {Sokolov}}}
  (\bibinfo {year} {2002}),\ \href@noop {} {\bibfield  {journal} {\bibinfo
  {journal} {Phys. Rev. E}\ }\textbf {\bibinfo {volume} {66}},\ \bibinfo
  {pages} {056105}}\BibitemShut {NoStop}%
\bibitem [{\citenamefont {Wasserman}\ and\ \citenamefont
  {Faust}(1994)}]{wass94}%
  \BibitemOpen
  \bibfield  {author} {\bibinfo {author} {\bibnamefont {Wasserman},
  \bibfnamefont {S.}}, \ and\ \bibinfo {author} {\bibfnamefont
  {K.}~\bibnamefont {Faust}}} (\bibinfo {year} {1994}),\ \href@noop {} {\emph
  {\bibinfo {title} {Social Network Analysis: Methods and Applications}}}\
  (\bibinfo  {publisher} {Cambridge University Press},\ \bibinfo {address}
  {Cambridge})\BibitemShut {NoStop}%
\bibitem [{\citenamefont {Watts}(2002)}]{Watts2002}%
  \BibitemOpen
  \bibfield  {author} {\bibinfo {author} {\bibnamefont {Watts}, \bibfnamefont
  {D.~J.}}} (\bibinfo {year} {2002}),\ \href
  {http://www.pubmedcentral.nih.gov/articlerender.fcgi?artid=122850\&tool=pmcentrez\&rendertype=abstract}
  {\bibfield  {journal} {\bibinfo  {journal} {Proceedings of the National
  Academy of Sciences of the United States of America}\ }\textbf {\bibinfo
  {volume} {99}}~(\bibinfo {number} {9}),\ \bibinfo {pages} {5766}}\BibitemShut
  {NoStop}%
\bibitem [{\citenamefont {Watts}\ \emph {et~al.}(2005)\citenamefont {Watts},
  \citenamefont {Muhamad}, \citenamefont {Medina},\ and\ \citenamefont
  {Dodds}}]{Wattsresurgent:2005}%
  \BibitemOpen
  \bibfield  {author} {\bibinfo {author} {\bibnamefont {Watts}, \bibfnamefont
  {D.~J.}}, \bibinfo {author} {\bibfnamefont {R.}~\bibnamefont {Muhamad}},
  \bibinfo {author} {\bibfnamefont {D.~C.}\ \bibnamefont {Medina}}, \ and\
  \bibinfo {author} {\bibfnamefont {P.~S.}\ \bibnamefont {Dodds}}} (\bibinfo
  {year} {2005}),\ \href@noop {} {\bibfield  {journal} {\bibinfo  {journal}
  {Proceedings of the National Academy of Sciences of the United States of
  America}\ }\textbf {\bibinfo {volume} {102}}~(\bibinfo {number} {32}),\
  \bibinfo {pages} {11157}}\BibitemShut {NoStop}%
\bibitem [{\citenamefont {Watts}\ and\ \citenamefont
  {Strogatz}(1998)}]{watts98}%
  \BibitemOpen
  \bibfield  {author} {\bibinfo {author} {\bibnamefont {Watts}, \bibfnamefont
  {D.~J.}}, \ and\ \bibinfo {author} {\bibfnamefont {S.~H.}\ \bibnamefont
  {Strogatz}}} (\bibinfo {year} {1998}),\ \href@noop {} {\bibfield  {journal}
  {\bibinfo  {journal} {Nature}\ }\textbf {\bibinfo {volume} {393}},\ \bibinfo
  {pages} {440}}\BibitemShut {NoStop}%
\bibitem [{\citenamefont {Webb}(2007)}]{gametheorywebb}%
  \BibitemOpen
  \bibfield  {author} {\bibinfo {author} {\bibnamefont {Webb}, \bibfnamefont
  {J.~N.}}} (\bibinfo {year} {2007}),\ \href@noop {} {\emph {\bibinfo {title}
  {Game Theory: Decisions, Interaction and Evolution}}},\ edited by\ \bibinfo
  {editor} {\bibnamefont {Springer}},\ Springer Undergraduate Mathematics
  Series\ (\bibinfo  {publisher} {Springer-Verlag},\ \bibinfo {address}
  {London})\BibitemShut {NoStop}%
\bibitem [{\citenamefont {Weber}\ and\ \citenamefont
  {Porto}(2007)}]{PhysRevE.76.046111}%
  \BibitemOpen
  \bibfield  {author} {\bibinfo {author} {\bibnamefont {Weber}, \bibfnamefont
  {S.}}, \ and\ \bibinfo {author} {\bibfnamefont {M.}~\bibnamefont {Porto}}}
  (\bibinfo {year} {2007}),\ \href {\doibase 10.1103/PhysRevE.76.046111}
  {\bibfield  {journal} {\bibinfo  {journal} {Phys. Rev. E}\ }\textbf {\bibinfo
  {volume} {76}},\ \bibinfo {pages} {046111}}\BibitemShut {NoStop}%
\bibitem [{\citenamefont {Wen}\ and\ \citenamefont {Zhong}(2012)}]{Wen2012967}%
  \BibitemOpen
  \bibfield  {author} {\bibinfo {author} {\bibnamefont {Wen}, \bibfnamefont
  {L.}}, \ and\ \bibinfo {author} {\bibfnamefont {J.}~\bibnamefont {Zhong}}}
  (\bibinfo {year} {2012}),\ \href {\doibase
  http://dx.doi.org/10.1016/j.nonrwa.2011.09.003} {\bibfield  {journal}
  {\bibinfo  {journal} {Nonlinear Analysis: Real World Applications}\ }\textbf
  {\bibinfo {volume} {13}}~(\bibinfo {number} {2}),\ \bibinfo {pages} {967
  }}\BibitemShut {NoStop}%
\bibitem [{\citenamefont {Weng}\ \emph {et~al.}(2013)\citenamefont {Weng},
  \citenamefont {Ratkiewicz}, \citenamefont {Perra}, \citenamefont
  {Gon\c{c}alves}, \citenamefont {Castillo}, \citenamefont {Bonchi},
  \citenamefont {Schifanella}, \citenamefont {Menczer},\ and\ \citenamefont
  {Flammini}}]{Weng2013}%
  \BibitemOpen
  \bibfield  {author} {\bibinfo {author} {\bibnamefont {Weng}, \bibfnamefont
  {L.}}, \bibinfo {author} {\bibfnamefont {J.}~\bibnamefont {Ratkiewicz}},
  \bibinfo {author} {\bibfnamefont {N.}~\bibnamefont {Perra}}, \bibinfo
  {author} {\bibfnamefont {B.}~\bibnamefont {Gon\c{c}alves}}, \bibinfo {author}
  {\bibfnamefont {C.}~\bibnamefont {Castillo}}, \bibinfo {author}
  {\bibfnamefont {F.}~\bibnamefont {Bonchi}}, \bibinfo {author} {\bibfnamefont
  {R.}~\bibnamefont {Schifanella}}, \bibinfo {author} {\bibfnamefont
  {F.}~\bibnamefont {Menczer}}, \ and\ \bibinfo {author} {\bibfnamefont
  {A.}~\bibnamefont {Flammini}}} (\bibinfo {year} {2013}),\ in\ \href@noop {}
  {\emph {\bibinfo {booktitle} {Proceedings of the 19th ACM SIGKDD
  International Conference on Knowledge Discovery and Data Mining}}},\ \bibinfo
  {series and number} {KDD '13}\ (\bibinfo  {publisher} {ACM},\ \bibinfo
  {address} {New York})\ pp.\ \bibinfo {pages} {356--364}\BibitemShut {NoStop}%
\bibitem [{\citenamefont {Wilf}(2006)}]{Wilf:2006:GEN:1204575}%
  \BibitemOpen
  \bibfield  {author} {\bibinfo {author} {\bibnamefont {Wilf}, \bibfnamefont
  {H.~S.}}} (\bibinfo {year} {2006}),\ \href@noop {} {\emph {\bibinfo {title}
  {Generatingfunctionology}}}\ (\bibinfo  {publisher} {A. K. Peters, Ltd.},\
  \bibinfo {address} {Natick, MA, USA})\BibitemShut {NoStop}%
\bibitem [{\citenamefont {Wilkinson}\ and\ \citenamefont
  {Sharkey}(2014)}]{Wilkinson2014}%
  \BibitemOpen
  \bibfield  {author} {\bibinfo {author} {\bibnamefont {Wilkinson},
  \bibfnamefont {R.~R.}}, \ and\ \bibinfo {author} {\bibfnamefont {K.~J.}\
  \bibnamefont {Sharkey}}} (\bibinfo {year} {2014}),\ \href {\doibase
  10.1103/PhysRevE.89.022808} {\bibfield  {journal} {\bibinfo  {journal} {Phys.
  Rev. E}\ }\textbf {\bibinfo {volume} {89}},\ \bibinfo {pages}
  {022808}}\BibitemShut {NoStop}%
\bibitem [{\citenamefont {Wu}\ and\ \citenamefont {Liu}(2008)}]{Wu2008}%
  \BibitemOpen
  \bibfield  {author} {\bibinfo {author} {\bibnamefont {Wu}, \bibfnamefont
  {X.}}, \ and\ \bibinfo {author} {\bibfnamefont {Z.}~\bibnamefont {Liu}}}
  (\bibinfo {year} {2008}),\ \href@noop {} {\bibfield  {journal} {\bibinfo
  {journal} {Physica A}\ }\textbf {\bibinfo {volume} {387}},\ \bibinfo {pages}
  {623}}\BibitemShut {NoStop}%
\bibitem [{\citenamefont {Yagan}\ \emph {et~al.}(2013)\citenamefont {Yagan},
  \citenamefont {Qian}, \citenamefont {Zhang},\ and\ \citenamefont
  {Cochran}}]{Yagan2013}%
  \BibitemOpen
  \bibfield  {author} {\bibinfo {author} {\bibnamefont {Yagan}, \bibfnamefont
  {O.}}, \bibinfo {author} {\bibfnamefont {D.}~\bibnamefont {Qian}}, \bibinfo
  {author} {\bibfnamefont {J.}~\bibnamefont {Zhang}}, \ and\ \bibinfo {author}
  {\bibfnamefont {D.}~\bibnamefont {Cochran}}} (\bibinfo {year} {2013}),\ \href
  {\doibase 10.1109/JSAC.2013.130606} {\bibfield  {journal} {\bibinfo
  {journal} {Selected Areas in Communications, IEEE Journal on}\ }\textbf
  {\bibinfo {volume} {31}}~(\bibinfo {number} {6}),\ \bibinfo {pages}
  {1038}}\BibitemShut {NoStop}%
\bibitem [{\citenamefont {Yan}\ \emph {et~al.}(2007)\citenamefont {Yan},
  \citenamefont {Fu}, \citenamefont {Ren},\ and\ \citenamefont
  {Wang}}]{Yan2007}%
  \BibitemOpen
  \bibfield  {author} {\bibinfo {author} {\bibnamefont {Yan}, \bibfnamefont
  {G.}}, \bibinfo {author} {\bibfnamefont {Z.-Q.}\ \bibnamefont {Fu}}, \bibinfo
  {author} {\bibfnamefont {J.}~\bibnamefont {Ren}}, \ and\ \bibinfo {author}
  {\bibfnamefont {W.-X.}\ \bibnamefont {Wang}}} (\bibinfo {year} {2007}),\
  \href@noop {} {\bibfield  {journal} {\bibinfo  {journal} {Physical Review E}\
  }\textbf {\bibinfo {volume} {75}}~(\bibinfo {number} {1}),\ \bibinfo {pages}
  {016108}}\BibitemShut {NoStop}%
\bibitem [{\citenamefont {Yang}\ and\ \citenamefont
  {Zhou}(2012)}]{PhysRevE.85.056106}%
  \BibitemOpen
  \bibfield  {author} {\bibinfo {author} {\bibnamefont {Yang}, \bibfnamefont
  {Z.}}, \ and\ \bibinfo {author} {\bibfnamefont {T.}~\bibnamefont {Zhou}}}
  (\bibinfo {year} {2012}),\ \href {\doibase 10.1103/PhysRevE.85.056106}
  {\bibfield  {journal} {\bibinfo  {journal} {Phys. Rev. E}\ }\textbf {\bibinfo
  {volume} {85}},\ \bibinfo {pages} {056106}}\BibitemShut {NoStop}%
\bibitem [{\citenamefont {Yeomans}(1992)}]{yeomans}%
  \BibitemOpen
  \bibfield  {author} {\bibinfo {author} {\bibnamefont {Yeomans}, \bibfnamefont
  {J.~M.}}} (\bibinfo {year} {1992}),\ \href@noop {} {\emph {\bibinfo {title}
  {Statistical mechanics of phase transitions}}}\ (\bibinfo  {publisher}
  {Oxford University Press},\ \bibinfo {address} {Oxford})\BibitemShut
  {NoStop}%
\bibitem [{\citenamefont {Youssef}\ \emph {et~al.}(2011)\citenamefont
  {Youssef}, \citenamefont {Kooij},\ and\ \citenamefont
  {Scoglio}}]{Mina_Caterina_VC2011}%
  \BibitemOpen
  \bibfield  {author} {\bibinfo {author} {\bibnamefont {Youssef}, \bibfnamefont
  {M.}}, \bibinfo {author} {\bibfnamefont {R.~E.}\ \bibnamefont {Kooij}}, \
  and\ \bibinfo {author} {\bibfnamefont {C.}~\bibnamefont {Scoglio}}} (\bibinfo
  {year} {2011}),\ \href@noop {} {\bibfield  {journal} {\bibinfo  {journal}
  {Journal of Computational Science}\ }\textbf {\bibinfo {volume} {2}},\
  \bibinfo {pages} {286}}\BibitemShut {NoStop}%
\bibitem [{\citenamefont {Youssef}\ and\ \citenamefont
  {Scoglio}(2011)}]{Youssef2011}%
  \BibitemOpen
  \bibfield  {author} {\bibinfo {author} {\bibnamefont {Youssef}, \bibfnamefont
  {M.}}, \ and\ \bibinfo {author} {\bibfnamefont {C.}~\bibnamefont {Scoglio}}}
  (\bibinfo {year} {2011}),\ \href {\doibase
  http://dx.doi.org/10.1016/j.jtbi.2011.05.029} {\bibfield  {journal} {\bibinfo
   {journal} {Journal of Theoretical Biology}\ }\textbf {\bibinfo {volume}
  {283}},\ \bibinfo {pages} {136 }}\BibitemShut {NoStop}%
\bibitem [{\citenamefont {Zanette}(2001)}]{Zanette2001}%
  \BibitemOpen
  \bibfield  {author} {\bibinfo {author} {\bibnamefont {Zanette}, \bibfnamefont
  {D.}}} (\bibinfo {year} {2001}),\ \href {\doibase 10.1103/PhysRevE.64.050901}
  {\bibfield  {journal} {\bibinfo  {journal} {Physical Review E}\ }\textbf
  {\bibinfo {volume} {64}}~(\bibinfo {number} {5}),\ \bibinfo {pages}
  {050901}}\BibitemShut {NoStop}%
\bibitem [{\citenamefont {Zanette}\ and\ \citenamefont
  {Risau-Gusm\'{a}n}(2008)}]{Zanette2008}%
  \BibitemOpen
  \bibfield  {author} {\bibinfo {author} {\bibnamefont {Zanette}, \bibfnamefont
  {D.~H.}}, \ and\ \bibinfo {author} {\bibfnamefont {S.}~\bibnamefont
  {Risau-Gusm\'{a}n}}} (\bibinfo {year} {2008}),\ \href {\doibase
  10.1007/s10867-008-9060-9} {\bibfield  {journal} {\bibinfo  {journal}
  {Journal of Biological Physics}\ }\textbf {\bibinfo {volume} {34}},\ \bibinfo
  {pages} {135}}\BibitemShut {NoStop}%
\bibitem [{\citenamefont {Zeng}\ and\ \citenamefont {Zhang}(2013)}]{Zeng2013}%
  \BibitemOpen
  \bibfield  {author} {\bibinfo {author} {\bibnamefont {Zeng}, \bibfnamefont
  {A.}}, \ and\ \bibinfo {author} {\bibfnamefont {C.-J.}\ \bibnamefont
  {Zhang}}} (\bibinfo {year} {2013}),\ \href {\doibase
  10.1016/j.physleta.2013.02.039} {\bibfield  {journal} {\bibinfo  {journal}
  {Physics Letters A}\ }\textbf {\bibinfo {volume} {377}}~(\bibinfo {number}
  {14}),\ \bibinfo {pages} {1031}}\BibitemShut {NoStop}%
\bibitem [{\citenamefont {Zhao}\ and\ \citenamefont {Gao}(2007)}]{Zhao2007}%
  \BibitemOpen
  \bibfield  {author} {\bibinfo {author} {\bibnamefont {Zhao}, \bibfnamefont
  {H.}}, \ and\ \bibinfo {author} {\bibfnamefont {Z.}~\bibnamefont {Gao}}}
  (\bibinfo {year} {2007}),\ \href@noop {} {\bibfield  {journal} {\bibinfo
  {journal} {Europhys. Lett.}\ }\textbf {\bibinfo {volume} {79}},\ \bibinfo
  {pages} {38002}}\BibitemShut {NoStop}%
\bibitem [{\citenamefont {Zhou}\ \emph {et~al.}(2007)\citenamefont {Zhou},
  \citenamefont {Liu},\ and\ \citenamefont {Li}}]{Zhou2007}%
  \BibitemOpen
  \bibfield  {author} {\bibinfo {author} {\bibnamefont {Zhou}, \bibfnamefont
  {J.}}, \bibinfo {author} {\bibfnamefont {Z.}~\bibnamefont {Liu}}, \ and\
  \bibinfo {author} {\bibfnamefont {B.}~\bibnamefont {Li}}} (\bibinfo {year}
  {2007}),\ \href {\doibase 10.1016/j.physleta.2007.01.094} {\bibfield
  {journal} {\bibinfo  {journal} {Physics Letters A}\ }\textbf {\bibinfo
  {volume} {368}}~(\bibinfo {number} {6}),\ \bibinfo {pages} {458}}\BibitemShut
  {NoStop}%
\end{thebibliography}

%

\end{document}